\documentclass[11pt]{article}

\usepackage[utf8]{inputenc}
\usepackage{jheppub} 
\usepackage[usenames,dvipsnames,table]{xcolor}
\usepackage{mathtools}
\usepackage{amsmath,amssymb}
\usepackage{tikz}
\usepackage{graphicx}
\usepackage{xspace}
\usetikzlibrary{positioning}
\usetikzlibrary{arrows}
\usetikzlibrary{decorations.pathreplacing}
\usetikzlibrary{shapes.geometric}
\usetikzlibrary{calc}
\usetikzlibrary{decorations.pathmorphing}
\usetikzlibrary{shapes.misc} 
\usepackage{tikz-cd}
\usetikzlibrary{snakes}
\usepackage{pdflscape}
\usepackage{sectsty} \allsectionsfont{\boldmath} 
\usepackage[detect-all]{siunitx}
\usepackage{cancel}
\usepackage{afterpage}

\hypersetup{hidelinks}
\hypersetup{linktoc=all}
\graphicspath{ {Figures/} } 

\DeclarePairedDelimiter\floor{\lfloor}{\rfloor}
\newcommand{\be}{\begin{eqnarray}}
\newcommand{\ee}{\end{eqnarray}}
\newcommand{\bea}{\begin{eqnarray}}
\newcommand{\eea}{\end{eqnarray}}
\newcommand{\nn}{\nonumber}
\newcommand{\bn}{\begin{enumerate}}
\newcommand{\en}{\end{enumerate}}

\def\Tr{\mathop{\mathrm{Tr}}\nolimits}

\def\e{\epsilon}

\newcommand{\udl}[1]{\mathrm{d} #1 \,}

\newcommand{\sbfunc}[1]{s_b\left( #1\right)}

\newcommand{\Gpq}[1]{\Gamma_e\left( #1\right)}

\def\ga{\alpha}

\def\Gc{\Gamma}
\def\Gd{\Delta}
\def\gd{\delta}

\def\gs{\sigma}

\newcommand{\bigphantomspace}{\phantom{\frac{\frac{\frac{1}{1}}{1}}{\frac{1}{1}}}}
\usepackage{array}
\newcommand\undermat[2]{
  \makebox[0pt][l]{$\smash{\underbrace{\phantom{
    \begin{matrix}#2\end{matrix}}}_{\text{$#1$}}}$}#2}
\preprint{\begin{flushright} USTC-ICTS/PCFT-23-24 \\ DMUS-MP-23/12 \end{flushright}}

\title{Probing bad theories with the dualization algorithm {\fontfamily{ppl}\selectfont I} }

\author[a,b]{Simone Giacomelli,}
\author[c,d]{Chiung Hwang,}
\author[a,b,e]{Fabio Marino,}
\author[a,b]{Sara Pasquetti,}
\author[f]{Matteo Sacchi}

\affiliation[a]{Dipartimento di Fisica, Università di Milano-Bicocca,
Piazza della Scienza 3, I-20126 Milano, Italy}
\affiliation[b]{INFN, sezione di Milano-Bicocca, Piazza della Scienza 3, I-20126 Milano, Italy}
\affiliation[c]{Interdisciplinary Center for Theoretical Study, University of Science and Technology of China, Hefei, Anhui 230026, China}
\affiliation[d]{Peng Huanwu Center for Fundamental Theory, Hefei, Anhui 230026, China}
\affiliation[e]{
Department of Mathematics, University of Surrey, Guildford, GU2 7XH, UK}
\affiliation[f]{Mathematical Institute, University of Oxford, Woodstock Road, Oxford, OX2 6GG, United Kingdom}

\emailAdd{simone.giacomelli@unimib.it}
\emailAdd{chiung@ustc.edu.cn}
\emailAdd{f.marino25@campus.unimib.it}
\emailAdd{sara.pasquetti@gmail.com} 
\emailAdd{matteo.sacchi@maths.ox.ac.uk}

\abstract{
Recently an algorithm to build $SL(2,\mathbb{Z})$ duals, including mirror duals, of 3d $\mathcal{N}=4$  quiver theories and their 4d $\mathcal{N}=1$ uplift has been introduced.
In this work we use this new tool to study the so-called \emph{bad theories}.
Our approach allows us to determine  exactly indices/partition functions for generic values of fugacities/real mass and FI parameters revealing their surprising feature:
 the  4d index/3d partition function of  a bad theory behaves as a sum of distributions rather than an ordinary function of the deformation parameters. We focus on the bad SQCD, with $U(N_c)$ gauge group in 3d and $USp(2N_c)$ in 4d, while in an upcoming paper we will consider linear quivers which, in the 3d case, have both unitary and special unitary bad nodes.
}

\begin{document} 

\maketitle
\flushbottom
\section{Introduction and summary}

Mirror symmetry \cite{Intriligator:1996ex,Hanany:1996ie} is one of the most famous examples of infrared (IR) duality. It relates pairs of 3d $\mathcal{N}=4$ theories, with the Higgs branch of one theory being mapped to the Coulomb branch of the other theory and vice-versa. This property, together with the fact that the Higgs branches of theories with eight supercharges in $3\leq d \leq 6$ dimensions are not corrected upon circle reduction \cite{Argyres:1996eh}, has allowed studying such Higgs branches, even for non-Lagrangian theories, as the Coulomb branches of 3d $\mathcal{N}=4$ quiver gauge theories. See \cite{Ferlito:2017xdq,Cabrera:2018jxt,Bourget:2019rtl,DelZotto:2014kka,Hanany:2018uhm,Cabrera:2019izd,Cabrera:2019dob,Bourget:2020asf,Bourget:2020gzi,Beratto:2020wmn,Closset:2020scj,Akhond:2020vhc,Bourget:2020mez,vanBeest:2020kou,Giacomelli:2020gee,Giacomelli:2020ryy,VanBeest:2020kxw,Closset:2020afy,Akhond:2021knl,Arias-Tamargo:2021ppf,Bourget:2021xex,vanBeest:2021xyt,Carta:2021dyx,Xie:2021ewm,Sperling:2021fcf,Nawata:2021nse,Closset:2021lwy,Akhond:2022jts,Carta:2022spy,Carta:2022fxc,Kang:2022zsl,Giacomelli:2022drw,Hanany:2022itc,Bourget:2023uhe,Carta:2023bqn,Bourget:2023cgs} for a partial list of references. 

Recently an algorithm to build $SL(2,\mathbb{Z})$ duals, including mirror duals, of 3d $\mathcal{N}=4$  quiver theories and their 4d $\mathcal{N}=1$ uplift has been introduced \cite{Bottini:2021vms,Hwang:2021ulb,Comi:2022aqo}.
The algorithm uses a set of basic duality moves which  can in turn be derived by iterative application of Seiberg-like dualities (Aharony \cite{Aharony:1997gp} in 3d and Intriligator--Pouliot  \cite{Intriligator:1995ne} in 4d). 
So thanks to the algorithm we can demonstrate mirror and $SL(2,\mathbb{Z})$ dualities for theories with 8 supercharges assuming only Seiberg-like dualities, effectively relying  only on the weaponry of theories with four supercharges.
 
In this work we initiate the study of  the so-called \emph{bad theories} in the sense of \cite{Gaiotto:2008ak} from this new   perspective.
As we will see, our approach allows us to determine  exactly indices/partition functions for generic values of fugacities/real mass and Fayet--Iliopoulos (FI) parameters revealing their surprising feature:
 the  4d index/3d partition function of  a bad theory behaves as a sum of distributions rather than an ordinary function of the deformation parameters.

A 3d $\mathcal{N}=4$ bad theory is usually characterized by the fact that some monopole operators have a scaling dimension, when computed with the ultraviolet (UV) R-symmetry, that is below the unitarity bound for chiral operators, $R\geq \tfrac{1}{2}$. For a linear quiver gauge theory with matter in the fundamental and bifundamental representations, the scaling dimensions of monopole operators are uniquely determined by the ranks of the gauge and of the flavor nodes, so the condition for a theory to be bad translates into a constraint between such ranks. For example, for an SQCD theory with gauge group $U(N_c)$ and  $N_f$ flavors, the constraint is $N_f<2N_c-1$, while the theory is said to be \emph{ugly} for $N_f=2N_c-1$.\footnote{In this case the fundamental monopole operators have exactly the free R-charge of $\tfrac{1}{2}$ and the theory flows to the $U(N_c-1)$ SQCD with $N_f$ plus a free twisted hypermultiplet corresponding to the decoupled monopole \cite{Gaiotto:2008ak}.}
The situation in which some chiral operators apparently seem to violate the unitarity bound is not so uncommon, especially for theories with four supercharges. The interpretation is usually that at low energies such operators become a free sector that is decoupled from the rest of the theory, which instead flows to a superconformal field theory (SCFT).\footnote{In order to isolate the interacting part, one can introduce a so-called \emph{flipping field} in the UV definition of the theory \cite{Benvenuti:2017lle}, which is a gauge singlet chiral field $S$ that couples to the free operator $O$ with a superpotential term $\gd \mathcal{W}=S\,O$. The F-term equation of $S$ then sets $O=0$ in the chiral ring. One should then perform $a$-maximization \cite{Intriligator:2003jj} to check if there are still operators below the bound, in which case the flipping procedure should be repeated. This process should be iterated until there are no operators below the bound.} The free sector is also associated with an emergent global symmetry that acts on it. Such a symmetry mixes with the R-symmetry so as to give a new, correct IR R-symmetry under which the free sector has the R-charge that saturates the unitarity bound, as expected.

Another key feature of bad theories is the intricate structure of their moduli spaces of vacua.
For a good theory the Higgs and the Coulomb branches are two hyper-Kähler cones meeting at their tips.  At this special point, which is usually taken to be the origin of the moduli space, the full UV gauge symmetry is preserved, and the theory flows to an SCFT.  For a bad theory, instead, as pointed out in  \cite{Assel:2017jgo}, 
typically the most singular locus in the moduli space, the one with the highest codimension, is not just a point, i.e.~the singularity is not isolated and there is no point where the full UV gauge symmetry is preserved. On this locus, instead, the gauge symmetry is broken to a subgroup of lower rank and around it the Coulomb branch looks like that of a theory with smaller gauge group times some flat directions (which for 3d theories with 8 supercharges are parametrized by free twisted hypermultiplets).
For us this will be the hallmark of a bad theory,  rather than looking at operators falling below the unitarity bound.
This feature will also define for us a 4d $\mathcal{N}=1$ bad theory.

Various properties of bad theories in 3d, including those discussed above, have been investigated in  \cite{Nanopoulos:2010bv,Kim:2012uz,Yaakov:2013fza,Bashkirov:2013dda,Hwang:2015wna,Hwang:2017kmk,Assel:2017jgo,Dey:2017fqs,Assel:2018exy}. In particular, in \cite{Assel:2017jgo}, the full quantum moduli space of the bad SQCD with unitary gauge group has been studied (see also \cite{Bourget:2021jwo}). 
Moreover, it has been observed, for example, in \cite{Razamat:2014pta,Closset:2020afy,Carta:2022spy,Kang:2022zsl,Akhond:2022jts} that bad theories can arise from the compactification to 3d of higher dimensional theories with eight supercharges. 
In particular, in \cite{Razamat:2014pta,Kang:2022zsl}, it was pointed out that for 4d $\mathcal{N}=2$ SCFTs, one gets a bad theory upon circle reduction to 3d whenever their Hall--Littlewood chiral ring is different from their Higgs branch chiral ring, which for class $\mathcal{S}$ theories \cite{Gaiotto:2009we,Gaiotto:2009hg} happens whenever the genus of the Riemann surface is greater than one \cite{Gadde:2011uv} or if there are multiple twist lines \cite{Kang:2022zsl}. Relatedly, bad theories also arise from torus compactification of 5d $\mathcal{N}=1$ SCFTs realized by compactifying M-theory on certain Calabi--Yau three-fold canonical hypersurface singularities \cite{Closset:2020scj,Closset:2020afy,Closset:2021lwy}.

In this work, we will focus on the example of the 3d $\mathcal{N}=4$ SQCD with  $U(N_c)$ gauge group and $N_f<2N_c$ fundamental hypermultiplets, as well as its 4d uplift introduced in \cite{Hwang:2020wpd}, which is the 4d $\mathcal{N}=1$ $USp(2 N_c)$ SQCD with one antisymmetric and $2N_f+4$ fundamental chirals (plus gauge singlet fields), represented on the left of Figure  \ref{SQCDzero} in Section \ref{sec:4dbadSQCD}, again for $N_f<2N_c$. In an upcoming work \cite{Giacomelli:2024laq} we will extend the analysis to quiver theories with bad nodes, which in 3d are of U/SU type.

One of our main results is the explicit  evaluation of the 4d supersymmetric indices \cite{Romelsberger:2005eg,Kinney:2005ej,Dolan:2008qi,Rastelli:2016tbz} and the 3d $S^3_b$ partition functions \cite{Kapustin:2009kz,Jafferis:2010un,Hama:2010av,Hama:2011ea} of the bad SQCD.\footnote{We expect a similar formula to hold for the 3d supersymmetric index due to the factorization of 3d partition functions into holomorphic blocks \cite{Pasquetti:2011fj,Beem:2012mb,Hwang:2012jh}. This can in principle be derived by repeating the application of the dualization algorithm we discuss here at the level of the 3d index.} In 3d, it is known that for bad theories the 3-sphere partition is not convergent (and the 3d index cannot be expanded because of the sum over monopoles with non-positive scaling dimension\footnote{Nevertheless, the index can be written in a factorized form, the product of vortex and anti-vortex partition functions summed over Higgs vacua, which is useful to test dualities with decoupled monopole operators as long as the dual theories have non-zero FI parameters, required to define the vortex partition function, because the vortex partition function can be expanded in the vorticity fugacity instead of the scaling dimension fugacity \cite{Hwang:2015wna,Hwang:2017kmk,Hwang:2018uyj}.}).
 Curiously, also the index of the 4d $\mathcal{N}=1$ uplift develops
some problematic features for $N_f<2N_c$, due to the presence of operators violating the 4d unitarity bound. 
We will see that also this theory presents the hallmark  of having the UV gauge symmetry broken to a subgroup in the IR, hence we will refer to this theory as the bad 4d SQCD

We find that  the schematic structure of the index/partition function of the  bad SQCD is given by a sum of terms each involving a Dirac delta distribution, enforcing a particular constraint on the fugacities/FI parameters, which multiplies the 4d index/3d partition function of an interacting SCFT and various singlets including a set of free fields. Such terms, which we will often call \emph{frames}, are labelled by $r=N_f-N_c+1, \cdots, \lfloor N_f/2\rfloor$, and for each $r$ the interacting part is given by a good SQCD, which is $U(r)$ with $N_f$ hypers in 3d and $USp(2r)$ with one antisymmetric and $2N_f+4$ fundamental chirals in 4d. 
In addition, there is an extra frame with no delta distribution, for which the interacting part is the good SQCD $U(N_f-N_c)$ with $N_f$ hypers in 3d and $USp(2N_f-2N_c)$ with one antisymmetric and $2N_f+4$ fundamental chirals in 4d.

In three dimensions, the contribution of the singlet fields together with the delta function can be understood as coming from a set of $N_c-r$ free twisted hypers. As we mentioned before, for bad theories some monopoles operators have UV R-charges less than 1/2, which violate the unitarity bound. This indicates that they become a decoupled free sector in the IR, which is associated with an emergent symmetry, or become massive and integrated out. The IR R-symmetry is obtained by mixing the UV one with a specific combination of the Cartan generators of this emergent symmetry in such a way that the free sector has precisely the free R-charge $R_{\text{IR}}=\frac{1}{2}$ \cite{Hwang:2015wna}.
In particular we will see that the fact that partition functions behave as delta distributions is due to the appearance of chiral multiplets  with vanishing UV R-charge.
This indicates that the divergent behavior of the 3d partition function, which should now be understood as a distribution, is due to the fact that we are evaluating it with a wrong R-symmetry making the contribution of some decoupled singlets divergent.

\begin{figure}[t]
	\centering
	\includegraphics[width=\textwidth]{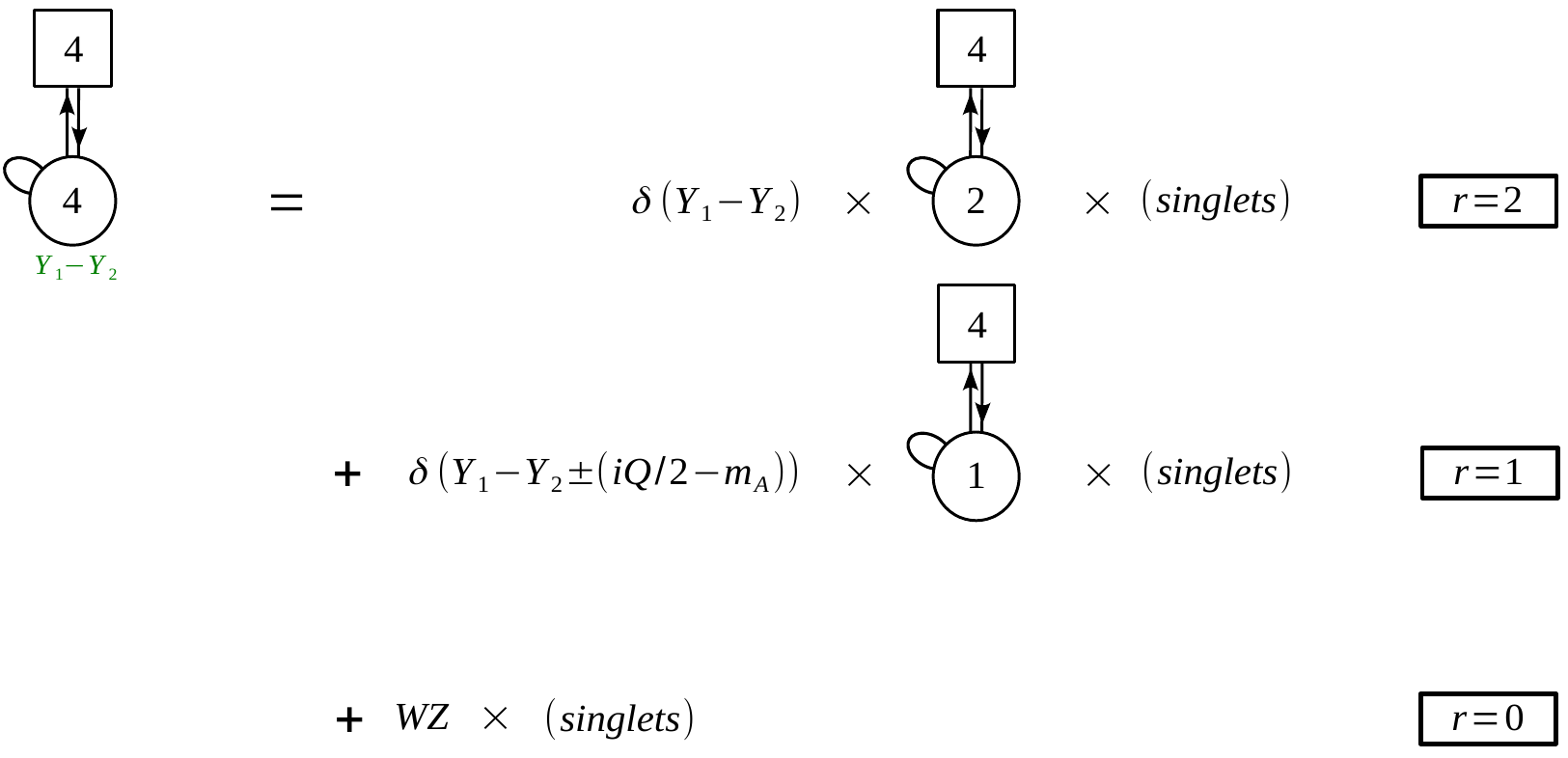}
	\caption{Schematic structure of the partition function of the 3d $\mathcal{N}=4$ $U(4)$ SQCD with $N_f=4$ flavors. The FI is parametrized by $Y_1-Y_2$, while $m_A$ is the real mass for the axial symmetry and $Q=b+b^{-1}$ with $b$ the squashing parameters of $S^3_b$.}
		 \label{3dNc4Nf4cartoon}
\end{figure}

For example, in the case of $U(4)$ SQCD with $N_f=4$ flavors, our result is schematically represented in Figure \ref{3dNc4Nf4cartoon}, where the FI parameter is parametrized as $Y_1-Y_2$, $m_A$ is the real mass for the $U(1)_A$ axial symmetry that is the commutant of the $U(1)_R$ $\mathcal{N}=2$ R-symmetry inside the $SU(2)_H\times SU(2)_C$ $\mathcal{N}=4$ R-symmetry, and $Q=b+b^{-1}$ with $b$ being the squashing parameter of $S^3_b$. 
In this case, we get three distinct frames:
\begin{enumerate}
\item when the FI is turned off, $Y_1-Y_2=0$, we get the partition function of the $U(2)$ SQCD with $N_f=4$ flavors, and the contribution of $\delta(Y_1-Y_2)$ plus that of the singlets can be understood as coming from two free twisted hypermultiplets after the mixing of the R-symmetry;
\item when the FI is instead tuned to the specific values {$Y_1-Y_2=\mp\left(i\frac{Q}{2}-m_A\right)$}, we get the partition function of the $U(1)$ SQCD with $N_f=4$ flavors, and the contribution of 
{$\delta\left(Y_1-Y_2\pm\left(i\frac{Q}{2}-m_A\right)\right)$} plus that of the singlets can be re-packaged into that of three free twisted hypermultiplets;
\item finally, when the FI is turned on and completely generic, we just get the partition function of four free twisted hypermultiplets  corresponding to the monopole operators in the original bad theory.
\end{enumerate}
The explicit expressions that we find for the $S^3_b$ partition function in the case of generic $N_f$ and $N_c$ are given in eqs.~\eqref{eq:3d pf}-\eqref{eq:3d pfNflNc}.

The 3-sphere partition function of the $U(N_c)$ SQCD with $N_f$ flavors for $N_c\leq N_f<2N_c$ has been previously studied in \cite{Yaakov:2013fza}. As we have already mentioned, the associated matrix integral that can be obtained from supersymmetric localization is divergent, but it can be regularized by turning on a non-vanishing FI parameter. In \cite{Yaakov:2013fza}, it was found that for generic non-vanishing FI this is identical to that of the good $U(N_f-N_c)$ SQCD with $N_f$ flavors and $2N_c-N_f$ twisted hypermultiplets. Our result is compatible with \cite{Yaakov:2013fza}, since if the FI is not tuned to one of the specific values dictated by the delta distributions, the only relevant contribution is the one from the term with no delta, which is indeed the partition function of the good $U(N_f-N_c)$ SQCD with $N_f$ flavors and $2N_c-N_f$ twisted hypermultiplets. Nevertheless, the expression that we find is more general than the one obtained in \cite{Yaakov:2013fza} since it is valid for any value of the FI and it characterizes completely the divergent behavior of the partition function.

Our result is also compatible with the findings of 
\cite{Assel:2017jgo}, where by studying the structure of the full quantum moduli space of the $U(N_c)$ SQCD with $N_f$ flavors for $N_f<2N_c$, in particular the quantum Coulomb branch using the techniques of \cite{Bullimore:2015lsa}, it was found that the good ``dual" low energy effective description proposed in \cite{Yaakov:2013fza} does not sit at the most singular locus of the moduli space. Instead, at the most singular locus, \cite{Assel:2017jgo} found a good $U\left(\lfloor N_f/2\rfloor\right)$ SQCD with $N_f$ flavors and $N_c-\lfloor N_f/2\rfloor$ twisted hypermultiplets.
This singular locus does not have the maximal possible codimension since the rank of the gauge group is reduced, as expected for a bad theory.
Our result is consistent with this result since by exploring the space of all possible values of the FI, we can at most get an SQCD theory with rank $\lfloor N_f/2\rfloor$ with the correct number of free twisted hypers.  Turning on a generic FI, this locus of the moduli space as well as those described by a $U(r)$ SQCD for $r=N_f-N_c+1, \cdots, \lfloor N_f/2\rfloor$ are lifted, and one is left only with the singular locus with the $U(N_f-N_c)$ gauge group found in \cite{Yaakov:2013fza}. Here the reason why we need a generic value of the FI different from those dictated by the delta distributions rather than just a non-zero value, which is the case of \cite{Assel:2017jgo}, is that our partition function is refined by the axial mass $m_A$, deforming the $\mathcal N=4$ theory to $\mathcal N = 2^*$. This aspect will be discussed more in Section \ref{sec:semi-classical}.

\begin{figure}[!ht]
	\centering
	\includegraphics[width=\textwidth]{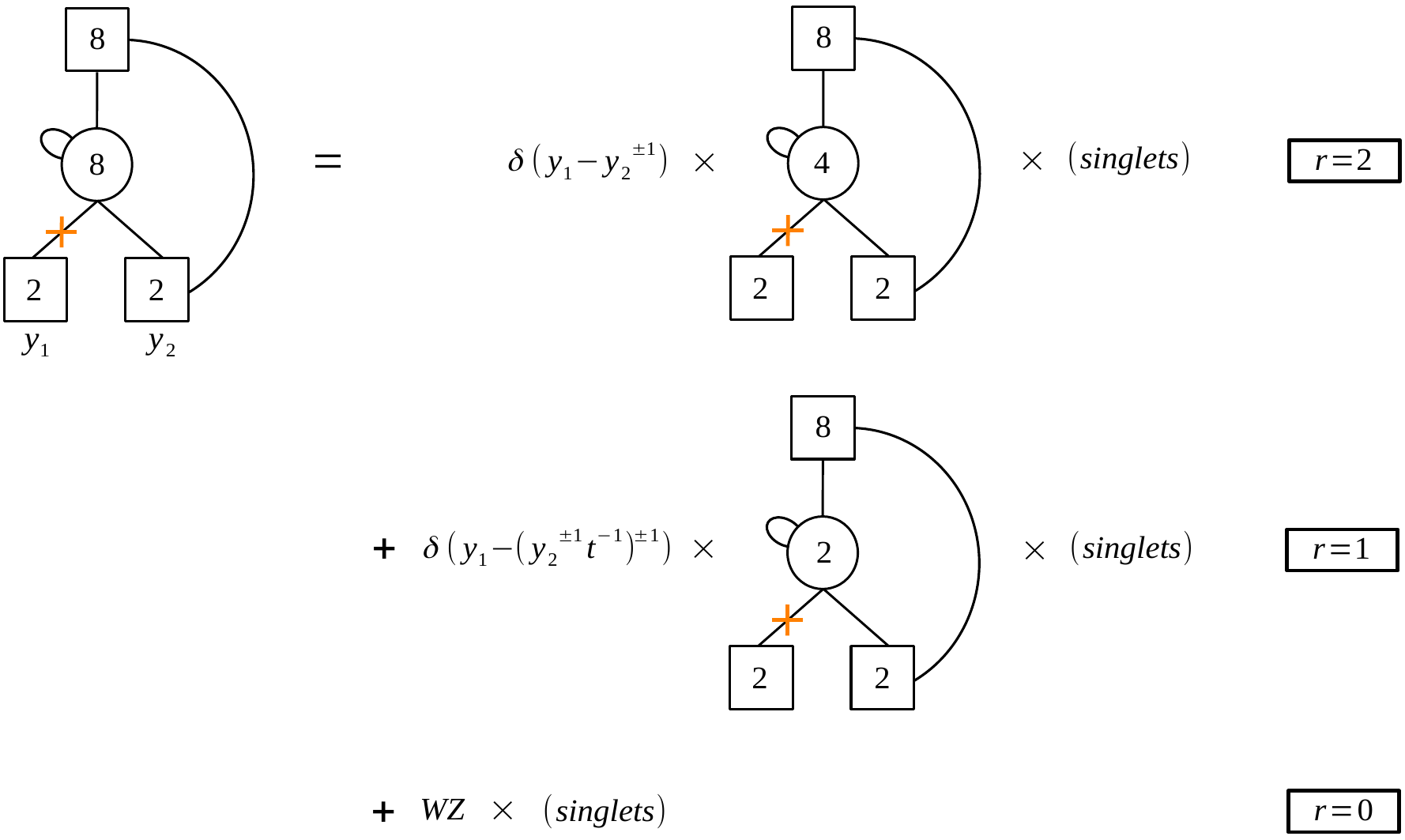}
	\caption{Schematic structure of the index of the 4d $\mathcal{N}=1$ $USp(8)$ SQCD with one antisymmetric and $2N_f+4=8+4$ fundamental chirals, where the 8 vertical chirals $Q_i$ couple to the antisymmetric $A$ as $\sum_i AQ_i^2$. $t$ is the fugacity for the only $U(1)$ symmetry under which the antisymmetric and the 8 fundamental chirals are charged.}
		 \label{4dNc4Nf4cartoon}
\end{figure}

Consider now the 4d set-up, in the specific example of the 4d $\mathcal{N}=1$ bad $USp(8)$ SQCD with $8+4$ fundamental chirals, which is the 4d uplift of the 3d one that we just considered. Our result is schematically represented in Figure \ref{4dNc4Nf4cartoon}, where $y_1$ and $y_2$ are the fugacities for the $SU(2)$ symmetries acting on the chirals which are not coupled to the antisymmetric  and $t$ is the fugacity for the only $U(1)$ symmetry under which the antisymmetric and the 8 fundamental chirals are charged (which is related to the axial symmetry in 3d).

The structure in 4d is completely analogous to the one in 3d, with  three distinct frames:
\begin{enumerate}
\item when the fugacities are set to $y_1=y_2^{\pm1}$ we get the index of the SQCD with gauge group $USp(4)$, plus some singlet fields;
\item when the fugacities are instead set to $y_1=(y_2^{\pm1}t^{-1})^{\pm1}$ we get the index of the SQCD with gauge group $USp(2)$, plus some singlet fields;
\item finally, when $y_1$ and $y_2$ are generic we just get the index of a Wess--Zumino (WZ) model with various singlet fields. 
\end{enumerate}
The explicit expressions that we find for the 4d index in the case of generic $N_f$ and $N_c$ are given in eqs.~\eqref{eq:indbad4d}-\eqref{eq:indbad4dNfoddlNc}.
As in the  3d case, a possible interpretation of this result is that in the moduli space of vacua of the theory there is no singular locus of maximal codimension where the full gauge symmetry is preserved. This is the sense in which we say that also the 4d theory is bad.

We  also propose an interpretation for the constraints on the parameters set by the delta functions. When the parameters are tuned to a value set by one of the delta constraints, one combination of the corresponding symmetries is killed and this is associated to a specific operator acquiring  a Vacuum Expectation Value (VEV). Specifically, in the frame $r$ such operator is one of the basic monopoles dressed with $N_c-r-1$ powers of the adjoint chiral in 3d, while in 4d it is the meson constructed with the fields charged under  $SU(2)_{y_1}\times SU(2)_{y_2}$  dressed with $N_c-r-1$ powers of the antisymmetric chiral. Such a VEV Higgses the gauge group of the bad SQCD reducing its rank by $N_c-r$ so that we indeed get the good SQCD with gauge group $U(r)$ in 3d or $USp(2r)$ in 4d. Equivalently, one can vice-versa think that by tuning the FI in 3d or mass deformation in 4d to the value corresponding to the delta constraint, most of the moduli space is lifted and one is left with only the singular locus where the corresponding operator is taking a VEV and where the associated effective description is in terms of the gauge theory of rank $r$ plus the free sector.
We show this both at the level of the index of the 4d theory and at the level of the equations of motion. 

Finally, we will show that both in 4d and in 3d we can consider a suitable deformation of the SQCD that has a single duality frame, consisting of the $\mathcal{N}=4$ $U(N_f-N_c)$ SQCD with $N_f$ flavors in 3d and the $\mathcal{N}=1$ $USp(2N_f-2N_c)$ SQCD with one antisymmetric and $2N_f+4$ chirals in 4d. Such deformation consists of flipping a set of operators that include those that get a VEV when the delta constraint is implemented which leads to one of the various frames. This prevents them from taking such a VEV and so the theory cannot flow to these frames. Mathematically, the contribution of the singlets gives, in all frames except the one without a delta, a zero that completely kills these terms. 

The rest of the paper is structured as follows. In Section \ref{sec:4dbadSQCD}, we 
apply the mirror dualization algorithm  \cite{Bottini:2021vms,Hwang:2021ulb,Comi:2022aqo}
to obtain our expression for the  index of the 4d bad SQCD.
 We demonstrate the dualization algorithm for some specific examples and also give the general formula for arbitrary $N_c$ and $N_f$. In Section \ref{sec:higgs}, we discuss the Higgsing of the 4d bad SQCD triggered by particular meson VEVs using both the index and the classical equations of motion. 
 In Section \ref{sec:3danalysis}, we move on to the 3d bad SQCD and discuss its IR properties using the squashed $S_b^3$ partition function. 
 Finally, in Section \ref{sec:semi-classical}, we analyze  the moduli spaces of bad theories both in 3d and in 4d in terms of the semi-classical equations of motion and, for the 3d case, the brane set-ups in Type IIB string theory. 

\section{The index of the 4d bad SQCD}
\label{sec:4dbadSQCD}

In this section we apply the dualization algorithm  \cite{Hwang:2021ulb} to the bad SQCD.
We will begin by quickly reviewing  the algorithm. We refer the reader  to \cite{Comi:2022aqo},
where all the steps of the dualization of the  good SQCD have been discussed in great detail.

The algorithm consists of various steps. In the first one, we chop a theory into certain building blocks by ungauging all the gauge nodes. In the second step, we dualize each block with the basic duality moves. After restoring the original gauge interactions, we typically obtain adjacent $\mathsf{S}$-wall theories in the quiver, which annihilate in pairs to give Identity-walls (see \cite{Bottini:2021vms} for more details about the wall theories). In the third step, we implement the identifications imposed by such Identity-walls, which results in a theory where some fields might have acquired a VEV. The fourth and last step is then to study the effect of such VEVs, which can be done with various methods as explained in \cite{Comi:2022aqo}. 

While the first three steps are identical for good and bad theories, the differences arise at the level of the VEV propagation which becomes considerably more intricate in the case of bad theories. Our main goal is to study the result of the VEV propagation in the case of the bad SQCD. We will do so in some explicit examples and then give the general result, while in Appendix \ref{app:general} we will make some considerations about its derivation.


\subsection{SQCD $\mathsf{S}$-dualization}

\begin{figure}[!ht]
	\includegraphics[width=\textwidth]{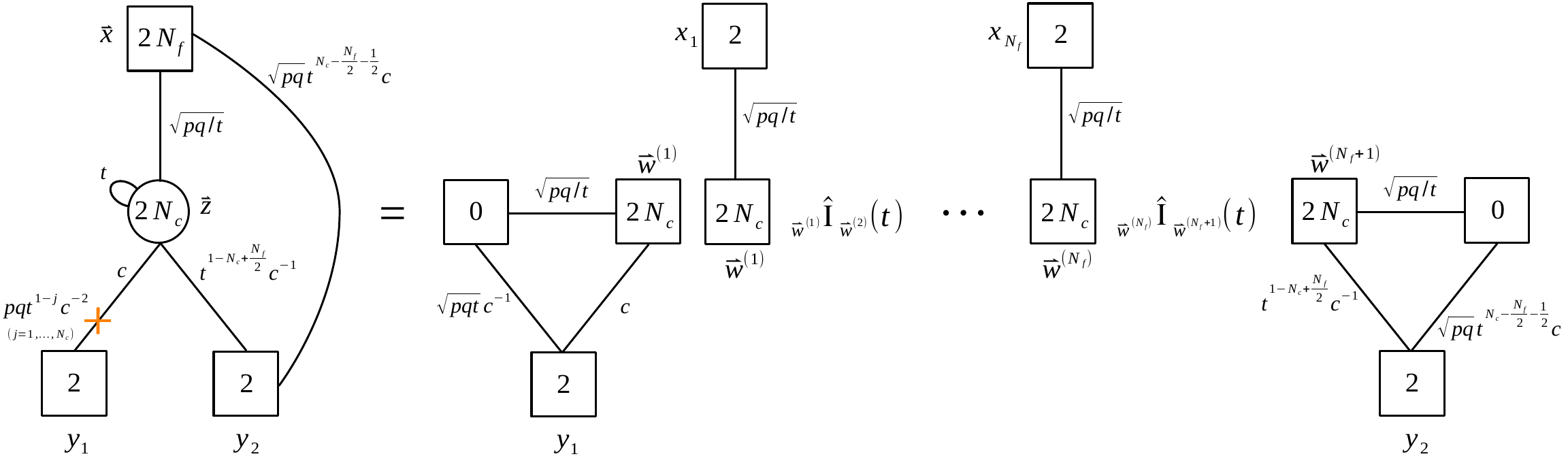}
	\caption{The SQCD and its decomposition into QFT blocks. The orange cross indicates a tower of singlets flipping the meson dressed with the $k$-th power of the antisymmetric  with $k=0,\dots,N_c-1$. The blocks are glued with the integration measure containing vector and antisymmetric chiral fields. 
To simplify the intermediate steps we remove the singlets of the original SQCD, the orange cross and the $SU(2)_{y_2}\times USp(2N_f)$ bifundamental
and reinsert them at the end of the dualization. }
		 \label{SQCDzero}
\end{figure}

We present what we call the 4d SQCD, namely the 4d $\mathcal{N}=1$ $USp(2N)$ gauge theory with one antisymmetric and $2N_f+4$ fundamental chirals, depicted on the left of Figure \ref{SQCDzero}, where the $2N_f$ (vertical chirals) $Q_i$ couple to the antisymmetric $A$ as $\mathcal{W}=\sum_i AQ_i^2$.
There are also some additional gauge singlet fields and a superpotential, we refer the reader to \cite{Hwang:2020wpd,Comi:2022aqo} for the details. The index of this theory is given by 
\be
&&\mathcal{I}_{\text{SQCD}(N_c,N_f)}(\vec{x};y_1,y_2;t;c)=\prod_{j=1}^{N_c}\Gpq{pq\,t^{1-j}c^{-2}}\prod_{a=1}^{N_f}\Gpq{(pq)^{\frac{1}{2}}t^{\frac{1}{2}+M}c\,x_a^{\pm1}y_2^{\pm1}}\nn\\
&&\qquad\times\oint\udl{\vec{z}_{N_c}}\Gd_{N_c}(\vec{z};t)\prod_{j=1}^{N_c}\Gpq{c\,y_1^{\pm1}z_j^{\pm1}}\Gpq{t^{-M}c^{-1}y_2^{\pm1}z_j^{\pm1}}\prod_{a=1}^{N_f}\Gpq{(pq)^{\frac{1}{2}}t^{-\frac{1}{2}}x_a^{\pm1}z_j^{\pm1}}\,,\nn\\
\label{isqcd}
\ee
where the power $M$ is set by the anomaly cancellation condition to be $M=N_c-\frac{N_f}{2}-1$, 
and the factors
\begin{align}
\Gd_{n}(\vec{x};t) =\Gd_{n}(\vec{x}) A_{n}(\vec{x};t)\, , \qquad  
A_{n}(\vec{x};t)=\Gpq{t}^{n} \prod_{i<j}^{n}\Gpq{t x_i^{\pm1} x_j^{\pm1}} \,, \label{intmes}
\end{align}
\begin{equation}
\Gd_{n}(\vec z_{n})=\frac{\left[(p;p)_\infty (q;q)_\infty\right]^n}{\prod_{i=1}^n\Gpq{x_i^{\pm2}}\prod_{i<j}^n\Gpq{x_i^{\pm1} x_j^{\pm1}}},\quad \udl{\vec{z}_{n}}=\frac{1}{2^n n!}\prod_{i=1}^n \frac{\udl{z_i}}{2\pi iz_i}
\end{equation}
encode the contributions of the antisymmetric chiral and of vector multiplets.

As we mentioned, the first step of the algorithm consists in the  decomposition of this theory into some standard basic building blocks, which we shall call \emph{QFT blocks} following \cite{Hwang:2021ulb,Comi:2022aqo}, as shown on the right of Figure \ref{SQCDzero}. More precisely, the theory is chopped into two  triangle shaped blocks, the bifundamental  $\mathsf{B}_{10}$ blocks, and  $N_f$ flavor  $\mathsf{B}_{01}$ blocks, associated to the  $2N_f$ chirals in the fundamental of $USp(2N_c)$.
It is understood that blocks are glued  by restoring the contribution of the antisymmetric and vector multiplets with measure 
$\oint\udl{\vec{z}_{N_c}}\Gd_{N_c}(\vec{z};t).$

\begin{figure}[t]
	\includegraphics[width=\textwidth]{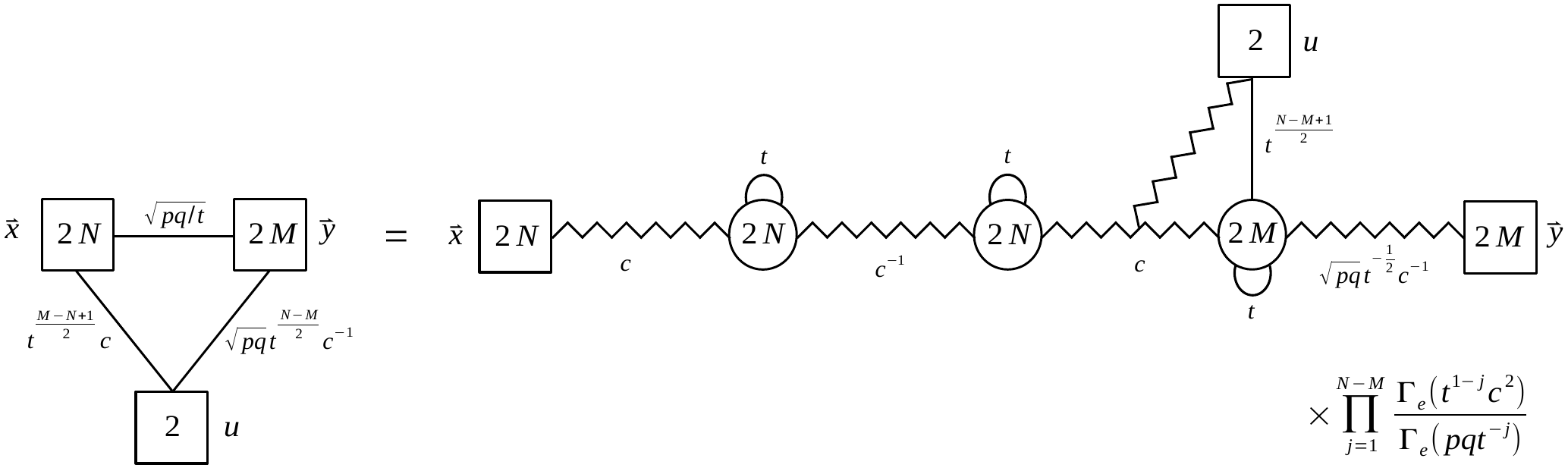}
	\caption{The $\mathsf{S}$-duality move for the $\mathsf{B}_{10}$ QFT block.}
	\label{10S01}
\end{figure}

\begin{figure}[t]
	\includegraphics[width=\textwidth]{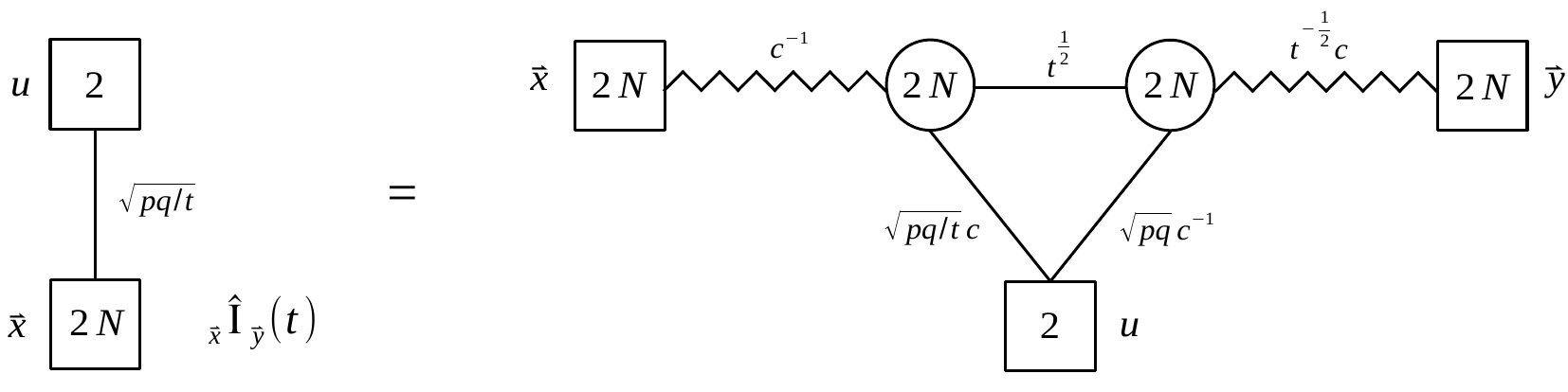}
	\caption{The $\mathsf{S}$-duality move for the $\mathsf{B}_{01}$ QFT block.}
	\label{01S10}
\end{figure}

\begin{figure}[t]
	\includegraphics[width=\textwidth]{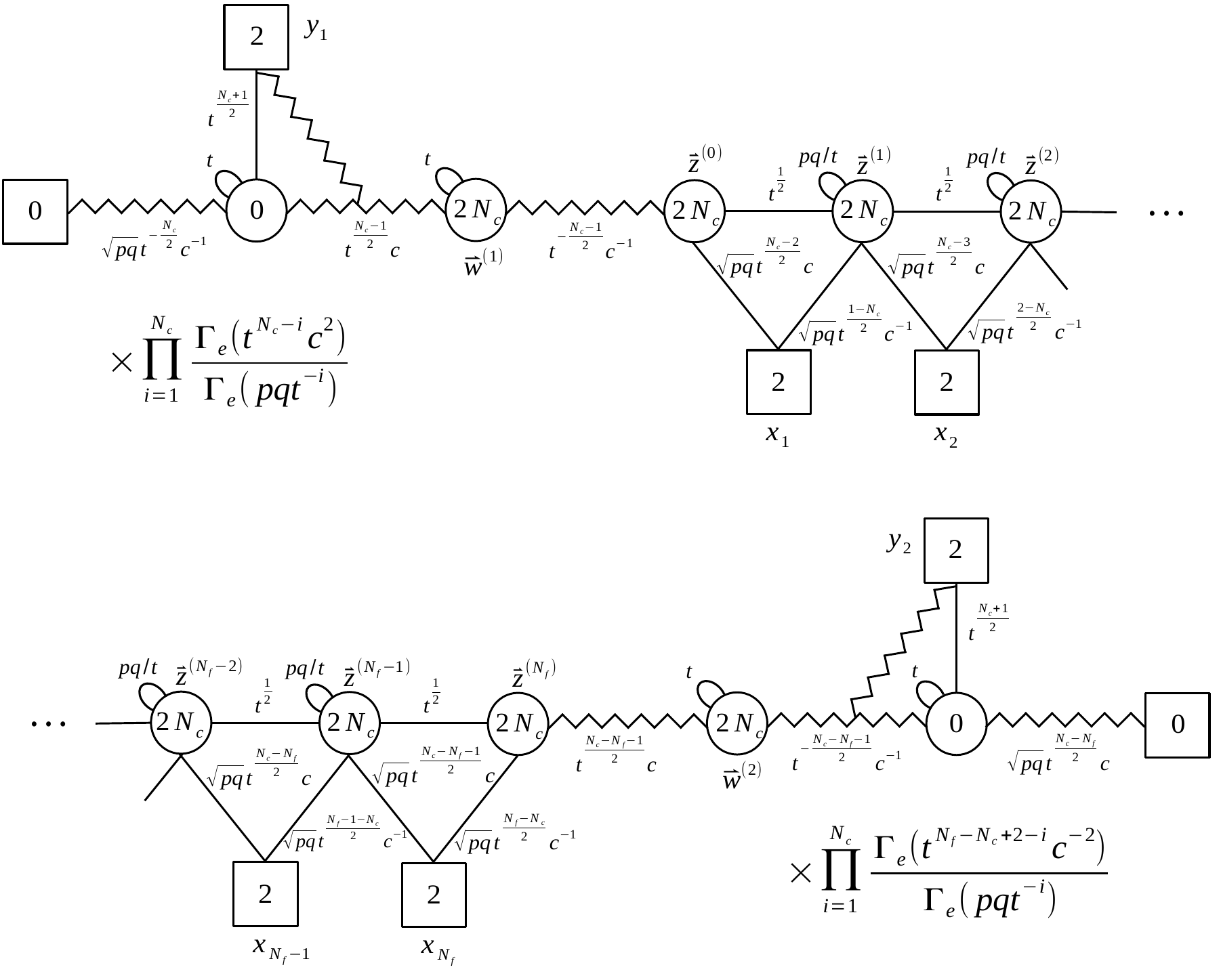}
	\caption{$\mathsf{S}$-dualized SQCD. }
	 \label{SQCDone}
\end{figure}

We now dualize each of these  types of QFT blocks using the basic $\mathsf{S}$-duality moves\footnote{The reason for this name is that these are dualities that mimic in field theory the $S$-duality operation at the level of the Type IIB brane set-up \cite{Hanany:1996ie} of the 3d $\mathcal{N}=4$ version of the theory, although at present we do not have a brane interpretation directly of the 4d theories.} of Figures \ref{10S01} and \ref{01S10} respectively, and glue them back by restoring the original gauge symmetries, which results in the dualized quiver shown in Figure \ref{SQCDone}. In these figures, a wiggle line connecting two $USp(2N)$ nodes represents the $FE[USp(2N)]$  $\mathsf{S}$-wall theory while a wiggle line connecting a $USp(2N)$ 
and a $USp(2M)\times USp(2)$ node represent an asymmetric $\mathsf{S}$-wall theory (see e.g.~\cite{Bottini:2021vms,Comi:2022aqo} for its definition and properties). The extra singlets are denoted in the figure by their index contributions.


\begin{figure}[!ht]
	\includegraphics[width=\textwidth]{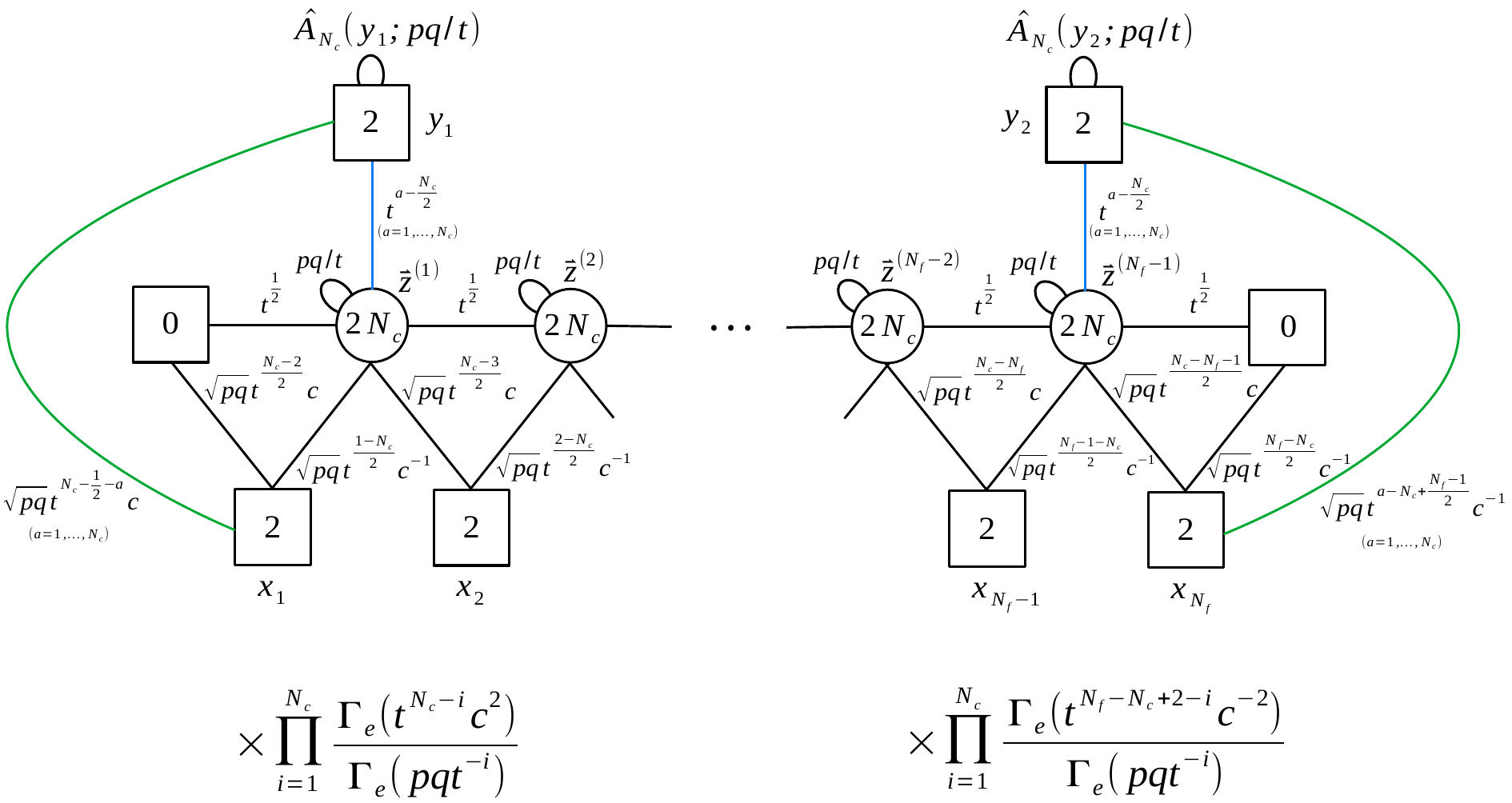}
	\caption{The theory obtained by implementing the identifications set by the Identity-walls. The blue and green legs denote sets of chirals labelled by  $a=1,\dots,N_c$. Mesons constructed with blue chirals acquire a VEV.}
	\label{SQCDtwo}
\end{figure}

An important property of the $\mathsf{S}$-wall theory is that concatenating two of them we get an Identity-wall. At the level of the index we have the following identity proven in \cite{Bottini:2021vms}:
\begin{eqnarray}
\label{eq:deltaNN}
&&\oint\udl{\vec{z}_N}\Gd_N(\vec{z};t)\mathcal{I}_{\mathsf{S}}^{(N)}(\vec{f};\vec{z};t;c)\mathcal{I}_{\mathsf{S}}^{(N)}(\vec{z};\vec{h};t;c^{-1})={}_{\vec{f}\,}\hat{\mathbb{I}}_{\vec{h}}(t) \,,
\end{eqnarray}
where $\mathcal{I}_{\mathsf{S}}^{(N)}$ is the index of the $\mathsf{S}$-wall theory (see e.g.~(2.1) of \cite{Comi:2022aqo}), while that of the Identity-wall is
\begin{equation}\label{eq:idopNN}
{}_{\vec f\,}\hat{\mathbb{I}}_{\vec h}(t)=\frac{\prod_{j=1}^N 2\pi if_j}{\Gd_N(\vec{f};t)}\sum_{\gs\in S_N}\sum_{\pm}\prod_{j=1}^N\gd\left(f_j- h_{\gs(j)}^\pm\right)\,.
\end{equation}
We used this property to fuse the $\mathsf{S}$-walls coming from the dualization of the flavors in Figure \ref{01S10} to arrive at the quiver in Figure \ref{SQCDone}. 
If we instead glue an $\mathsf{S}$ and an asymmetric $\mathsf{S}$-wall theory we obtain an asymmetric Identity-wall
\begin{equation}
\label{eq:idop}
{}_{\vec{f}\,}\hat{\mathbb{I}}_{\vec{h},w}(t)=\frac{{ \prod_{j=1}^{N} 2\pi i f_j }}{\Gd_N (\vec f;t) }\left.\sum_{\sigma \in S_N,\pm}\,\prod_{i=1}^{N}\delta\left(f_i-h_{\sigma(i)}^{\pm1}\right)\right|_{h_{M+j}=t^{\frac{N-M+1-2j}{2}}w} \,.
\end{equation}
We use this property at the two sides of the quiver in Figure \ref{SQCDone}. This leads to identifications that specialize the gauge fugacities $\vec{z}^{\,(0)}$ and $\vec{z}^{\,(N_f)}$ of the leftmost and rightmost nodes in the following geometric progression:
\begin{align}
z^{(0)}_j=t^{\tfrac{N_c+1-2j}{2}} y_1\,, \qquad z^{(N_f)}_j=t^{\tfrac{N_c+1-2j}{2}} y_2\,, \qquad j=1,\cdots,N_c 
\end{align}
so as to arrive at the quiver in Figure \ref{SQCDtwo}. 
The blue lines indicate two sets of $N_c$ chirals in the bifundamental of $SU(2)_{y_1} \times USp(2N_c)_{z^{(1)}}$ and that of $SU(2)_{y_2} \times USp(2N_c)_{z^{(N_f-1)}}$, respectively. They contribute to the index by
\begin{align}
  \prod_{i=1}^{N_c}\Gpq{t^{a-\tfrac{N_c}{2}}z^{(1)\pm}_i y_1^\pm  }, \qquad 
  \prod_{i=1}^{N_c}\Gpq{t^{a-\tfrac{N_c}{2}}z^{(N_f-1)\pm}_i y_2^\pm  },
 \qquad a=0,\, 1, \, \cdots, \, N_c \,.
 \label{blue}
 \end{align}
In addition, we have the following singlets  coming from the dualization:
\begin{align}
\nonumber& \widehat{A}_{N_c}(y_1;pq/t )\prod_{a=1}^{N_c}\Gpq{\sqrt{pq} t^{N_c-\frac{1}{2}-a} c x_1^\pm y_1^\pm } 
 \frac{ \Gpq{t^{N_c-a}c^2 }}{\Gpq{p q t^{-a}}} \\
&\times \widehat{A}_{N_c}(y_2;pq/t ) \prod_{a=1}^{N_c} \Gpq{\sqrt{pq} t^{a-N_c+\frac{N_f-1}{2}} c^{-1} x_{N_f}^\pm y_2^\pm }
  \frac{  \Gpq{t^{N_f-N_c+2-a} c^{-2}}}{\Gpq{p q t^{-a}}} \,,
\end{align}
where we defined
\begin{align}
\label{eq:spec_antisymm}
\widehat{A}_{N_c}(v;pq/t)={A}_{N_c}\left(t^{\tfrac{N_c-1}{2}} v,\cdots ,t^{\tfrac{1-N_c}{2}} v;pq/t\right)  \,.
\end{align}
Now we have reached a quiver theory with no duality-walls left. As we are going to see shortly, there are some VEVs that trigger an RG flow, which we study in several ways in the following subsections.

\begin{figure}[!ht]
	\includegraphics[width=\textwidth]{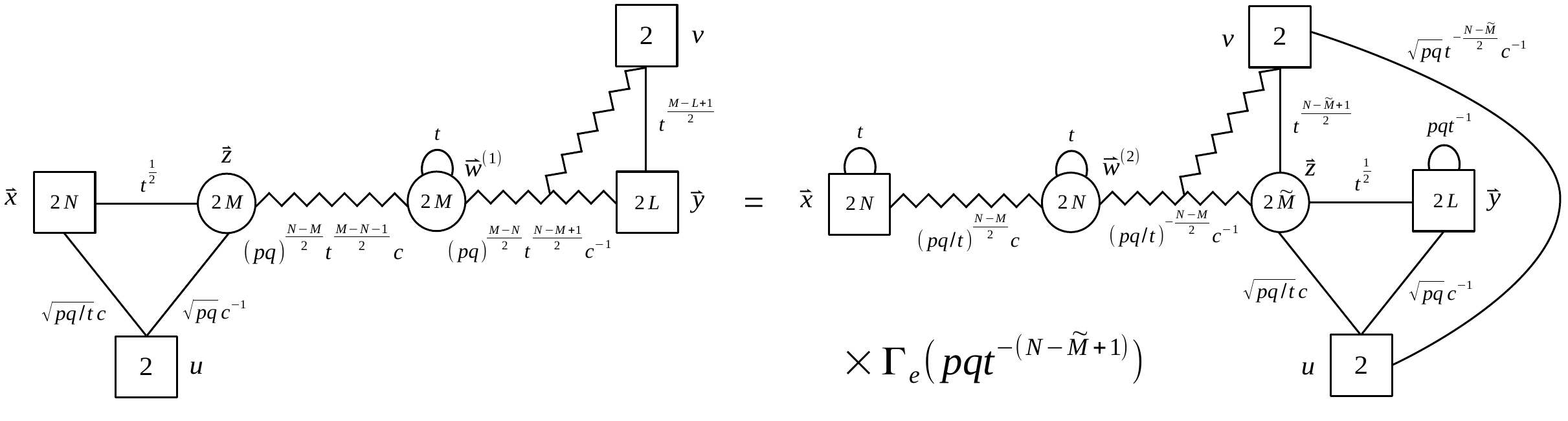}
	\caption{Hanany--Witten duality move. On the r.h.s.~the rank of the second gauge node becomes $\widetilde{M}=N+L-M+1$.}
	 \label{nHWmove}
\end{figure}

One way to study the VEV propagation is to apply a  sequence of Hanany--Witten (HW) duality moves,\footnote{Again the name is by analogy with the Hanany--Witten brane move \cite{Hanany:1996ie} swapping a D5 and an NS5-brane with D3-branes stretched in between, which this duality mimics in field theory, that is typically needed to re-order branes after applying $S$-duality in order to read off a gauge theory.} which is an IR duality that we schematically depict in Figure \ref{nHWmove}.

For $N\geq\widetilde{M}\geq 0$, 
the associated index identity is given by \cite{Comi:2022aqo}
\begin{align}\label{eq:HWid}
    & \oint \udl{\Vec{z}_M} \,\, \Delta_M\left(\Vec{z}\right) 
    \mathcal{I}^{(N,M)}_{(1,0)}\left( \vec{x};\vec{z};u;pqt^{-1};c (pqt^{-1})^\frac{ N-M}{2}\right)
    \,_{\vec{z}}\hat{\mathbb{I}}_{\Vec{y},\,v}(t)
    \prod_{k=1}^L \Gamma_e\left( t^{\frac{M-L+1}{2}} y_k^{\pm} v^{\pm} \right) \nonumber\\
    & \quad = 
    \Gamma_e\left( pqt^{-(N-\widetilde{M}+1)} \right) \Gamma_e\left( (pq)^{\frac{1}{2}} t^{-\frac{N-\widetilde{M}}{2}} c^{-1} u^{\pm} v^{\pm} \right)  A_N\left(\vec{x};t\right) A_L\left(\vec{y};pq/t\right) \nonumber\\
    & \qquad \times 
    \oint \udl{\Vec{z}_{\widetilde{M}}}\,\, \Delta_{\widetilde{M}}\left( \Vec{z} \right) \,_{\vec{x}}\hat{\mathbb{I}}_{\Vec{z},\,v}(t) 
    \prod_{j=1}^{\widetilde{M}} \Gamma_e\left( t^{\frac{N-\widetilde{M}+1}{2}} z_j^{\pm} v^{\pm} \right)  
    \mathcal{I}^{(\widetilde{M},L)}_{(1,0)}\left( \vec{z};\vec{y};u;pqt^{-1};c (pqt^{-1})^\frac{ \widetilde{M}-L}{2}\right)\,,
\end{align}
where $\mathcal{I}^{(N,M)}_{(1,0)}$ is the contribution of the $\mathsf{B}_{10}$ block (see eq.~(2.22) of \cite{Comi:2022aqo} for its definition).

\begin{figure}[h]
\centering
\includegraphics[scale=.39]{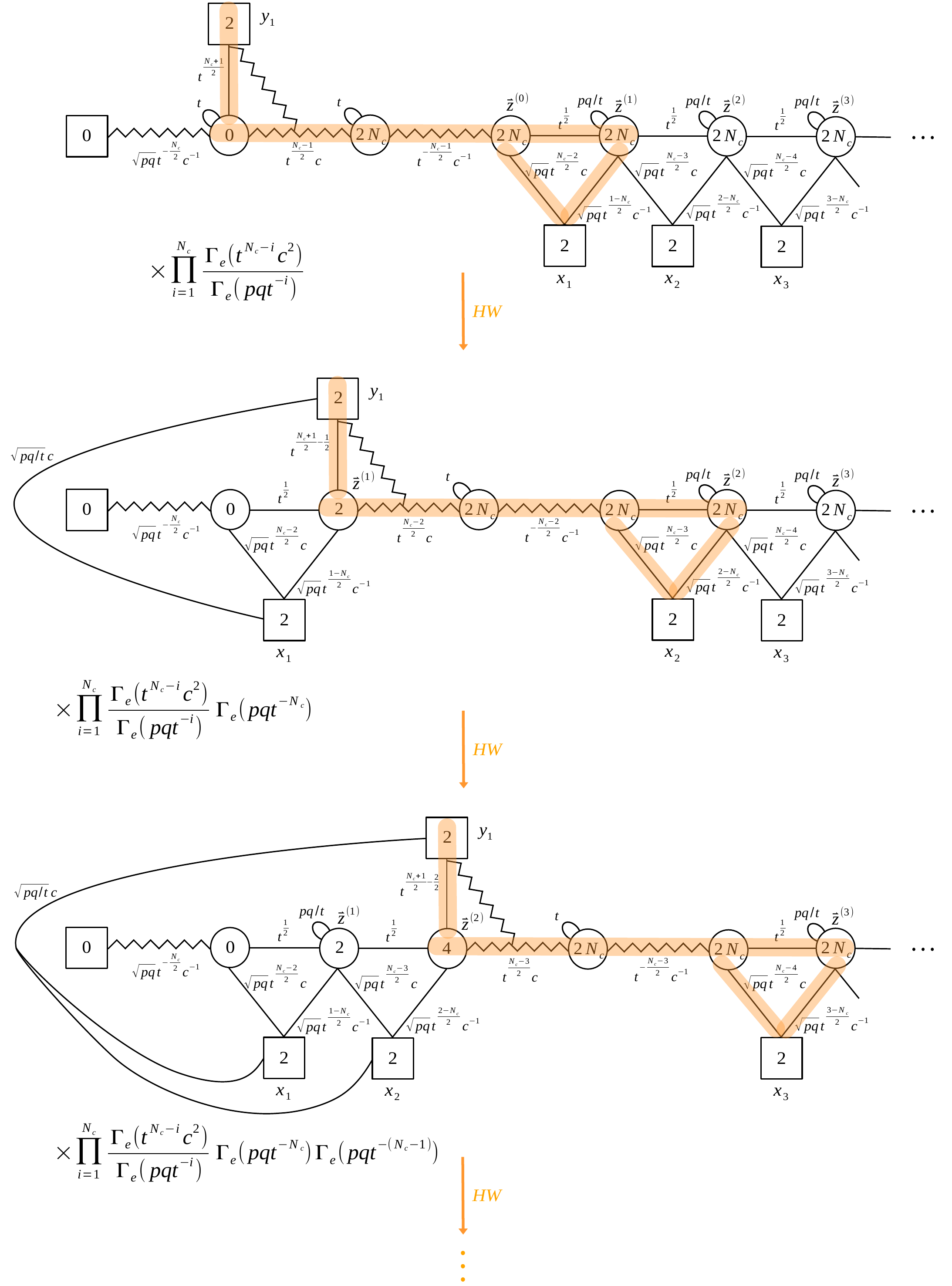}
\caption{First iterations  of the HW duality move.}
\label{fig:HW_0_1_2}
\end{figure}

Consider the $\mathsf{S}$-dualized SQCD in Figure \ref{SQCDone}, focusing on the left part of the quiver to begin with.
We apply the HW move to the sequence of $\mathsf{B}_{01}$-$\mathsf{B}_{10}$ blocks highlighted in orange in the first quiver in Figure
\ref{fig:HW_0_1_2} and reach the  second quiver.
Notice that  the  $USp(2N_c)_{z^{(1)}}$ gauge node becomes a $USp(2)_{z^{(1)}}$ node and two singlets have been produced:  the $SU(2)_{y_1}\times SU(2)_{x_1}$ bifundamental and the one corresponding to $\Gamma_e(pqt^{-N_c})$, which actually cancels one of the singlets of Figure \ref{SQCDone}. We then apply again the HW move to the new highlighted sequence of $\mathsf{B}_{01}$-$\mathsf{B}_{10}$ blocks in the second quiver of Figure
\ref{fig:HW_0_1_2} to reach the third quiver and continue moving to the right.

We can implement the HW moves also on the right side of the quiver propagating to the left. For a good quiver, such propagations of the HW moves, or equivalently Higgsing, from both ends of the quiver stop before colliding to each other. On the other hand, as we will see shortly, the propagating HW moves for a bad theory with $N_f<2N_c$ meet at some point before the VEVs are completely extinguished. This is what gives these theories the complicated structure we mentioned in the Introduction. In the rest of the section we will tackle this subtle problem, which will result in the general expression for the index of a bad SQCD.

We split the analysis between the two cases of $N_f$ even and odd. For each of these, we will start by discussing some specific example in detail and then we will give the general result, on which we further comment in Appendix \ref{app:general}.

\subsection{Even $N_f$}

\subsubsection{Example: $N_c=N_f=2$}
\label{22sec}

\begin{figure}[!ht]
	\centering
    \includegraphics[width=1\textwidth]{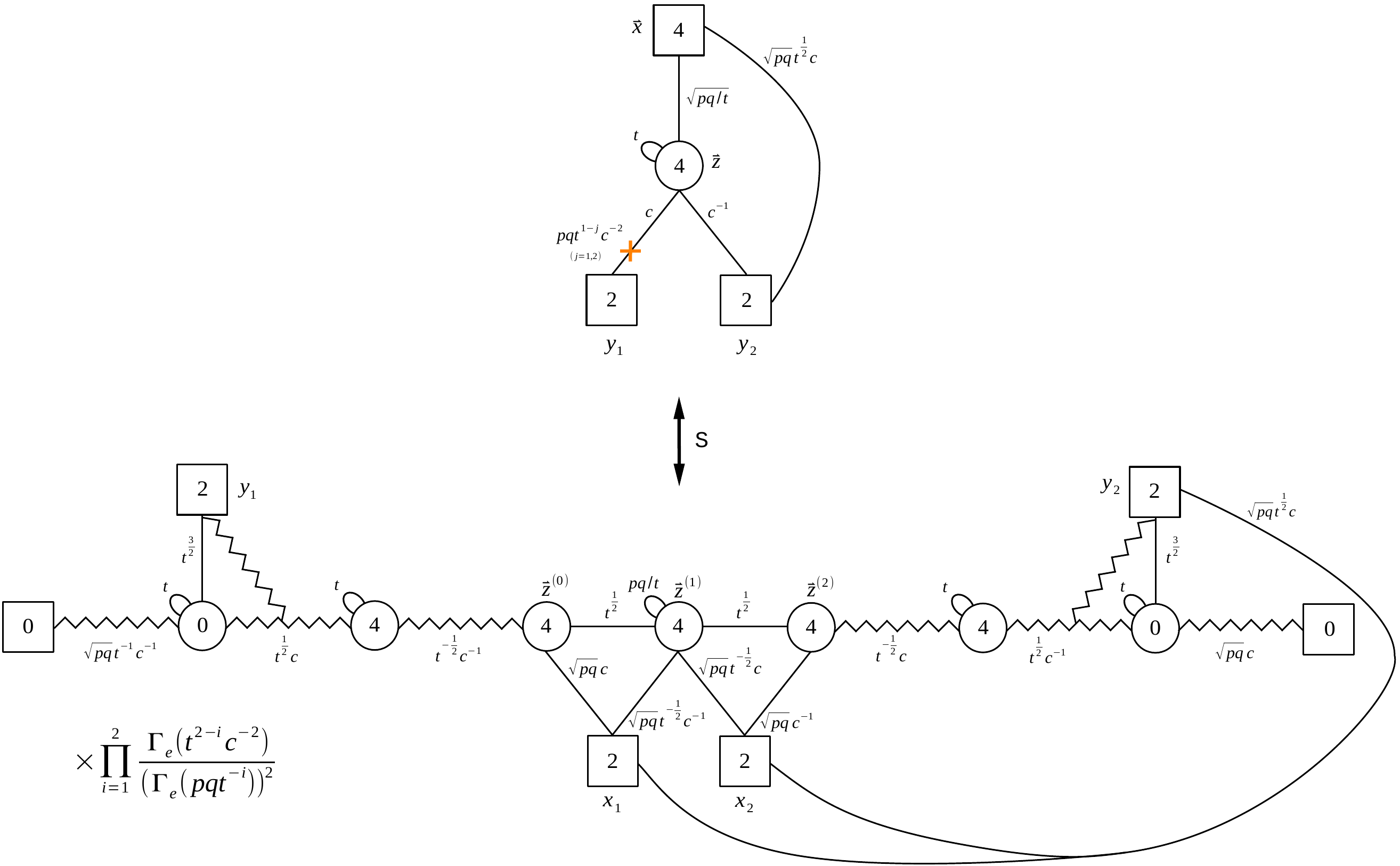}
    \caption{The $N_c=N_f=2$ SQCD and the result of its $\mathsf{S}$-dualization before propagating the VEVs.}
    \label{fig:SQCD_Nc=Nf=2_Sdualised}
\end{figure}

We first consider the $N_c=N_f=2$ SQCD. 
After the standard procedure of the $\mathsf S$-dualization, we get the theory on the bottom of Figure \ref{fig:SQCD_Nc=Nf=2_Sdualised}.

\begin{figure}[!ht]
	\centering
    \includegraphics[width=1\textwidth]{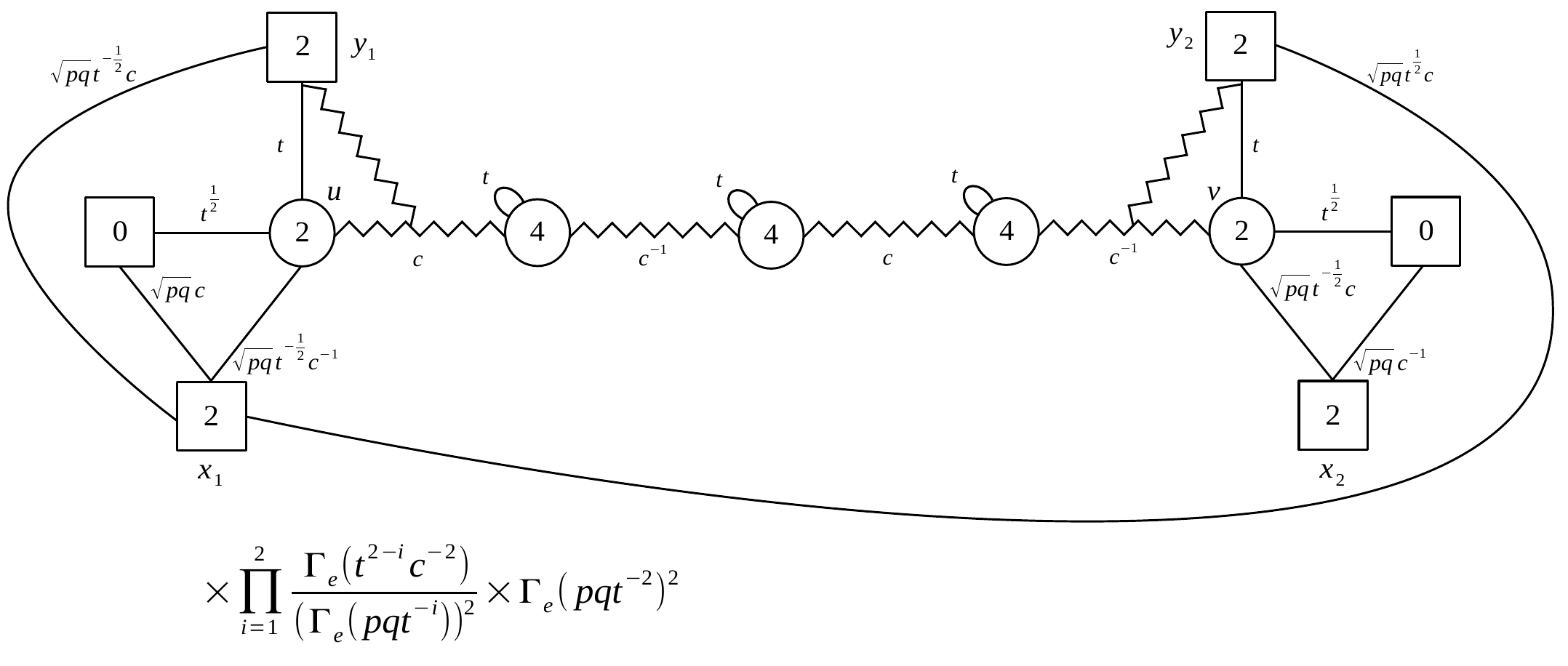}
    \caption{Quiver for the $\mathsf{S}$-dual of the $N_f=N_c=2$ SQCD after applying the HW moves.}
    \label{fig:SQCD_Nc=Nf=2_dual_HW_setI}
\end{figure}

We now apply the HW duality move, for example, once on the left side and once on the right side of the quiver to arrive at the configuration shown in Figure \ref{fig:SQCD_Nc=Nf=2_dual_HW_setI}. One should note that this is merely one convenient choice and we can also apply the HW move twice only on the same side of the quiver, which should result in the same conclusion. We will discuss such multiple choices in Appendix \ref{moreframe} for the case of $N_c=N_f=4$.

\begin{figure}[!ht] 
	\centering
	\includegraphics[width=.8\textwidth]{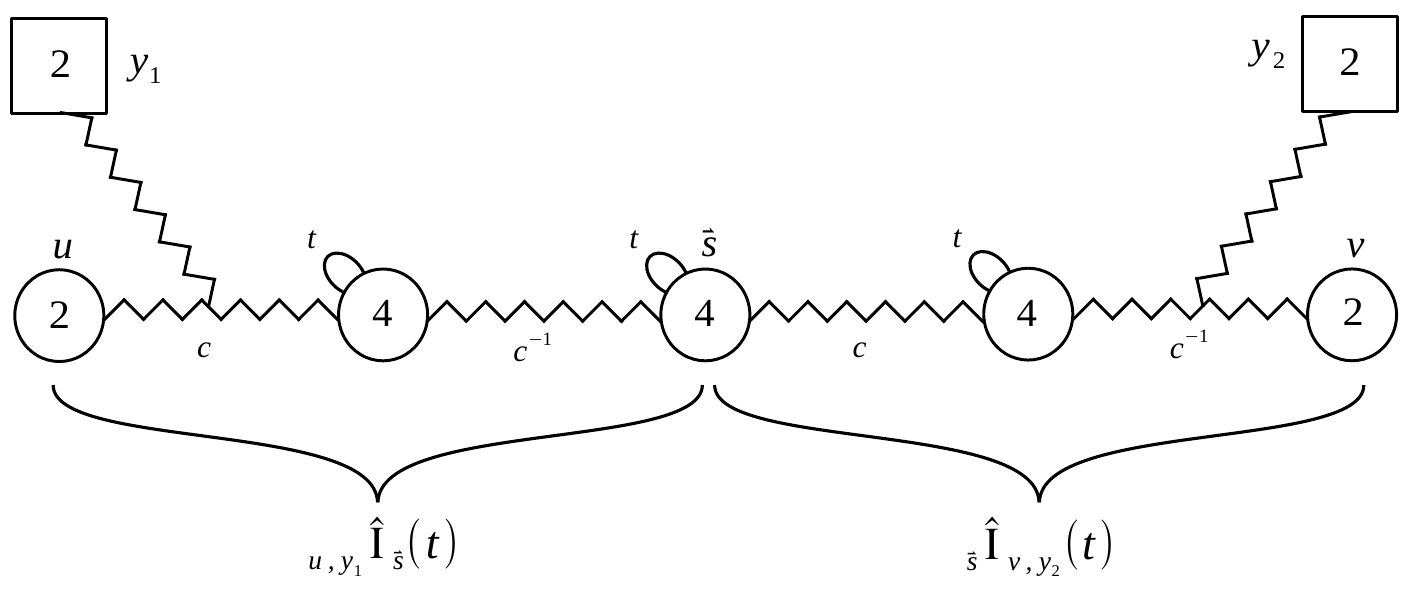}
	\caption{We isolate the part of the $N_f=N_c=2$ quiver after applying the HW moves that contains the two colliding Identity-walls.}
	\label{fig:SQCD_Nc=Nf=2_dual_setI_deltaNorm}
\end{figure}
It is useful to isolate the index contribution of the colliding Identity-walls shown in Figure \ref{fig:SQCD_Nc=Nf=2_dual_setI_deltaNorm}, containing
\begin{align}\nonumber
&{\Delta_2\left( \vec{s}, t \right)}\Delta_1\left( u \right) 
	\Delta_1\left( v \right)
{}_{\vec{s}}\hat{\mathbb{I}}_{u,y_1}(t) {}_{\vec{s}}\hat{\mathbb{I}}_{v,y_2}(t)=
	{\Delta_2\left( \vec{s}, t \right)}
	\Delta_1\left( u \right) 
	\Delta_1\left( v \right)
\frac{{ \prod_{j=1}^{2} 2\pi i s_j }}{\Gd_2 (\vec s;t) }\nn\\
&\qquad\qquad\qquad\times\left.\sum_{\sigma \in S_2,\pm}\prod_{i=1}^{2}\delta\left(s_i-f_{\sigma(i)}^{\pm1}\right)\right|_{\substack{f_1=u,\\ f_2=y_1}}
\frac{{ \prod_{i=1}^{2} 2\pi i s_i }}{\Gd_2 (\vec s;t) }\left.\sum_{\sigma \in S_2,\pm}\prod_{i=1}^{2}\delta\left(s_i-h_{\sigma(i)}^{\pm1}\right)\right|_{\substack{h_1=v,\\ h_2=y_2}}\,,
\label{eqdelta2}
\end{align}
where ${\Delta_2\left( \vec{s}, t \right)}$ is the contribution of the antisymmetric and the vector of the middle 
$USp(4)_{\vec s}$ while  $\Delta_1\left( u \right)\Delta_1\left( v \right)$ is the contribution of the vector multiplets of the 
left and right $USp(2)$ nodes without antisymmetric fields. 

\begin{figure}[!ht] 
	\centering
	\includegraphics[width=.5\textwidth]{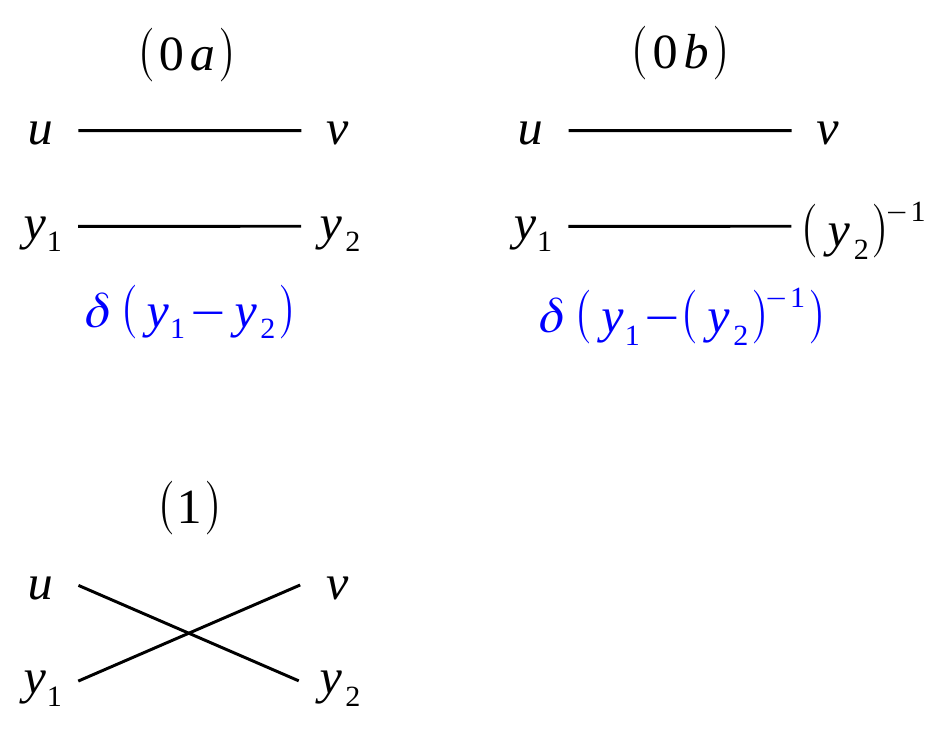}
	\caption{Possible inequivalent identifications of fugacities enforced by the colliding Identity-walls for $N_c=N_f=2$.}
	\label{22id}
\end{figure}

The product of delta functions in eq.~\eqref{eqdelta2} forces us to identify each component of the two vector of fugacities\footnote{We do not consider as independent identifications those differing by a sign flip of the power of the gauged fugacities as these are equivalent up to a transformation of the Weyl of the gauge group. In other words, these apparently different identifications would give equal result, whose multiplicity then cancels against the Weyl factor in the integration measure.}
\be
\vec{f}&=&\{ u,y_1 \}\,,\nn\\
\vec{h}&=&\{ v,y_2 \}\,.
\ee
The delta function involving a gauge fugacity can be then used to remove the corresponding integration, while a delta function identifying the two global fugacities remains as a factor in the final index. The possible inequivalent identifications are summarized in Figure \ref{22id} and we shall next study their result in turn.
%

\begin{figure}[!ht]
	\centering
    \includegraphics[width=\textwidth]{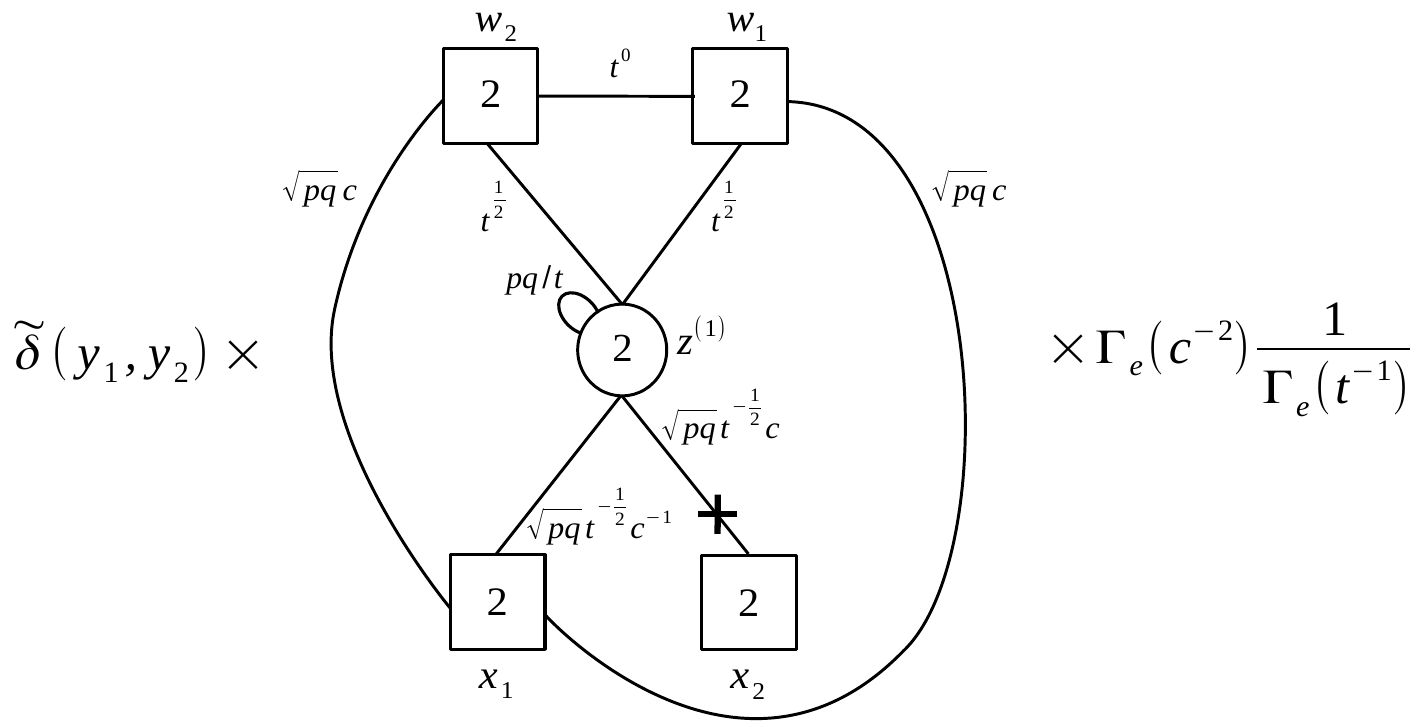}
    \caption{Frame zero of the $N_f=N_c=2$ SQCD. We restored all the singlet flipping fields.}
    \label{fig:SQCD_Nc=Nf=2_ACdual}
\end{figure}

\paragraph{Frame zero.}  We first consider the identifications dubbed as $(0a)$ and $(0b)$ in Figure \ref{22id}, where the two gauge fugacities $u=v$ are identified and we have two identifications of the global fugacities $y_1$ and $y_2$, namely $y_1=y_2^\pm$. 
Both choices lead to the same result, which we call frame zero and display it in Figure \ref{fig:SQCD_Nc=Nf=2_ACdual}, up to different identification conditions between $y_1$ and $y_2$.

Let's first consider the identification $(0a)$  in Figure \ref{22id}.
\begin{itemize}
\item 
The contribution of vector and antisymmetric  chirals in eq.~\eqref{eqdelta2} becomes
\begingroup\allowdisplaybreaks
\begin{align}
	& \cancel{\Delta_2\left( \vec{s}, t \right)}
	\frac{1}{\cancel{\Delta_2\left( \vec{s}, t \right)}}
	\frac{1}{\Delta_2\left( \vec{s}, t \right)} 
	\Delta_1\left( u \right) 
	\Delta_1\left( v \right) = \nonumber\\[7pt]
	& \quad = 
	\prod_{i<j}^{2}\Gamma_e\left( s_i^{\pm}s_j^{\pm} \right)
	\prod_{i=1}^{2}\Gamma_e\left( s_i^{\pm 2} \right)
	\left[\Gamma_e\left( t \right)^2
	\prod_{i<j}^{2}\Gamma_e\left( t s_i^{\pm}s_j^{\pm} \right)
	\right]^{-1} 
	\left[ \Gamma_e\left( u^{\pm 2} \right) \right]^{-1}
	\left[ \Gamma_e\left( v^{\pm 2} \right) \right]^{-1} \nonumber\\[7pt]
	& \quad \xrightarrow{\,\,(0a)\,\,} 
	\frac{1}{(p;p)_\infty (q;q)_\infty}\Delta_1\left({z}^{(1)},pqt^{-1}\right) 
	\Gamma_e\left( z^{(1)\pm} y_1^{\pm} \right)
	\Gamma_e\left( y_1^{\pm 2} \right)
	\Bigg[ \Gamma_e\left( t \right)
	\Gamma_e\left( t z^{(1)\pm} y_1^{\pm} \right)
	\Bigg]^{-1} \,. 
	\label{eq:Nc=Nf=2_frame0_deltanormalisation_result}
\end{align}
\endgroup
\item
The contributions of the $USp(2)_u\times USp(2)_{y_1}$  and $USp(2)_v\times USp(2)_{y_2}$  bifundamentals  in Figure \ref{fig:SQCD_Nc=Nf=2_dual_HW_setI} give
\begin{align}
	\Gamma_e\left( t u^{\pm} y_1^{\pm} \right) 
	 \xrightarrow{(0a)}  
	\Gamma_e\left( t z^{(1)\pm} y_1^{\pm} \right) \,, \quad
	\Gamma_e\left( t v^{\pm} y_2^{\pm} \right) 
	 \xrightarrow{(0a)}  
	\Gamma_e\left( t z^{(1)\pm} y_2^{\pm} \right) \,.
\end{align}
\item We can then combine the remaining chirals in Figure \ref{fig:SQCD_Nc=Nf=2_dual_HW_setI} by introducing the variables\footnote{Notice that the choice $w_1=y_1t^{\frac{1}{2}}$, $w_2=y_2t^{-\frac{1}{2}}$ is equivalent.
}
\begin{equation}
	\begin{cases}
		w_1=y_1t^{-\frac{1}{2}}\,,\\
		w_2=y_2t^{\frac{1}{2}}\,,
	\end{cases} 
	\label{eq:Nc=Nf=2_frame0_wRedef}
\end{equation}
as
\begin{align}
	& \Gamma_e\left(\sqrt{pq}t^{-\frac{1}{2}}c x_1^{\pm} y_1^{\pm} \right)
	\Gamma_e\left(\sqrt{pq}t^{\frac{1}{2}}c x_1^{\pm} y_2^{\pm} \right) 
	\Gamma_e\left(\sqrt{pq}t^{-\frac{1}{2}}c^{-1} u^{\pm} x_1^{\pm} \right)
	\Gamma_e\left(\sqrt{pq}t^{-\frac{1}{2}}c \,v^{\pm} x_2^{\pm} \right)
	\nonumber\\
	& \xrightarrow{\,\,(0a),\,(\ref{eq:Nc=Nf=2_frame0_wRedef})\,\,}
	\Gamma_e\left(\sqrt{pq}c x_1^{\pm} w_1^{\pm} \right)
	\Gamma_e\left(\sqrt{pq}c x_1^{\pm} w_2^{\pm} \right) \nonumber\\
	& \quad\qquad\qquad\times
	\Gamma_e\left(\sqrt{pq}t^{-\frac{1}{2}}c^{-1} z^{(1)\pm} x_1^{\pm} \right)
	\Gamma_e\left(\sqrt{pq}t^{-\frac{1}{2}}c \, z^{(1)\pm} x_2^{\pm} \right) \,.
\end{align}
\end{itemize}

Collecting all the contributions and implementing the redefinition \eqref{eq:Nc=Nf=2_frame0_wRedef}
we obtain the theory in Figure \ref{fig:SQCD_Nc=Nf=2_ACdual} with index

\begin{align}
\label{eq:ind22-0}
 \tilde{\delta}\left(y_1,y_2\right)
    \frac{\Gamma_e\left(c^{-2}\right)}{\Gamma_e\left(t^{-1}\right)}  
    \Gamma_e\left( w_1^\pm w_2^\pm \right)    
    \mathcal{I}_{\text{SQCD}(1,2)}\left(\vec x;w_2,w_1;t;t^{\frac{1}{2}}c\right)\Big|_{w_{1,2}=y_{1,2}t^{\mp \frac{1}{2}}}\,,
\end{align}
where we defined
\begin{align}
	\tilde{\delta}\left(x,y\right) = \frac{2 \pi i x}{(p;p)_\infty (q;q)_\infty}\delta\left(x-y\right)\,.
\end{align}
Notice that above, using the result in Appendix \ref{The 4d mirror pair}, we rewrote the interacting part in terms of its $\mathsf S$-dual (in this case the theory is actually self-dual) that is the good SQCD with $N_c=1$ and $N_f=2$ with a shifted $c$ fugacity. We then have an
$SU(2)_{w_1}\times SU(2)_{w_2}$ bifundamental  chiral and two more singlets.\footnote{
Notice that by imposing \eqref{eq:Nc=Nf=2_frame0_wRedef} on \eqref{eq:Nc=Nf=2_frame0_deltanormalisation_result}, the term $\Gamma_e\left( y_1^{\pm 2} \right)$ produces only $\Gamma_e\left( w_1^{+}w_2^{+} \right)\Gamma_e\left( w_1^{-}w_2^{-} \right)$. In order to reconstruct the whole $SU(2)_{w_1}\times SU(2)_{w_2}$ bifundamental $\Gamma_e\left( w_1^{\pm}w_2^{\pm} \right)$ we have to multiply and divide by the missing pieces. This operation contributes to the singlets shown in Figure \ref{fig:SQCD_Nc=Nf=2_ACdual}.}

The identification $(0b)$ in  Figure \ref{22id}  produces the same result but with $\tilde{\delta}\left(y_1,y_2^{-1}\right)$  in front of it.

\paragraph{Frame one.} Let's now consider the identification $(1)$  in Figure \ref{22id}. In this case 
each gauge fugacity is identified  with one of the global fugacities. This allows to remove all of the integrals and there is no constraint on the global $y_1$ and $y_2$ fugacities. 
Let us then see the effect of these identifications on the theory.
\begin{itemize}
\item 
The contribution of vector and antisymmetric  chirals in eq.~\eqref{eqdelta2} gives
\begingroup\allowdisplaybreaks
\begin{align}
	& \cancel{\Delta_2\left( \vec{s}, t \right)}
	\frac{1}{\cancel{\Delta_2\left( \vec{s}, t \right)}}
	\frac{1}{\Delta_2\left( \vec{s}, t \right)} 
	\Delta_1\left( u \right) 
	\Delta_1\left( v \right) = \nonumber\\[7pt]
	& \quad = 
	\prod_{i<j}^{2}\Gamma_e\left( s_i^{\pm}s_j^{\pm} \right)
	\prod_{i=1}^{2}\Gamma_e\left( s_i^{\pm 2} \right)
	\left[\Gamma_e\left( t \right)^2
	\prod_{i<j}^{2}\Gamma_e\left( t s_i^{\pm}s_j^{\pm} \right)
	\right]^{-1} 
	\left[ \Gamma_e\left( u^{\pm 2} \right) \right]^{-1}
	\left[ \Gamma_e\left( v^{\pm 2} \right) \right]^{-1} \nonumber\\[7pt]
	& \quad \xrightarrow{\,\,(1a)\,\,} 
	\Gamma_e\left( y_1^{\pm} y_2^{\pm} \right)
	\Bigg[ \Gamma_e\left( t \right)^2
	\Gamma_e\left( t  y_1^{\pm} y_2^{\pm} \right)
	\Bigg]^{-1} \,.
\end{align}
\endgroup
\item
The contributions of the $USp(2)_u\times USp(2)_{y_1}$  and $USp(2)_v\times USp(2)_{y_2}$  bifundamentals  in Figure \ref{fig:SQCD_Nc=Nf=2_dual_HW_setI} give
\begin{align}
	\Gamma_e\left( t u^{\pm} y_1^{\pm} \right) 
	 \xrightarrow{(1a)}  
	\Gamma_e\left( t y_2^{\pm} y_1^{\pm} \right) \,,\quad
	\Gamma_e\left( t v^{\pm} y_2^{\pm} \right) 
	 \xrightarrow{(1a)}  
	\Gamma_e\left( t y_1^{\pm} y_2^{\pm} \right) \,.
\end{align}
\item
The remaining chirals become
\begin{align}
	& \Gamma_e\left(\sqrt{pq}t^{-\frac{1}{2}}c x_1^{\pm} y_1^{\pm} \right)
	\Gamma_e\left(\sqrt{pq}t^{\frac{1}{2}}c x_1^{\pm} y_2^{\pm} \right) 
	\Gamma_e\left(\sqrt{pq}t^{-\frac{1}{2}}c^{-1} u^{\pm} x_1^{\pm} \right)
	\Gamma_e\left(\sqrt{pq}t^{-\frac{1}{2}}c \,v^{\pm} x_2^{\pm} \right)
	\nonumber\\
	& \xrightarrow{\,\,(1a)\,\,}
	\prod_{i=1}^2 \Gamma_e\left(\sqrt{pq}t^{-\frac{1}{2}}c x_i^{\pm} y_1^{\pm} \right) \,.
\end{align}
\end{itemize}

\begin{figure}[!ht]
	\centering
    \includegraphics[width=.7\textwidth]{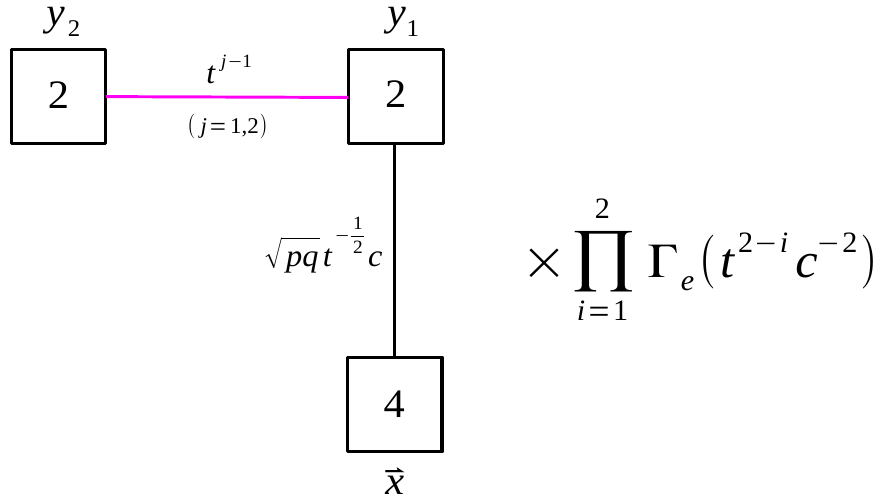}
     \caption{Frame zero of the $N_f=N_c=2$ SQCD. We restored all the singlet flipping fields.}
    \label{fig:SQCD_Nc=Nf=2_ITAdual}
\end{figure}

Collecting all the contribution we obtain the theory in Figure \ref{fig:SQCD_Nc=Nf=2_ITAdual}, with index
\begin{align}
  \prod_{j=1}^2\Gamma_e\left(c^{-2} t^{2-j}\right)  
    \prod_{j=1}^2\Gamma_e\left(t^{j-1} y_1^\pm y_2^\pm\right) 
    \mathcal{I}^{(4d)}_{\text{SQCD}(0,2)}\left(\vec x;y_2,y_1;t;tc\right)
    \,.
\end{align}
This is a WZ model which, up to singlets including an
$SU(2)_{y_1}\times SU(2)_{y_2}$ bifundamental, 
 we rewrote in terms of its $\mathsf S$-dual
  the good SQCD with $N_c=0$ and $N_f=2$ with a shifted $c$ fugacity, again using the result in Appendix \ref{The 4d mirror pair}.

\begin{figure}[!p] 
	\centering
	\includegraphics[width=.75\textwidth]{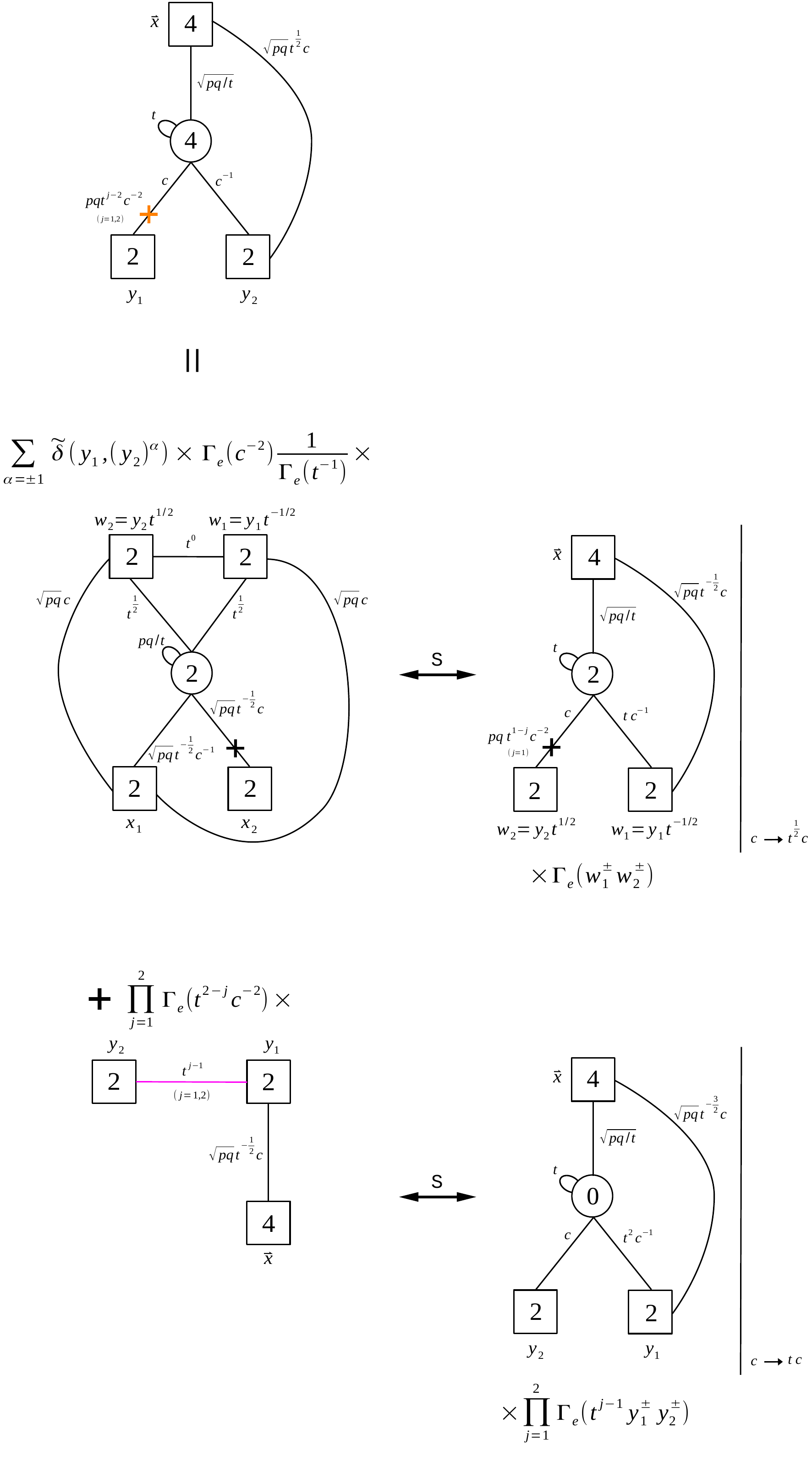}
	\caption{The final result for the $N_c=N_f=2$ SQCD.}
	\label{fig:SQCD_Nc=Nf=2_All_In_One}
\end{figure}

Collecting the results from the frames zero and one, we find that the index of the bad $N_c=N_f=2$ SQCD is given by 
\begin{align}
    & \mathcal{I}_{\text{SQCD}(2,2)}\left(\vec x;y_1,y_2;t;c\right) = \nonumber\\
    & \, = \quad
    \sum_{\alpha=\pm 1}\Bigg[
    \tilde{\delta}\left(y_1,(y_2)^{\alpha}\right)
    \frac{\Gamma_e\left(c^{-2}\right)}{\Gamma_e\left(t^{-1}\right)}  
    \Gamma_e\left( w_1^\pm w_2^\pm \right)    
    \mathcal{I}_{\text{SQCD}(1,2)}\left(\vec x;w_2,w_1;t;t^{\frac{1}{2}}c\right)\Big|_{w_{1,2}=y_{1,2}t^{\mp \frac{1}{2}}} \Bigg]
    \nonumber\\
    & \quad\,\, + 
    \prod_{j=1}^2\Gamma_e\left(c^{-2} t^{2-j}\right)  
    \prod_{j=1}^2\Gamma_e\left(t^{j-1} y_1^\pm y_2^\pm\right) 
    \mathcal{I}_{\text{SQCD}(0,2)}\left(\vec x;y_2,y_1;t;tc\right)
    \,.
\end{align}
This result is summarized in Figure \ref{fig:SQCD_Nc=Nf=2_All_In_One}.

\subsubsection{Example: $N_c=N_f=4$}
\label{44sec}

Let us now consider the $N_c=N_f=4$ SQCD. This second example exhibits some subtleties that were not present in the previous one, which we shall discuss.

\begin{figure}[!ht]
	\centering
	\makebox[\linewidth][c]{
    \includegraphics[width=\textwidth]{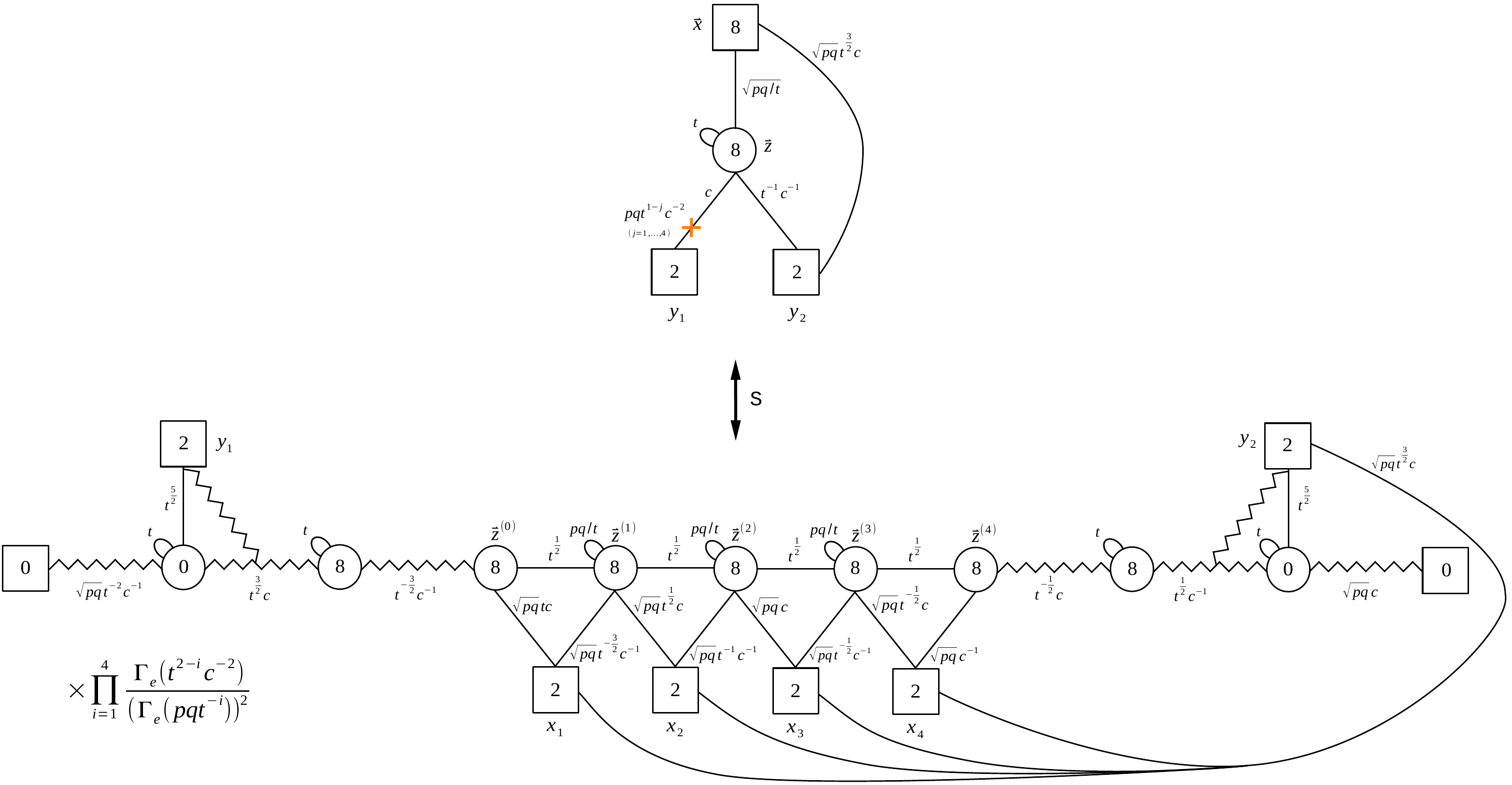}}
    \caption{The $N_c=N_f=4$ SQCD and the result of its $\mathsf{S}$-dualization before propagating the VEVs.}
    \label{fig:SQCD_Nc=Nf=4_Sdualised}
\end{figure}

After the $\mathsf S$-dualization to the quiver shown on the bottom of Figure \ref{fig:SQCD_Nc=Nf=4_Sdualised}, we have various possibilities for propagating the VEV.  The propagating VEVs can indeed meet at many points in the quiver, depending on how many HW moves we choose to perform on the left and on the right. We consider, for example, the case where we perform two HW moves on each side,
whose result is shown in Figure \ref{fig:SQCD_Nc=Nf=4_dual_HW_setI}. Other possibilities yield the same final result  and are discussed in Appendix \ref{app:alternativeHWcollisions}.
\begin{figure}[!ht]
    \includegraphics[width=\textwidth]{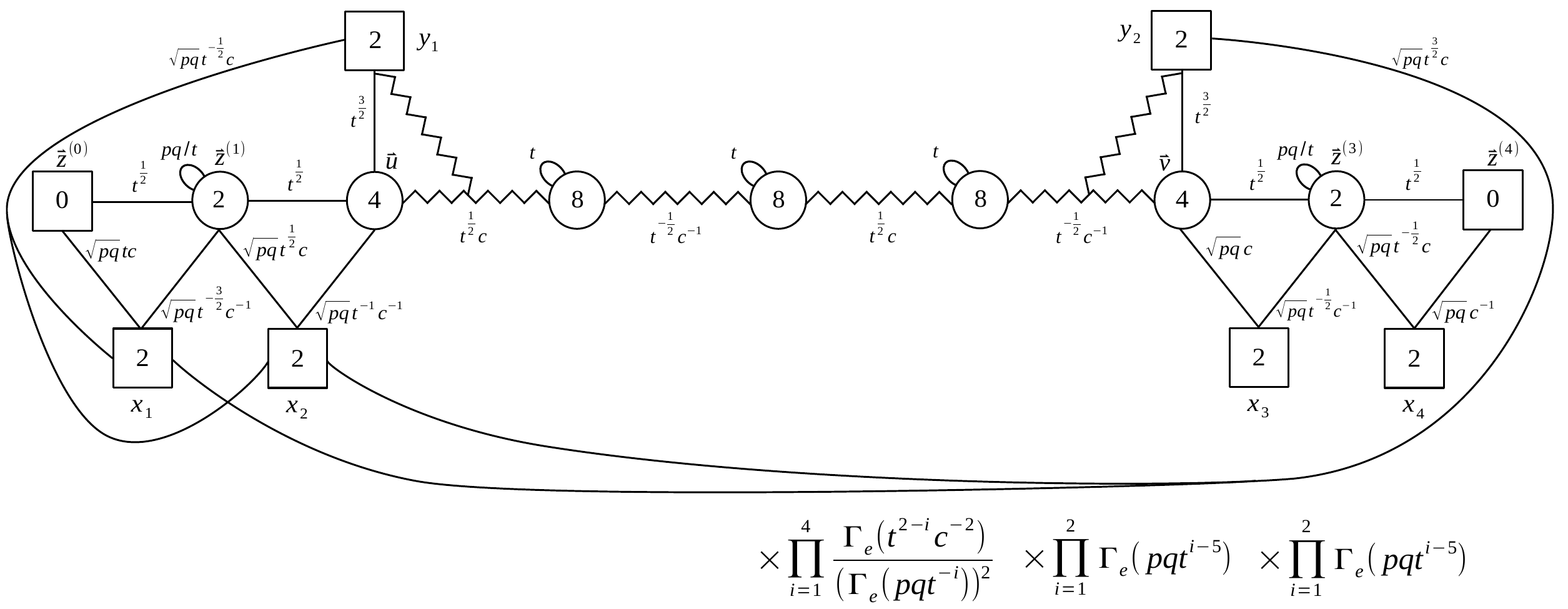}
    \caption{Quiver for the $\mathsf{S}$-dual of the $N_f=N_c=4$ SQCD after applying the HW moves.}
    \label{fig:SQCD_Nc=Nf=4_dual_HW_setI}
\end{figure}

\begin{figure}[!ht] 
	\centering
	\includegraphics[width=.7\textwidth]{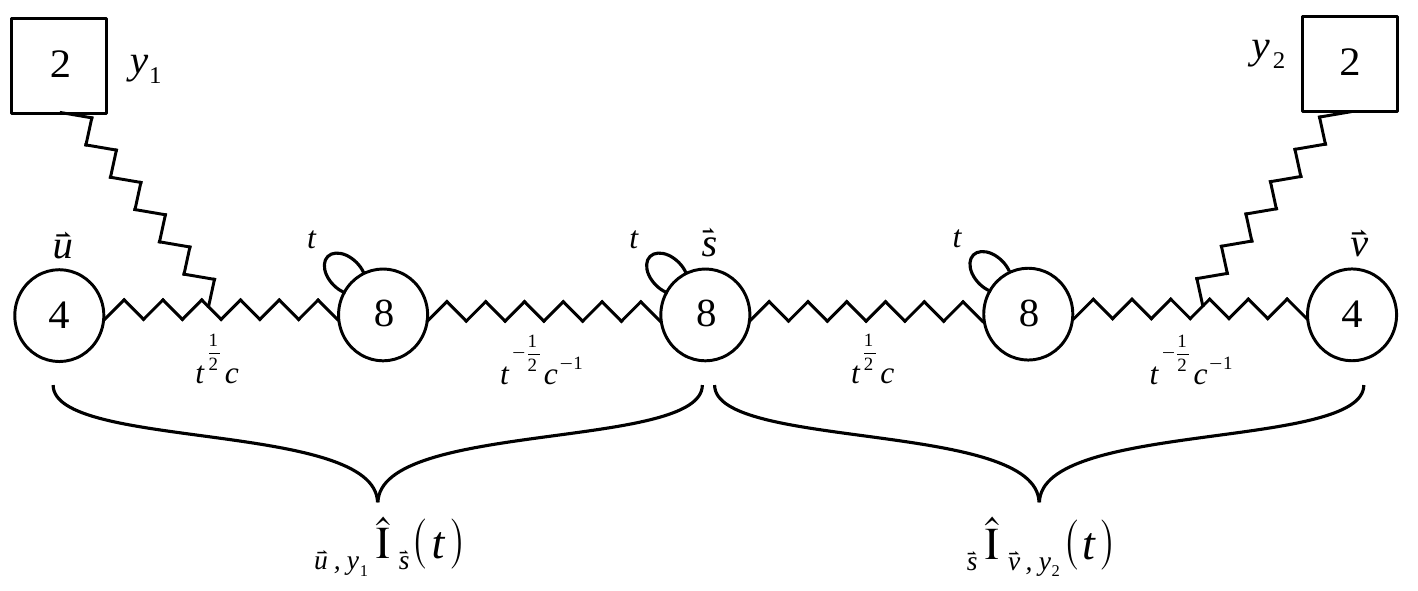}
	\caption{We isolate the part of the $N_f=N_c=4$ quiver after applying the HW moves that contains the two colliding Identity-walls.}
	\label{fig:SQCD_Nc=Nf=4_dual_setI_deltaNorm}
\end{figure}

We find again a configuration with two colliding asymmetric Identity-walls, as highlighted in Figure \ref{fig:SQCD_Nc=Nf=4_dual_setI_deltaNorm}, which contribute to the index as 

\begin{align}\nonumber
&{\Delta_4\left( \vec{s}, t \right)}\Delta_2\left( \vec{u} \right) 
	\Delta_2\left( \vec{v} \right)
{}_{\vec{s}}\hat{\mathbb{I}}_{\vec{u},y_1}(t) {}_{\vec{s}}\hat{\mathbb{I}}_{\vec{v},y_2}(t)=\\
&=
{\Delta_4\left( \vec{s}, t \right)}\Delta_2\left( \vec{u} \right) 
	\Delta_2\left( \vec{v} \right)
	\frac{{ \prod_{j=1}^{4} 2\pi i s_j }}{\Gd_4 (\vec s;t) }\left.\sum_{\sigma \in S_4,\pm}\prod_{i=1}^{4}\delta\left(s_i-f_{\sigma(i)}^{\pm1}\right)\right|_{\substack{f_1=u_1,\\f_2=u_2,\\f_3=y_1t^{1/2} \\ f_4=y_1t^{-1/2} }} \nonumber \\
&\quad\times \frac{{ \prod_{j=1}^{4} 2\pi i s_j }}{\Gd_4 (\vec s;t) }\left.\sum_{\sigma \in S_4,\pm}\prod_{i=1}^{4}\delta\left(s_i-h_{\sigma(i)}^{\pm1}\right)\right|_{\substack{h_1=v_1,\\h_2=v_2,\\h_3=y_2t^{1/2} \\ h_4=y_2t^{-1/2} }}\,.
\label{eqdelta4}
\end{align}

We can see that in this case the product of delta functions in \eqref{eqdelta4} forces us to identify the two sets of fugacities on each side of the colliding Identity-walls\footnote{As before we do not consider as independent identifications those that differ for permutation and sign flip of the power of the gauge fugacities as these multiplicities are taken into account by the Weyl factor in the integration measure.}
\be
\vec{f}&=&\left\{ u_1, u_2, y_1t^{\frac{1}{2}}, y_1t^{-\frac{1}{2}} \right\} \,, \label{eq:var441} \\
\vec{h}&=&\left\{ v_1, v_2, y_2t^{\frac{1}{2}}, y_2t^{-\frac{1}{2}} \right\} \label{eq:var442} \,.
\ee
As we saw in the previous example, two colliding Identity-walls lead to multiple frames that originate from different identifications between the two sets $\vec{f}$ and $\vec{h}$. Nevertheless, in general not all identifications lead to physical frames. We propose a simple criterion for which identifications should be considered. These are those for which after the identifications the variables $f_i$ satisfy the following conditions:
\begin{enumerate}
\item $f_i\neq f_j^{\pm1}$ for any pair of $i$ and $j$;
\item $f_i\neq 1$ for any $i$.
\end{enumerate}
These conditions exclude in particular the fixed points of the action of the $USp(8)$ Weyl group on the variables $\vec{f}$. This is because the property of the Identity-wall \eqref{eq:deltaNN} was derived assuming the variables $f_i$ satisfy the above conditions.
We have tested this criterion in many examples and we expect it indeed provides the correct way to isolate the physical frames.

\begin{figure}[!ht] 
	\centering
	\includegraphics[width=.8\textwidth]{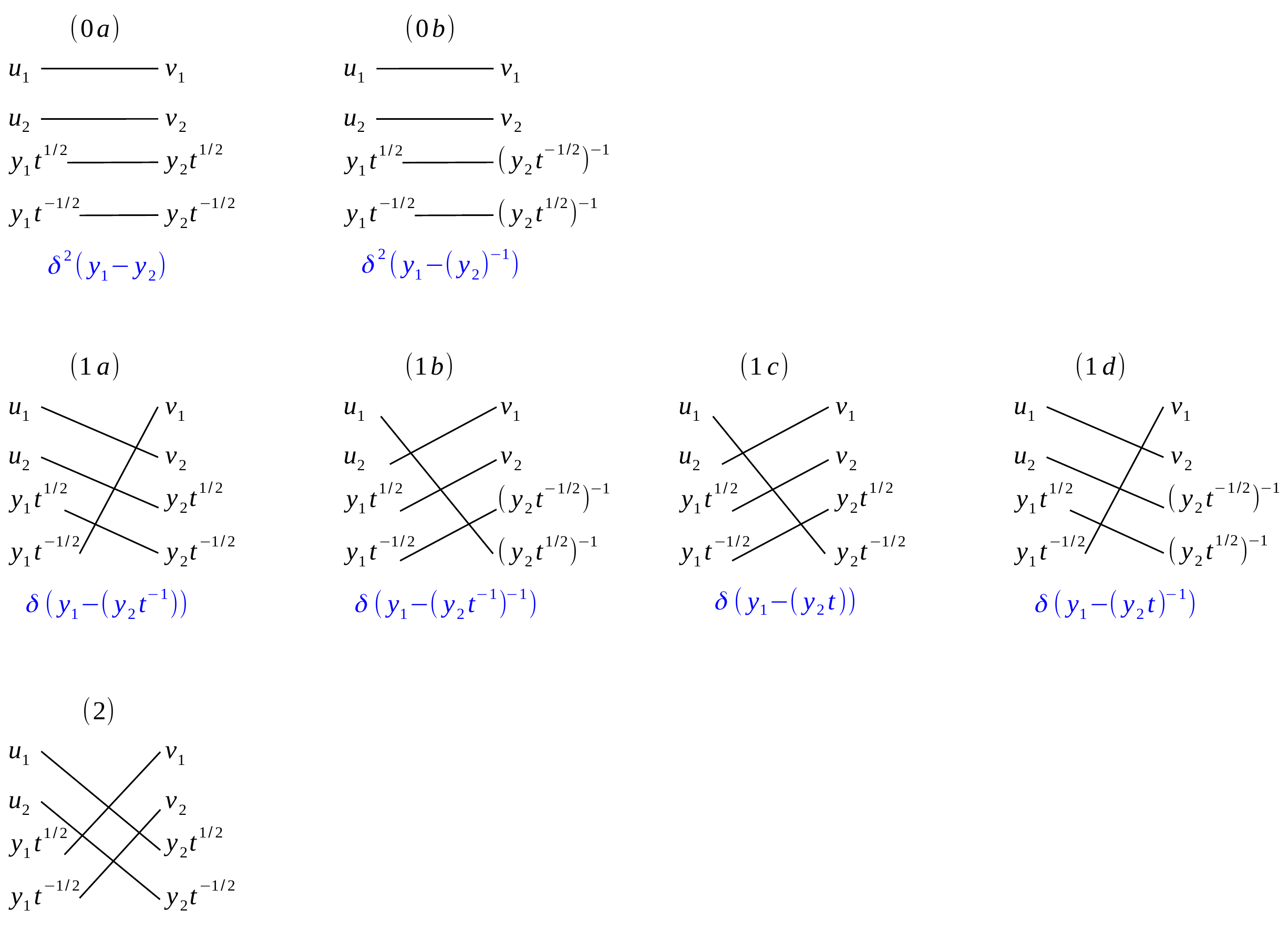}
	\caption{Possible inequivalent identifications of fugacities enforced by the colliding Identity-walls for $N_c=N_f=4$.}
	\label{44id}
\end{figure}

For $N_c = N_f = 4$, the inequivalent physical identifications are summarized in Figure \ref{44id}, where they are grouped in accordance with the three distinct frames that they lead to, which we shall now study in turn. There are more possible identifications that violate the criterion above  and  do not yield physical frames, in Appendix \ref{moreframe} we argue indeed that they should not be considered. 

\paragraph{Frame zero.} Consider the identification $(0a)$  in Figure \ref{44id}
\begin{align}
	\begin{cases}
		u_1=v_1=z^{(2)}_1\,,\\
		u_2=v_2=z^{(2)}_2\,,\\
		y_1t^{\frac{1}{2}}=y_2t^{\frac{1}{2}}\,,\\
		y_1t^{-\frac{1}{2}}=y_2t^{-\frac{1}{2}}\,.
	\end{cases} 
	\label{eq:Nc=Nf=4_frame1_identif}
\end{align}
So we get twice the constraint $y_1=y_2$, yielding $\delta^2(y_1-y_2)$.
Moreover, after the identifications we get the following contributions.

\begin{itemize}
\item The contribution of vector and antisymmetric  chirals in eq.~\eqref{eqdelta4} becomes
\begingroup\allowdisplaybreaks
\begin{align}
\label{eq:va}
	& \cancel{\Delta_4\left( \vec{s}, t \right)}
	\frac{1}{\cancel{\Delta_4\left( \vec{s}, t \right)}}
	\frac{1}{\Delta_4\left( \vec{s}, t \right)} 
	\Delta_2\left( \vec{u} \right) 
	\Delta_2\left( \vec{v} \right) = \nonumber\\[7pt]
	& \quad \xrightarrow{\,\,(\ref{eq:Nc=Nf=4_frame1_identif})} 
	\frac{1}{[(p;p)_\infty (q;q)_\infty]^2}\Delta_2\left(\vec{z}^{\,(2)},pqt^{-1}\right) 
	\prod_{i=1}^2\prod_{j=1}^2\Gamma_e\left( z_i^{(2)\pm} \left( y_1t^{\frac{3-2j}{2}} \right)^{\pm} \right)\nonumber\\
	&\quad\quad\times
	\prod_{i<j}^2\Gamma_e\left( \left( y_1t^{\frac{3-2i}{2}} \right)^{\pm}\left( y_1t^{\frac{3-2j}{2}} \right)^{\pm}\right)
	\prod_{i=1}^2\Gamma_e\left( \left ( y_1t^{\frac{3-2i}{2}} \right)^{\pm 2} \right) \nonumber\\
	& \quad\quad\times
	\Bigg[ \Gamma_e\left( t \right)^2 
	\prod_{i=1}^2\prod_{j=1}^2\Gamma_e\left( t z_i^{(2)\pm} \left( y_1t^{\frac{3-2j}{2}} \right)^{\pm} \right) \times
	\prod_{i<j}^2\Gamma_e\left( t \left( y_1t^{\frac{3-2i}{2}} \right)^{\pm}\left( y_1t^{\frac{3-2j}{2}} \right)^{\pm}\right)
	\Bigg]^{-1} \,.
\end{align}
\endgroup

\item
The $USp(4)_{\vec{u}} \times SU(2)_{y_1} $ and  $USp(4)_{\vec{v}}\times SU(2)_{y_2} $ bifundamentals in Figure \ref{fig:SQCD_Nc=Nf=4_dual_HW_setI} become
\begin{align}
	\prod_{i=1}^2\Gamma_e\left( t^{\frac{3}{2}} u_i^{\pm} y_1^{\pm} \right) 
	 \xrightarrow{\,\,(\ref{eq:Nc=Nf=4_frame1_identif})}  
	\prod_{i=1}^2\Gamma_e\left( t^{\frac{3}{2}} z_i^{(2)\pm} y_1^{\pm} \right) \,, \\[3pt]
	\prod_{i=1}^2\Gamma_e\left( t^{\frac{3}{2}} v_i^{\pm} y_2^{\pm} \right) 
	 \xrightarrow{\,\,(\ref{eq:Nc=Nf=4_frame1_identif})}  
	\prod_{i=1}^2\Gamma_e\left( t^{\frac{3}{2}} z_i^{(2)\pm} y_2^{\pm} \right) \,.
\end{align}

\item
The $USp(4)_{\vec{u}} \times USp(2)_{z^{(1)}} $ and  $USp(4)_{\vec{v}}\times USp(2)_{z^{ (3)}} $ bifundamentals in Figure \ref{fig:SQCD_Nc=Nf=4_dual_HW_setI} become
\begin{align}
	\prod_{i=1}^2\Gamma_e\left( t^{\frac{1}{2}} z^{(1)\pm} u_i^{\pm} \right) 
	 \xrightarrow{\,\,(\ref{eq:Nc=Nf=4_frame1_identif})}  
	\prod_{i=1}^2\Gamma_e\left( t^{\frac{1}{2}} z^{(1)\pm} z_i^{(2)\pm} \right) \,,\\[3pt]
	\prod_{i=1}^2\Gamma_e\left( t^{\frac{1}{2}} v_i^{\pm} z^{(3)\pm} \right) 
	 \xrightarrow{\,\,(\ref{eq:Nc=Nf=4_frame1_identif})}  
	\prod_{i=1}^2\Gamma_e\left( t^{\frac{1}{2}} z_i^{(2)\pm} z^{(3)\pm} \right) \,.
\end{align}
\item
Finally, by defining
\begin{equation}
	\begin{cases}
		w_1=y_1t^{-1}\,,\\
		w_2=y_2t\,,
	\end{cases} 
	\label{eq:Nc=Nf=4_frame1_wRedef}
\end{equation}

we can combine the remaining chirals as
\begingroup\allowdisplaybreaks
\begin{align}
	& \prod_{i=1}^2\Gamma_e\left(\sqrt{pq}t^{-\frac{1}{2}}c x_i^{\pm} y_1^{\pm} \right)
	\prod_{i=1}^2\Gamma_e\left(\sqrt{pq}t^{\frac{3}{2}}c x_i^{\pm} y_2^{\pm} \right) \nonumber\\
	& \,\,\times
	\prod_{i=1}^2\Gamma_e\left(\sqrt{pq}t^{-1}c^{-1} u_i^{\pm} x_2^{\pm} \right)
	\prod_{i=1}^2\Gamma_e\left(\sqrt{pq}c \,v_i^{\pm} x_3^{\pm} \right)
	\nonumber\\
	& \xrightarrow{\,\,(\ref{eq:Nc=Nf=4_frame1_identif}),\,(\ref{eq:Nc=Nf=4_frame1_wRedef})\,\,}
	\prod_{i=1}^2\Gamma_e\left(\sqrt{pq}t^{\frac{1}{2}}c x_i^{\pm} w_1^{\pm} \right)
	\prod_{i=1}^2\Gamma_e\left(\sqrt{pq}t^{\frac{1}{2}}c x_i^{\pm} w_2^{\pm} \right) \nonumber\\
	& \quad\qquad\qquad\quad\times
	\prod_{i=1}^2\Gamma_e\left(\sqrt{pq}t^{-1}c^{-1} z_i^{(2)\pm} x_2^{\pm} \right)
	\prod_{i=1}^2\Gamma_e\left(\sqrt{pq}c \, z_i^{(2)\pm} x_3^{\pm} \right) \,.
\end{align}
\endgroup
\end{itemize}

One should note that \eqref{eq:va} includes a zero factor in the last line
\begin{align}
\label{eq:zero}
\Gpq{t \left(y_1 t^\frac12\right)^{-1} \left(y_1 t^{-\frac12}\right)}^{-1} =\Gpq{ 1}^{-1} = \Gpq{pq} = 0 \,,
\end{align}
which may imply that the index vanishes. However, this zero must be treated carefully because of $\delta^2(y_1-y_2) = \delta(y_1-y_2) \, \delta(0)$, which includes a divergent factor $\delta(0)$. In Appendix \ref{moreframe} we argue that indeed the zero and the delta divergence cancel out and in the rest of the paper we will adopt this prescription when $\delta(0)$ and $\Gpq{pq}$ are encountered.

\begin{figure}[!ht] 
	\centering
	\includegraphics[width=.8\textwidth]{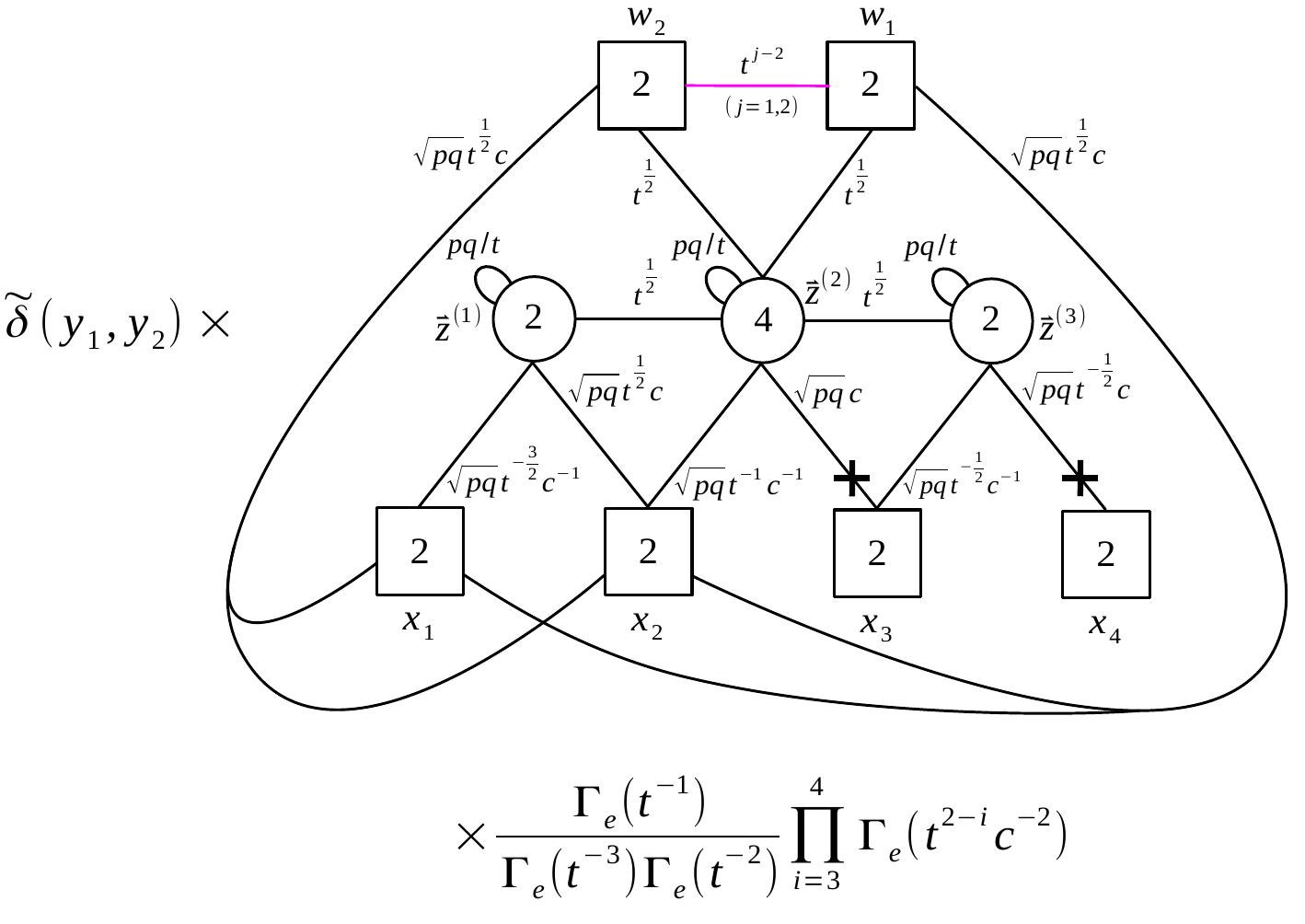}
	\caption{Frame zero of the $N_f=N_c=4$ SQCD. We restored all the singlet flipping fields.}
	\label{fig:SQCD_Nc=Nf=4_ACdual}
\end{figure}

Collecting all the contributions and implementing the redefinition \eqref{eq:Nc=Nf=4_frame1_wRedef}, we obtain the  theory in Figure \ref{fig:SQCD_Nc=Nf=4_ACdual} with index
\begin{align}
\label{eq:ind44-0}
&
	\tilde\delta\left(y_1,y_2 \right)
    \frac{\Gamma_e\left(t^{-1}\right)}{\Gamma_e\left(t^{-3}\right)\Gamma_e\left(t^{-2}\right)} 
    \prod_{j=3}^4\Gamma_e\left(c^{-2} t^{2-j}\right) \nn\\
    &\quad\times
    \prod_{j=1}^2\Gamma_e\left(t^{j-2} w_1^\pm w_2^\pm\right)    
    \mathcal{I}_{\text{SQCD}(2,4)}\left(\vec x;w_2,w_1;t;tc\right)\Big|_{w_{1,2}=y_{1,2}t^{\mp 1}} \,.
\end{align}
Notice that above,  using the result in Appendix \ref{The 4d mirror pair},  we rewrote the interacting part in terms of its $\mathsf S$-dual, the good SQCD with $N_c=2$ and $N_f=4$ with a shifted $c$ fugacity. We then have two
$SU(2)_{w_1}\times SU(2)_{w_2}$ bifundamental  chiral (denoted in pink in the figure) and some extra singlets.

The identification $(0b)$ in  Figure \ref{44id}  produces the same result but with $\tilde{\delta}\left(y_1,y_2^{-1}\right)$ in front of it.

\paragraph{Frame one.}
Consider now the identification $(1a)$  in Figure \ref{44id}
\begin{align}
	\begin{cases}
		u_1=v_1=z^{(2)}\,,\\
		u_2=y_2t^{\frac{1}{2}}\,,\\
		v_2=y_1t^{-\frac{1}{2}}\,,\\
		y_1t^{\frac{1}{2}}=y_2t^{-\frac{1}{2}}\,,
		\end{cases} 
		\label{eq:Nc=Nf=4_frame2_identif}
\end{align}
which leads to the overall delta function
\begin{equation}
	\delta\left(y_1-y_2t^{-1}\right) \,.
	\label{eq:Nc=Nf=4_frame2_delta}
\end{equation}
Let us then see the effect of these identifications on the theory.

\begin{itemize}
\item The contribution of vector and antisymmetric  chirals in eq.~\eqref{eqdelta4} becomes
\begingroup\allowdisplaybreaks
\begin{align}
	& \cancel{\Delta_4\left( \vec{s}, t \right)}
	\frac{1}{\cancel{\Delta_4\left( \vec{s}, t \right)}}
	\frac{1}{\Delta_4\left( \vec{s}, t \right)} 
	\Delta_2\left( \vec{u} \right) 
	\Delta_2\left( \vec{v} \right) = \nonumber\\[7pt]
	& \qquad \xrightarrow{\,\,(\ref{eq:Nc=Nf=4_frame2_identif}),\,(\ref{eq:Nc=Nf=4_frame2_delta})\,\,} 
	\frac{1}{(p;p)_\infty (q;q)_\infty}\Delta_1\left(z^{(2)},pqt^{-1}\right) 
	\prod_{i=1}^3\Gamma_e\left( z^{(2)\pm} \left( y_1t^{\frac{5-2i}{2}} \right)^{\pm} \right)\nonumber\\
	&\qquad\quad\times
	\prod_{i<j}^3\Gamma_e\left( \left( y_1t^{\frac{5-2i}{2}} \right)^{\pm}\left( y_1t^{\frac{5-2j}{2}} \right)^{\pm}\right)
	\prod_{i=1}^3\Gamma_e\left( \left( y_1t^{\frac{5-2i}{2}} \right)^{\pm 2} \right) \nonumber\\
	& \qquad\quad\times
	\Bigg[ \Gamma_e\left( t \right)^3 
	\prod_{i=1}^3\Gamma_e\left( t z^{(2)\pm} \left( y_1t^{\frac{5-2i}{2}} \right)^{\pm} \right)
	\prod_{i<j}^3\Gamma_e\left( t \left( y_1t^{\frac{5-2i}{2}} \right)^{\pm}\left( y_1t^{\frac{5-2j}{2}} \right)^{\pm}\right) \nonumber\\
	& \qquad\qquad\quad\times 
	\Gamma_e\left( z^{(2)\pm} \left(y_2t^{\frac{1}{2}} \right)^{\pm} \right)
	\Gamma_e\left( \left(y_2t^{\frac{1}{2}} \right)^{\pm 2} \right) \nonumber\\
	& \qquad\qquad\quad\times 
	\Gamma_e\left( z^{(2)\pm} \left(y_1t^{-\frac{1}{2}} \right)^{\pm} \right)
	\Gamma_e\left( \left(y_1t^{-\frac{1}{2}} \right)^{\pm 2} \right)
	\Bigg]^{-1} \,.
\end{align}
\endgroup

\item 
The $USp(4)_{\vec{u}} \times SU(2)_{y_1} $ and  $USp(4)_{\vec{v}}\times SU(2)_{y_2} $ bifundamentals in Figure \ref{fig:SQCD_Nc=Nf=4_dual_HW_setI} become
\begin{align}
	\prod_{i=1}^2\Gamma_e\left( t^{\frac{3}{2}} u_i^{\pm} y_1^{\pm} \right) 
	& \xrightarrow{\,\,(\ref{eq:Nc=Nf=4_frame2_identif}),\,(\ref{eq:Nc=Nf=4_frame2_delta})\,\,}  
	\Gamma_e\left( t^{\frac{3}{2}} z^{(2)\pm} y_1^{\pm} \right) 
	\Gamma_e\left( t^{\frac{3}{2}} \left(y_2t^{\frac{1}{2}}\right)^{\pm} y_1^{\pm} \right) \,,\\[7pt]
	\prod_{i=1}^2\Gamma_e\left( t^{\frac{3}{2}} v_i^{\pm} y_2^{\pm} \right) 
	& \xrightarrow{\,\,(\ref{eq:Nc=Nf=4_frame2_identif}),\,(\ref{eq:Nc=Nf=4_frame2_delta})\,\,}  
	\Gamma_e\left( t^{\frac{3}{2}} z^{(2)\pm} y_2^{\pm} \right) 
	\Gamma_e\left( t^{\frac{3}{2}} \left(y_1t^{-\frac{1}{2}}\right)^{\pm} y_2^{\pm} \right) \,.
\end{align}
\item
The $USp(4)_{\vec{u}} \times USp(2)_{z^{(1)}} $ and  $USp(4)_{\vec{v}}\times USp(2)_{z^{ (3)}} $ bifundamentals in Figure \ref{fig:SQCD_Nc=Nf=4_dual_HW_setI} become
\begin{align}
	\prod_{i=1}^2\Gamma_e\left( t^{\frac{1}{2}} z^{(1)\pm} u_i^{\pm} \right) 
	& \xrightarrow{\,\,(\ref{eq:Nc=Nf=4_frame2_identif}),\,(\ref{eq:Nc=Nf=4_frame2_delta})\,\,}  
	\Gamma_e\left( t^{\frac{1}{2}} z^{(1)\pm} z^{(2)\pm} \right)
	\Gamma_e\left( t^{\frac{1}{2}} z^{(1)\pm} \left(y_2t^{\frac{1}{2}}\right)^{\pm} \right) \,,\\[7pt]
	\prod_{i=1}^2\Gamma_e\left( t^{\frac{1}{2}} v_i^{\pm} z^{(3)\pm} \right) 
	& \xrightarrow{\,\,(\ref{eq:Nc=Nf=4_frame2_identif}),\,(\ref{eq:Nc=Nf=4_frame2_delta})\,\,}  
	\Gamma_e\left( t^{\frac{1}{2}} z^{(2)\pm} z^{(3)\pm} \right)
	\Gamma_e\left( t^{\frac{1}{2}} \left(y_1t^{-\frac{1}{2}}\right)^{\pm} z^{(3)\pm} \right) \,.
\end{align}
\item
We can then combine the remaining chirals in Figure \ref{fig:SQCD_Nc=Nf=4_dual_HW_setI} by introducing the variables
\begin{equation}
	\begin{cases}
		w_1=y_1t^{-\frac{1}{2}}\,,\\
		w_2=y_2t^{\frac{1}{2}}
	\end{cases} 
	\label{eq:Nc=Nf=4_frame2_wRedef}
\end{equation}
and employing the condition \eqref{eq:Nc=Nf=4_frame2_delta} as
\begin{align}
	& \prod_{i=1}^2\Gamma_e\left(\sqrt{pq}t^{-\frac{1}{2}}c x_i^{\pm} y_1^{\pm} \right)
	\prod_{i=1}^2\Gamma_e\left(\sqrt{pq}t^{\frac{3}{2}}c x_i^{\pm} y_2^{\pm} \right) \nonumber\\
	& \,\,\times
	\prod_{i=1}^2\Gamma_e\left(\sqrt{pq}t^{-1}c^{-1} u_i^{\pm} x_2^{\pm} \right)
	\prod_{i=1}^2\Gamma_e\left(\sqrt{pq}c \,v_i^{\pm} x_3^{\pm} \right)
	\nonumber\\
	& \xrightarrow{\,\,(\ref{eq:Nc=Nf=4_frame2_identif}),\,(\ref{eq:Nc=Nf=4_frame2_delta}),\,(\ref{eq:Nc=Nf=4_frame2_wRedef})\,\,}
	\prod_{i=1}^3\Gamma_e\left(\sqrt{pq}c x_i^{\pm} w_1^{\pm} \right)
	\Gamma_e\left(\sqrt{pq}tc x_1^{\pm} w_2^{\pm} \right) \nonumber\\
	& \qquad\qquad\qquad\qquad\times
	\Gamma_e\left(\sqrt{pq}t^{-1}c^{-1} z^{(2)\pm} x_2^{\pm} \right)
	\Gamma_e\left(\sqrt{pq}c \, z^{(2)\pm} x_3^{\pm} \right) \,,
\end{align}
where the last row is the saw of the resulting quiver.
\end{itemize}

\begin{figure}[!ht] 
	\centering
	\includegraphics[width=.8\textwidth]{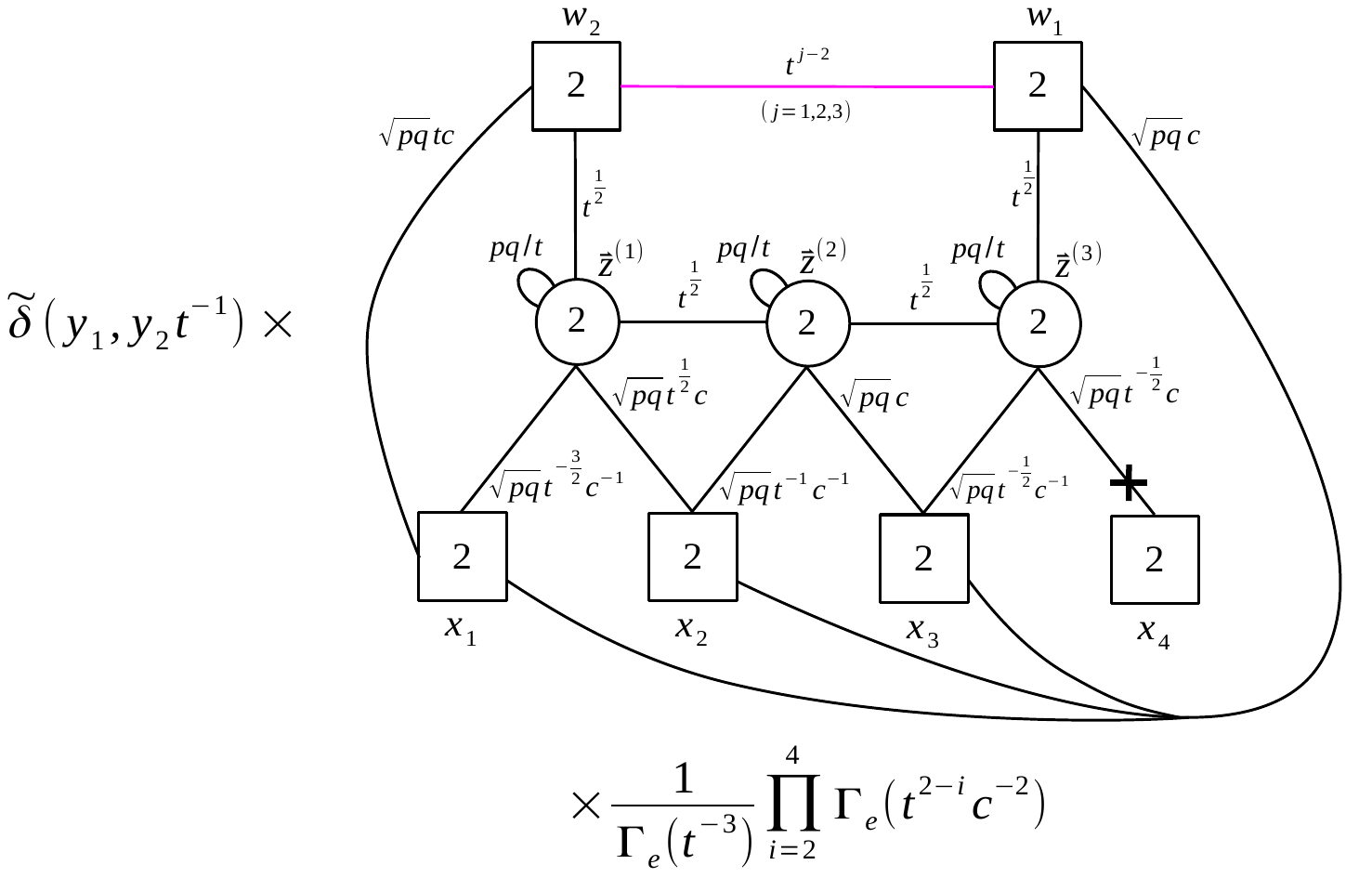}
	\caption{Frame one of the $N_f=N_c=4$ SQCD. We restored all the singlet flipping fields.}
	\label{fig:SQCD_Nc=Nf=4_INTdual}
\end{figure}

Collecting all the contributions and using the redefinition \eqref{eq:Nc=Nf=4_frame2_wRedef} we get the  theory in Figure \ref{fig:SQCD_Nc=Nf=4_INTdual} with index
\begin{align}
\label{eq:ind44-1}
   &
    \tilde\delta\left(y_1,y_2 t^{-1}\right)
    \frac{1}{\Gamma_e\left(t^{-3}\right)} 
    \prod_{j=2}^4\Gamma_e\left(c^{-2} t^{2-j}\right)\nn\\
    &\qquad\qquad\times
    \prod_{j=1}^3\Gamma_e\left(t^{j-2} w_1^\pm w_2^\pm\right)    
    \mathcal{I}_{\text{SQCD}(1,4)}\left(\vec x;w_2,w_1;t;t^{\frac{3}{2}}c\right)\Big|_{w_{1,2}=y_{1,2}t^{\mp \frac{1}{2}}} \,.
\end{align}

Again,  using the result in Appendix \ref{The 4d mirror pair},   we rewrote the interacting part in terms of its $\mathsf S$-dual, the good SQCD with $N_c=1$ and $N_f=4$ with a shifted $c$ fugacity. We then have three 
$SU(2)_{w_1}\times SU(2)_{w_2}$ bifundamental chirals (in pink in the figure) and some extra singlets.

Notice that if we had instead implemented the identification $(1b)$  in Figure \ref{44id}, we would have obtained the same theory but with a $\tilde{\delta}\left(y_1,y_2^{-1}t\right)$ in front of it.
On the other hand, the identifications $(1c)$ and $(1d)$ lead again to the same theory, but with a $\tilde{\delta}\left(y_1,(y_2t)^{\pm1}\right)$ in front of it and with a different definition of the $w_1$ and $w_2$ variables
\begin{align}
	\begin{cases}
		w_1=y_1t^{\frac{1}{2}}\,,\\
		w_2=y_2t^{-\frac{1}{2}}\,.
	\end{cases}
\end{align}

\paragraph{Frame two.}
Finally, we consider the identification $(2)$ in Figure \ref{44id} 
\begin{align}
	\begin{cases}
		u_1=y_2t^{\frac{1}{2}}\,,\\
		u_2=y_2t^{-\frac{1}{2}}\,,\\
		v_1=y_1t^{\frac{1}{2}}\,,\\
		v_2=y_1t^{-\frac{1}{2}}\,.
		\end{cases} 
		\label{eq:Nc=Nf=4_frame3_identif}
\end{align}
In this case  each gauge fugacity is identified  with one of the global fugacities, while there is no delta relating the two the global $y_1$ and $y_2$ fugacities. 
Let us then see the effect of these identifications on the theory.
\begin{itemize}
\item 
The contribution of vector and antisymmetric  chirals in eq.~\eqref{eqdelta4} becomes
\begingroup\allowdisplaybreaks
\begin{align}
	& \cancel{\Delta_4\left( \vec{s}, t \right)}
	\frac{1}{\cancel{\Delta_4\left( \vec{s}, t \right)}}
	\frac{1}{\Delta_4\left( \vec{s}, t \right)} 
	\Delta_2\left( \vec{u} \right) 
	\Delta_2\left( \vec{v} \right) = \nonumber\\[7pt]
	& \qquad \xrightarrow{\,\,(\ref{eq:Nc=Nf=4_frame3_identif})\,\,} 
	\prod_{i=1}^2\prod_{j=1}^2\Gamma_e\left( \left( y_1t^{\frac{3-2i}{2}} \right)^{\pm} \left( y_2t^{\frac{3-2j}{2}} \right)^{\pm} \right) \nonumber\\
	& \qquad\qquad\quad\times
	\Bigg[ \Gamma_e\left( t \right)^4 
	\prod_{i<j}^2\Gamma_e\left( t \left( y_1t^{\frac{3-2i}{2}} \right)^{\pm}\left( y_1t^{\frac{3-2j}{2}} \right)^{\pm}\right) \nonumber\\
	& \qquad\qquad\qquad\quad\times 
	\prod_{i<j}^2\Gamma_e\left( t \left( y_2t^{\frac{3-2i}{2}} \right)^{\pm}\left( y_2t^{\frac{3-2j}{2}} \right)^{\pm}\right) \nonumber\\
	& \qquad\qquad\qquad\quad\times
	\prod_{i=1}^2\prod_{j=1}^2\Gamma_e\left( t \left( y_1t^{\frac{3-2i}{2}} \right)^{\pm} \left( y_2t^{\frac{3-2j}{2}} \right)^{\pm} \right)
	\Bigg]^{-1} \,.
\end{align}
\endgroup
\item
The $USp(4)_{\vec{u}} \times SU(2)_{y_1} $ and  $USp(4)_{\vec{v}}\times SU(2)_{y_2} $ bifundamentals in Figure (\ref{fig:SQCD_Nc=Nf=4_dual_HW_setI}) become
\begin{align}
	\prod_{i=1}^2\Gamma_e\left( t^{\frac{3}{2}} u_i^{\pm} y_1^{\pm} \right) 
	& \xrightarrow{\,\,(\ref{eq:Nc=Nf=4_frame3_identif})\,\,}  
	\Gamma_e\left( t^{\frac{3}{2}} \left(y_2t^{\frac{1}{2}}\right)^{\pm}  y_1^{\pm} \right) 
	\Gamma_e\left( t^{\frac{3}{2}} \left(y_2t^{-\frac{1}{2}}\right)^{\pm} y_1^{\pm} \right) \,,\\[7pt]
	\prod_{i=1}^2\Gamma_e\left( t^{\frac{3}{2}} v_i^{\pm} y_2^{\pm} \right) 
	& \xrightarrow{\,\,(\ref{eq:Nc=Nf=4_frame3_identif})\,\,}  
	\Gamma_e\left( t^{\frac{3}{2}} \left(y_1t^{\frac{1}{2}}\right)^{\pm} y_2^{\pm} \right) 
	\Gamma_e\left( t^{\frac{3}{2}} \left(y_1t^{-\frac{1}{2}}\right)^{\pm} y_2^{\pm} \right) \,.
\end{align}
\item
The $USp(4)_{\vec{u}} \times USp(2)_{z^{(1)}} $ and  $USp(4)_{\vec{v}}\times USp(2)_{z^{ (3)}} $ bifundamentals in Figure (\ref{fig:SQCD_Nc=Nf=4_dual_HW_setI}) become
\begin{align}
	\prod_{i=1}^2\Gamma_e\left( t^{\frac{1}{2}} z^{(1)\pm} u_i^{\pm} \right) 
	& \xrightarrow{\,\,(\ref{eq:Nc=Nf=4_frame3_identif})\,\,}  
	\Gamma_e\left( t^{\frac{1}{2}} z^{(1)\pm} \left(y_2t^{\frac{1}{2}}\right)^{\pm} \right)
	\Gamma_e\left( t^{\frac{1}{2}} z^{(1)\pm} \left(y_2t^{-\frac{1}{2}}\right)^{\pm} \right) \nonumber\\
	& \qquad\quad =
	\Gamma_e\left( t z^{(1)\pm} y_2^{\pm} \right) 
	\Gamma_e\left( z^{(1)\pm} y_2^{\pm} \right)  
	\label{eq:Nc=Nf=4_frame3_pinching_z1}
	\,,\\[7pt]
	\prod_{i=1}^2\Gamma_e\left( t^{\frac{1}{2}} v_i^{\pm} z^{(3)\pm} \right) 
	& \xrightarrow{\,\,(\ref{eq:Nc=Nf=4_frame3_identif})\,\,}  
	\Gamma_e\left( t^{\frac{1}{2}} \left(y_1t^{\frac{1}{2}}\right)^{\pm} z^{(3)\pm} \right)
	\Gamma_e\left( t^{\frac{1}{2}} \left(y_1t^{-\frac{1}{2}}\right)^{\pm} z^{(3)\pm} \right) \nonumber\\
	& \qquad\quad =
	\Gamma_e\left( t z^{(3)\pm} y_1^{\pm} \right) 
	\Gamma_e\left( z^{(3)\pm} y_1^{\pm} \right)  
	\,. \label{eq:Nc=Nf=4_frame3_pinching_z3}
\end{align}
\end{itemize}

By inspecting eqs.~\eqref{eq:Nc=Nf=4_frame3_pinching_z1} and \eqref{eq:Nc=Nf=4_frame3_pinching_z3} we notice that the mesons built with the chirals whose index contributions are
\be
\Gamma_e\left( z^{(1)\pm} y_2^{\pm} \right)\Gamma_e\left( z^{(3)\pm} y_1^{\pm} \right)
\ee
acquire a VEV which Higgses completely the $USp(2)_ {z^{(1)}}$ and
$USp(2)_ {z^{(3)}}$ gauge nodes. Indeed their poles pinch the integration contour at the points
\begin{align}
	z^{(1)} &\xrightarrow{\,\,(\ref{eq:Nc=Nf=4_frame3_pinching_z1})\,\,}  y_2 \,,\label{eq:eq:Nc=Nf=4_frame3_higgsing_z1}\\
	z^{(3)}& \xrightarrow{\,\,(\ref{eq:Nc=Nf=4_frame3_pinching_z3})\,\,}  y_1 \,. \label{eq:eq:Nc=Nf=4_frame3_higgsing_z3}
\end{align}
Following \cite{Gaiotto:2012xa} we should then take the residue at such poles, which eliminates the integrations corresponding to the $USp(2)_ {z^{(1)}}$ and
$USp(2)_ {z^{(3)}}$ gauge nodes that are Higgsed.

After taking the residue and various simplifications to select only the fields that survive the Higgsing, we obtain the theory shown in Figure \ref{fig:SQCD_Nc=Nf=4_ITAdual}. 
This is a WZ model where,  using the result in Appendix \ref{The 4d mirror pair},  we recognize, up to singlets including four
$SU(2)_{y_1}\times SU(2)_{y_2}$ bifundamentals,  the $\mathsf S$-dual of the good SQCD with $N_c=0$ and $N_f=4$ with a shifted $c$ fugacity. Its index is
\begin{align}
  \prod_{j=1}^4\Gamma_e\left(c^{-2} t^{2-j}\right)  
    \prod_{j=1}^4\Gamma_e\left(t^{j-2} y_1^\pm y_2^\pm\right) 
    \mathcal{I}_{\text{SQCD}(0,4)}\left(\vec x;y_2,y_1;t;t^{2}c\right)\,.
\end{align}

\begin{figure}[!ht] 
	\centering
	\includegraphics[width=.5\textwidth]{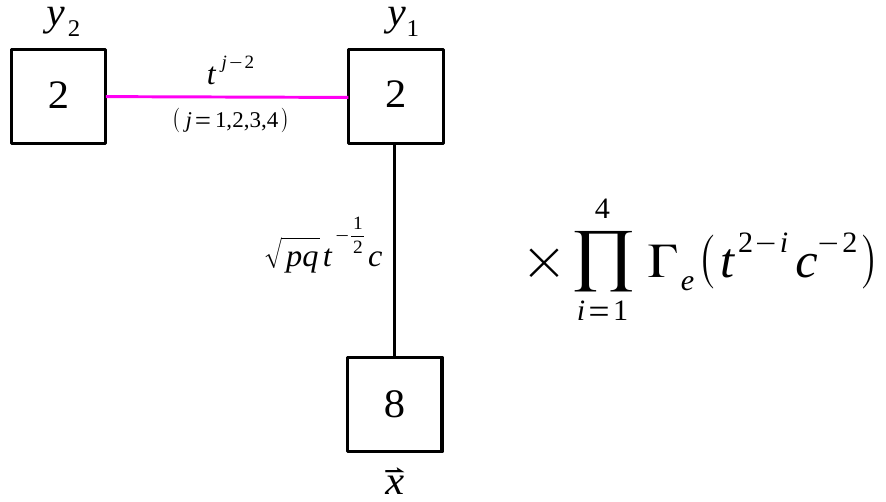}
	\caption{Frame two of the $N_f=N_c=4$ SQCD. We restored all the singlet flipping fields.}
	\label{fig:SQCD_Nc=Nf=4_ITAdual}
\end{figure}

\begin{figure}[!p] 
	\centering
	\includegraphics[width=.7\textwidth]{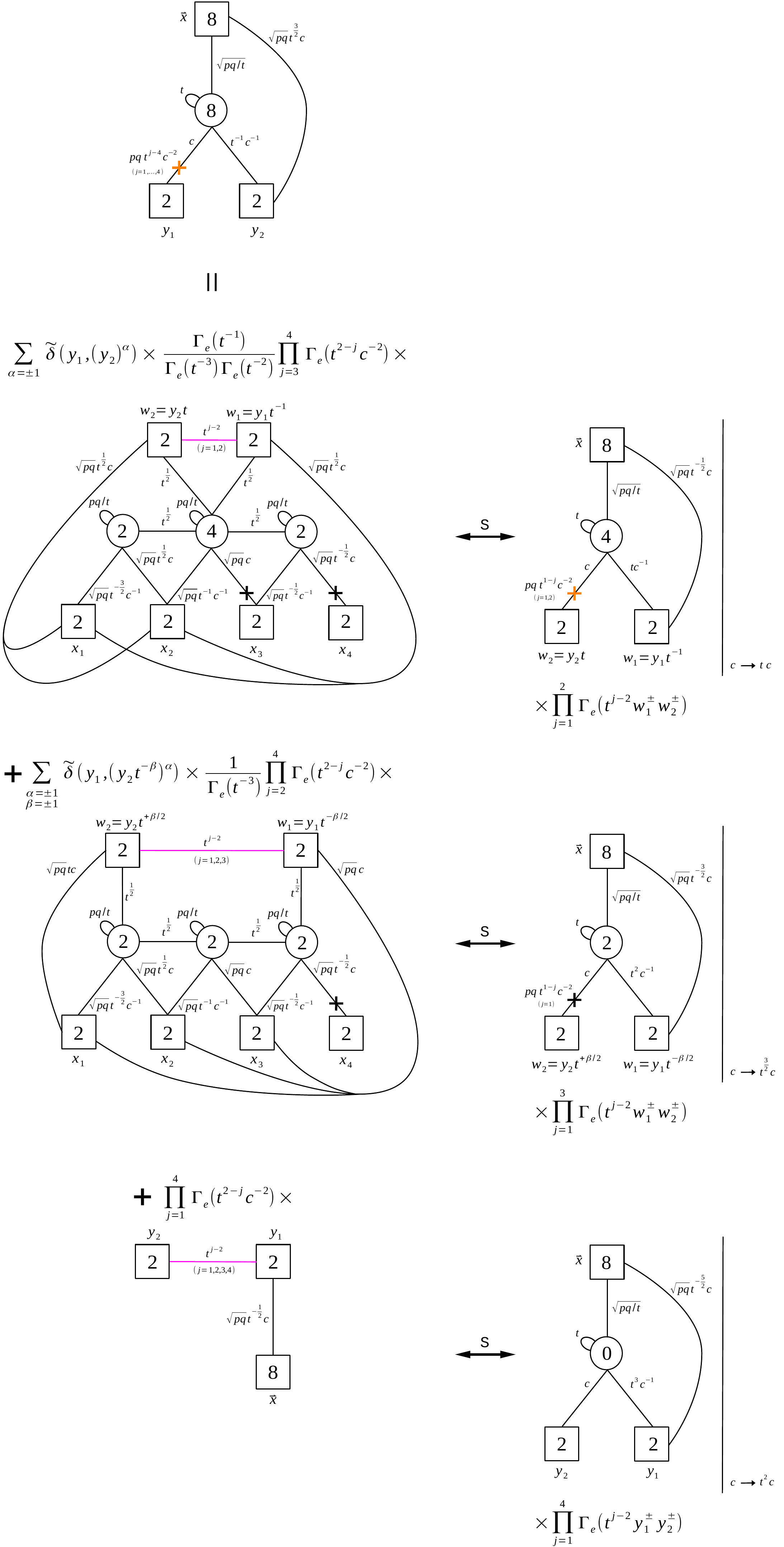}
	\caption{The final result for the $N_c=N_f=4$ SQCD.}
	\label{fig:SQCD_Nc=Nf=4_All_In_One}
\end{figure}

Collecting the results from the frames zero, one and two, we find that the index of the bad $N_c=N_f=4$ SQCD is given by 
\begin{align}
\label{eq:ind44}
    & \mathcal{I}_{\text{SQCD}(4,4)}\left(\vec x;y_1,y_2;t;c\right) = \nonumber\\
    & \, = \quad
    \sum_{\alpha=\pm 1}\Bigg[
    \tilde\delta\left(y_1,(y_2)^{\alpha}\right)
    \frac{\Gamma_e\left(t^{-1}\right)}{\Gamma_e\left(t^{-3}\right)\Gamma_e\left(t^{-2}\right)} 
    \prod_{j=3}^4\Gamma_e\left(c^{-2} t^{2-j}\right) \nonumber\\
    & \qquad\qquad\qquad \times
    \prod_{j=1}^2\Gamma_e\left(t^{j-2} w_1^\pm w_2^\pm\right)    
    \mathcal{I}_{\text{SQCD}(2,4)}\left(\vec x;w_2,w_1;t;tc\right)\Big|_{w_{1,2}=y_{1,2}t^{\mp 1}} \Bigg]
    \nonumber\\
    & \quad\,\, + \sum_{\substack{\alpha=\pm 1 \\ \beta=\pm 1}}\Bigg[
    \tilde\delta\left(y_1,\left(y_2 t^{-\beta}\right)^{\alpha}\right)
    \frac{1}{\Gamma_e\left(t^{-3}\right)} 
    \prod_{j=2}^4\Gamma_e\left(c^{-2} t^{2-j}\right)\nonumber\\
    & \qquad\qquad\qquad\times
    \prod_{j=1}^3\Gamma_e\left(t^{j-2} w_1^\pm w_2^\pm\right)    
    \mathcal{I}_{\text{SQCD}(1,4)}\left(\vec x;w_2,w_1;t;t^{\frac{3}{2}}c\right)\Big|_{w_{1,2}=y_{1,2}t^{\mp \frac{1}{2}\beta}} \Bigg] 
    \nonumber\\
    & \quad\,\, + 
    \prod_{j=1}^4\Gamma_e\left(c^{-2} t^{2-j}\right)  
    \prod_{j=1}^4\Gamma_e\left(t^{j-2} y_1^\pm y_2^\pm\right) 
    \mathcal{I}_{\text{SQCD}(0,4)}\left(\vec x;y_2,y_1;t;t^{2}c\right)
    \,.
\end{align}
This result is summarized in Figure \ref{fig:SQCD_Nc=Nf=4_All_In_One}.

\subsubsection{General result}
\label{subsubsec:genNfeven}

We now give the general result for the case in which $N_f$ is even. As mentioned in the Introduction, we find a sum of terms corresponding to $\Bigl\lfloor\frac{N_f}{2}\Bigr\rfloor-N_f+N_c=N_c-\frac{N_f}{2}$ distinct frames weighted by a delta and one separate frame with no delta. Each frame with the delta actually appears four times, corresponding to all the possible signs of the powers of $y_2$ and $t$ in the relation with $y_1$, except when the power of $t$ is zero in which case we only have two terms corresponding to $y_2$ and $y_2^{-1}$.

We find that every frame corresponds to the good SQCD with gauge group $USp(2r)$ for $r=N_f-N_c,\cdots,\frac{N_f}{2}$, plus a free sector of $4(N_c-r)$ chirals and some extra singlets. If we define $M = N_c-\frac{N_f}{2}-1$ the general formula when $N_c \leq N_f<2N_c$ becomes 
\begingroup\allowdisplaybreaks
\begin{align}\label{eq:indbad4dNfevengNc}
    & 
    \mathcal{I}_{\text{SQCD}\left(N_c,\,N_c \leq N_f < 2N_c\right)}(\vec x;y_1,y_2;t;c) = 
    \nonumber\\[10pt]
    & = 
        \sum_{\substack{\alpha=\pm 1 }} \left\{
    \tilde\delta\left(y_1,\left(y_2\right)^{\alpha}\right)
    \frac{\prod_{j=1}^{M}\Gamma_e\left(t^{-j}\right)}{\prod_{j=0}^{M} \Gamma_e\left(t^{\, j-2M-1}\right)}
    \prod_{j=0}^{M}\Gamma_e\left(c^{-2} t^{\, j-2M}\right)
    \prod_{j=1}^{N_c-\frac{N_f}{2}}\Gamma_e\left(t^{\, j-M-1} w_1^\pm w_2^\pm\right)
    \right. \nonumber\\
    & \qquad\qquad\qquad \times \left.\left.
    \mathcal{I}_{\text{SQCD}\left(\frac{N_f}{2},\,N_f\right)}\left(\vec x;w_2,w_1;t;t^{\frac{N_c-\frac{N_f}{2}}{2}}c\right) \right|_{w_{1,2}=y_{1,2} t^{\mp \frac{1}{2}\left(M+1\right)}} 
    \right\} \nonumber\\[10pt]
&    
+    \sum_{n=1}^{M} 
     \sum_{\substack{\alpha=\pm 1 \\ \beta=\pm 1}} \left\{
    \tilde\delta\left(y_1,\left(y_2 t^{-n\beta}\right)^{\alpha}\right)
    \frac{\prod_{j=1}^{M-n}\Gamma_e\left(t^{-j}\right)}{\prod_{j=0}^{M-n} \Gamma_e\left(t^{\, j-2M-1}\right)}
    \right. \nonumber\\
    & \qquad\qquad\qquad \times \left. 
    \prod_{j=0}^{M+n}\Gamma_e\left(c^{-2} t^{\, j-2M}\right)
    \prod_{j=1}^{N_c-\frac{N_f}{2}+n}\Gamma_e\left(t^{\, j-M-1} w_1^\pm w_2^\pm\right)
    \right. \nonumber\\
    & \qquad\qquad\qquad \times \left.\left.
    \mathcal{I}_{\text{SQCD}\left(\frac{N_f}{2}-n,\,N_f\right)}\left(\vec x;w_2,w_1;t;t^{\frac{N_c-\frac{N_f}{2}+n}{2}}c\right) \right|_{w_{1,2}=y_{1,2} t^{\mp \frac{1}{2}\left(M+1-n\right)\beta}} 
    \right\} \nonumber\\[10pt]
    & + \left\{
    \prod_{j=0}^{2M+1}\Gamma_e\left(c^{-2} t^{\, j-2M}\right) 
    \prod_{j=1}^{2M+2}\Gamma_e\left(t^{\, j-M-1} y_1^\pm y_2^\pm\right) 
    \mathcal{I}_{\text{SQCD}\left(N_f-N_c,\,N_f\right)}\left(\vec x;y_2,y_1;t;t^{\frac{2N_c-N_f}{2}}c\right) \right\} \,,
\end{align}
\endgroup
where the label $n$ in the summation is related to the previous $r$ by $r=\frac{N_f}{2}-n$. Instead when $N_f<N_c$ we find
\begingroup\allowdisplaybreaks
\begin{align}\label{eq:indbad4dNfevenlNc}
    & 
    \mathcal{I}_{\text{SQCD}\left(N_c,\,N_f < N_c\right)}(\vec x;y_1,y_2;t;c) = 
    \nonumber\\[10pt]
    & = \quad 
    \sum_{n=0}^{\frac{N_f}{2}} 
    \sum_{\alpha=\pm 1} \sum_{\substack{\beta = 1 \text{ if } n = 0, \\ \beta = \pm1 \text{ otherwise}}}
    \left\{
    \tilde\delta\left(y_1,\left(y_2 t^{-n\beta}\right)^{\alpha}\right)
    \frac{\prod_{j=1}^{M-n}\Gamma_e\left(t^{-j}\right)}{\prod_{j=0}^{M-n} \Gamma_e\left(t^{j-2M-1}\right)} 
    \right.
    \\
    & \qquad\qquad\qquad \times \left. 
    \prod_{j=0}^{M+n}\Gamma_e\left(c^{-2} t^{j-2M}\right)
    \prod_{j=1}^{M+n+1}\Gamma_e\left(t^{j-M-1} w_1^\pm w_2^\pm\right)
    \right. \nonumber\\
    & \qquad\qquad\qquad \times \left.\left.
    \mathcal{I}_{\text{SQCD}\left(\frac{N_f}{2}-n,\,N_f\right)}\left(\vec x;w_2,w_1;t;t^{\frac{M+n+1}{2}}c\right) 
    \right|_{w_{1,2}=y_{1,2} t^{\mp \frac{1}{2}\left(M-n+1
    \right)\beta}} 
    \right\} \,.\nonumber
\end{align}
\endgroup
Notice that in this case all frames come with a delta distribution. 
In Appendix \ref{app:generalNfeven} we comment on the derivation of this general result.


\subsection{Odd $N_f$}

\subsubsection{Example: $N_c=6$, $N_f=7$}
\label{67sec}
\begin{figure}[!ht]
	\centering
	\makebox[\linewidth][c]{
    \includegraphics[scale=.3]{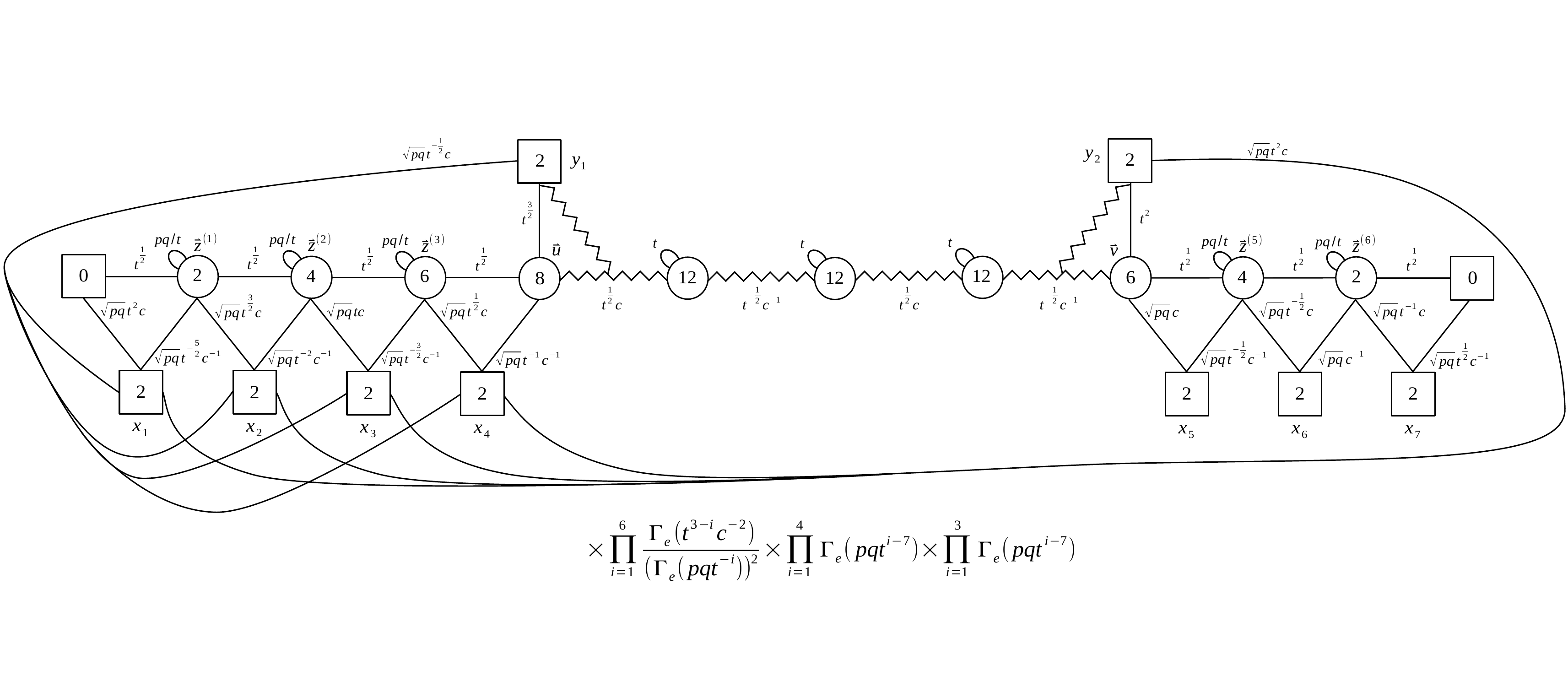}
    }
    \caption{The result of the $\mathsf{S}$-dualization of the $N_c=6$, $N_f=7$ SQCD and of the application of the HW duality moves.}
    \label{fig:SQCD_Nc=6_Nf=7_dual_HW_setI}
\end{figure}

We next discuss the case with odd $N_f$, starting again from a particular example and then considering the general case. Specifically, we consider the $N_c=6$, $N_f=7$ SQCD. Here we will be more brief and give less details, since the analysis is completely analogous to the one we discussed extensively for even $N_f$.
As before, after the $\mathsf S$-dualization we apply the HW duality move until we get two colliding Identity-walls.
We consider applying the HW duality move four times on the left side of the quiver and three times on the right side so to reach the configuration in Figure \ref{fig:SQCD_Nc=6_Nf=7_dual_HW_setI}.

\begin{figure}[!ht] 
	\centering
	\makebox[\linewidth][c]{
	\includegraphics[width=.85\textwidth]{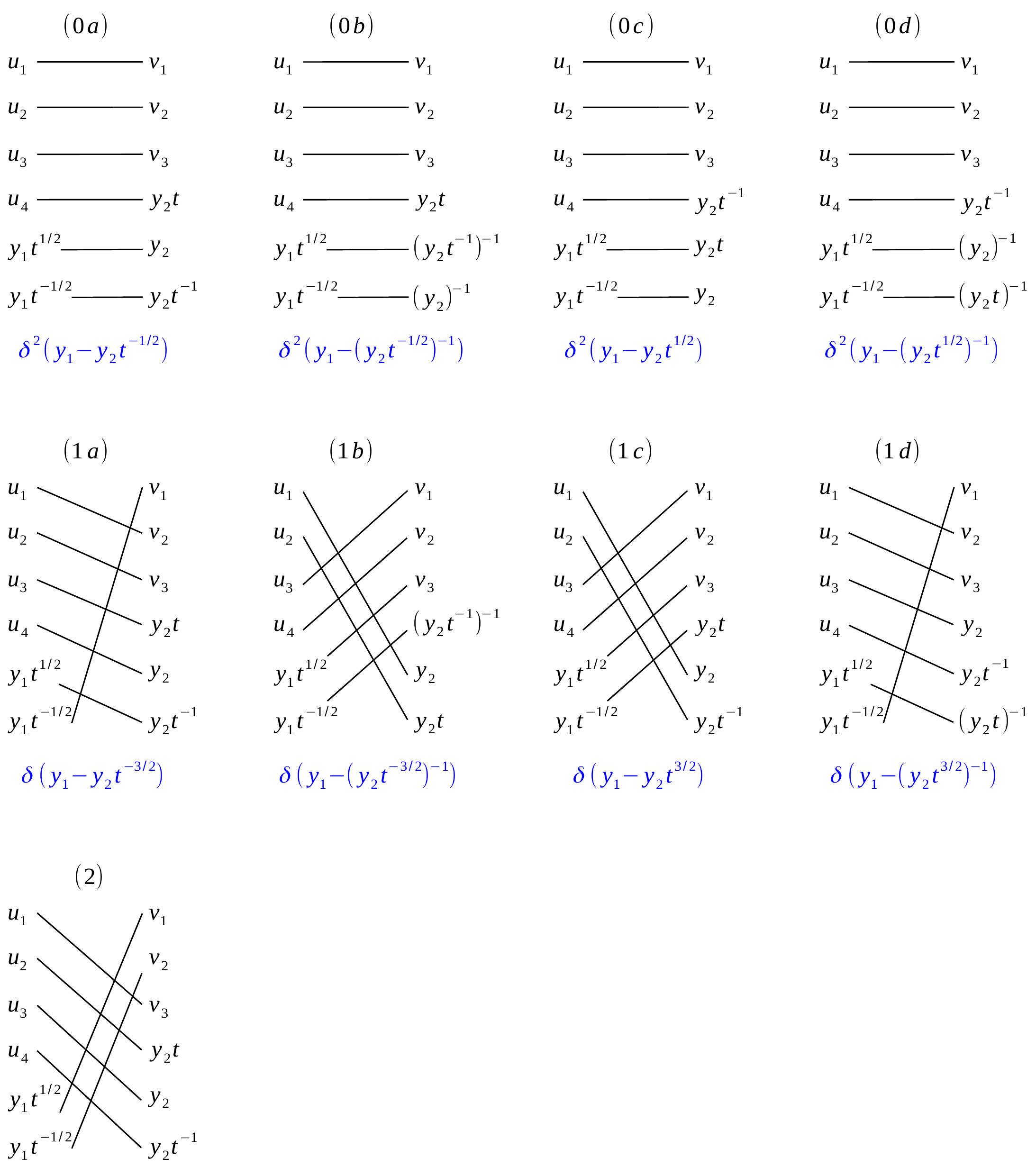}
	}
	\caption{Possible inequivalent identifications of fugacities enforced by the colliding Identity-walls for $N_c=6$, $N_f=7$.}
	\label{67id}
\end{figure}

The two colliding asymmetric Identity-walls force the identification between the following two sets of Cartans:
\begin{align}
	\vec{f}&=\left\{ u_1, u_2, u_3, u_4, y_1t^{\frac{1}{2}}, y_1t^{-\frac{1}{2}} \right\} \,, \nonumber\\
	\vec{h}&=\left\{ v_1, v_2, v_3, y_2t, y_2, y_2t^{-1} \right\} \,.
\end{align}
All the physical identifications, which we remind are those for which $f_i\neq f_j^{\pm1}$ and $f_i\neq 1$, are listed in Figure \ref{67id}, where they are grouped in accordance with the three distinct frames that they lead to, which we shall now study in turn. Again we neglect additional identifications that do not lead to physical frames as discussed in Appendix \ref{moreframe}.

\begin{figure}[!ht] 
	\centering
	\includegraphics[width=\textwidth]{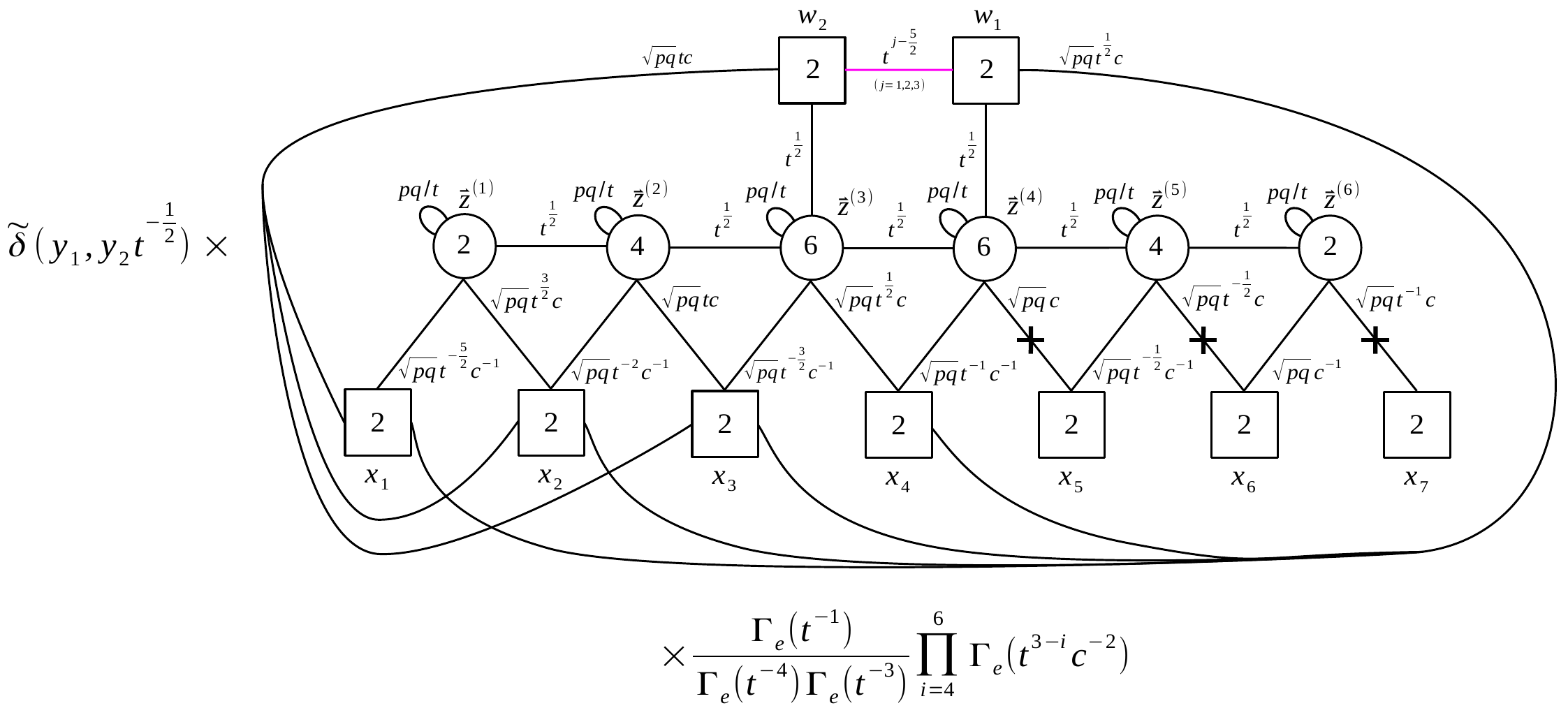}
	\caption{Frame zero of the $N_c=6$, $N_f=7$ SQCD.}
	\label{fig:SQCD_Nc=6_Nf=7_ACdual}
\end{figure}

\paragraph{Frame zero.} Let us consider the identification  $(0a)$ in Figure \ref{67id} which takes us to the frame zero shown in Figure \ref{fig:SQCD_Nc=6_Nf=7_ACdual}, where we defined
\begin{align}
	\begin{cases}
		w_1=y_1t^{-1}\,,\\
		w_2=y_2t\,.
	\end{cases} 
\end{align}
The result we find for the index is
\begin{align}
    &
  \tilde\delta\left(y_1,y_2t^{-\frac{1}{2}}\right)
    \frac{\Gamma_e\left(t^{-1}\right)}{\Gamma_e\left(t^{-4}\right)\Gamma_e\left(t^{-3}\right)} 
    \prod_{j=4}^6\Gamma_e\left(c^{-2} t^{3-j}\right) \nn\\
    &\quad\times
    \prod_{j=1}^3\Gamma_e\left(t^{j-\frac{5}{2}} w_1^\pm w_2^\pm\right)    
    \mathcal{I}_{\text{SQCD}(3,7)}\left(\vec x;w_2,w_1;t;t^{\frac{3}{2}}c\right)\Big|_{w_{1,2}=y_{1,2}t^{\mp 1}} \,.
\end{align}
Above,  using the result in Appendix \ref{The 4d mirror pair},  we rewrote the interacting part in terms of its $\mathsf S$-dual,  the good SQCD with $N_c=3$ and $N_f=7$ with a shifted $c$ fugacity. We then have three
$SU(2)_{w_1}\times SU(2)_{w_2}$ bifundamental chiral (in pink in the figure) and some extra  singlets.

If we had instead implemented the identification $(0b)$ in Figure \ref{67id} we would have obtained the same theory but with a $\tilde{\delta}(y_1,y_2^{-1}t^{\frac{1}{2}})$ in front of it.
On the other hand, implementing the identification $(0c)$ or $(0d)$ in Figure \ref{67id}, we would have obtained the same theory but with a 
$\tilde{\delta}(y_1,(y_2t^{\frac{1}{2}})^{\pm1})$
in front of it and with a different definition of the $w_1$ and $w_2$ variables
\begin{align}
	\begin{cases}
		w_1=y_1t\,,\\
		w_2=y_2t^{-1}\,.
	\end{cases} 
\end{align}

\begin{figure}[!ht] 
	\centering
	\includegraphics[width=\textwidth]{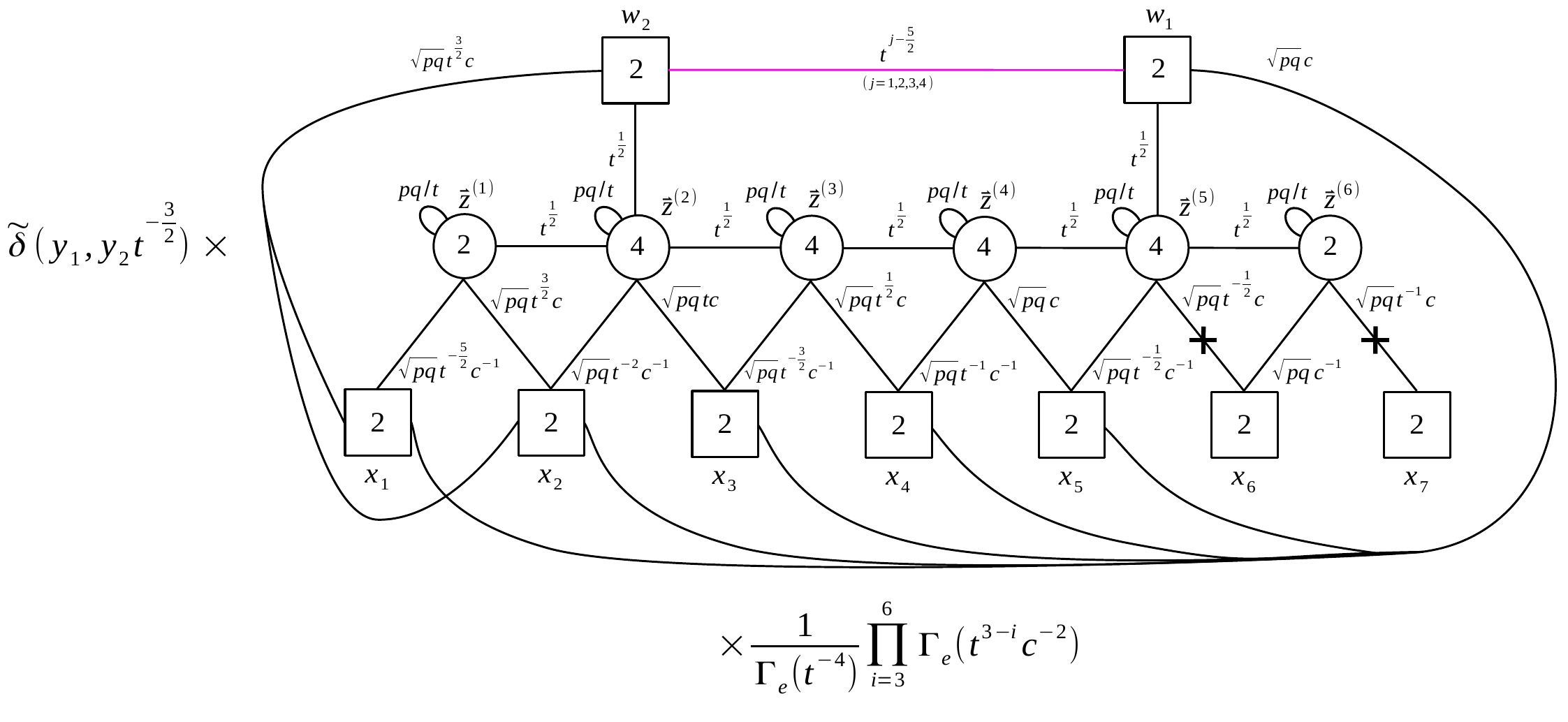}
	\caption{Frame one of the $N_c=6$, $N_f=7$ SQCD.}
	\label{fig:SQCD_Nc=6_Nf=7_INTdual}
\end{figure}

\paragraph{Frame one.} We then consider the identification $(1a)$ in Figure \ref{67id}, which takes us to frame one shown in Figure \ref{fig:SQCD_Nc=6_Nf=7_INTdual}, where we defined
\begin{align}
	\begin{cases}
		w_1=y_1t^{-\frac{1}{2}}\,,\\
		w_2=y_2t^{\frac{1}{2}}\,.
	\end{cases} 
\end{align}
The result we find for the index is
\begin{align}
    &
    \tilde\delta\left(y_1,y_2 t^{-\frac{3}{2}}\right)
    \frac{1}{\Gamma_e\left(t^{-4}\right)} 
    \prod_{j=3}^6\Gamma_e\left(c^{-2} t^{3-j}\right)\nn\\
    &\quad\times
    \prod_{j=1}^4\Gamma_e\left(t^{j-\frac{5}{2}} w_1^\pm w_2^\pm\right)    
    \mathcal{I}_{\text{SQCD}(2,7)}\left(\vec x;w_2,w_1;t;t^{2}c\right)\Big|_{w_{1,2}=y_{1,2}t^{\mp \frac{1}{2}}} \,.
\end{align}
Again,  using the result in Appendix \ref{The 4d mirror pair},  we rewrote the interacting part in terms of its $\mathsf S$-dual, the good SQCD with $N_c=2$ and $N_f=7$ with a shifted $c$ fugacity. We then have four
$SU(2)_{w_1}\times SU(2)_{w_2}$ bifundamental chiral (in pink in the figure) and some extra  singlets.

If we had instead implemented the identification $(1b)$ in Figure \ref{67id} we would have obtained the same theory but with a $\tilde{\delta}(y_1,y_2^{-1}t^{\frac{3}{2}})$ in front of it.
On the other hand, implementing the identification $(1c)$ or $(1d)$ in Figure \ref{67id} we would have obtained the same theory but with a $\tilde{\delta}(y_1,(y_2t^{\frac{3}{2}})^{\pm1})$ in front of it and with a different definition of the $w_1$ and $w_2$ variables
\begin{align}
	\begin{cases}
		w_1=y_1t^{\frac{1}{2}}\,,\\
		w_2=y_2t^{-\frac{1}{2}}\,.
	\end{cases} 
\end{align}

\begin{figure}[!ht] 
	\centering
	\includegraphics[width=\textwidth]{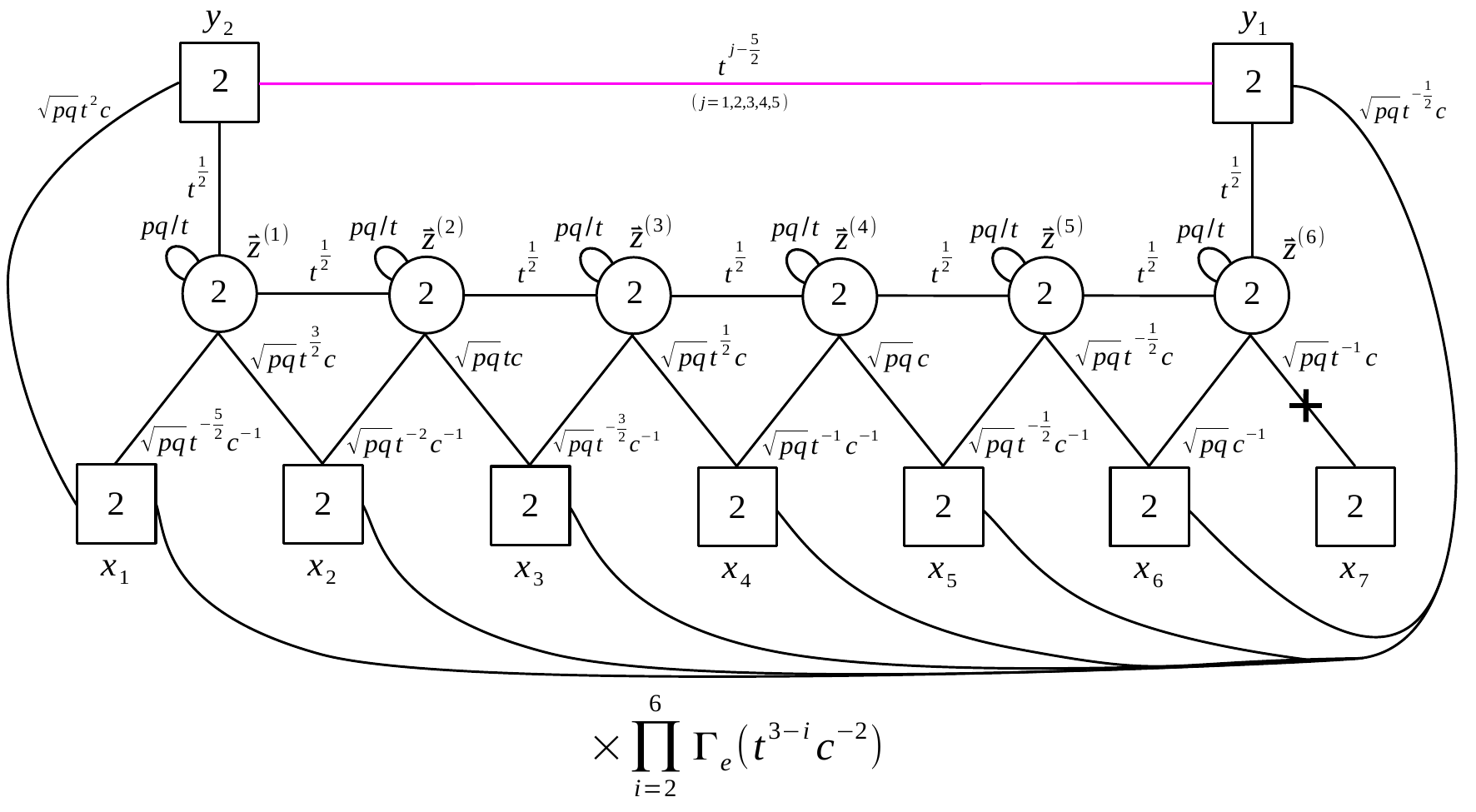}
	\caption{Frame two of the $N_c=6$, $N_f=7$ SQCD.}
	\label{fig:SQCD_Nc=6_Nf=7_ITAdual}
\end{figure}

\paragraph{Frame two.} Finally, let us consider the identification $(2)$ in Figure \ref{67id} leading to frame two in Figure \ref{fig:SQCD_Nc=6_Nf=7_ITAdual}. The index is given by
\begin{align}
    \prod_{j=2}^6\Gamma_e\left(c^{-2} t^{3-j}\right)  
    \prod_{j=1}^5\Gamma_e\left(t^{j-\frac{5}{2}} y_1^\pm y_2^\pm\right) 
    \mathcal{I}_{\text{SQCD}(1,7)}\left(\vec x;y_2,y_1;t;t^{\frac{5}{2}}c\right)\,.
\end{align}
Now,  using the result in Appendix \ref{The 4d mirror pair},  we rewrite  the interacting part as its
  $\mathsf S$-dual, the good SQCD with $N_c=1$ and $N_f=7$ with a shifted $c$ fugacity. We then have five
$SU(2)_{w_1}\times SU(2)_{w_2}$ bifundamental chiral (in pink in the figure) and some extra  singlets.

\begin{figure}[!p] 
	\centering
	\includegraphics[width=.9\textwidth]{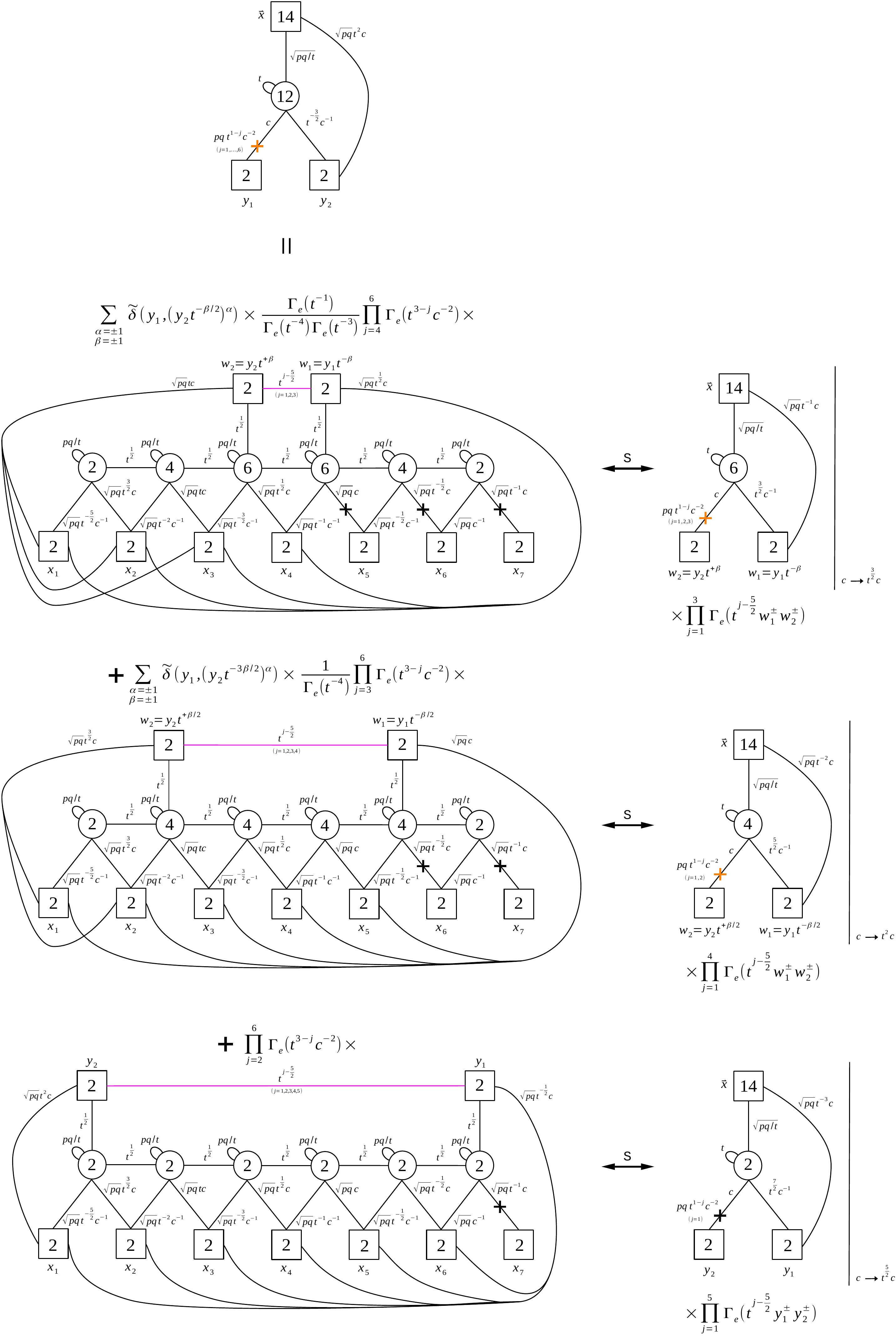}
	\caption{The final result for the $N_c=6$, $N_f=7$ SQCD.}
	\label{fig:SQCD_Nc=6_Nf=7_All_In_One}
\end{figure}

Collecting the results from the frames zero and one, we find that the index of the bad $N_c=6$, $N_f=7$ SQCD is given by
\begin{align}
	& \mathcal{I}_{\text{SQCD}(6,7)}\left(\vec x;y_1,y_2;t;c\right) = \nonumber\\
    & \, = \quad
    \sum_{\substack{\alpha=\pm 1 \\ \beta=\pm 1 }}\Bigg[
    \tilde\delta\left(y_1,\left(y_2t^{-\frac{1}{2}\beta}\right)^{\alpha}\right)
    \frac{\Gamma_e\left(t^{-1}\right)}{\Gamma_e\left(t^{-4}\right)\Gamma_e\left(t^{-3}\right)} 
   	\prod_{j=4}^6\Gamma_e\left(c^{-2} t^{3-j}\right) \nonumber\\
    &\qquad\qquad\qquad \times
    \prod_{j=1}^3\Gamma_e\left(t^{j-\frac{5}{2}} w_1^\pm w_2^\pm\right)    
    \mathcal{I}_{\text{SQCD}(3,7)}\left(\vec x;w_2,w_1;t;t^{\frac{3}{2}}c\right)\Big|_{w_{1,2}=y_{1,2}t^{\mp \beta}} \Bigg]
    \nonumber\\
    & \quad\,\, + \sum_{\substack{\alpha=\pm 1 \\ \beta=\pm 1 }}\Bigg[
    \tilde\delta\left(y_1,\left(y_2 t^{-\frac{3}{2}\beta}\right)^{\alpha}\right)
    \frac{1}{\Gamma_e\left(t^{-4}\right)} 
    \prod_{j=3}^6\Gamma_e\left(c^{-2} t^{3-j}\right)\nonumber\\
    & \qquad\qquad\qquad\times
    \prod_{j=1}^4\Gamma_e\left(t^{j-\frac{5}{2}} w_1^\pm w_2^\pm\right)    
    \mathcal{I}_{\text{SQCD}(2,7)}\left(\vec x;w_2,w_1;t;t^{2}c\right)\Big|_{w_{1,2}=y_{1,2}t^{\mp \frac{1}{2}\beta}} \Bigg]
    \nonumber\\
    & \quad\,\, + 
    \prod_{j=2}^6\Gamma_e\left(c^{-2} t^{3-j}\right)  
    \prod_{j=1}^5\Gamma_e\left(t^{j-\frac{5}{2}} y_1^\pm y_2^\pm\right) 
    \mathcal{I}_{\text{SQCD}(1,7)}\left(\vec x;y_2,y_1;t;t^{\frac{5}{2}}c\right)
    \,.
\end{align}
This result is summarized in Figure \ref{fig:SQCD_Nc=6_Nf=7_All_In_One}.

\subsubsection{General result}
\label{subsubsec:genNfodd}

We now give the general result for the case in which $N_f$ is odd. As mentioned in the Introduction and similarly to the even $N_f$ case, we find a sum of terms corresponding to $\left\lfloor\frac{N_f}{2}\right\rfloor-N_f+N_c$ distinct frames weighted by a delta and one separate frame with no delta. Each frame with the delta actually appears four times, corresponding to all the possible signs of the powers of $y_2$ and $t$ in the relation with $y_1$.\footnote{For odd $N_f$, unlike the case of even $N_f$, there is no delta with a trivial power of $t$.}

We find that every frame corresponds to the good SQCD with gauge group $USp(2r)$ for $r=N_f-N_c,\cdots ,k$, plus a sector of $4(N_c-r)$ bifundamental chirals of $SU(2)_{w_1}\times SU(2)_{w_2}$ and some extra singlets. If we define $M = N_c-\frac{N_f}{2}-1$ the general formula for $N_c \leq N_f < 2N_c$ becomes
\begingroup\allowdisplaybreaks
\begin{align}\label{eq:indbad4dNfoddgNc}
    & 
    \mathcal{I}_{\text{SQCD}\left(N_c,\,N_c \leq N_f < 2N_c\right)}(\vec x;y_1,y_2;t;c) = 
    \nonumber\\[10pt]
    & = \quad 
    \sum_{n=0}^{M-\frac{1}{2}} 
    \sum_{\substack{\alpha=\pm 1 \\ \beta=\pm 1}} \left\{
    \tilde\delta\left(y_1,\left(y_2 t^{-(n+\frac{1}{2})\beta}\right)^{\alpha}\right)
    \frac{\prod_{j=1}^{M-\frac{1}{2}-n}\Gamma_e\left(t^{-j}\right)}{\prod_{j=0}^{M-\frac{1}{2}-n} \Gamma_e\left(t^{\, j-2M-1}\right)}
    \right. \nonumber\\
    & \qquad\qquad\qquad\quad \times 
    \prod_{j=0}^{M+\frac{1}{2}+n}\Gamma_e\left(c^{-2} t^{\, j-2M}\right)
    \prod_{j=1}^{M+\frac{1}{2}+n+1}\Gamma_e\left(t^{\, j-M-1} w_1^\pm w_2^\pm\right)
    \nonumber\\
    & \qquad\qquad\qquad\quad \times \left.\left.
    \mathcal{I}_{\text{SQCD}\left(\frac{N_f-1}{2}-n,\,N_f\right)}\left(\vec x;w_2,w_1;t;t^{\frac{M+\frac{1}{2}+n+1}{2}}c\right) \right|_{w_{1,2}=y_{1,2} t^{\mp \frac{1}{2}\left(M-n+\frac{1}{2}\right)\beta}} 
    \right\} \nonumber\\[10pt]
    & \qquad\quad + \left\{
    \prod_{j=0}^{2M+1}\Gamma_e\left(c^{-2} t^{\, j-2M}\right) 
    \prod_{j=1}^{2M+2}\Gamma_e\left(t^{\, j-M-1} y_1^\pm y_2^\pm\right) 
    \right. \nonumber\\
    & \qquad\qquad\qquad \left.
    \phantom{\frac{\frac{\frac{1}{1}}{1}}{\frac{\frac{1}{1}}{1}}}
    \times
    \mathcal{I}_{\text{SQCD}\left(N_f-N_c,\,N_f\right)}\left(\vec x;y_2,y_1;t;t^{\frac{2N_c-N_f}{2}}c\right) \right\}\,,
\end{align}
\endgroup
while for $N_f<N_c$ we find
\begingroup\allowdisplaybreaks
\begin{align}
    & 
    \mathcal{I}_{\text{SQCD}\left(N_c,\,N_f < N_c\right)}(\vec x;y_1,y_2;t;c) = 
    \nonumber\\[10pt]
    & = \quad 
    \sum_{n=0}^{\frac{N_f-1}{2}} 
    \sum_{\substack{\alpha=\pm1 \\ \beta=\pm1}}
    \left\{
    \tilde\delta\left(y_1,\left(y_2 t^{-\left(n+\frac{1}{2}\right)\beta}\right)^{\alpha}\right)
    \frac{\prod_{j=1}^{M-\frac{1}{2}-n}\Gamma_e\left(t^{-j}\right)}{\prod_{j=0}^{M-\frac{1}{2}-n} \Gamma_e\left(t^{j-2M-1}\right)} 
    \right. \nonumber
    \\
    & \qquad\qquad\qquad \times \left. 
    \prod_{j=0}^{M+\frac{1}{2}+n}\Gamma_e\left(c^{-2} t^{j-2M}\right)
    \prod_{j=1}^{M+n+\frac{3}{2}}\Gamma_e\left(t^{j-M-1} w_1^\pm w_2^\pm\right)
    \right. \nonumber\\
    & \qquad\qquad\qquad \times \left.\left.
    \mathcal{I}_{\text{SQCD}\left(\frac{N_f-1}{2}-n,\,N_f\right)}\left(\vec x;w_2,w_1;t;t^{\frac{M+n+\frac{3}{2}}{2}}c\right) 
    \right|_{w_{1,2}=y_{1,2} t^{\mp \frac{1}{2}\left(M-n+\frac{1}{2}
    \right)\beta}} 
    \right\}\,.
\end{align}
\endgroup
In Appendix \ref{app:generalNfodd} we comment on the derivation of this general result.

\subsection{Generic $N_f$}
We close this section  giving the general result by combining even and odd case.
The general formula for $N_c \leq N_f < 2N_c$ reads as follows:
\begingroup\allowdisplaybreaks
\begin{align}\label{eq:indbad4d}
    & 
    \mathcal{I}_{\text{SQCD}\left(N_c,\,N_c \leq N_f < 2N_c\right)}(\vec x;y_1,y_2;t;c) = 
    \nonumber\\[10pt]
    & = \quad 
    \sum_{n=0}^{M+\epsilon} 
    \sum_{\alpha=\pm 1} \sum_{\substack{\beta = 1 \text{ if } n-\epsilon = 0, \\ \beta = \pm1 \text{ otherwise}}} \left\{
    \tilde\delta\left(y_1,\left(y_2 t^{-(n-\epsilon)\beta}\right)^{\alpha}\right)
    \frac{\prod_{j=1}^{M+\epsilon-n}\Gamma_e\left(t^{-j}\right)}{\prod_{j=0}^{M+\epsilon-n} \Gamma_e\left(t^{\, j-2M-1}\right)} 
    \right. \nonumber\\
    & \qquad\qquad\qquad\quad \times 
    \prod_{j=0}^{M-\epsilon+n}\Gamma_e\left(c^{-2} t^{\, j-2M}\right)
    \prod_{j=1}^{M-\epsilon+n+1}\Gamma_e\left(t^{\, j-M-1} w_1^\pm w_2^\pm\right)
    \nonumber\\
    & \qquad\qquad\qquad\quad \times \left.\left.
    \mathcal{I}_{\text{SQCD}\left(\frac{N_f}{2}+\epsilon-n,\,N_f\right)}\left(\vec x;w_2,w_1;t;t^{\frac{M-\epsilon+n+1}{2}}c\right) \right|_{w_{1,2}=y_{1,2} t^{\mp \frac{1}{2}\left(M+\epsilon-n+1\right)\beta}} 
    \right\} \nonumber\\[10pt]
    & \qquad\quad + \left\{
    \prod_{j=0}^{2M+1}\Gamma_e\left(c^{-2} t^{\, j-2M}\right) 
    \prod_{j=1}^{2M+2}\Gamma_e\left(t^{\, j-M-1} y_1^\pm y_2^\pm\right) 
    \phantom{\frac{\frac{\frac{1}{1}}{1}}{\frac{\frac{1}{1}}{1}}}
    \right. \nonumber\\
    & \qquad\qquad\qquad \left.
    \phantom{\frac{\frac{\frac{1}{1}}{1}}{\frac{\frac{1}{1}}{1}}}
    \times
    \mathcal{I}_{\text{SQCD}\left(N_f-N_c,\,N_f\right)}\left(\vec x;y_2,y_1;t;t^{\frac{2N_c-N_f}{2}}c\right) \right\}\,,
    \end{align}
\endgroup
where we defined
\be
\epsilon=\begin{cases}0&N_f\text{ even}\,,\\-\frac{1}{2}&N_f\text{ odd}\,.\end{cases}
\ee
Note that, for $N_f = 2 N_c-1$, i.e., the ugly case, only the last term with no delta constraint contributes to the index since $M+\epsilon = -1$.
On the other hand, for $N_f<N_c$ we have
\begingroup\allowdisplaybreaks
\begin{align}\label{eq:indbad4dNfoddlNc}
    & 
    \mathcal{I}_{\text{SQCD}\left(N_c,\,N_f < N_c\right)}(\vec x;y_1,y_2;t;c) = 
    \nonumber\\[10pt]
    & = \quad 
    \sum_{n=0}^{\frac{N_f}{2}+\epsilon} 
    \sum_{\alpha=\pm 1} \sum_{\substack{\beta = 1 \text{ if } n-\epsilon = 0, \\ \beta = \pm1 \text{ otherwise}}}
    \left\{
    \tilde\delta\left(y_1,\left(y_2 t^{-\left(n-\epsilon\right)\beta}\right)^{\alpha}\right)
    \frac{\prod_{j=1}^{M+\epsilon-n}\Gamma_e\left(t^{-j}\right)}{\prod_{j=0}^{M+\epsilon-n} \Gamma_e\left(t^{j-2M-1}\right)} 
    \right. \nonumber
    \\
    & \qquad\qquad\qquad \times 
    \prod_{j=0}^{M-\epsilon+n}\Gamma_e\left(c^{-2} t^{j-2M}\right)
    \prod_{j=1}^{M-\epsilon+n+1}\Gamma_e\left(t^{j-M-1} w_1^\pm w_2^\pm\right)
    \nonumber\\
    & \qquad\qquad\qquad \times \left.\left.
    \mathcal{I}_{\text{SQCD}\left(\frac{N_f}{2}+\epsilon-n,\,N_f\right)}\left(\vec x;w_2,w_1;t;t^{\frac{M-\epsilon+n+1}{2}}c\right) 
    \right|_{w_{1,2}=y_{1,2} t^{\mp \frac{1}{2}\left(M+\epsilon-n+1
    \right)\beta}} 
    \right\}\,.
\end{align}
\endgroup

\section{Higgsing of the 4d bad SQCD} 
\label{sec:higgs}

In this section we study more in details the physical implications of the constraints coming from the delta functions that we found in our result of the previous section. We will show how each constraint breaks one combination of the global symmetries of the theory and  interpret this  as due to a specific operator acquiring a non-trivial VEV. This VEV triggers a Higgs mechanism which makes the theory flow to the good SQCD we associate with the delta constraint considered, including all the singlet fields that we found in the previous section.

We will show this from two different but equivalent perspectives. First, we will study the effect of the delta constraints at the level of the index. These will imply a collision of poles that contribute to the integrand of the index, whose effect is precisely to implement the Higgsing as in \cite{Gaiotto:2012xa}. The index perspective is useful to efficiently keep track of all the massless contributions after the Higgsing, so as to precisely match the result of the previous section. We will first make some considerations on the general case and then discuss in details the examples $N_c=N_f=2$ and $N_c=N_f=4$. 

Then we will revisit the same problem  from the perspective of the equations of motions. The analysis is identical in spirit to the one with the index, but it is phrased in a more field theoretic language.

Finally, we discuss how by considering a deformed version of the SQCD where we  flip all the operators acquiring  a  VEV in each frame we can obtain a more conventional duality with a single dual frame.

\subsection{The index perspective}
\label{sec:index higgsing}

We begin with the index perspective, focusing on the SQCD index which we write here again for convenience
\be
&&\mathcal{I}_{\text{SQCD}(N_c,N_f)}(\vec{x};y_1,y_2;t;c)=\prod_{a=1}^{N_c}\Gpq{pq\,t^{1-a}c^{-2}}\prod_{i=1}^{N_f}\Gpq{(pq)^{\frac{1}{2}}t^{M+\frac{1}{2}}c\,x_i^{\pm1}y_2^{\pm1}}\nn\\
&&\qquad\times\oint\udl{\vec{z}_{N_c}}\Gd_{N_c}(\vec{z};t)\prod_{a=1}^{N_c}\Gpq{c\,y_1^{\pm1}z_a^{\pm1}}\Gpq{t^{-M}c^{-1}y_2^{\pm1}z_a^{\pm1}}\prod_{i=1}^{N_f}\Gpq{(pq)^{\frac{1}{2}}t^{-\frac{1}{2}}x_i^{\pm1}z_a^{\pm1}}\,,\nn\\
\ee
where $M=N_c-\frac{N_f}{2}-1\geq 0$ since we are working with a bad theory.

In the previous section we found the following constraints on the fugacities imposed by the delta functions:
\be\label{eq:deltaconstr}
y_1=\left(y_2^{\pm1} t^{\epsilon-n}\right)^{\pm1},\qquad n=0,\cdots,\Big\lfloor\frac{N_f}{2}\Big\rfloor-(N_f-N_c)-1\,,
\ee
where we defined
\be
\epsilon=\begin{cases}0&N_f\text{ even}\,,\\-\frac{1}{2}&N_f\text{ odd}\end{cases}
\ee
and when $\epsilon=n=0$, so that the power of $t$ is zero, we only have the two constraints
\be
y_1=y_2^\pm\,.
\ee

We are now going to argue that the four different constraints that we have for a fixed $n$ correspond to a different component of the meson constructed with the fields of the saw and possibly dressed with the antisymmetric chiral, which is an $SU(2)_{y_1}\times SU(2)_{y_2}$ bifundamental, taking a VEV. In this subsection, in particular, we are interested in studying the behavior of the index when such constraints are imposed, so to understand the effect of the Higgsing triggered by this VEV.
Notice that the two constraints $y_1=y_2 t^{\epsilon-n}$ and $y_1=\left(y_2 t^{\epsilon-n}\right)^{-1}$ are actually equivalent thanks to the invariance $y_1\to y_1^{-1}$ due to the $SU(2)_{y_1}$ symmetry, and similarly for $y_1=y_2^{-1} t^{\epsilon-n}$ and $y_1=\left(y_2^{-1} t^{\epsilon-n}\right)^{-1}$. In the following we then focus on the two constraints $y_1=y_2^{\pm1} t^{\epsilon-n}$. 

Plugging $y_1=y_2 t^{\epsilon-n}$ into the index of the SQCD we get
\be
&&\mathcal{I}_{\text{SQCD}(N_c,N_f)}(\vec{x};t^{\epsilon-n}y_2,y_2;t;c)=\prod_{a=1}^{N_c}\Gpq{pq\,t^{1-a}c^{-2}}\prod_{i=1}^{N_f}\Gpq{(pq)^{\frac{1}{2}}t^{M+\frac{1}{2}}c\,x_i^{\pm1}y_2^{\pm1}}\nn\\
&&\qquad\times\oint\udl{\vec{z}_{N_c}}\Gd_{N_c}(\vec{z})\Gpq{t}^{N_c}\prod_{a<b}^{N_c}\Gpq{t\,z_a^{\pm1}z_b^{\pm1}}\prod_{a=1}^{N_c}\Gpq{t^{\epsilon-n}c\,y_2z_a^{\pm1}}\nn\\
&&\qquad\times\prod_{a=1}^{N_c}\Gpq{t^{n-\epsilon}c\,y_2^{-1}z_a^{\pm1}}\Gpq{t^{-M}c^{-1}y_2^{\pm1}z_a^{\pm1}}\prod_{i=1}^{N_f}\Gpq{(pq)^{\frac{1}{2}}t^{-\frac{1}{2}}x_i^{\pm1}z_a^{\pm1}}\,,\nn\\
\ee
where now we are writing explicitly the contribution of the antisymmetric chiral field. Let us focus on the following combination of gamma functions:
\be\label{eq:gammahiggs}
\Gpq{t^{\epsilon-n}c\,y_2z_{N_c-k+1}^{-1}}\prod_{a=1}^{k-1}\Gpq{t\,z_{N_c-k+a}z_{N_c-k+a+1}^{-1}}\Gpq{t^{-M}c^{-1}y_2^{-1}z_{N_c}}\,,
\ee
where $k=n-\epsilon+M+1$. One could equivalently choose any other combination of the gauge fugacities $z_a$ or their inverses $z_a^{-1}$, but since these are just related by a Weyl reflection the result of the next discussion would be the same.\footnote{Specifically, contributions coming from Weyl equivalent configurations will cancel part of the Weyl symmetry factor in front of the integral, in accordance with how the gauge group has been Higgsed.} These Gamma functions contribute with poles to the integral at the points defined by the following set of conditions:
\be\label{eq:poles}
&&t^{\epsilon-n}c\,y_2z_{N_c-k+1}^{-1}p^{m_1}q^{n_1}=1\,,\nn\\
&&t\,z_{N_c-k+a}z_{N_c-k+a+1}^{-1}p^{m_{a+1}}q^{n_{a+1}}=1\,,\quad a=1,\cdots,k-1\,,\nn\\
&&t^{-M}c^{-1}y_2^{-1}z_{N_c}p^{m_{k+1}}q^{n_{k+1}}=1\,,
\ee
where $m_i,n_i=0,\cdots,\infty$. For each of the integration variables $z_{N_c-k+1},\cdots ,z_{N_c}$ the corresponding integration contour is pinched by a pair of these sets of poles. Consider for example the variable $z_{N_c-k+1}$. One set of poles in this variable comes from the first line of \eqref{eq:poles}
\be\label{eq:pinchpoints}
z_{N_c-k+1}=t^{\epsilon-n}c\,y_2p^{m_1}q^{n_1}\,,
\ee
while the second set of poles can be determined combining the remaining lines of \eqref{eq:poles}
\be 
z_{N_c-k+1}&=&t^{-1}z_{N_c-k+2}p^{-m_2}q^{-n_2}=t^{-2}z_{N_c-k+3}p^{-m_2-m_3}q^{-n_2-n_3}=\cdots\nn\\
&=&t^{1-k}z_{N_c}p^{-\sum_{a=1}^{k-1}m_{a+1}}q^{-\sum_{a=1}^{k-1}n_{a+1}}=t^{M+1-k}c\,y_2p^{-\sum_{a=1}^{k}m_{a+1}}q^{-\sum_{a=1}^{k}n_{a+1}}\nn\\
&=&t^{\epsilon-n}c\,y_2p^{-\sum_{a=1}^{k}m_{a+1}}q^{-\sum_{a=1}^{k}n_{a+1}}\,.
\ee
We then see that the integration contour of $z_{N_c-k+1}$ is pinched at the point $z_{N_c-k+1}=t^{\epsilon-n}c\,y_2$ when all $m_i,n_i=0$. In general, we have that the integration contours of each of the $k$ variables are pinched at the points
\be
z_{N_c-k+a}=t^{\epsilon-n+a-1}c\,y_2\,,\qquad a=1,\cdots,k\,.
\ee

Following \cite{Gaiotto:2012xa}, we should then pick the residue of the index at these poles. This is the manifestation at the level of the index of the fact that the constraint \eqref{eq:deltaconstr} signals that the operator constructed from the chirals in \eqref{eq:gammahiggs}, which is a meson  in the $SU(2)_{y_1}\times SU(2)_{y_2}$ bifundamental possibly dressed with powers of the antisymmetric, induces a Higgs mechanism that breaks the gauge group from $USp(2N_c)$ to $USp(2N_c-2k)$. A similar Higgsing of a dressed meson corresponding to a multi-dimensional pinching of the integration contour of the index has also been recently considered in \cite{Bajeot:2023gyl}.

By looking at \eqref{eq:gammahiggs} it is clear that our analysis applies only if $k\geq 1$. Remembering that we defined $k=M+n-\epsilon-1$, this happens only for bad theories for which $M\geq 0$. In other words, for good theories the Gamma functions in \eqref{eq:gammahiggs} arising after imposing the constraint \eqref{eq:deltaconstr} do not in general lead to any pinching of the integration contour in the index.

If we considered instead the constraint $y_1=y_2^{-1} t^{\epsilon-n}$, all the previous equations would have been identical but with the replacement $y_2\to y_2^{-1}$. Hence, the result of the Higgsing in the two cases is equivalent, but with a different identification of the parameters. 

In the following we consider some examples to show the explicit result of the Higgsing at the level of the index. We will see in particular that the UV fugacities $c$, $y_1$ and $y_2$ will be redefined in the IR, in accordance with our general result from the previous subsection. We discuss the mapping of UV and IR fugacities in the case of generic $N_c$ and $N_f$ in Appendix \ref{app:UVIRfug}.

\subsubsection*{\boldmath $N_c=N_f=2$}

For $N_c=N_f=2$ we have that $M=0$ and the index of this SQCD is given by
\be
\label{eq:indx22}
&&\mathcal{I}_{\text{SQCD}(2,2)}(\vec{x};y_1,y_2;t;c)=\prod_{a=1}^{2}\Gpq{pq\,t^{1-a}c^{-2}}\prod_{i=1}^{2}\Gpq{(pq)^{\frac{1}{2}}t^{\frac{1}{2}}c\,x_i^{\pm1}y_2^{\pm1}}\nn\\
&&\times\oint\udl{\vec{z}_{2}} \Gd_2(\vec z) \Gpq{t}^2\Gpq{t\,z_1^{\pm1}z_2^{\pm1}}\prod_{a=1}^{2}\Gpq{c\,y_1^{\pm1}z_a^{\pm1}}\Gpq{c^{-1}y_2^{\pm1}z_a^{\pm1}}\prod_{i=1}^{2}\Gpq{(pq)^{\frac{1}{2}}t^{-\frac{1}{2}}x_i^{\pm1}z_a^{\pm1}}\,.\nn\\
\ee
In this case $\epsilon=0$ and the only possible value of $n$ is $n=0$, so we have only one constraint
\be
y_1=y_2
\ee
and we want to study the behavior of the index once this is imposed
\be
&&\mathcal{I}_{\text{SQCD}(2,2)}(\vec{x};y_2,y_2;t;c)=\prod_{a=1}^{2}\Gpq{pq\,t^{1-a}c^{-2}}\prod_{i=1}^{2}\Gpq{(pq)^{\frac{1}{2}}t^{\frac{1}{2}}c\,x_i^{\pm1}y_2^{\pm1}}\nn\\
&&\times\oint\udl{\vec{z}_{2}} \Gd_2(\vec z) \Gpq{t}^2\Gpq{t\,z_1^{\pm1}z_2^{\pm1}}\prod_{a=1}^{2}\Gpq{c\,y_2^{\pm1}z_a^{\pm1}}\Gpq{c^{-1}y_2^{\pm1}z_a^{\pm1}}\prod_{i=1}^{2}\Gpq{(pq)^{\frac{1}{2}}t^{-\frac{1}{2}}x_i^{\pm1}z_a^{\pm1}}\,.\nn\\
\ee

In this case the constraint is encoding a VEV for one component of the  $SU(2)_{y_1}\times SU(2)_{y_2}$ meson not dressed with the antisymmetric, and the chiral fields that compose it contribute to the index with the following gamma functions:
\be
\Gpq{c\,y_2z_2^{-1}}\Gpq{c^{-1}y_2^{-1}z_2}\,.
\ee
These provide two sets of poles
\be
z_2&=&c\,y_2\,p^{m_1}q^{n_1}\,,\nn\\
z_2&=&c\,y_2\,p^{-m_2}q^{-n_2}\,,\qquad m_i,n_i\geq 0\,,
\ee
which collide for $m_1=n_1=m_2=n_2=0$ and pinch the integration contour of the variable $z_2$ at\footnote{There are also other combinations of gamma functions that pinch the integration contour and whose residue one should consider. For example, replacing $z_2$ with $z_2^{-1}$ or with $z_1^{\pm1}$ one would obtain exactly the same result after taking the residue. This gives 4 equivalent possibilities, which take into account the difference between the $USp(4)$ Weyl group before the Higgsing and the $USp(2)$ Weyl group after the Higgsing. Moreover, still keeping $z_2$ there are other Gamma functions that contribute with a pole at $z_2=c\,y_2^{-1}$, but it turns out that we can choose an integration contour, or equivalently choose a parameter region, such that only the pole at $z_2 = c y_2$ is enclosed by the contour while the pole at $z_2 = c y_2^{-1}$ is not. Thus, we only consider the residue of the former pole.}
\be
z_2=c\,y_2\,.
\ee
The physical interpretation of this fact is that the VEV for the meson is Higgsing the $USp(4)$ gauge group down to $USp(2)$ and this Higgsing is realized in the index by taking the residue of the integral at this point.

After taking the residue we get
\be
&&\mathcal{I}_{\text{SQCD}(2,2)}(\vec{x};y_2,y_2;t;c)=\Gpq{1}\Gpq{t}\Gpq{c^{\pm2}}\Gpq{y_2^{\pm2}}\prod_{a=1}^{2}\Gpq{pq\,t^{1-a}c^{-2}}\nn\\
&&\times\prod_{i=1}^{2}\Gpq{(pq)^{\frac{1}{2}}t^{\frac{1}{2}}c\,x_i^{\pm1}y_2^{-1}}\Gpq{(pq)^{\frac{1}{2}}t^{-\frac{1}{2}}c\,x_i^{\pm1}y_2}\oint {\frac{1}{2} \frac{\udl{z_{1}}}{2 \pi i z_1}} \Gd_1(z)\Gpq{z_1^{\pm1}(c\,y_2)^{-1}}^{-1}\nn\\
&&\times\Gpq{t}\Gpq{t\,z_1^{\pm1}(c\,y_2)^{\pm1}}\Gpq{c\,y_2^{-1}z_1^{\pm1}}\Gpq{c^{-1}y_2^{\pm1}z_1^{\pm1}}\prod_{i=1}^{2}\Gpq{(pq)^{\frac{1}{2}}t^{-\frac{1}{2}}x_i^{\pm1}z_1^{\pm1}}\,.\nn\\
\ee
Introducing the variables $w_1=y_1t^{-\frac{1}{2}}=y_2t^{-\frac{1}{2}}$ and $w_2=y_2t^{\frac{1}{2}}$ and simplifying some contributions, we can rewrite this result as
\be
\mathcal{I}_{(2,2)}(\vec{x};y_2,y_2;t;c)&=&\Gpq{1}\Gpq{t^{-1}}^{-1}\Gpq{c^{-2}}\Gpq{w_1^{\pm1}w_2^{\pm1}}\left.\mathcal{I}_{(1,2)}\left(\vec{x};w_2,w_1;t;t^{\frac{1}{2}}c\right)\right|_{w_{1,2}=y_{1,2} t^{\mp \frac{1}{2}}} \,,\nn\\
\ee
which, up to the divergent factor $\Gpq{1}$,  matches perfectly with the coefficient of the delta function $\tilde\gd(y_1;y_2)$ that we have in \eqref{eq:ind22-0}. We point out that the parameters $w_1$, $w_2$ in the final good SQCD are different from the analogous $y_1$, $y_2$ in the original bad SQCD and also $c$ got redefined to $c'=ct^{\frac{1}{2}}$, again in accordance with \eqref{eq:ind22-0}. We discuss the general mapping between the UV fugacities $c$, $y_1$ and $y_2$ and the IR fugacities $c'$, $w_1$, $w_2$ for arbitrary $N_c$ and $N_f$ in Appendix \ref{app:UVIRfug}.

Finally we would like to interpret the divergent factor of $\Gpq{1}$ in terms of the delta distribution. To see this, let us go back to \eqref{eq:indx22} and relabel $y_1$ and $y_2$ as
\begin{align}
y_1 = y s^{-1}\,, \qquad y_2 = y s \,,
\end{align}
respectively, to explore the  limit $y_1 \rightarrow y_2$ corresponding to  $s \rightarrow 1$. In particular, we rewrite the following factors in terms of $y$ and $s$:
\begin{align}
\label{eq:pinching factors}
\Gpq{c\,y_1 z_2^{\pm1}}\Gpq{c^{-1}y_2^{-1}z_2^{\pm1}} = \Gpq{s^{-2}} \Gpq{pq s^2} \Gpq{s^{-1} (c\,y)^\pm z_2^{\pm1}} \,,
\end{align}
where we have inserted $1 = \Gpq{s^{-2}} \Gpq{pq s^2}$ on the right hand side.
Now the last two factors in the $s \rightarrow 1$ limit give (see e.g.~\cite{Spiridonov:2014cxa,Bottini:2021vms})
\begin{align}
\lim_{s \rightarrow 1}  \Gpq{pq s^2} \Gpq{s^{-1} (c\,y)^\pm z_2^{\pm1}} =\frac{ 2\pi i z_2 \Gpq{ z_2^{\pm2}}}{(p;p)_\infty (q;q)_\infty} \left[\delta(z_2-y c)+\delta(z_2-(y c)^{-1})\right]\,.
\end{align}
These delta distributions restrict $z_2$ to be $(yc)^\pm$ once integrated, which is exactly the Higgsing effect explained above.
Now we are left with the first factor $\Gpq{s^2} = \Gpq{y_2/y_1}$ which diverges in the limit $s \rightarrow 1$.
We can actually argue that as a distribution $\lim_{y_2\to y_1}  \Gpq{y_2/y_1} \sim \delta (y_2-y_1)$.\footnote{For example we can consider the sequence $f_k(X)=\Gpq{ e^{-1/k}}$ for $\frac{1}{k} \leq X\leq \frac{2}{k}$ and zero otherwise, which is $\delta(X)$ in the limit $k\rightarrow \infty$ because 
$$
\lim_{k\to \infty} \int_{-\infty}^\infty f_k(X) \Phi(X) dX=\lim_{k\to \infty} \Gpq{ e^{-1/k}} \frac{1}{k} \Phi(X_k) = \Phi(0)
$$
for some $X_k$ in the range $\frac{1}{k} \leq X_k\leq \frac{2}{k}$, going to zero when $k \rightarrow \infty$
since for small $X$ we have $\Gamma(e^{-X})\sim \frac{1}{(p;p)_\infty (q;q)_\infty X}$
}
Hence, from now on when we write a $\Gpq{1}$ this should be understood as the delta function evaluated at the origin due to the constraint imposed on the fugacities $y_1$ and $y_2$.

\subsubsection*{\boldmath $N_c=N_f=4$}

Let us consider now an example in which the operator getting a VEV is a dressed meson. For $N_c=N_f=4$ we have that $M=1$ and the index of this SQCD is given by
\be
&&\mathcal{I}_{\text{SQCD}(4,4)}(\vec{x};y_1,y_2;t;c)=\nn\\
&&=\prod_{a=1}^{4}\Gpq{pq\,t^{1-a}c^{-2}}\prod_{i=1}^{4}\Gpq{(pq)^{\frac{1}{2}}t^{\frac{1}{2}}c\,x_i^{\pm1}y_2^{\pm1}}\oint\udl{\vec{z}_{4}} \Gd_4(\vec z) \Gpq{t}^4\prod_{a<b}^4\Gpq{t\,z_a^{\pm1}z_b^{\pm1}}\nn\\
&&\times\prod_{a=1}^{4}\Gpq{c\,y_1^{\pm1}z_a^{\pm1}}\Gpq{t^{-1}c^{-1}y_2^{\pm1}z_a^{\pm1}}\prod_{i=1}^{4}\Gpq{(pq)^{\frac{1}{2}}t^{-\frac{1}{2}}x_i^{\pm1}z_a^{\pm1}}\,.\nn\\
\ee
Again $\epsilon=0$ since the number of flavors is even, but now we can have two possible values of $n$
\be
y_1=y_2t^{-n}\,,\qquad n=0,1\,.
\ee

We will consider the case $n=0$ first, for which the constraint is
\be
y_1=y_2
\ee
and so the index becomes
\be
&&\mathcal{I}_{\text{SQCD}(4,4)}(\vec{x};y_2,y_2;t;c)=\nn\\
&&=\prod_{a=1}^{4}\Gpq{pq\,t^{1-a}c^{-2}}\prod_{i=1}^{4}\Gpq{(pq)^{\frac{1}{2}}t^{\frac{1}{2}}c\,x_i^{\pm1}y_2^{\pm1}}\oint\udl{\vec{z}_{4}} \Gd_4(\vec z) \Gpq{t}^4\prod_{a<b}^4\Gpq{t\,z_a^{\pm1}z_b^{\pm1}}\nn\\
&&\times\prod_{a=1}^{4}\Gpq{c\,y_2^{\pm1}z_a^{\pm1}}\Gpq{t^{-1}c^{-1}y_2^{\pm1}z_a^{\pm1}}\prod_{i=1}^{4}\Gpq{(pq)^{\frac{1}{2}}t^{-\frac{1}{2}}x_i^{\pm1}z_a^{\pm1}}\,.\nn\\
\ee
Let us focus on the following combination of gamma functions:
\be
\Gpq{c\,y_2z_3^{-1}}\Gpq{t\,z_3z_4^{-1}}\Gpq{t^{-1}c^{-1}y_2^{-1}z_4}\,.
\ee
These provide three sets of poles
\be\label{eq:poles44k0}
&&z_3=c\,y_2p^{m_1}q^{n_1}\,,\nn\\
&&z_4z_3^{-1}=t\,p^{m_2}q^{n_2}\,,\nn\\
&&z_4=t\,c\,y_2p^{-m_3}q^{-n_3}\,,\qquad m_i,n_i\geq 0\,,
\ee
which pinch the integration contours of the variables $z_3$ and $z_4$ at the points
\be\label{eq:pinch44k0}
z_3=c\,y_2\,,\qquad z_4=t\,z_3=t\,c\,y_2\,.
\ee
This fact indicates that the component of the    $SU(2)_{y_1}\times SU(2)_{y_2}$ meson dressed with the antisymmetric once,
contributing to the index with the three gamma functions above, is taking a VEV which Higgses the $USp(8)$ gauge group down to $USp(4)$.

In order to see the simultaneous pinching of two of the integration variables, it is useful to restrict the fugacities to be such that $\mathrm{Arg}(p)=\mathrm{Arg}(q)=\mathrm{Arg}(t)=\mathrm{Arg}(c)=\mathrm{Arg}(y_2)=0$, so that the poles lie on $\mathrm{Arg}(z_3)=\mathrm{Arg}(z_4)=0$ and we can then only focus on $|z_3|$ and $|z_4|$. Then, if we first set $|z_3|=c\,y_2$ we can see that the last two sets of poles in \eqref{eq:poles44k0}
\be
&&|z_4|=t\,c\,y_2p^{m_2}q^{n_2}\,,\nn\\
&&|z_4|=t\,c\,y_2p^{-m_3}q^{-n_3}\,,\qquad m_i,n_i\geq 0\,,
\ee
pinch the integration contour at $|z_4|=t\,c\,y_2$ similarly to the more conventional previously analyzed case. Instead, for other values of $|z_3|$ there is no pinching. Similarly, if we set $|z_4|=t\,c\,y_2$ and look at the first two sets of poles in \eqref{eq:poles44k0}
\be
&&|z_3|=c\,y_2p^{m_1}q^{n_1}\,,\nn\\
&&|z_3|=c\,y_2p^{-m_2}q^{-n_2}\,,\qquad m_i,n_i\geq 0\,,
\ee
we can see that these pinch the integration contour at $|z_3|=c\,y_2$, while for other values of $|z_4|$ there is no pinching. This shows that the three sets of poles \eqref{eq:poles44k0} simultaneously pinch the integration contours of $z_3$ and $z_4$ at the points \eqref{eq:pinch44k0}.

We then take the residues of the integral of the index at these points. After some simplifications and introducing the new variables $w_1=y_1t^{-1}=y_2t^{-1}$ and $w_2=y_2t$ we can rewrite the result as
\be
\mathcal{I}_{\text{SQCD}(4,4)}(\vec{x};y_2,y_2;t;c)&=&\Gpq{1}\frac{\Gpq{t^{-1}}\Gpq{t^{-1}c^{-2}}\Gpq{t^{-2}c^{-2}}}{\Gpq{t^{-2}}\Gpq{t^{-3}}}\Gpq{t^{-1}w_1^{\pm1}w_2^{\pm1}}\Gpq{w_1^{\pm1}w_2^{\pm1}}\nn\\
&\times&\left.\mathcal{I}_{(2,4)}\left(\vec{x};w_2,w_1;t;t\,c\right)\right|_{w_{1,2}=y_{1,2} t^{\mp 1}} \,,\nn\\
\ee
as expected from \eqref{eq:ind44-0}, where as before we should think of $\Gpq{1}$ as $\lim_{y_1\to y_2} \Gpq{y_2/y_1} \sim \delta (y_2-y_1)$.

We next consider the case $n=1$, where the constraint is
\be
y_1=y_2t^{-1}
\ee
and so the index becomes
\be
&&\mathcal{I}_{\text{SQCD}(4,4)}(\vec{x};y_2t^{-1},y_2;t;c)=\nn\\
&&=\prod_{a=1}^{4}\Gpq{pq\,t^{1-a}c^{-2}}\prod_{i=1}^{4}\Gpq{(pq)^{\frac{1}{2}}t^{\frac{1}{2}}c\,x_i^{\pm1}y_2^{\pm1}}\oint\udl{\vec{z}_{4}} \Gd_4(\vec z) \Gpq{t}^4\prod_{a<b}^4\Gpq{t\,z_a^{\pm1}z_b^{\pm1}}\nn\\
&&\times\prod_{a=1}^{4}\Gpq{c(t^{-1}y_2)^{\pm1}z_a^{\pm1}}\Gpq{t^{-1}c^{-1}y_2^{\pm1}z_a^{\pm1}}\prod_{i=1}^{4}\Gpq{(pq)^{\frac{1}{2}}t^{-\frac{1}{2}}x_i^{\pm1}z_a^{\pm1}}\,.\nn\\
\ee
Here we should focus on the following gamma functions:
\be
\Gpq{t^{-1}c\,y_2z_2^{-1}}\Gpq{t\,z_2z_3^{-1}}\Gpq{t\,z_3z_4^{-1}}\Gpq{t^{-1}c^{-1}y_2^{-1}z_4}\,,
\ee
which provide four sets of poles
\be
&&z_2=t^{-1}c\,y_2p^{m_1}q^{n_1}\,,\nn\\
&&z_3z_2^{-1}=t\,p^{m_2}q^{n_2}\,,\nn\\
&&z_4z_3^{-1}=t\,p^{m_3}q^{n_3}\,,\nn\\
&&z_4=t\,c\,y_2p^{-m_4}q^{-n_4}\,,\qquad m_i,n_i\geq 0\,.
\ee
By generalizing the reasoning we did for $n=0$, we can see that these pinch the integration contours of the variables $z_2$, $z_3$ and $z_4$ at the points
\be
&&z_2=t^{-1}c\,y_2\,,\nn\\
&&z_3=t\,z_2=c\,y_2\,,\nn\\
&&z_4=t\,z_3=t\,c\,y_2\,.
\ee
The physical interpretation in this case is that the    $SU(2)_{y_1}\times SU(2)_{y_2}$ meson dressed twice with the antisymmetric is taking a VEV, which Higgses the $USp(8)$ gauge group down to $USp(2)$. To implement this Higgsing in the index we take again the residue of the integral at these points. After some simplifications and introducing the new variables $w_1=y_1t^{-\frac{1}{2}}=y_2t^{-\frac{3}{2}}$ and $w_2=y_2t^{\frac{1}{2}}$ we can rewrite the result as
\be
\mathcal{I}_{\text{SQCD}(4,4)}(\vec{x};y_2 t^{-1},y_2;t;c)&=&\Gpq{1}\frac{\prod_{j=2}^4\Gpq{c^{-2}t^{2-j}}}{\Gpq{t^{-3}}}\Gpq{t^{-1}w_1^{\pm1}w_2^{\pm1}}\Gpq{w_1^{\pm1}w_2^{\pm1}}\Gpq{t\,w_1^{\pm1}w_2^{\pm1}}\nn\\
&\times&\left.\mathcal{I}_{(1,4)}\left(\vec{x};w_2,w_1;t;t^{\frac{3}{2}}c\right)\right|_{w_{1,2}=y_{1,2} t^{\mp \frac{1}{2}}} \,,\nn\\
\ee
as expected from \eqref{eq:ind44-1}, where again we think of $\Gpq{1}$ as $\lim_{y_1\to y_2 t^{-1}} \Gpq{y_2 t^{-1}/y_1} \sim \delta (y_2 t^{-1}-y_1)$.

Finally, again for $n=1$ we also consider the other constraint
\be
y_1=y_2^{-1}t^{-1}\,,
\ee
for which the index becomes
\be
&&\mathcal{I}_{\text{SQCD}(4,4)}(\vec{x};y_2^{-1}t^{-1},y_2;t;c)=\nn\\
&&=\prod_{a=1}^{4}\Gpq{pq\,t^{1-a}c^{-2}}\prod_{i=1}^{4}\Gpq{(pq)^{\frac{1}{2}}t^{\frac{1}{2}}c\,x_i^{\pm1}y_2^{\pm1}}\oint\udl{\vec{z}_{4}} \Gd_4(\vec z) \Gpq{t}^4\prod_{a<b}^4\Gpq{t\,z_a^{\pm1}z_b^{\pm1}}\nn\\
&&\times\prod_{a=1}^{4}\Gpq{c(t^{-1}y_2^{-1})^{\pm1}z_a^{\pm1}}\Gpq{t^{-1}c^{-1}y_2^{\pm1}z_a^{\pm1}}\prod_{i=1}^{4}\Gpq{(pq)^{\frac{1}{2}}t^{-\frac{1}{2}}x_i^{\pm1}z_a^{\pm1}}\,.\nn\\
\ee
Here we should focus on the following gamma functions:
\be
\Gpq{t^{-1}c\,y_2^{-1}z_2^{-1}}\Gpq{t\,z_2z_3^{-1}}\Gpq{t\,z_3z_4^{-1}}\Gpq{t^{-1}c^{-1}y_2z_4}\,,
\ee
which provide four sets of poles
\be
&&z_2=t^{-1}c\,y_2^{-1}p^{m_1}q^{n_1}\,,\nn\\
&&z_3z_2^{-1}=t\,p^{m_2}q^{n_2}\,,\nn\\
&&z_4z_3^{-1}=t\,p^{m_3}q^{n_3}\,,\nn\\
&&z_4=t\,c\,y_2^{-1}p^{-m_4}q^{-n_4}\,,\qquad m_i,n_i\geq 0\,.
\ee
Again these pinch the integration contours of the variables $z_2$, $z_3$ and $z_4$ at the points
\be
&&z_2=t^{-1}c\,y_2^{-1}\,,\nn\\
&&z_3=t\,z_2=c\,y_2^{-1}\,,\nn\\
&&z_4=t\,z_3=t\,c\,y_2^{-1}\,.
\ee

In this case the operator that is taking a VEV corresponds to another component of the    $SU(2)_{y_1}\times SU(2)_{y_2}$ meson dressed twice with the antisymmetric,  but the effect of the VEV is the same, namely the $USp(8)$ gauge group is Higgsed down to $USp(2)$. The difference is in the identifications of the parameters. Specifically, in this case we introduce the new variables $w_1=y_1t^{\frac{1}{2}}=y_2^{-1}t^{-\frac{1}{2}}$  and $w_2=y_2t^{-\frac{1}{2}}$, so that after some simplifications we can rewrite the result as 
\be
\mathcal{I}_{(4,4)}(\vec{x};y_2^{-1} t^{-1},y_2;t;c)&=&\Gpq{1}\frac{\prod_{j=2}^4\Gpq{c^{-2}t^{2-j}}}{\Gpq{t^{-3}}}\Gpq{t^{-1}w_1^{\pm1}w_2^{\pm1}}\Gpq{w_1^{\pm1}w_2^{\pm1}}\Gpq{t\,w_1^{\pm1}w_2^{\pm1}}\nn\\
&\times&\left.\mathcal{I}_{(1,4)}\left(\vec{x};w_2,w_1;t;t^{\frac{3}{2}}c\right)\right|_{w_{1,2}=y_{1,2} t^{\pm \frac{1}{2} }} \,.\nn\\
\ee

This again matches nicely with the second term in \eqref{eq:ind44} with $(\alpha,\beta) = (-1,-1)$, where $\Gpq{1}$ is interpreted as $\lim_{y_1\to y_2^{-1} t^{-1}}  \Gpq{y_2^{-1} t^{-1}/y_1} \sim \delta (y_2^{-1} t^{-1}-y_1)$.

\subsection{The classical equations of motion perspective}

Let us now analyze the Higgsing in the 4d SQCD theory from the perspective of the equations of motion. The discussion is conceptually equivalent to the one with the index that we discussed in the previous subsection, just rephrased in a different, possibly more familiar language. Moreover, we will be more general by keeping $N_c$ and $N_f$ arbitrary and we will identify the field components which remain massless after the Higgsing.

We label the $USp(2N_c)$ antisymmetric chiral by $A$ and the $2N_f$ flavors by $Q$ while the $USp(2N_c)$ fundamentals charged under $SU(2)_{y_{1}}$ and $SU(2)_{y_{2}}$ will be denoted as $P_L$ and $P_R$ respectively.
The superpotential of the theory reads 
\be\label{suppot}
\mathcal{W} = AQQ + QP_RS + \sum_{i=0}^{N_c}\beta_i A^iP_LP_L\,,
\ee
where we contract indices using the symplectic matrix $J_{ij}$, $S$ is a gauge singlet in the bifundamental of $USp(2N_f)\times SU(2)_{y_2}$ 
and $\beta_i$ are the singlets represented by the orange cross in Figure \ref{SQCDzero}.

Introducing as before the parameters $M=N_c-\frac{N_f}{2}-1$ for (minus) the $U(1)_t$ charge of $P_R$ and $n$ labelling the various frames (the rank of the gauge group is $\floor*{\frac{N_f}{2}}-n$), we will now see that as expected the constraint $y_1=y_2t^{\epsilon-n}$ can be implemented by giving a non-vanishing expectation value to the  $SU(2)_{y_{1}}\times SU(2)_{y_{2}}$ meson, dressed by the antisymmetric $A$ $M+n-\epsilon$ times. This corresponds to a dressed monopole operator in the three-dimensional $U(N_c)$ SQCD theory we will discuss in Section \ref{sec:3danalysis}. 

\subsubsection{Expectation value for the elementary fields}

We will turn on an expectation value for $A$, $P_{L,1}$ and $P_{R,2}$ only, where the subscripts $1,2$ denote the $SU(2)_{y_i}$ components, and therefore from \eqref{suppot} we easily see that F-terms do not provide any constraints. This corresponds to the fact that the meson dressed with the antisymmetric $M+n-\epsilon$ times is the only gauge invariant operator with a non-vanishing VEV and in the infrared we recover the SQCD theory with gauge group $USp\left(2\floor*{\frac{N_f}{2}}-2n \right)$ plus a collection of free chiral multiplets. 

In order to simplify the formulas in the rest of this section, it is convenient to introduce the parameter $k\equiv M+n+1-\epsilon$ and
choose a basis in which the symplectic form $J$ has the block-diagonal form 
\be\label{simplectic} J=\left(\begin{array}{c|c}
J_{N_c-k} & 0\\
\hline
0 & J_k \\
\end{array}\right) \,,
\ee 
where 
\be J_m=i\sigma_2\otimes I_m=\left(\begin{array}{c|c}
0 & I_m\\
\hline
-I_m & 0 \\
\end{array}\right)\,. 
\ee 
In order to reproduce the results of the index analysis presented above we make the following choice for the VEV of the fields:
\be\label{vevsol}
A=\left(\begin{array}{c|c}
0_{2N_c-2k} & 0\\
\hline
0 & \begin{array}{c|c} 0 &\mathbb{J}_k\\
\hline
-\mathbb{J}_k^T& 0\\\end{array}\\
\end{array}\right)\,,\quad P_{L,1}=e_{2N_c-k}\,,\quad P_{R,2}=e_{2N_c-k+1}\,,
\ee
where $0_{2N_c-2k}$ denotes the $(2N_c-2k)\times (2N_c-2k)$ trivial matrix, $e_i$ is the vector whose $i$-th entry is equal to one and the other components are vanishing, and $\mathbb{J}_k$ denotes the $k$-dimensional Jordan block. Our convention is that $P_{L,1}$ has fugacity $y_1$ while $P_{L,2}$ has fugacity $1/y_1$ and analogously for $P_{R,i=1,2}$ . We set all other fields to zero. As we have already noticed, the F-terms are automatically satisfied and one can check that \eqref{vevsol} satisfies the D-flatness condition as well. 

The solution \eqref{vevsol} breaks the gauge group to $USp(2N_c-2k)$ and the only gauge invariant operator with a non-vanishing VEV is the meson dressed with the antisymmetric $A$ $k-1$ times. In order to see this it is convenient to multiply the fields by the symplectic form $J$
\be\label{vevsol2}
JA=\left(\begin{array}{c|c}
0_{2N_c-2k} & 0\\
\hline
0 & \begin{array}{c|c} -\mathbb{J}_k^T & 0\\
\hline
0& -\mathbb{J}_k\\\end{array}\\
\end{array}\right)\,,\quad JP_{L,1}=-e_{2N_c}\,,\quad JP_{R,2}=e_{2N_c-2k+1}\,.
\ee
In this basis it is straightforward to see that the matrix $JA$ is nilpotent and satisfies the relation $(JA)^k=0$. If we contract indices via ordinary matrix multiplication (viewing $P_{L,1}$ and $P_{R,2}$ as $2N_c$ by $1$ matrices), the $USp$-invariant dressed mesons read 
$$(P_{L,1})^T(JA)^m JP_{R,2}=-(P_{R,2})^T(JA)^m JP_{L,1}$$
and it can be easily checked from \eqref{vevsol2} that they vanish unless $m=k-1$. 

Let us now check that the non-vanishing VEV for $(P_{L,1})^T(JA)^{k-1} JP_{R,2}$ imposes the expected constraint on fugacities. Denoting the fugacities associated with the broken $USp$ Cartan generators as $z_i$ with $i=1,\dots, n-\epsilon+M+1$, we have that the VEV for the antisymmetric $A$ in \eqref{vevsol} imposes the constraints  
\be
\label{Aconstr2} z_{i+1}=tz_i\,,\quad i=1,\dots, n-\epsilon+M\,,
\ee 
whereas the VEVs for $P_{L,1}$ and $P_{R,2}$ lead instead to 
\be
\label{Pconstr2} y_1cz_{n-\epsilon+M+1}=1\,,\quad y_2cz_1=t^{-M}\,.
\ee 
By combining \eqref{Aconstr2} and \eqref{Pconstr2} we indeed recover the desired relation \eqref{eq:deltaconstr}
\be
\frac{y_1}{y_2}t^{k-1-M}=\frac{y_1}{y_2}t^{n-\epsilon}=1\,.
\ee

\paragraph{Giving a VEV to other meson components.} 

Before proceeding with the detailed analysis of the Higgsing induced by the VEV \eqref{vevsol}, we would like to point out that we can alternatively give VEV to the meson components with charge $y_1y_2$, $y_2/y_1$ or $1/y_1y_2$. Let us focus for definiteness on the component with charge $y_2/y_1$. Giving it a VEV simply corresponds to replacing \eqref{vevsol} with 
\be\label{vevsol3}
A=\left(\begin{array}{c|c}
0_{2N_c-2k} & 0\\
\hline
0 & \begin{array}{c|c} 0 & \mathbb{J}_{k}\\
\hline
-\mathbb{J}_{k}^T& 0\\\end{array}\\
\end{array}\right)\,,\quad P_{L,2}=e_{2N_c-k}\,,\quad P_{R,1}=e_{2N_c-k+1}\,,
\ee
while the rest is unchanged. This again breaks the gauge group down to $USp(2N_c-2k)$ and imposes the same constraint 
\be 
y_1=y_2t^{k-1-M}\,.
\ee 

The VEV for the other two meson components can be treated analogously. The case $y_1y_2$ corresponds to giving a VEV to $P_{L,1}$ and $P_{R,1}$ in \eqref{vevsol3}, whereas $1/y_1y_2$ requires a VEV for $P_{L,2}$ and $P_{R,2}$. These lead respectively to the fugacity constraints 
\be
 y_1y_2t^{k-1-M}=1
 \ee 
and
\be 
y_1y_2=t^{k-1-M}\,.
\ee 
This reproduces precisely the pattern of delta functions we have seen before at the level of the index.
The analysis which follows can be repeated in these cases as well without major modifications.

\subsubsection{Low-energy effective theory} 

In order to determine the effect of the Higgsing we should first of all identify the fields which remain massless in the IR. As is well known there are two ways in which matter fields can be removed from the spectrum: they can combine with vector multiplets associated with broken gauge generators into long multiplets (Higgs mechanism) or they acquire a mass due to the expansion of the superpotential (in our case \eqref{suppot}) around the VEV. We will now discuss these two effects separately. 

In order to simplify the equations it will sometimes be convenient to work in terms of submatrices and we will adopt the following notation: if $M$ is a $N\times N$ matrix, we use $M|_k$ ($k<N$) to denote the $k\times k$ matrix obtained by keeping the last $k$ rows and columns of $M$. Analogously, for a $N$-dimensional vector $V$ we denote by $V|_k$ its last $k$ components.

\paragraph{Higgs mechanism.} 

As reviewed e.g.~in \cite{Agarwal:2014rua}, when discussing the Higgs mechanism for supersymmetric theories, we can identify the components of matter chiral multiplets which combine with vectors into long multiplets by acting on the expectation values with an infinitesimal complexified gauge transformation. In the case at hand we have to consider the variation of $A$, $P_{L,1}$ and $P_{R,2}$, which are the only fields with a non-zero VEV. We find the equations 
\be\label{variation}\delta A = U\langle A\rangle+\langle A\rangle U^T;\quad \delta P_{L,R}=U\langle P_{L,R}\rangle,\ee 
where the matrix $U$ is a linear combination with complex coefficients of $\mathfrak{usp}(2N_c)$ generators. This is defined by the condition\footnote{$USp(2N_c)$ matrices $M$ satisfy the relation $MJM^T=J$, from which the relation \eqref{Liegen} for the Lie algebra elements follows.} \be\label{Liegen} U^TJ+JU=0.\ee From \eqref{simplectic} and \eqref{Liegen} we find that $U$ is of the form \be\label{transform} U=\left(\begin{array}{c|c} U_{2N_c-2k}& B\\ \hline C & U_{2k}\end{array}\right)\,,\ee 
where $B$ is an unconstrained $(2N_c-2k)\times 2k$ complex matrix, $C$ can be written in terms of $B$ as $C=(J_{N_c-k}BJ_k)^T$ and $U_{2N_c-2k}$, $U_{2k}$ are $\mathfrak{usp}(2N_c-2k)$ and $\mathfrak{usp}(2k)$ matrices respectively. Clearly, since the gauge group is Higgsed to $USp(2N_c-2k)$ the transformations $U_{2N_c-2k}$ act trivially on the VEVs and therefore we will set them to zero from now on. The other generators instead act non-trivially on the VEVs and we will now consider them.

Let us start from $B$ and $C$ transformations in \eqref{transform}. When acting on $\langle A\rangle$, these produce components $\delta A_{Ij}=-\delta A_{jI}$ with $I\leq 2N_c-2k$ and $j>2N_c-2k$. Since the unbroken gauge group acts on gauge indices $I\leq 2N_c-2k$ only, these are fundamentals of the unbroken gauge group. By plugging in \eqref{variation}-\eqref{transform} with $U_{2N_c-2k}= U_{2k}=0$ and using \eqref{vevsol}, we find that $\delta A_{Ij}$ is trivial for $j=2N_c-k$ and $j=2N_c-k+1$ whereas the other components are unconstrained. As a result, only the components
\be
V_{1,I}=A_{I,j=2N_c-k}\,,\quad V_{2,I}=A_{I,j=2N_c-k+1}\,,\quad I=1,\cdots, 2N_c-2k
\ee
belong to the massless spectrum in the IR and become a pair of fundamentals of the unbroken $USp(2N_c-2k)$ gauge group, whereas the other components fit into long multiplets. Analogously, when we consider the action of $B$ and $C$ transformations on the VEV of $P_{L,1}$ and $P_{R,2}$ we find that all the components of $P_{L,1}^I$ and $P_{R,2}^I$ with $I\leq 2N_c-2k$ become part of long multiplets and disappear from the spectrum. 

The analysis of $U_{2k}$ transformations is a bit more involved. These generate only components of $A$, $P_{L,1}$ and $P_{R,2}$ with all gauge indices in the range $2N_c-2k+1,\dots, 2N_c$ and therefore correspond to gauge singlets in the infrared. Since other components do not play any role, in this discussion we will restrict the gauge indices to be in the above mentioned range and therefore we work in terms of the $2k$-dimensional submatrices and subvectors $A|_{2k}$ and $P_{L,R}|_{2k}$. 

Let us start from the antisymmetric multiplet. Most of its components combine with vector multiplets associated with the broken generators of the gauge group into long multiplets and therefore disappear from the low energy spectrum. There are however $k$ components which survive the Higgsing and become free chiral multiplets in the infrared. These can be chosen to have charge 
\be
t\frac{z_j}{z_1}\,,\quad j=1,\dots k
\ee
and therefore, using \eqref{Aconstr2}, we end up with chirals having charge $t^j$. Let us explain how to derive the above statement. We have to consider again the variation of $A|_{2k}$ around its VEV induced by infinitesimal complexified $USp(2k)$ gauge transformations. This leads to the following matrix equation (see \eqref{Liegen})
\be
\label{var1} \delta A_{2k}=U\langle A\rangle|_{2k}+\langle A\rangle|_{2k} U^T\,,\quad U=\left(\begin{array}{c|c}\alpha & \beta\\ \hline \gamma & -\alpha^T\\\end{array}\right)\,,
\ee 
where $\alpha$ is a generic $k\times k$ complex matrix while $\beta$ and $\gamma$ are symmetric and remember from \eqref{vevsol} that 
\be
\langle A\rangle|_{2k}= \left(\begin{array}{c|c}0 & \mathbb{J}_k\\ \hline -\mathbb{J}^T_k & 0\\\end{array}\right)\,.
\ee
From \eqref{var1} we can see that the diagonal $k\times k$ blocks of $\delta A|_{2k}$ are generic antisymmetric matrices, and therefore all the components of $A|_{2k}$ inside these blocks combine with vector multiplets. On the other hand, we observe that $\delta A|_{2k}$ in \eqref{var1} can never be of the following form 
\be\label{antyvar} \left(\begin{array}{c|c}0 & D\\ \hline -D^T & 0\\\end{array}\right)\,,\quad D=\left(\begin{array}{cccc}
d_1 & 0&\dots&0\\
\vdots & \ddots &\ddots &\\
d_{k-1} &&\ddots&0\\
d_k&d_{k-1}&\dots & d_1
\end{array}\right)\,.\ee 
This is because the upper-right block of $\delta A|_{2k}$ is given by $[\alpha, \mathbb J_k]\equiv UR$, which satisfies $\mathrm{Tr} [UR (\mathbb J_k)^n] = 0$ for $n = 0,\dots,k$, indicating that the sum of the components of $UR$ along each diagonal line below the main diagonal vanishes. Notice that all the entries along the diagonals of the matrix $D$ in \eqref{antyvar} are identical. These are the $k$ components of $A|_{2k}$ which remain in the low-energy spectrum and do not combine with vector multiplets. Using \eqref{Aconstr2} one can easily see that $d_j$ has charge $j$ under $U(1)_t$. 

The above argument shows that most of the components of $A|_{2k}$ (more precisely $2k^2-2k$ of them) combine with vector multiplets associated with broken generators of the gauge group into long multiplets. We can now observe that $USp(2k)$ has dimension $2k^2+k$ and therefore there are $3k$ generators of the gauge group which do not leave the VEV invariant and yet the corresponding vector multiplets do not combine with components of $A$. They combine instead with the same number of components of  $P_{L,1}|_{2k}$ and $P_{R,2}|_{2k}$. In order to identify them, we can first identify the transformations $U$ which leave $\langle A\rangle|_{2k}$ invariant and then consider how they act on the VEV of $P_{L,1}|_{2k}$ and $P_{R,2}|_{2k}$.
The relevant gauge transformations are given by components of $U$ such that $\delta A|_{2k}$ in \eqref{var1} vanishes. It can be checked that there are precisely $3k$ of them and they all act non-trivially on the VEV of  $P_{L,1}|_{2k}$ and $P_{R,2}|_{2k}$. The corresponding variations span $3k$ out of the $4k$ components of $P_{L,1}|_{2k}$ and $P_{R,2}|_{2k}$. These combine with the corresponding vector multiplets into massive long multiplets, leaving $k$ components which persist in the low-energy spectrum. From this computation we find that the surviving $k$ massless components have charges 
\be
cy_1z_j\,,\quad \frac{t^{-M}}{cz_{k+1-j}y_2}\,,\quad j=1,\cdots,k\,.
\ee 
By making use of \eqref{Aconstr2} we conclude that these have charge $j-k$ under $U(1)_t$. Notice that one of the chirals has charge zero under $U(1)_t$. This corresponds to a linear combination of the components of $P_{L,1}$ and $P_{R,2}$ which are given a VEV. Overall, we find that from the fluctuations around the VEV of $A$, $P_{L,1}$ and $P_{R,2}$ we get $2k$ gauge singlets charged under $U(1)_t$ only. These become free fields in the IR and are the only free fields which survive the 3d limit. We will discuss this in detail in Section \ref{sec:3danalysis}. 

Let us summarize our findings so far: 
\begin{itemize}
\item From the antisymmetric chiral $A$ we get in the IR an antisymmetric of the unbroken $USp(2N_c-2k)$ gauge group, two fundamentals $V_1$ and $V_2$ and $k$ singlets; 
\item From the fundamentals $P_{L,1}$ and $P_{R,2}$ we get only $k$ gauge singlets. 
\end{itemize}
As a result of this analysis we see that the spectrum of the infrared theory lacks the analogs of $P_{L,1}$ and $P_{R,2}$ since the Higgs mechanism has removed them. However, these are replaced by the fields $V_1$ and $V_2$ as we will now see. In order to show this, we need to look in detail at the expansion of the superpotential around the VEV of the fields which is what we discuss next.

\paragraph{Mass terms from the superpotential.}

We now study the expansion of the superpotential \eqref{suppot} for the fluctuations around the VEV \eqref{vevsol} in order to determine possible mass terms and the resulting superpotential for the massless IR fields.

Let us start from the superpotential terms in \eqref{suppot} involving the flavors $Q$. When the gauge index for the flavors is in the range $1,\dots, 2N_c-2k$ the corresponding components form fundamentals of the unbroken gauge group in the infrared, whereas the other components describe gauge singlets. As we will now see, most of these become massive when we expand the superpotential. Using matrix multiplication notation the relevant terms in \eqref{suppot} can be written as $(Q_i)^TJAJQ^i$,\footnote{Notice that the flavor indices are contracted using the antisymmetric symplectic matrix.} $(Q_i)^TJP_{R,2}S_1^i$ and $(Q_i)^TJP_{R,1}S_2^i$. Indeed only the first two lead to mass terms upon expanding around the VEV. By replacing $A$ with its VEV in the first superpotential term, we get 
\be \sum_{a=2N_c-2k+2}^{2N_c-k}Q_{a,i}Q_{a+k-1}^{i},\ee
 implying that all the $Q$ singlets with gauge index from $2N_c-2k+2$ to $2N_c-1$ become massive and can be integrated out. As a result only $Q_{2N_c-2k+1}$ and $Q_{2N_c}$ survive in the infrared. This is not the end of the story though, since if we now replace $P_{R,2}$ with its VEV in the second term we find the further mass term 
 \be Q_{2N_c-2k+1,i}S_1^i\ee 
 and therefore we conclude that both $S_1$ \footnote{For simplicity we suppress the flavor index of $S$.} and the $(2N_c-2k+1)$-th color component of $Q$ become massive due to this superpotential term. As a result, only $Q_{2N_c}$ remains massless and will replace in the IR the multiplet $S_1$. This can be seen by expanding the term $(Q_i)^TJAJQ^i$ around the VEV. From this procedure we get an analogous term involving the antisymmetric and fundamental flavors for the unbroken $USp(2N_c-2k)$ gauge group and also a term involving $Q_{2N_c}$ and the fundamental $V_1$ we have mentioned before. This term explicitly reads 
 \be Q_{i}^TJV_1Q_{2N_c},\ee 
 where $Q_i$ denote the flavors of the infrared gauge theory. We can now notice that this superpotential term has the same structure as  $(Q_i)^TJP_{R,2}S_1^i$, where $V_1$ plays the role of $P_{R,2}$ and $Q_{2N_c}$ that of $S_1$. In summary we find that: 
 \begin{itemize}
 \item From the flavors $Q$ of the UV theory we get $2N_f$ fundamentals of the IR gauge group and one singlet in the fundamental of the flavor symmetry group, all other components acquire a mass due to the expansion of the superpotential around the VEV. 
 \item The singlet $S_1$ charged under the flavor symmetry group acquires a mass and disappears from the spectrum however, in the IR theory it is replaced by the only surviving $Q$ singlet. Finally, the fundamental $V_1$ of the IR gauge group arising from the decomposition of the antisymmetric $A$ replaces the multiplet $P_{R,2}$ which disappears from the IR spectrum due to the Higgs mechanism. 
 \end{itemize}

Finally, we need to discuss the fate of the fields $V_2$ and $P_{L,1}$. In order to understand this, let us look at the last part of \eqref{suppot} involving the singlets $\beta_i$. In matrix notation this can be rewritten as 
\be\label{betaterm} \sum_i\beta_i (P_{L,\gamma})^T(JA)^iJP_{L,\delta}\epsilon^{\gamma \delta}=\sum_i\beta_i\left[(P_{L,1})^T(JA)^iJP_{L,2}-(P_{L,2})^T(JA)^iJP_{L,1}\right]\,.\ee
Since in all these terms both $P_{L,1}$ and $P_{L,2}$ always appear and we are not giving VEV to $P_{L,2}$, nor to any of the $\beta_i$’s, the only way to generate a mass term is to set both $A$ and $P_{L,1}$ to their expectation value and ignore their fluctuations around the VEV, since these can only induce higher order terms. The nilpotency of $JA$ now tells us that we can get a mass term only for $i<k$. The corresponding $\beta_i$ singlets become massive, together with $k$ of the components of $P_{L,2}$ which are singlets of the infrared $USp(2N_c-2k)$ gauge group. 

The remaining $\beta_i$ singlets with $i\geq k$ then become the singlets of the infrared theory and the corresponding leading superpotential terms we find upon expanding \eqref{betaterm} around the VEV are 
\be\label{supexp} \sum_{i\geq k}\beta_i\left[ \langle P_{L,1}\rangle^T\langle JA\rangle^{k-1}(JA)^{i+1-k}(JP_{L,2})-(P_{L,2})^T(JA)^{i+1-k}\langle JA\rangle^{k-1}\langle JP_{L,1}\rangle\right]\,.\ee 
From \eqref{vevsol2} we can now notice that $\langle JA\rangle^{k-1}\langle JP_{L,1}\rangle=e_{2N_c-k+1}$ and therefore we find
\be\label{exps} A\langle JA\rangle^{k-1}\langle JP_{L,1}\rangle=V_2.\ee 
Using \eqref{exps} we can rewrite \eqref{supexp} as 
\be\label{supp2}\sum_{i\geq 0}\beta_i\left[ (V_2)^T(JA)^{i}(JP_{L,2})-(P_{L,2})^T(JA)^{i}(JV_2)\right]\,.\ee 
Notice that \eqref{supp2} has the same structure as \eqref{betaterm}, where $V_2$ plays the role of $P_{L,1}$ in the infrared. 
From this discussion we conclude that 
\begin{itemize} 
\item The fundamental $P_{L,1}$, which is removed from the low-energy spectrum due to the Higgs mechanism, is replaced in the IR by the multiplet $V_2$ coming from the expansion of the antisymmetric chiral. 
\item Out of the $4k$ gauge singlets coming from the UV fields $P_{L,2}$ and $P_{R,1}$, $3k$ of them become free fields in the infrared while the other $k$ (coming from the field $P_{L,2}$) acquire a mass, together with some of the $\beta$ singlets.
\end{itemize}

From the above analysis we conclude that the low-energy spectrum is consistent with the expectation that a VEV for the meson dressed $k-1$ times leads to the electric theory with rank $N_c-k$. We also recover for the Higgsed theory the superpotential \eqref{suppot} from the expansion of the UV superpotential around the expectation value of the various fields. Furthermore, the theory is accompanied by a collection of free fields: $2k$ of them arise from the components of the fields acquiring a VEV and are only charged under $U(1)_t$, so they survive upon compactifying the theory to 3d and mass deforming it to make it $\mathcal{N}=4$. The other $3k$ free chirals arise from the other fields and are removed from the spectrum upon compactification since they are also charged under $U(1)_c$, which is the symmetry for which we perform a real mass deformation in 3d. 

With the results obtained so far it is possible to connect the UV and IR fugacities, as we did from the index analysis of the two examples analyzed in the previous section, by identifying which role is played in the IR by the UV fields that remain massless. We perform this analysis for generic $N_c$ and $N_f$ in Appendix \ref{app:UVIRfug}.

\subsection{Flips and a single frame duality}
\label{sec:single frame duality}

Before closing this section we would like to comment on the IR behavior of an interesting variant of the SQCD model with $N_c\leq N_f<2N_c$ discussed so far. Recall that the bad SQCD models we have considered include dressed mesonic operators $P_L A^k P_R$, whose VEV plays a significant role in determining the IR dynamics of the theory, and the dressed diagonal mesons $A^k (P_R)^2$. Now we want to flip these mesons dressed with up to $2N_c-N_f-1$ powers of the antisymmetric, resulting in a theory with a reduced moduli space and a conventional IR duality with a single dual frame.

Specifically, let us introduce extra gauge singlets $T_k$ and $W_k$ for $k = 1, \dots, 2 N_c-N_f-1$ in the bifundamental and singlet representation of $SU(2)_{y_1} \times SU(2)_{y_2}$, respectively, coupling them to the original operators as follows:
\begin{align}
\label{eq:flip}
\sum_{k = 1}^{2 N_c-N_f} \left[T_k P_L A^{k-1} P_R + W_k A^{k-1} (P_R)^2\right] \,.
\end{align}
We have omitted the gauge and $SU(2)_{y_i}$ indices for simplicity, which should be contracted properly to form an invariant of those symmetries.
Notice that the equation of motion for $T_k$ demands that
\begin{align}
P_L A^{k-1} P_R = 0 \,.
\end{align}
Namely, these mesons are now removed from the chiral ring, and we cannot activate their VEV Higgsing the theory down to one of  the multiple dual frames associated with a particular delta constraint, which as we have shown determines which meson gets a VEV. Instead, we expect the resulting theory to be dual to a single good theory, the last frame of the multiple duals with no delta constraint. In this frame, the original mesons $P_L A^k P_R$ and $A_k (P_R)^2$ are mapped to gauge singlets, which become massive because of the superpotential terms in \eqref{eq:flip}. Thus, the remaining theory is the good $USp(2 N_f-2 N_c)$ SQCD theory without extra singlets.

Indeed,  at the level of the index, it is  easy to check how this is realized. First note that the index contribution of $T_k$ and $W_k$ are given by
\begin{gather}
\prod_{k = 1}^{2 N_c-N_f} \Gpq{p q t^{-k+N_c-\frac{N_f}{2}} y_1^\pm y_2^\pm} , \label{tk} \\
\prod_{k = 1}^{2 N_c-N_f} \Gpq{p q t^{-k-1+2 N_c-N_f} c^2} .
\end{gather}
When evaluated on the delta constraint in each frame, the contribution of $T_k$ in eq. ~\eqref{tk} includes $\Gpq{pq}^2$, producing a zero of second order. We can think that one of these factors cancels the divergence $\Gamma_e(1)$ corresponding to the delta while the other zero kills the contribution of that  frame completely. Only the last frame survives in the end, and we get the following identity:
\begin{align}
\label{eq:4dflipped}
&\mathcal{I}_{\text{SQCD}\left(N_c,\,N_c \leq N_f < 2N_c\right)}(\vec x,y_1,y_2,t,c) \nonumber \\
&\times \prod_{k = 1}^{2 N_c-N_f} \Gpq{p q t^{-k+N_c-\frac{N_f}{2}} y_1^\pm y_2^\pm} \Gpq{p q t^{-k-1+2 N_c-N_f} c^2} \nonumber \\
&= \mathcal{I}_{\text{SQCD}\left(N_f-N_c,\,N_f\right)}\left(\vec x,y_2,y_1,t,t^{\frac{2N_c-N_f}{2}}c\right) ,
\end{align}
indicating a more conventional IR duality for a bad SQCD with a single good dual theory.

The cancelation between the $\Gpq{pq}$ and $\Gpq{1}$, of course, needs care; one should refine the restricted fugacities, such as \eqref{eq:var441} and \eqref{eq:var442} for the $N_c=N_f=4$ case, and take a suitable limit to see if they indeed cancel.  In Appendix \ref{appsingleframe} we check this for the $N_f=N_c=4$ case, where we observe that the delta associated with each frame is canceled by one of the extra $\Gpq{pq}$ factors also refined properly, and the other $\Gpq{pq}$ gives a zero killing the whole contribution of the frame.

\section{The 3d bad SQCD}
\label{sec:3danalysis}

We will now move to the study of the 3d bad SQCD with $U(N_c)$ gauge group and $N_f<2N_c-1$ fundamental hypers,  where monopole operators  have  scaling dimension falling below the unitarity bound for chiral operators. We will present our result for its $S^3_b$ partition function for arbitrary value of the FI and real mass parameters. 

As in the 4d case, the partition function is a distribution rather than an ordinary function.
More precisely it is a sum of frames, each containing a delta distribution, enforcing  a particular constraint on the FI parameter, multiplied by 
 the partition functions of a given good SQCD, plus extra singlets. Each of these frames is reached by giving a VEV to a different dressed monopole operator.  In addition, there is an extra frame with no delta distribution, for which the interacting part is the good $U(N_f - N_c)$ SQCD.
 
We will then rewrite the partition function of the bad SQCD in a form which makes it easier to compare with previous results in the literature, e.g.~\cite{Yaakov:2013fza,Assel:2017jgo}, and allows us to determine the correct IR R-symmetry in each dual frame. Finally, we will discuss how we can obtain a more conventional duality with a single dual frame by flipping the monopoles whose VEVs trigger the Higgsing to each frame.

\subsection{Partition function and IR properties}

In order to study the partition function of the 3d bad SQCD, we could retrace all the steps  of the previous sections, applying the 3d version of the  local dualization algorithm \cite{Bottini:2021vms,Hwang:2021ulb,Comi:2022aqo}. As in the 4d case, for bad theories the algorithm brings us to multiple dual frames equipped with different constraints on parameters, which in this case are real masses appearing in the $S^3_b$ partition function, where such constraints reflect the symmetry breaking due to VEVs of monopole operators. 

However, knowing already the 4d result, we can just implement  the standard 4d to 3d reduction procedure  (see for example \cite{Pasquetti:2019tix,Pasquetti:2019hxf,Bottini:2021vms,Comi:2022aqo}) to reduce the $S^3\times S^1$ index of a 4d $USp(2N_c)$ SQCD to the $S^3_b$ partition function of a 3d $U(N_c)$ SQCD. This can be done by using the asymptotic behavior of the elliptic gamma function
\be
\lim_{r\to0}\Gc_e\left(\e^{2\pi irs};p=\e^{-2\pi rb},q=\e^{-2\pi rb^{-1}}\right)=\e^{-\frac{i\pi}{6r}\left(i\frac{Q}{2}-s\right)}\sbfunc{i\frac{Q}{2}-s}\,,
\ee
where $Q = b+b^{-1}$ and $r$ is the radius of the circle which we shrink to zero. Indeed $s_b$ is the special function in terms of which the contribution of the 3d $\mathcal{N}=2$ vector and chiral multiplets to the $S^3_b$ partition function can be expressed. In particular, the contribution of a chiral of R-charge $r$ and charged under a $U(1)$ symmetry whose corresponding real mass is $m$ is given by
\be\label{eq:sbchir}
\sbfunc{i\frac{Q}{2}(1-r)-m}\,.
\ee
Since our 4d index contains delta distributions, their 3d limits should also be taken into account as follows:
\begin{align}
\lim_{r \rightarrow 0} \frac{2 \pi i y_1 \delta(y_1-y_2)}{(p;p)_\infty (q;q)_\infty} = \lim_{r \rightarrow 0} \frac{\delta(r Y_1-r Y_2)}{(p;p)_\infty (q;q)_\infty} = e^{\frac{\pi Q}{12 r}} \, \delta(Y_1-Y_2) \,,
\end{align}
where $y_{1,2} = e^{2 \pi i r Y_{1,2}}$, $p = e^{-2 \pi r b}$ and $q = e^{-2 \pi r b^{-1}}$. Note that the extra exponential factor is crucial to have a finite 3d partition function identity when taking the $r \rightarrow 0$ limit. This arises from the following asymptotic behavior of the q-Pochhammer symbol:
\be
\lim_{\ga\to 0}(e^{-2\ga};e^{-2\ga})_\infty=\sqrt{\frac{\pi}{\ga}}\exp\left[-\frac{\pi^2}{12\ga}\right]\,.
\ee

On top of the dimensional reduction limit, in order to flow to the 3d $\mathcal{N}=4$ theory we also need to perform a series of deformations. The first one is a combination of real mass deformation for the non-abelian flavor symmetries and Coulomb branch VEV for the gauge group which breaks all of them from symplectic to their unitary subgroup of the same rank. The second one is instead a real mass deformation for the $U(1)_c$ symmetry, which gets rid of the extra fields compared to the $\mathcal{N}=4$ theory. These deformations are implemented at the level of the $S^3_b$ partition function by using that
\be
\lim_{s\to\pm\infty}\sbfunc{s}=\e^{\pm i\frac{\pi}{2}s^2}\,.
\ee
We refer the reader to \cite{Pasquetti:2019tix,Pasquetti:2019hxf,Bottini:2021vms,Comi:2022aqo} for more details on these limits.
The 3d result for  the $N_f=7$, $N_c=6$ case is shown in Figure \ref{fig:bad_SQCD_3d_Nc=6_Nf=7_all_1}.
\afterpage{
\begin{landscape}
\begin{figure}[tbp]
\centering
\includegraphics[width=1.4\textwidth]{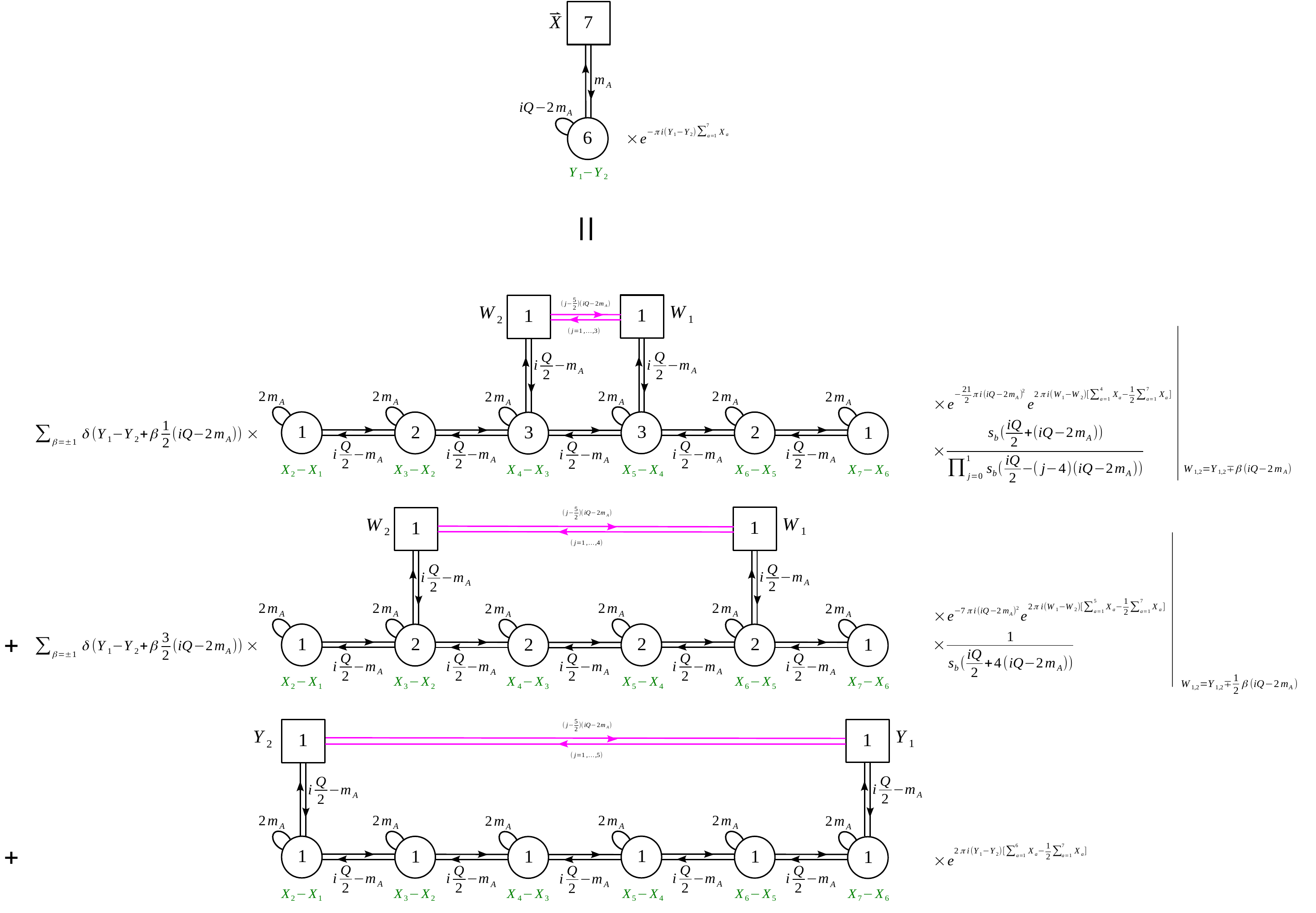}
\caption{The final result of the bad 3d $\mathcal{N}=4$ SQCD with $N_c=6$ and $N_f=7$.}
\label{fig:bad_SQCD_3d_Nc=6_Nf=7_all_1}
\end{figure}
\end{landscape}
}

The general result for  $N_c\leq N_f < 2N_c$, where using the result in Appendix \ref{3dgoodres} we re-express each frame as a 3d good SQCD with extra singlets, is given by\footnote{For convenience we have added a mixed CS coupling  $e^{-\pi i (Y_1+Y_2) \sum_{i = 1}^{N_f} X_i}$  on both sides so that each SQCD frame has a neat
  mixed CS coupling between the topological $U(1)$ and the diagonal $U(1)$ of the flavor symmetry.}

\begin{align}
\label{eq:3d pf}
	& 
	\mathcal{Z}^{3d}_{\text{SQCD}_U(N_c,N_c \leq N_f < 2N_c)}\left(\vec{X};Y_1-Y_2\right) \times
	e^{-\pi i (Y_1-Y_2) \sum_{i=1}^{N_f}X_i}
	= \nonumber\\[10pt]
	& \quad = 
	\sum_{n=0}^{\floor*{\frac{N_f}{2}}-(N_f-N_c)-1} 
	\sum_{\substack{\beta = 1 \text{ if } n-\epsilon = 0, \\ \beta = \pm1 \text{ otherwise}}}
	\left[
	\delta\Big(Y_1-\left(Y_2-\beta(n-\epsilon)(iQ-2m_A)\right) \Big) 
	\times 
	\bigphantomspace\right.
	\nonumber \\
	&\qquad \quad \times
	e^{-\pi i (W_2-W_1)\sum_{i=1}^{N_f}X_i} 
	\nonumber\\
	&\qquad \quad \times
	\frac{\prod_{j=1}^{N_c-N_f+\floor*{\frac{N_f}{2}}-1-n}s_b\left(\frac{iQ}{2}+j(iQ-2m_A)\right)}{\prod_{j=0}^{N_c-N_f+\floor*{\frac{N_f}{2}}-1-n} s_b\left(\frac{iQ}{2}-(j-(2N_c-N_f-1))(iQ-2m_A)\right)} 
	\nonumber\\ 
	&\qquad \quad \times
	\prod_{j=1}^{N_c-\floor*{\frac{N_f}{2}}+n}s_b\left(\frac{iQ}{2}-\left(j-N_c+\frac{N_f}{2}\right)(iQ-2m_A)\pm (W_1- W_2)\right) 
	\nonumber\\
	&\left.\qquad \quad \times
	\mathcal{Z}^{3d}_{\text{SQCD}_U\left(\floor*{\frac{N_f}{2}}-n,N_f\right)}\left(\vec{X};W_2-W_1\right)\Bigg|_{W_{1,2}=Y_{1,2}\mp\frac{1}{2}\beta\left(N_c-N_f+\floor*{\frac{N_f}{2}}-n\right)(iQ-2m_A)}
	 \right] \nonumber\\[10pt]
	 &\qquad+
	e^{-\pi i (Y_2-Y_1)\sum_{i=1}^{N_f}X_i} 
	 \times
	\prod_{j=1}^{2 N_c-N_f}s_b\left(\frac{iQ}{2}-\left(j-N_c+\frac{N_f}{2}\right)(iQ-2m_A)\pm (Y_1- Y_2)\right) 
	\nonumber\\
	&\qquad \quad \times
	\mathcal{Z}^{3d}_{\text{SQCD}_U\left(N_f-N_c,N_f\right)}\left(\vec{X};Y_2-Y_1\right)
	 \,.
\end{align}

As in the 4d case we have a frame, the last one, with no delta constraint. This is  reached when the value of the FI parameter is generic and the dual is the good theory with $U(N_f-N_c)$ theory, in accordance with the finding of \cite{Kim:2012uz,Yaakov:2013fza}.

All other frames come with a delta distribution implementing a specific breaking of the topological symmetry corresponding to a given dressed  monopole operator  acquiring a VEV and Higgsing the theory down to the expected dual good  $U(r)$ SQCD with $r=N_f-N_c+1, \cdots, \lfloor N_f/2\rfloor$, where we recall that $r=\lfloor\tfrac{N_f}{2}\rfloor-n$. In particular, the $n$-th frame is reached at the end of the RG flow triggered by the VEV of fundamental monopoles $V_{i}^\pm$ dressed with the adjoint scalar in the vector multiplet $i=M+n-\epsilon=k+1$ times (with the sign $\beta$ related to the topological charge of the monopoles).
  
Similarly, taking the 3d reduction  we obtain the result for the  $N_f < N_c$ case
\begin{align}\label{eq:3d pfNflNc}
	& 
	\mathcal{Z}^{3d}_{\text{SQCD}_U(N_c,N_f < N_c)}\left(\vec{X};Y_1-Y_2\right) \times
	e^{-\pi i (Y_1-Y_2) \sum_{i=1}^{N_f}X_i}
	= \nonumber\\[10pt]
	& \quad = 
	\sum_{n=0}^{\floor*{\frac{N_f}{2}}} 
	\sum_{\substack{\beta = 1 \text{ if } n-\epsilon = 0, \\ \beta = \pm1 \text{ otherwise}}}
	\left[
	\delta\Big(Y_1-\left(Y_2-\beta(n-\epsilon)(iQ-2m_A)\right) \Big) 
	\times
	\bigphantomspace\right. \nonumber \\
	&\qquad \quad \times
	e^{-\pi i (W_2-W_1)\sum_{i=1}^{N_f}X_i} 
	\nonumber\\
	&\qquad \quad \times
	\frac{\prod_{j=1}^{N_c-N_f+\floor*{\frac{N_f}{2}}-1-n}s_b\left(\frac{iQ}{2}+j(iQ-2m_A)\right)}{\prod_{j=0}^{N_c-N_f+\floor*{\frac{N_f}{2}}-1-n} s_b\left(\frac{iQ}{2}-(j-(2N_c-N_f-1))(iQ-2m_A)\right)} 
	\nonumber\\ 
	&\qquad \quad \times
	\prod_{j=1}^{N_c-\floor*{\frac{N_f}{2}}+n}s_b\left(\frac{iQ}{2}-\left(j-N_c+\frac{N_f}{2}\right)(iQ-2m_A)\pm (W_1- W_2)\right) 
	\nonumber\\
	&\left.\qquad \quad \times
	\mathcal{Z}^{3d}_{\text{SQCD}_U\left(\floor*{\frac{N_f}{2}}-n,N_f\right)}\left(\vec{X};W_2-W_1\right)\Bigg|_{W_{1,2}=Y_{1,2}\mp\frac{1}{2}\beta\left(N_c-N_f+\floor*{\frac{N_f}{2}}-n\right)(iQ-2m_A)}
	 \right] .
\end{align} 
Notice that as in  the 4d case, for this range of the number of flavors, there is no dual frame with  unbroken topological $U(1)$ symmetry and 
each frame comes with a delta function.

We can  further simplify the above results, in particular the contribution of the singlet fields, by implementing the constraints from the delta distributions. Such simplifications reflect the effect of integrating out  fields acquiring mass due to symmetry breaking corresponding to the delta constraints. Furthermore, we can also read information about the IR physics of bad theories from the simplified partition functions, which give consistent results with the literature. This is what we shall show next.


\subsubsection{$N_f = 0, 1$}

The SQCD partition function \eqref{eq:3d pfNflNc} for $N_f = 0$ is given by
\begin{align}
\label{eq:Nf=0}
	& \mathcal{Z}^{3d}_{\text{SQCD}_U(N_c,N_f=0)}\left(Y_1-Y_2\right)
	= \nonumber\\
	&= \delta\Big(Y_1-Y_2\Big) \frac{\prod_{j=1}^{N_c-1}s_b\left(\frac{iQ}{2}+j(iQ-2m_A)\right)}{\prod_{j=0}^{N_c-1} s_b\left(\frac{iQ}{2}-(j-(2N_c-1))(iQ-2m_A)\right)} 
	\nonumber\\ 
	&\quad \times
	\prod_{j=1}^{N_c}s_b\left(\frac{iQ}{2}-\left(j-N_c\right)(iQ-2m_A)\pm (W_1- W_2)\right)\Bigg|_{W_{1,2}=Y_{1,2}\mp\frac{1}{2} N_c (iQ-2m_A)} \\
	&= \delta\Big(Y_1-Y_2\Big) s_b\left(\frac{iQ}{2}-N_c (iQ-2m_A)\right) \prod_{j=1}^{N_c-1}s_b\left(\frac{iQ}{2}\pm j(iQ-2m_A)\right) \,,
\end{align}
where in the second line we simplified the contributions of the chirals for $Y_1 \rightarrow Y_2$.
Note that this expression diverges when $Y_1 \rightarrow Y_2$ due to $\delta(Y_1-Y_2)$. In 4d, we argued that such divergent delta distribution is captured  by an elliptic gamma function with argument 1, which can be interpreted as the contribution of a chiral with zero $R$-charge. This is still true in 3d; namely, $\delta(0) \sim s_b\left(\frac{i Q}{2}\right)$.
Thus, the partition function \eqref{eq:Nf=0} of the pure Yang-Mills theory can be written as
\begin{align}
	\lim_{Y_1 \rightarrow Y_2} \mathcal{Z}^{3d}_{\text{SQCD}_U(N_c,N_f=0)}\left(Y_1-Y_2\right) = \prod_{j=1}^{N_c}s_b\left(\frac{iQ}{2}-\left(\frac12 \pm \left(j-\frac12\right)\right)(iQ-2m_A)\right) ,
\label{ffc}	
\end{align}
corresponding to $ N_c$ pairs of  chirals with
\be\label{eq:UVR}
R^{\pm,j}_{UV}=\left(\frac12 \pm \left(j-\frac12\right)\right)\,,
\ee
which can be read using the definition   \eqref{eq:sbchir} after  redefining
\be
m_A=\tilde m_A+ i \frac{Q}{4}\,,
\ee
as in  the usual conventions  for good theories.
Since these chirals are all supposed to be free in the IR we expect that there will be accidental emergent $U(1)_j$ symmetries, for $j = 1, \dots, N_c$, acting on the free fields with the $k$-th pair of chirals in \eqref{ffc} carrying  charge $\pm \delta_{j,k}$.
In the IR the $R$- and axial symmetries are then related to those in the UV by the mixing with $U(1)_j$  (with  mixing coefficient $\left(k-\frac12\right) $) as follows:
\begin{align}
\begin{aligned}
\label{eq:IR sym 0}
R^{\pm,j}_{IR} &=R^{\pm,j}_{UV}-\sum_{k=1}^{N_c} \left(k-\frac12\right) (\pm \delta_{j,k}) \,, \\
A^{\pm,j}_{IR} &= A^{\pm,j}_{UV}+2 \sum_{k= 1}^{N_c} \left(k-\frac12\right) (\pm \delta_{j,k}) \,,
\end{aligned}
\end{align}
which implies that all the chirals have $R_{IR} = \tfrac12$ and $A_{IR} = -1$ in the IR, as expected for free fields constituting twisted hypers. 
If we set $\tilde m_A = 0$ or equivalently $m_a=i\tfrac{Q}{4}$, those symmetries combine into $\mathcal N = 4$ supersymmetry, and the free chirals form $N_c$ free twisted hypers. The IR global symmetry is also enhanced to $USp(2 N_c)$ whose Cartans are given by $U(1)_j$.
This is consistent with the analysis of  \cite{Assel:2017jgo} where this theory was argued  to have a smooth moduli space 
and described at each point of the moduli space by $N_c$ free twisted hypers.

Similarly, the 3d $\mathcal N=4$ SQCD with a single fundamental hyper was also argued by \cite{Assel:2017jgo}  to have a smooth moduli space, and indeed  we find that the  partition function \eqref{eq:3d pf} for $N_f = 1$ is given by

\begin{align}
\label{eq:Nf=1}
	& \mathcal{Z}^{3d}_{\text{SQCD}_U(N_c,N_f=1)}\left(Y_1-Y_2\right)
	= \nonumber\\
	&= \delta\left(Y_1-\left(Y_2-\frac{\beta}{2} (i Q-2 m_A)\right)\right) \frac{\prod_{j=1}^{N_c-2}s_b\left(\frac{iQ}{2}+j(iQ-2m_A)\right)}{\prod_{j=0}^{N_c-2} s_b\left(\frac{iQ}{2}-(j-2 (N_c-1))(iQ-2m_A)\right)} 
	\nonumber\\ 
	&\quad \times
	\prod_{j=1}^{N_c}s_b\left(\frac{iQ}{2}-\left(j-N_c+\frac12\right)(iQ-2m_A)\pm (W_1- W_2)\right)\Bigg|_{W_{1,2}=Y_{1,2}\mp\frac{1}{2} \beta (N_c-1) (iQ-2m_A)} \\
	&= \delta\left(Y_1-\left(Y_2-\frac{\beta}{2} (i Q-2 m_A)\right)\right) s_b\left(\frac{iQ}{2}-N_c (iQ-2m_A)\right) \prod_{j=1}^{N_c-1}s_b\left(\frac{iQ}{2}\pm j(iQ-2m_A)\right) \,,
\end{align}
which in the $Y_1 \rightarrow Y_2-\frac{\beta}{2} (i Q-2 m_A)$ limit can be written as
\begin{align}
	\lim_{Y_1 \rightarrow Y_2-\frac{\beta}{2} (i Q-2 m_A)} \mathcal{Z}^{3d}_{\text{SQCD}_U(N_c,N_f=1)}\left(Y_1-Y_2\right) = \prod_{j=1}^{N_c}s_b\left(\frac{iQ}{2}-\left(\frac12 \pm \left(j-\frac12\right)\right)(iQ-2m_A)\right) \,.
\end{align}
 corresponding to $N_c$ free twisted hypers with IR symmetries given again by \eqref{eq:IR sym 0}.

\subsubsection{$2 \leq N_f < N_c$}

For $N_f <  N_c$  the  SQCD  has  $\floor*{\frac{N_f}{2}}$ dual frames, each of which is described by a dual good theory with decoupled twisted hypers. The monopole VEV leading to each dual frame imposes the following  constraint on the FI parameter:

\begin{align}
\label{eq:FI cond}
Y_1-Y_2 = -\beta(n-\epsilon)(iQ-2 m_A)
\end{align}
for $n = 0, \dots, \floor*{\frac{N_f}{2}}$.
By taking the limit $Y_1 \rightarrow Y_2-\beta(n-\epsilon)(iQ-2 m_A)$,  in the partition function \eqref{eq:3d pf} 
we obtain the $S^3_b$ partition function of the $n$-th frame
\begin{align}
\label{eq:3d dual pf}
	& 
	\lim_{Y_1 \rightarrow Y_2-\beta(n-\epsilon)(iQ-2 m_A)} \mathcal{Z}^{3d}_{\text{SQCD}_U(N_c,N_f < 2N_c)}\left(\vec{X};Y_1-Y_2\right) \times
	e^{-\pi i (Y_1-Y_2) \sum_{i=1}^{N_f}X_i}
	= \nonumber\\[10pt]
	& \quad = 
	e^{-\pi i \beta\left(N_c-\frac{N_f}{2}\right) (iQ-2 m_A)\sum_{i=1}^{N_f}X_i} 
	\nonumber\\
	& \qquad\quad\times
	\prod_{j=1}^{N_c-\floor*{\frac{N_f}{2}}+n}s_b\left(\frac{iQ}{2}-\left(\frac12\pm\left(j-\frac12\right)\right) (iQ-2m_A)\right) 
	\nonumber\\
	& \qquad\quad\times
	\mathcal{Z}^{3d}_{\text{SQCD}_U\left(\floor*{\frac{N_f}{2}}-n,N_f\right)}\left(\vec{X};\beta\left(N_c-\frac{N_f}{2}\right) (iQ-2 m_A)\right) ,
\end{align}
which is given by  the good $U(\tilde N_c)$ theory with $\tilde N_c = \floor*{\frac{N_f}{2}}-n$, multiplied by the contribution of $N_c-\tilde N_c$ pairs of chirals with

\be
R^{\pm,j}_{UV}=\frac12\pm\left(j-\frac12\right)\,.
\ee
We expect these to form $N_c-\tilde N_c$ free twisted hypers in the IR. For this to happen there should be $U(1)_j$ symmetries for $j=1,\cdots ,N_c-\tilde N_c$ under which the $k$-th pair of chirals have charges $\pm \delta_{j,k}$ and these should mix (with coefficient $\left(k-\frac12\right)$)  with the UV R- and axial symmetries to give the IR ones as follows:
\begin{align}
\begin{aligned}
R^{\pm,j}_{IR} &=R^{\pm,j}_{UV}-\sum_{k=1}^{N_c-\tilde N_c} \left(k-\frac12\right) (\pm \delta_{j,k}) \,, \\
A^{\pm,j}_{IR} &= A^{\pm,j}_{UV}+2 \sum_{k= 1}^{N_c-\tilde N_c} \left(k-\frac12\right) (\pm \delta_{j,k}) \,,
\end{aligned}
\end{align}
In this way, all the chirals have $R_{IR} = \frac12$ and $A_{IR} = -1$ and form twisted hypers in the IR. The $U(1)_j$ symmetries are then enhanced to $USp(2 N_c-2 \tilde N_c)$. 
Note that the partition function vanishes if the FI condition \eqref{eq:FI cond} is not met.


\subsubsection{$N_c \leq N_f < 2 N_c$}

Lastly, we consider the theories with $N_c \leq N_f < 2 N_c$, in which case, there are $\floor*{\frac{N_f}{2}}-N_f+N_c+1$ dual frames. Of these, the $\floor*{\frac{N_f}{2}}-N_f+N_c$ which appear with a delta are treated in the same way as the previous case. Specifically, if the FI parameter $Y_1-Y_2$ saturates to the specific value
\begin{align}
Y_1-Y_2 = -\beta(n-\epsilon)(iQ-2 m_A)
\end{align}
for $n = 0, \dots, \floor*{\frac{N_f}{2}}-N_f+N_c-1$, the $S^3_b$ partition function is given by \eqref{eq:3d dual pf}.

On the other hand, if the FI parameter has a generic value other than those, the dual theory is given by the good $U(N_f-N_c)$ theory with  unbroken topological symmetry and $2 N_c-N_f$ decoupled twisted hypers \cite{Kim:2012uz,Yaakov:2013fza}. The $S^3_b$ partition function is given by
\begin{align}
	&\mathcal{Z}^{3d}_{\text{SQCD}_U(N_c,N_c \leq N_f < 2N_c)}\left(\vec{X};Y_1-Y_2\right) \times
	e^{-\pi i (Y_1-Y_2) \sum_{i=1}^{N_f}X_i} \nonumber \\
	&=e^{-\pi i (Y_2-Y_1)\sum_{i=1}^{N_f}X_i} 
	\nonumber\\
	& \times
	\prod_{j=1}^{2 N_c-N_f}s_b\left(\frac{iQ}{2}-\left(j-N_c+\frac{N_f}{2}\right)(iQ-2m_A)\pm (Y_1- Y_2)\right) 
	\nonumber\\
	& \times
	\mathcal{Z}^{3d}_{\text{SQCD}_U\left(N_f-N_c,N_f\right)}\left(\vec{X};Y_2-Y_1\right) ,
\end{align}
which was also observed in \cite{Yaakov:2013fza} assuming a generic FI parameter. Similarly, the same relation was derived for the superconformal index in \cite{Hwang:2015wna,Hwang:2017kmk}.

Note that the right hand side is given by the partition function of the good $U(N_f-N_c)$ theory multiplied by the contribution of $2 N_c-N_f$ pairs of chirals, which we expect to form $2 N_c-N_f$ free twisted hypers in the IR. \cite{Hwang:2015wna} shows that it is convenient to relabel their contribution as follows to see how they form twisted hypers:
\begin{align}\label{eq:free twisted hypers}
&\prod_{j=1}^{2 N_c-N_f}s_b\left(\frac{iQ}{2}-\left(j-N_c+\frac{N_f}{2}\right)(iQ-2m_A)\pm (Y_1- Y_2)\right) \nonumber \\
&= \prod_{j=1}^{2 N_c-N_f}s_b\left(\frac{iQ}{2}-\left(\frac12 \pm \left(j-N_c+\frac{N_f}{2}-\frac12\right)\right)(iQ-2m_A)\pm (Y_1- Y_2)\right) \,,
\end{align}
where two $\pm$ signs on the right hand side are correlated.
These correspond to $2 N_c-N_f$ pairs of chirals with UV R-charge
\be
R^{\pm,j}_{UV}=\frac12 \pm \left(j-N_c+\frac{N_f}{2}-\frac12\right)\,.
\ee
Then there is the emergent $USp(4 N_c-2 N_f)$ symmetry acting on those twisted hypers in the IR, whose Cartans $U(1)_j$ act on the $k$-th pair of chirals above with charges $\pm \delta_{j,k}$. This mixes (with coefficient $\left(j-N_c+\frac{N_f}{2}-\frac12\right)$) with the UV $R$- and axial symmetries in the following way:
\begin{align}
R^{\pm,j}_{IR} &=R^{\pm,j}_{UV}-\sum_{k= 1}^{2 N_c-N_f} \left(j-N_c+\frac{N_f}{2}-\frac12\right)  (\pm \delta_{j,k}) \,, \\
A^{\pm,j}_{IR} &= A^{\pm,j}_{UV}+2 \sum_{k= 1}^{2 N_c-N_f} \left(j-N_c+\frac{N_f}{2}-\frac12\right) (\pm \delta_{j,k})  \,.
\end{align}


Even though the mixing coefficients are different from those in the previous cases, all the chirals again have $R_{IR} = \frac12$ and $A_{IR} = -1$ and form twisted hypers in the IR. 

One may also recall that this theory has an unbroken topological $U(1)$ with associated real mass parameter $Y_1-Y_2$, which is now identified with the diagonal combination of $U(1)_j$. Namely, those twisted hypers are charged under the topological $U(1)$ symmetry of the original bad theory and mapped to its monopole operators decoupled in the IR.

 As found in \cite{Hwang:2015wna}, the precise map of the monopole operators is as follows:\footnote{The identity for this map can also be derived from the wall-crossing phenomenon of vortex world-volume quantum mechanics \cite{Hwang:2017kmk}.}
\begin{align}
\begin{aligned}
\label{eq:monopole map}
(V_i^+,V_{2 N_c-N_f-1-i}^-) \quad &\longleftrightarrow \quad \text{free twisted hypers} \,, \qquad \, i = 0, \dots, 2N_c-N_f-1 \,, \\
V_i^\pm \quad &\longleftrightarrow \quad \tilde V_{N_f-2 N_c+i}^\pm \,, \qquad \qquad \qquad i = 2 N_c-N_f, \dots, N_c-1 \,,
\end{aligned}
\end{align}
where $V_i^\pm$ is the monopole operator of the original $U(N_c)$ bad theory dressed with the adjoint scalar in the vector multiplet $i$ times, whose UV charges are given by
\begin{align}
R(V_i^\pm) &= \frac{N_f}{2}-N_c+1+i \,, \\
A(V_i^\pm) &= -N_f+2 N_c-2-2 i \,,
\end{align}
and $\tilde V_i^\pm$ is that of the dual $U(N_f-N_c)$ theory with charges
\begin{align}
R(\tilde V_i^\pm) &= -\frac{N_f}{2}+N_c+1+i \,, \\
A(\tilde V_i^\pm) &= N_f-2 N_c-2-2 i \,.
\end{align}
The charges of $V_i^\pm$ and those of $\tilde V_i^\pm$ and the free twisted hypers, read from \eqref{eq:free twisted hypers}, exactly match under the map \eqref{eq:monopole map}.

Notice that we are mapping to free hypers not only monopoles that  using the UV R-charge assignment fall below the unitarity bound, but also some that are above. Indeed the number of monopoles violating the unitarity bound is equal to the number of twisted hypers in the $n=0$ frame.
 This is non-trivial and related to the observation in the next subsection.

\subsection{Flips and a single frame duality}

Similarly to the 4d case discussed in Subsection \ref{sec:single frame duality}, we can consider the 3d SQCD with $N_c\leq N_f<2N_c$  where we flip the tower of monopoles dressed  up to $2N_c-N_f-1$ powers of the adjoint by introducing the following superpotential
\begin{align}
\sum_{i = 0}^{2 N_c-N_f-1} V_i^\pm v_i^\mp\,,
\end{align}
where $v_i^\pm$ are flipping fields.
Since  these monopoles are removed from the chiral ring, we cannot activate the VEV for them that Higgses the theory to one of  the multiple dual frames
corresponding to a particular delta constraint. We expect to have a single dual frame, the last one which has no delta. Moreover, since as we have previously discussed in this frame the monopoles that we are flipping are all mapped to singlet fields, the latter are completely integrated out. If instead we flip monopoles up to a lower level of dressing, some of the other frames  survive.

Notice also, following the comment at the end of the previous subsection, that we are not only flipping the monopoles falling below the unitarity bound, but we are also flipping monopoles which are above the bound so we are deforming the theory non-trivially with an $\mathcal{N}=2$ deformation that triggers a new RG flow. Nevertheless this deformed theory flows in the IR to the good $\mathcal{N}=4$ SQCD with gauge group $U(N_f-N_c)$ and $N_f$ flavors.

At the level of the partition function we can easily check that the extra flipping fields we introduced, when evaluated on the delta constraint in each frame, produce a zero of second order which both cancels the divergence of the delta and  kills completely the contribution of that frame. In the end only the last frame survives. Equivalently, we can think of taking the 3d limit of the identity \eqref{eq:4dflipped} for the flipped 4d SQCD. The 3d partition function identity for the single frame duality of the flipped bad SQCD is
\begin{align}
	&\mathcal{Z}^{3d}_{\text{SQCD}_U(N_c,N_c \leq N_f < 2N_c)}\left(\vec{X};Y_1-Y_2\right) \times
	e^{-\pi i (Y_1-Y_2) \sum_{i=1}^{N_f}X_i} \nonumber \\
	& \times
	\prod_{j=1}^{2 N_c-N_f}s_b\left(-\frac{iQ}{2}+\left(\frac12 \pm \left(j-N_c+\frac{N_f}{2}-\frac12\right)\right)(iQ-2m_A)\pm (Y_1- Y_2)\right)
	\nonumber\\
	&=e^{-\pi i (Y_2-Y_1)\sum_{i=1}^{N_f}X_i} \times \mathcal{Z}^{3d}_{\text{SQCD}_U\left(N_f-N_c,N_f\right)}\left(\vec{X};Y_2-Y_1\right) \,.
\end{align}

\section{Semi-classical analysis in 3d and 4d}
\label{sec:semi-classical}

\subsection{Analysis of the equations of motion in 3d}
\label{subsec:3dN=4eom}

Let us now see how to recover the various frames for the 3d theory at the classical level using just its equations of motion. The analysis is based on the results presented in \cite{Argyres:1996eh,deBoer:1996mp}. We start by considering the moduli space of the 3d theory with gauge group $U(N_c)$. For 3d $\mathcal N=4$ theories with $U(1)$ factors, one can turn on three FI parameters for each $U(1)$, forming a triplet of the $SU(2)_H \in SU(2)_H \times SU(2)_C$ R-symmetry. Sometimes it is convenient to decompose this triplet into one complex FI and one real FI, which appear in the F-term and D-term equations, respectively, when expressed in terms of the $\mathcal N=2$ subgroup. With a complex FI parameter $\lambda$ and a real FI parameter $\zeta$, 
the F- and D-terms read\footnote{Together with \eqref{f2}, the equation \eqref{f3} is equivalent to the following condition
\begin{gather}
[\Phi,\Phi^\dagger] = 0 \,, \\
Q Q^\dagger-\widetilde Q^\dagger \widetilde Q=\zeta \, \mathbb{I}_{N_c} \,, \\
Q^\dagger \Phi = 0 \,, \quad \Phi \widetilde Q^\dagger = 0 \,.
\end{gather}
which can be seen by squaring \eqref{f3}.}
\be
Q \widetilde Q=\lambda \mathbb{I}_{N_c} \,,
\label{f1}
\ee
\be
\widetilde Q \Phi=0 \,, \quad \Phi Q=0 \,,
\label{f2}
\ee
\be
[\Phi,\Phi^\dagger]+Q Q^\dagger-\widetilde Q^\dagger \widetilde Q=\zeta \, \mathbb{I}_{N_c} \,,
\label{f3}
\ee
where numerical coefficients, which may depend on the gauge coupling, are omitted for simplicity. Here each field represents a matrix. Specifically, $Q$ is an $N_c \times N_f$ matrix with color index $i = 1, \dots, N_c$ and flavor index $a = 1,\dots, N_f$. $\widetilde Q$ is an $N_f \times N_c$ matrix with flavor index $a = 1,\dots, N_f$ and color index $i = 1, \dots, N_c$. Lastly, $\Phi$ is an $N_c \times N_c$ matrix with two color indices $i,j = 1, \dots, N_c$. In addition, the standard covariant derivative term includes the effective real mass for matter fields, giving additional vacuum equations
\begin{align}
\label{eq:mass}
\Sigma Q = 0 \,, \qquad -\Sigma \widetilde Q = 0 \,, \qquad [\Sigma,\Phi] = 0 \,,
\end{align}
where $\Sigma$ is an $N_c \times N_c$ matrix representing the real scalar in the vector multiplet.

Let us consider first the case of a good theory $N_f\geq 2N_c$. Using an $SU(2)_H$ rotation, one can align the FI parameters in a suitable way, e.g.~$\lambda \neq 0$ and $\zeta = 0$, which is a convenient choice to uplift to the 4d case that we will discuss in Subsection \ref{subsec:4deom}.
The VEV of the matter fields which parametrize the Higgs branch 
and solve the e.o.m.~\eqref{f1}-\eqref{f3}   can be put with a gauge and flavor rotation in the following form: 
\begin{align}
Q =
\left (
\begin{array}{ccc|c|c}
q_1 & & & & \\
& \ddots & & \,\,\,0\,\,\,\, & \,\,\,0\,\,\, \\
\undermat{N_c}{& \,\,\,\,\,\, & q_{N_c} &} \undermat{N_c}{\,\,\,\,\,\,\,\,\,\,\,\,\,\,\,\,\,\,\,\,\,} & \undermat{N_f-2N_c}{\,\,\,\,\,\,\,\,\,\,\,\,\,\,\,\,\,\,\,\,} \\
\end{array}
\right ) 
\,, \qquad
\widetilde{Q}^T =
\left (
\begin{array}{ccc|ccc|c}
\tilde{q}_1 & & & \alpha_1 & & & \\
& \ddots & & & \ddots & & 0 \\
\undermat{N_c}{& \,\,\,\,\,\, & \tilde{q}_{N_c} &} \undermat{N_c}{\,\,\,\,\,\,& \,\,\,\,\, & \alpha_{N_c} &} \undermat{N_f-2N_c}{\,\,\,\,\,\,\,\,\,\,\,\,\,\,\,\,\,\,\,\,} \\
\end{array}
\right ) \,, \label{quarkvev1} \\
\nonumber
\end{align}
with\footnote{Strictly speaking the second equation reads $|q_i|^2=|\tilde{q}_i|^2+|\alpha_i|^2$ however, using color and flavor rotations, we can set the phase of all the parameters to the same value and therefore we can drop the modulus.} 
\be\label{eom} 
\left\lbrace\begin{array}{l}
\tilde{q}_iq_i=\lambda\;\;\forall i\,.\\ 
q_i^2=\tilde{q}_i^2+\alpha_i^2 \;\;\forall i.\\ 
\end{array}\right.
\ee
To have non-zero $Q$ and $\widetilde Q$, equations \eqref{f2} and \eqref{eq:mass} demand $\Phi = \Sigma = 0$.


If we now turn off the FI parameter, we can set to zero all $\tilde{q}_i$ coefficients and the solution reduces to 
\begin{align}
Q =
\left (
\begin{array}{ccc|c|c}
q_1 & & & & \\
& \ddots & & \,\,\,0\,\,\,\, & \,\,\,0\,\,\, \\
\undermat{N_c}{& \,\,\,\,\,\,\,\,\,\,\,& q_{N_c} } & \undermat{N_c}{\,\,\,\,\,\,\,\,\,\,\,\,\,\,\,\,\,\,\,\,} & \undermat{N_f-2N_c}{\,\,\,\,\,\,\,\,\,\,\,\,\,\,\,\,\,\,\,\,} \\
\end{array}
\right ) 
\,, \qquad
\widetilde{Q}^T =
\left (
\begin{array}{c|ccc|c}
& q_1 & & & \\
0\,\,\,& & \ddots & & 0 \\
\undermat{N_c}{\,\,\,\,\,\,\,\,\,\,\,\,\,\,\,\,\,\,} & \undermat{N_c}{\,\,\,\,\,\,& \,\,\,\,\,\, & q_{N_c} } &\undermat{N_f-2N_c}{\,\,\,\,\,\,\,\,\,\,\,\,\,\,\,\,\,\,\,\,} \\
\end{array}
\right ) \,. \label{quarkvev2} \\
\nonumber
\end{align}
We can also set to zero some of the $q_i$ coefficients, but this is just a special case of \eqref{quarkvev2}. In general the VEV we have just described breaks the gauge group completely, indicating that we are in the Higgs phase as expected. 

If we now try to move to the root of the Higgs branch (where Higgs and Coulomb branches intersect), which can be implemented by setting to zero all the $q_i$ parameters in \eqref{quarkvev2}, we find that the full $U(N_c)$ gauge symmetry is restored, indicating that we are at the origin of the moduli space where the SCFT lives. 
From the e.o.m.~\eqref{f2} we also see that  if we are on the HB with VEVs parameterized by \eqref{quarkvev2} we can't turn on VEVs for $\Phi$ and $\Sigma$.

So for the good theories we see the following familiar facts:
\begin{itemize}
\item At a generic point of the HB $\Phi$ and $\Sigma$ cannot take the VEV, so there is no residual CB direction.
\item At the origin of the HB the gauge group is unbroken and equal to $U(N_c)$.
\end{itemize}

Let us now see how this picture changes in the bad case $N_c<N_f<2N_c$. The solution \eqref{quarkvev1} with the FI parameter turned on now reads 
\begin{align}
Q  =
\left (
\begin{array}{ccccc|c}
q_1 & & & & & \\
& \ddots & & & & \\
& & \ddots & & & 0 \\
& & & \ddots & & \\
\undermat{N_c}{\,\,\,\,\,\,& \,\,\,\,\,\,& \,\,\,\,\,\,\,\,\,\,\,\,\,\,\,& \,\,\,\,\,\, & q_{N_c}} & \undermat{N_f-N_c}{\,\,\,\,\,\,\,\,\,\,\,\,\,\,\,\,\,} \\
\end{array}
\right ), \quad
\widetilde{Q}^T =
\left (
\begin{array}{ccccc|ccc}
\tilde{q}_1 & & & & & \alpha_1 & & \\
& \ddots & & & & & \ddots & \\
& & \tilde{q}_{N_f-N_c} & & & & & \alpha_{N_f-N_c} \\
& & & \ddots & & & \,\,\,\,0 & \\
\undermat{N_c}{\,\,\,\,\,\,& \,\,\,\,\,\,& \,\,\,\,\,\,\,\,\,\,\,\, \,\,\,\,\,\,\,\,\,\,& \,\,\,\,\,\, &\tilde{q}_{N_c}} & \undermat{N_f-N_c}{\,\,\,\,\,\,&\,\,\,\,\,\,&\,\,\,\,\,\,\,\,\,\,\,\,\,\,\,\,\,\,\,} \\
\end{array}
\right ) \,. \nonumber \\
\nonumber \\
\label{quarkvev3}
\end{align}
When we switch off the FI parameter this reduces, modulo a flavor rotation, to 
\begin{align}
Q =
\left (
\begin{array}{ccc|c|c}
q_1 & & & & \\
& \ddots & & & \\
& & q_{N_f-N_c} & 0 & \,0 \\
\undermat{N_f-N_c}{\,\,\,\,\,\,& \,\,\,\,\, \,\,\,\,\,0 & \,\,\,\,\,\,\,\,\,\,\,\,\,\,\,\,\,& } \undermat{N_f-N_c}{\,\,\,\,\,\,\,\,\,\,\,\,\,\,\,\,\,\,\,\,\,\,\,\,} & \undermat{2N_c-N_f}{\,\,\,\,\,\,\,\,\,\,\,\,\,\,\,\,\,\,\,\,} \\
\end{array}
\right )
\,, \quad
\widetilde{Q}^T =
\left (
\begin{array}{c|ccc|c}
& q_1 & & & \\
0\,\, & & \ddots & & 0 \\
& & & q_{N_f-N_c} & \\
\undermat{N_f-N_c}{\,\,\,\,\,\,\,\,\,\,\,\,\,\,\,\,\,\,\,\,\,\,\,\,} & \undermat{N_f-N_c}{\,\,\,\,\,\,& \,\,\,\,\, 0 &\,\,\,\,\,\,\,\,\,\,\,\,\,\,\,\,\,\,\,} & \undermat{2N_c-N_f}{\,\,\,\,\,\,\,\,\,\,\,\,\,\,\,\,\,\,\,\,\,\,\,\,} \\
\end{array}
\right ) \,. \nonumber \\
\nonumber \\
\label{quarkvev4}
\end{align}

To begin with we notice that if we take the first $N_f-N_c$ rows of the matrices in eq.~\eqref{quarkvev4} they coincide with the VEVs (with FI turned off) parameterizing the HB of good theory in \eqref{quarkvev2} with the replacement $N_c\to N_f-N_c$.

We also notice that, contrary to the good case, we can now turn on VEV for the adjoint chiral $\Phi$ and the real scalar $\Sigma$,
at a generic point on the HB (that is without turning off the coefficients $q_i$).
More precisely, with a non-zero VEV for the quarks of the form \eqref{quarkvev4}, the e.o.m.~\eqref{f1}, \eqref{f2}, \eqref{f3} allow VEVs of the form 
\begin{align}
\Phi &=\text{Diag}(0,\dots,0,\phi_1,\dots,\phi_{2N_c-N_f}) \,, \label{vectvev} \\
\Sigma &= \text{Diag}(0,\dots,0,\sigma_1,\dots,\sigma_{2N_c-N_f}) \,. \label{vectvev_sigma}
\end{align}
So  contrary to the good case at each point in the HB there is a residual CB direction.
Note that the abelian real scalars $\sigma_i$ together with the dualized photons $a_i$ can combine into monopole chiral operators \cite{deBoer:1997kr,Aharony:1997bx}
\begin{align}
\label{eq:monopole}
V^\pm_i \sim \exp\left[\pm\left(2 \pi \sigma_i/e_\text{eff}^2+i a_i\right)\right] \,,
\end{align}
which are the proper coordinates of the Coulomb branch near the fixed point in the moduli space.

We also see that, contrary to the good case,  as we move to the root of the Higgs branch by sending $q_i$ to zero in \eqref{quarkvev4}, we do not land on the origin of the moduli space, but rather on a $(2N_c-N_f)$-dimensional submanifold of the Coulomb branch. As the $q_i$ parameters go to zero a $U(N_f-N_c)$ subgroup of the gauge group becomes unbroken, leading to the conclusion that on each point of the submanifold the low-energy effective theory has a $U(N_f-N_c)$ gauge group. Here we recognize our last frame with no delta distribution associated.

Another important difference with respect to the good case is that we can now find extra solutions for the squark VEVs besides \eqref{quarkvev4} when the FI parameter is set zero from the beginning (so we are looking at a set of  solutions which are  not connected to the $\lambda\neq 0$ case). We can indeed observe that for any $0\leq r\leq \lfloor N_f/2\rfloor$ the VEVs 
\begin{align}
Q =
\left (
\begin{array}{ccc|c|c}
q_1 & & & & \\
& \ddots & & & \\
& & q_{r} & 0 & \,0 \\
\undermat{r}{\,\,\,& \,\,\,\,\,0\,\,\,\,\, & & } \undermat{r}{\,\,\,\,\,\,\,\,\,\,\,\,\,\,\,\,\,\,\,\,\,\,\,\,} & \undermat{N_f-2r}{\,\,\,\,\,\,\,\,\,\,\,\,\,\,\,\,\,\,\,\,} \\
\end{array}
\right )
\,,\qquad
\widetilde{Q}^T =
\left (
\begin{array}{c|ccc|c}
& q_1 & & & \\
0\,\, & & \ddots & & 0 \\
& & & q_{r} & \\
\undermat{r}{\,\,\,\,\,\,\,\,\,\,\,\,\,\,\,\,\,\,\,\,\,\,\,\,} & \undermat{r}{& \,\, \,\, 0 \,\,\,\,\, &\,\,\,\,\,\,} & \undermat{N_f-2r}{\,\,\,\,\,\,\,\,\,\,\,\,\,\,\,\,\,\,\,\,\,\,\,\,} \\
\end{array}
\right ) 
\label{quarkvev5} 
\nonumber\\[20pt]
\end{align}

\noindent do solve the equations of motion. While for $r\leq N_f-N_c$ \eqref{quarkvev5} is just a special case of \eqref{quarkvev4}, when $r$ lies in the range 
$N_f-N_c<r\leq \lfloor N_f/2\rfloor$ we find new solutions corresponding to the other frames we have discussed in the previous sections. Again we are allowed to turn on a VEV for the adjoint chiral $\Phi$ of the form 
\be\label{vectvev2} \Phi=\text{Diag}(0,\dots,0,\phi_1,\dots,\phi_{N_c-r})\ee
and a VEV for the real scalar $\Sigma$ in the vector multiplet of the form
\begin{align}
\label{eq:Sigma}
\Sigma = \text{Diag}(0,\dots,0,\sigma_1,\dots,\sigma_{N_c-r}) \,.
\end{align}
We can argue that at the corresponding root of the Higgs branch (when we send in \eqref{quarkvev5} all the $q_i$’s to zero) the unbroken gauge group is $U(r)$. The analysis of the partition function precisely reproduces this structure with multiple frames.

Notice that in the bad case, because of the residual CB  at each point of the HB, when we go to the origin of the HB the gauge group is still broken, and there is no point in the moduli space where the full $U(N_c)$ is preserved.\footnote{This is a quantum
statement, the classical analysis we just performed cannot probe the regions of small VEVs for  $\Phi$ and $\Sigma$ (basically turning off the VEVs in \eqref{vectvev} and \eqref{vectvev2}).}

At this stage it is worth pausing for a moment and discuss the relation between the above classical discussion and the partition function argument we are proposing. One simple difference between the two approaches is that in the classical analysis presented above we have characterized the various frames, for vanishing squark VEV, using the expectation value for the three adjoint scalars in the $\mathcal{N}=4$ vector multiplet as in \eqref{vectvev}-\eqref{vectvev_sigma} and \eqref{vectvev2}-\eqref{eq:Sigma} whereas in Section \ref{sec:3danalysis} we have considered VEVs for monopole operators. 

As is well known (see \cite{Gaiotto:2008ak, Assel:2017jgo}), at a generic point of the Coulomb branch the theory abelianizes and locally in a neighborhood of each point we can parametrize the Coulomb branch using the VEV of the three of real scalars in the abelian vector multiplets and the dual photons. The entire Coulomb branch of the non-abelian theory is then parametrized by considering Weyl invariant combinations of these fields. The three real scalars transform as triplets of the $SU(2)_C$ factor of the R-symmetry, while the dual photons are uncharged. We can then describe the Coulomb branch in a fixed complex structure by combining two out of the three scalars into a complex field and pairing up the third with the dual photon to form the (bare) monopole operators. As remarked in \cite{Gaiotto:2008ak}, in doing this we are implicitly selecting a specific $U(1)$ subgroup of $SU(2)_C$ (corresponding to a choice of $\mathcal{N}=2$ subalgebra) under which the complex field has charge 1. Once such a choice has been made, only the $U(1)$ subgroup is a manifest symmetry and indeed all such choices are physically equivalent up to an $SU(2)_C$ transformation. In that sense, the VEVs of monopole operators discussed in Section \ref{sec:higgs} can be regarded as the quantum description of the Coulomb branch parameterized by the adjoints scalars in the vector multiplet considered in this section. 

Notice also that both descriptions are consistent with the analysis of \cite{Assel:2017jgo}, which shows that the locus on the Coulomb branch corresponding to the frame with a $U(r)$ gauge group, which is isomorphic to the Coulomb branch of a $U(N_c-r)$ SQCD with $N_f-2r$ hypers, is characterized by a vanishing VEV for all Casimirs of the complex field $\Tr\Phi^{k}$ of degree $k>N_c-r$ and for all monopoles dressed more that $N_c-r-1$ times. In \eqref{vectvev2} the complex field has only $N_c-r$ non-vanishing components, saying that Casimirs of order higher than $N_c-r$ are not independent chiral ring operators and in our partition function analysis we enforce the condition that the monopole dressed $N_c-r-1$ times has a non zero VEV.

Lastly, let us comment on the solution with a non-zero real FI parameter, which is useful for the comparison with the $\mathcal N=2^*$ case discussed in the next subsection. While we have assumed  vanishing real FI, $\zeta = 0$, one can rotate $\lambda$ to $\zeta$ using the $SU(2)_H$ rotation so that we have $\lambda = 0$ and $\zeta \neq 0$. In that case, the VEVs of $Q$ and $\widetilde Q$ in \eqref{quarkvev3} are deformed to
\begin{align}
Q  =
\left(
\begin{array}{cccccc|ccc}
q_1 & & & & & & & & \\
& \ddots & & & & & & & \\
& & q_{N_f-N_c} & & & & & 0 & \\
& & & \sqrt{\zeta} & & & & & \\
& & & & \ddots & & & & \\
\undermat{N_c}{\,\,\,\,\,\,&\,\,\,\,\,\,\,\,&\,\,\,\,\,\,\,\,\,\,\,\,\,\,\,\,\,\,\,&\,\,\,\,\,\,\,\,\,&\,\,\,\,\,\,\,\,& \sqrt{\zeta}} & \undermat{N_f-N_c}{\,\,\,\,\,\,\,&\,\,\,\,&\,\,\,\,\,\,\,}
\end{array}
\right), \quad
\widetilde{Q}^T =
\left(
\begin{array}{cccccc|ccc}
& & & & & & \alpha_1 & & \\
& & & & & & & \ddots & \\
& & 0 & & & & & & \alpha_{N_f-N_c} \\
& & & & & & & & \\
& & & & & & & & \\
\undermat{N_c}{\,\,\,&\,\,\,\,&\,\,\,\,\,\,\,\,\,\,&\,\,\,\,\,&\,\,\,\,&\,\,\,\,\,} & \undermat{N_f-N_c}{\,\,\,\,\,\,\,&\,\,\,\,\,\,\,\,&\,\,\,\,\,\,\,\,\,\,\,\,\,\,\,\,\,\,\,\,\,}
\end{array}
\right) \,. \nonumber \\
\nonumber \\
\label{eq:vev with real FI}
\end{align}
with
\begin{align}
q_i^2-\alpha_i^2 = \zeta \,,
\end{align}
which is a physically equivalent solution to \eqref{quarkvev3}.

\subsubsection{Analysis with the $\mathcal N=2^*$ deformation}
\label{sec:N=2}
In the previous subsection, we considered two types of vacuum solutions: \eqref{eq:vev with real FI}, or equivalently \eqref{quarkvev3}, parametrizing the Higgs branch of the moduli space and \eqref{quarkvev5} parametrizing the mixed branch, together with the Coulomb moduli \eqref{vectvev2} and \eqref{eq:Sigma}. While \eqref{eq:vev with real FI} exists for an arbitrary value of the FI parameter, the solution \eqref{quarkvev5} is allowed only for  vanishing FI. In particular the latter solution, labeled by the integer $r \leq \floor*{\frac{N_f}{2}}$, 
consists of two parts: the Higgs part and the Coulomb part. Looking at the origin of the Higgs branch part, the effective theory is the $U(r)$ theory with $N_c-r$ free twisted hypers obtained by dualizing the Coulomb moduli, which is indeed consistent with the structure of the multiple duals found in the 4d index and 3d partition function analysis.

In the partition function analysis, we have seen that those multiple duals are actually distinguished by different values of the FI parameter. In this subsection, we demonstrate that such FI conditions can be derived from the $\mathcal N = 2$ deformation of the moduli space by turning on a real mass for the $U(1)_A$ symmetry defined by the charge assignment in Table \ref{tab:rep}.
\begin{table}[bp]
\centering
\begin{tabular}{c|c|c|c}
\hline
 & $U(N_c)$ & $SU(N_f)$ & $U(1)_A$ \\
\hline
$Q$ & $\mathbf{N_c}$ & $\overline{\mathbf{N_f}}$ & 1 \\
$\widetilde Q$ & $\overline{\mathbf{N_c}}$ & $\mathbf{N_f}$ & 1 \\
$\Phi$ & $\mathbf{adj}$ & $\mathbf{1}$ & $-2$ \\
\hline
\end{tabular}
\caption{\label{tab:rep} The representations of the matter fields.}
\end{table}
We will see that when we turn on the $U(1)_A$ real mass, the FI parameter is quantum corrected because the theory now has only four supercharges, and
a non-compact Coulomb direction of the mixed branch  can exist only when the \emph{effective} FI parameter, rather than the bare one, becomes zero. In particular we will show that  the condition for having a vanishing effective FI is exactly the delta distribution condition associated with each dual frame in the partition function analysis.

Let us discuss how the moduli space is deformed when we turn on real mass for $U(1)_A$.
The representations of the fields are summarized in Table \ref{tab:rep}.
First note that the F-term conditions remain the same
\begin{align}
\label{eq:F-term}
Q \widetilde Q= 0 \,, \qquad \Phi Q = 0 \,, \qquad \widetilde Q \Phi = 0 \,,
\end{align}
where the matrix indices are all suppressed. 
The D-term condition, on the other hand, is given by the following matrix equation (see for example \cite{Intriligator:2013lca}\footnote{Although \cite{Intriligator:2013lca} considers the fundamental representation only, the contribution of the adjoint field can easily be added.})
\begin{align}
\label{eq:D-term}
[\Phi,\Phi^\dagger]+Q Q^\dagger{}-\widetilde Q^\dagger \widetilde Q-\frac{\zeta_\text{eff}}{2 \pi} \, \mathbb I_{N_c}-\frac{k_\text{eff}}{2\pi} \, \Sigma = 0 \,.
\end{align}
Since now only the $\mathcal N=2$ subgroup is preserved, there can be quantum corrections to FI and CS terms. Namely, $\zeta_\text{eff}$ is the effective FI for the $U(1)$ factor inside the $U(N_c)$ gauge group
 given by
\begin{align}
\zeta_\text{eff} &= \zeta+\frac12 \sum_{\alpha} n_\alpha m_\alpha \, \mathrm{sgn}(m_\alpha+n_\alpha \sigma) \,,
\end{align}
where $\zeta$ is a bare FI parameter and $\sigma$ is the real scalar in the $U(1)$ vector multiplet. $\alpha$ labels each charged field with charge $n_\alpha$ and effective real mass $m_\alpha+n_\alpha \sigma$. In addition, even if a bare CS coupling is absent, given various real masses, the effective IR theory may have an effective CS term, which is given by
\begin{align}
\label{eq:CS}
k_\text{eff} &= \frac12 \sum_\alpha T_2(R_\alpha) \, \mathrm{sgn}(m_\alpha(\Sigma)) \,,
\end{align}
where $\alpha$ labels each charged field with effective real mass $m_\alpha(\Sigma)$, whose explicit form depends on the representation, and in the representation $R_\alpha$ with quadratic index $T_2(R_\alpha)$ normalized such that $T_2(\text{Fund}) = 1$ for the fundamental representation. For an abelian  gauge group the effective CS simplifies to
\begin{align}
k_\text{eff} = \frac12 \sum_{\alpha} n_\alpha^2 \, \mathrm{sgn}(m_\alpha+n_\alpha \sigma) \,.
\end{align}

Let us comment on possible types of solution to \eqref{eq:D-term} \cite{Intriligator:2013lca}. Firstly, for the $U(1)$ part, we have three types of solutions as follows.
\begin{itemize}
\item There are Higgs vacua parameterized by the VEVs of matter fields satisfying the vacuum equations, which break the $U(1)$ gauge group. Especially, if $\zeta_\text{eff}+k_\text{eff} \, \sigma \neq 0$, some of the matter fields must have non-zero VEVs to satisfy the D-term equation.
\item If $\zeta_\text{eff}+k_\text{eff} \, \sigma = 0$ with $k_\text{eff} \neq 0$, there is a topological vacuum at $\sigma = -\frac{\zeta_\text{eff}}{k_\text{eff}}$, described by the low-energy $U(1)_{k_\text{eff}}$ CS theory where all the matter fields become massive and are integrated out.
\item If $\zeta_\text{eff} = k_\text{eff} = 0$, usually for a particular range of $\sigma$, there are a continuous set of solutions parameterized by $\sigma$ in this range with vanishing VEV of the matter fields, called the Coulomb branch of vacua.
\end{itemize}

For the $SU(N)$ part, the types of solutions are similar, but one has to take into account some additional effects, such as the monopole instanton effect. Firstly, as in the $U(1)$ case, if matter fields can have non-zero VEVs, there are Higgs vacua parametrized by these VEVs, breaking the $SU(N)$ gauge group; on the other hand, if matter fields do not have VEVs, $\Sigma$ is such that $\zeta_\text{eff} \, \mathbb I_{N_c}+k_\text{eff} \, \Sigma = 0$ to satisfy the D-term equation. If $k_\text{eff} \neq 0$, there is a topological vacuum whose effective low energy theory is a Chern-Simons theory as in the $U(1)$ case. If $k_\text{eff} = 0$, there can be Coulomb branches parameterized by the Cartan part of $\Sigma$, breaking $SU(N)$ into $U(1)^{N-1}$. The effective low energy theory on the Coulomb branches is then governed by the above $U(1)$ solution. In this case, one should take into account quantum corrections due to monopole instanton superpotentials, lifting (part of) semi-classical Coulomb branches. Moreover, since $SU(N)$ can be partially broken, the combination of theses solutions, called mixed branches, are also allowed.

Lastly, the conditions from the mass terms are given by\footnote{These equations are the generalization of \eqref{eq:mass} in the presence of non-trivial real masses, which can be understood as the real scalar components of background vector multiplets for the corresponding global symmetries.}
\begin{align}
\begin{aligned}
\label{eq:mass*}
\left[(\mathsf m_A-m_a) \, \mathbb I_{N_c}+\Sigma\right] Q_a &= 0 \qquad \text{for each $a$}, \\
\left[(\mathsf m_A+m_a) \, \mathbb I_{N_c}-\Sigma\right] \widetilde Q^a &= 0 \qquad \text{for each $a$}, \\
-2 \mathsf m_A \Phi+[\Sigma,\Phi] &= 0 \,,
\end{aligned}
\end{align}
where $\mathsf m_A$ is the real mass for $U(1)_A$ and $m_a$ is the real mass for (the Cartan part of) the flavor $SU(N_f)$ symmetry. As usual, one can diagonalize $\Sigma$ using a gauge rotation so that $\Sigma$ can be written as
\begin{align}
\Sigma = \mathrm{Diag} \left(\sigma_1, \dots, \sigma_{N_c}\right) .
\end{align}
Then, component-wise, equation \eqref{eq:mass*} can be written as
\begin{align}
\begin{aligned}
 \label{eq:mass**}
M_{\text{eff}}(Q^i{}_a)&\equiv\left(\mathsf m_A-m_a+\sigma_i\right) Q^i{}_a = 0 \,, \\
M_{\text{eff}}(\widetilde{Q}^a{}_i)&\equiv\left(\mathsf m_A+m_a-\sigma_i\right) \widetilde Q^a{}_i = 0 \,, \\
M_{\text{eff}}(\Phi^i{}_j)&\equiv\left(-2 \mathsf m_A+\sigma_i-\sigma_j\right) \Phi^i{}_j = 0 \,.
\end{aligned}
\end{align}

Here, we are interested in how the mixed branch solution \eqref{quarkvev5} of the $\mathcal N=4$ case is deformed when the $U(1)_A$ real mass is turned on. For non-zero $\mathsf m_A$, the mixed branch solution is given by
\begin{align}
Q = \left(\begin{array}{ccc|c}
q & & &  \\
& \ddots & & \, \\
& & q & 0 \\
& & & \\
\undermat{r}{\,\,\,\,\,\,&\,\,\,\,\,\,\,\,\,\,\,&\,\,\,\,\,\,} & \undermat{N_f-r}{\,\,\,\,\,\,\,\,\,\,\,\,\,\,\,\,\,\,\,\,\,\,\,\,\,\,\,\,\,} \\
\end{array}\right)
\end{align}
and
\begin{align}
\label{eq:Sigma*}
\Sigma = \mathrm{Diag} \left(\undermat{r}{-\mathsf m_A, \dots, -\mathsf m_A},\undermat{N_c-r}{\sigma, \dots, \sigma}\right) , \\
\nonumber
\end{align}
where $\Sigma$ breaks the $U(N_c)$ gauge group into $U(r) \times U(N_c-r)$ with two separate effective FI parameters. The parameter $q$ is now  fixed by the D-term equation for the $U(r)$ part. On the other hand, the D-term equation for the $U(N_c-r)$ part is satisfied only when $\zeta_\text{eff}+k_\text{eff} \, \Sigma = 0$ in this sector because the components of $Q$ charged under $U(N_c-r)$ do not have the VEV. Especially, if both $\zeta_\text{eff}$ and $k_\text{eff}$ vanish for a certain range of $\sigma$, we have a continuous set of  Coulomb branch solutions parametrized by $\sigma$ in this range. We will see that this vanishing condition exactly gives the constraint on the FI parameter found in the partition function analysis in Section \ref{sec:3danalysis}. Note that here we take $\sigma_{r+1} = \dots = \sigma_{N_c} = \sigma$, whereas, in principle, $\sigma_i$ can be all different, further breaking $U(N_c-r)$ into $U(1)^{N_c-r}$. We comment on this case shortly.

To evaluate $\zeta_\text{eff}+k_\text{eff} \, \Sigma$ for the $U(N_c-r)$ part, first we notice that the $SU(N_c-r)$ part and the diagonal $U(1)$ part receive different quantum corrections because the adjoint field is only charged under $SU(N_c-r)$ and neutral under $U(1)$. Since the effective real mass of the adjoint field $\Phi$ restricted to the $U(N_c-r)$ part is
\begin{align}
M_{\text{eff}}(\Phi) &= -2 \mathsf m_A \,,
\end{align}
it generates a CS term of level
\begin{align}
-\mathrm{sgn}(\mathsf m_A) (N_c-r)
\end{align}
only for $SU(N_c-r)$ after integrated out, as explained in \eqref{eq:CS}. On the other hand, it doesn't yield any shift of the FI parameter since it is neutral under the diagonal $U(1)$. Next, the corrections to $\zeta_\text{eff}+k_\text{eff} \, \Sigma$
for $U(N_c-r)$ from the other fields read as follows:
\begin{align}
\label{eq:F}
\frac12 N_f |\mathsf m_A+\sigma|-\frac12 N_f |\mathsf m_A-\sigma| + \frac12 \sum_{j = 1}^r \Big(|-2 \mathsf m_A+\sigma-\sigma_j|-|-2 \mathsf m_A+\sigma_j-\sigma|\Big)
\end{align}
with $\sigma_j = -\mathsf m_A$ for $j = 1, \dots, r$, as given in \eqref{eq:Sigma*}. Each term is contributed by the fundamental, anti-fundamental, and the components of the original $U(N_c)$ adjoint field that are now in the bifundamental representation of $U(r) \times U(N_c-r)$. \eqref{eq:F} is a piece-wise constant function in $\sigma$. Assuming $|\sigma| \gg |\mathsf m_A|$ because we are interested in a non-compact Coulomb branch, one can evaluate \eqref{eq:F} as follows:
\begin{align}
\mathrm{sgn}(\sigma) (N_f-2 r) \mathsf m_A \,,
\end{align}
indicating the one-loop correction to the FI parameter and the absence of the contribution to the CS level.

In conclusion, we find that the 
effective theory in the IR is the product of an $\mathcal N = 2$ $U(1)$ Maxwell theory with FI parameter
\begin{align}
\zeta+\mathrm{sgn}(\sigma) (N_f-2 r) \mathsf m_A
\end{align}
and an $\mathcal N=2$ $SU(N_c-r)$ CS theory with level $-(N_c-r) \, \mathrm{sgn}(\mathsf m_A)$.

The latter CS theory has a unique vacuum, and the Coulomb directions parameterized by distinct $\sigma_i$ are all lifted \cite{Intriligator:2013lca,Witten:1999ds}, at least locally, which is why we focus on the case of equal $\sigma_i$.
Most importantly, the diagonal $\mathcal N = 2$ $U(1)$ Maxwell  theory has a supersymmetric vacuum, a non-compact Coulomb branch in this case, only when $\zeta_\text{eff}$ vanishes.

This condition is exactly the delta constraint associated with the dual $U(r)$ theory with $\beta = -\mathrm{sgn}(\sigma)$ found in the partition function analysis (see \eqref{eq:3d pf}) once we identify\footnote{The shift by $-\frac{iQ}{2}$ appears due to our choice of the UV R-symmetry in the partition function computation.}
\begin{align}
\zeta &= Y_1-Y_2 \,, \\
\mathsf m_A &= m_A-\frac{iQ}{2} \,.
\end{align}
Notice also that this $U(1)$ vector multiplet can be dualized into a chiral multiplet, which corresponds to the chiral with vanishing charges represented by a delta distribution in the partition function.
While we have considered the case with vanishing flavor masses, one can also analyze the moduli space with other choices of the flavor masses, which gives the same FI condition to have the mixed branch with a non-compact Coulomb direction. 

In  Appendix \ref{app:N=2} we will discuss a different choice of  flavors masses which contains also  non-compact Higgs branch solutions   from which we can for example extract the bound of $r \leq \floor*{\frac{N_f}{2}}$.

\subsection{Analysis of the equations of motion in 4d}
\label{subsec:4deom}

The properties we have found for the 3d $\mathcal N=4$ theory in Subsection \ref{subsec:3dN=4eom} can also be recovered in the 4d uplifted theory with gauge group $USp(2N_c)$. In this case we have a single squark matrix which is $2N_c\times 2N_f$ and we can set its VEV to be of the form 
\be
Q=\left(\begin{array}{c|c}
\begin{array}{c|c}B&0\\\end{array}&\\ 
\hline 
&\begin{array}{c|c} C & D\\\end{array}\\
\end{array}\right) \,,
\ee 
where $B$ and $C$ are $N_c\times N_c$ diagonal matrices with eigenvalues $b_i$ and $c_i$ respectively. The matrix $D$ is $N_c\times (N_f-N_c)$ and can be taken to be in diagonal form with eigenvalues $d_i$. In the good case we have $N_c$ different eigenvalues and $N_f-N_c$ in the bad case, when $N_f<2N_c$. The eigenvalues are subject to the constraints 
\be\label{eom2} 
\left\lbrace\begin{array}{l}
c_ib_i=\lambda\;\;\forall i\,,\\ 
b_i^2=c_i^2+d_i^2 \;\;\forall i\,,\\ 
\end{array}\right.
\ee 
imposed by F and D-terms. In \eqref{eom2} $\lambda$ is the 4d uplift of the FI parameter, associated with the superpotential term 
\be \delta\mathcal{W}=\lambda\Tr(JA)\,,\ee
where $J$ is the symplectic matrix and $A$ is the $USp(2N_c)$ antisymmetric chiral.

Let us now focus on the bad case, which is the most interesting for us. As in the 3d case, when we switch off $\lambda=0$ we can solve the equations of motion by setting the matrix $C$ to zero. Since the matrix $D$ has only $N_f-N_c$ non-trivial eigenvalues, the system \eqref{eom2} can be solved for $\lambda=0$ only by restricting the rank of $B$ to be at most $N_f-N_c$. This is nothing but the 4d counterpart of the last frame with no delta distribution associated.

As in the 3d case, we can find other solutions for $\lambda=0$ of the form 
\be\label{quark1}
Q=\left(\begin{array}{c|c}
\begin{array}{c|c}B&0\\\end{array}&\\ 
\hline 
&\begin{array}{c|c} 0 & D\\\end{array}\\
\end{array}\right) \,,
\ee 
where both $B$ and $D$ are diagonal and have rank $r\leq \lfloor\frac{N_f}{2}\rfloor$. Indeed when $r\leq N_f-N_c$ these are just special cases of the last frame with no delta distribution associated, whereas the other values of $r$ provide the missing frames. Notice that we can still solve the equations of motion by turning on the VEV for other fields (other than $Q$) in the remaining $N_c-r$ color directions. This is consistent with the structure of the 3d bad theory, in which at every point of the Higgs branch the Coulomb branch is never entirely lifted. 

We can extract information about the residual Coulomb branch in each frame from the 4d equations of motion by activating a VEV for the antisymmetric chiral $A$ and the fundamentals $P_{L,R}$. If in \eqref{quark1} $B$ and $D$ have the form 
\be B=\text{Diag}(b_1,\dots, b_r,0,\dots 0),\quad  D=\text{Diag}(d_1,\dots, d_r,0,\dots 0)\, ,\ee
one allowed solution is to take $A$ of the form 
\be A=\left(\begin{array}{c|c}0 & \Phi\\ \hline  -\Phi & 0\end{array}\right),\quad \Phi=(0,\dots 0,\phi_1,\dots,\phi_{N_c-r})\, ,\ee 
where $\Phi$ indeed plays the role of the $U(N_c)$ adjoint chiral in 3d we have discussed in Subsection \ref{subsec:3dN=4eom}. 
Another option is to turn on the VEV for $P_{L,R}$ and take $\langle JA\rangle$ to be nilpotent as in \eqref{vevsol} with $k=N_c-r$, which is consistent with the result of Section \ref{sec:higgs} where we turned on VEVs of dressed mesons constructed from the fields in the saw and the antisymmetric chiral, which are exactly those having VEVs along these remaining color directions. 

The two options we have just described correspond to submanifolds of the “Coulomb branch lifted to 4d’’ and correspond to giving a VEV to the Casimirs of $A$ (the counterpart of the Casimirs built out of the adjoint chiral in the 3d theory) or to the dressed mesons (which become monopoles in 3d). Notice that in the 3d $\mathcal N=4$ theory we can simply describe the residual Coulomb branch in terms of the triplet of adjoint scalars $\Phi$ and $\Sigma$, as a consequence of the hyperk\"ahler structure of the moduli space. In the 4d theory, where we do not have extended supersymmetry, this structure is not present and from the equations of motion we only see a collection of branches with a K\"ahler structure. We leave the understanding of quantum chiral ring relations in 4d and their connection with the hyperk\"ahler structure and enhanced R-symmetry of the 3d $\mathcal N=4$ theory for future work.


\subsection{Brane analysis of the magnetic frame}
We will now discuss the perspective of the Type IIB brane realization of our result about the 3d $\mathcal{N}=4$ SQCD. We consider the magnetic S-dual configuration, where we  have two D5-branes and $N_f$ NS5-branes. 
The analysis which follows is a generalization of the argument presented in \cite{Bourget:2021jwo} (although in the context of the mirror dual theory), where just the $n=0$ and the last frame with no delta distribution associated were discussed. Here we extend the analysis to include all the frames, recovering the result we have found at the level of the partition function.

\begin{figure}[!ht]
\centering
\resizebox{0.5\textwidth}{!}{
\begin{tikzpicture}
 \draw[red] (3.5,-1)--(3.5,1) (4,-1)--(4,1) (4.5,-1)--(4.5,1) (5,-1)--(5,1) (5.5,-1)--(5.5,1) (6,-1)--(6,1) (6.5,-1)--(6.5,1);
   \draw[blue]  (2.3,-0.1)--(3.3,0.9) (6.7,-0.1)--(7.7,0.9);
   \draw (3.15,0.75)--(6,0.75) (3,0.6)--(5.5,0.6) (2.85,0.45)--(5,0.45) (2.7,0.3)--(4.5,0.3) (2.55,0.15)--(4,0.15) (2.4,0)--(3.5,0) ; 
   \draw  (6.5,0.75)--(7.5,0.75) (6,0.6)--(7.35,0.6) (5.5,0.45)--(7.2,0.45) (5,0.3)--(7.05,0.3) (4.5,0.15)--(6.9,0.15) (4,0)--(6.75,0); 
   \draw (3.5,-0.5)--(6.5,-0.5); 
\end{tikzpicture} 	
}
\caption{The brane setup describing $U(6)$ SQCD with 7 flavors. Red lines are NS5-branes, blue lines are D5-branes and black lines are D3-branes.}
\label{tikzfig:step0}
\end{figure}
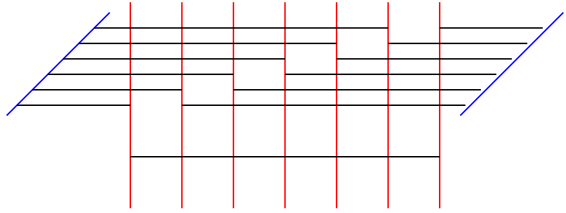

\begin{figure}[!ht]
\centering
\resizebox{0.5\textwidth}{!}{
\begin{tikzpicture}
 \draw[red] (2.5,-1)--(2.5,1) (3,-1)--(3,1) (3.5,-1)--(3.5,1) (5,-1)--(5,1) (6.5,-1)--(6.5,1) (7,-1)--(7,1) (7.5,-1)--(7.5,1);
   \draw[blue]  (3.8,-0.1)--(4.8,0.9) (5.2,-0.1)--(6.2,0.9);
   \draw (4.65,0.75)--(7,0.75) (4.5,0.6)--(6.5,0.6) (4.35,0.45)--(5,0.45); 
   \draw  (5,0.3)--(5.55,0.3) (3.5,0.15)--(5.4,0.15) (3,0)--(5.25,0); 
   \draw (2.5,-0.5)--(7.5,-0.5); 
\end{tikzpicture}
}
\caption{The brane setup after the HW moves.}
\label{tikzfig:step1}
\end{figure}
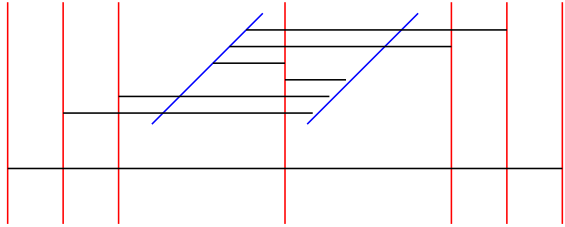

Focusing on the bad case $N_f<2N_c$, we start by suspending D3-branes between D5 and NS5 as much as we can. In the example of $U(6)$ SQCD with 7 flavors this means going to the configuration in Figure \ref{tikzfig:step0}.
If we now move the D5-branes using HW moves we can get to the configuration in Figure \ref{tikzfig:step1}.

\begin{figure}[!ht]
\centering
\resizebox{0.5\textwidth}{!}{
\begin{tikzpicture}
 \draw[red] (2.5,-1)--(2.5,1) (4,-1)--(4,1) (4.5,-1)--(4.5,1) (5,-1)--(5,1) (5.5,-1)--(5.5,1) (6,-1)--(6,1) (7.5,-1)--(7.5,1);
   \draw[blue]  (2.8,-0.1)--(3.8,0.9) (6.2,-0.1)--(7.2,0.9);
   \draw (2.5,-0.5)--(7.5,-0.5); 
\end{tikzpicture}
}
\caption{The brane setup describing the mirror dual of $U(1)$ with 7 flavors.}
\label{tikzfig:Itamar}
\end{figure}
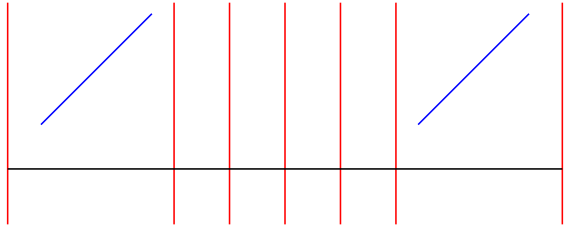

Close to the D5-branes the theory locally looks like the magnetic dual of $U(5)$ with 5 flavors. By moving along the Higgs branch of this theory (by sending to infinity all the D3 segments suspended between the two D5’s) we are left with the setup in Figure \ref{tikzfig:step2}, and with a further sequence of HW moves we end up with the configuration in Figure \ref{tikzfig:Itamar}, where we recognize the brane setup describing the mirror dual of SQED with 7 flavors. For generic $N_c$ and $N_f$ we end up with the mirror dual of $U(N_f-N_c)$ SQCD with $N_f$ flavors, i.e. the last frame with no delta distribution associated discussed in \cite{Yaakov:2013fza}. 
\begin{figure}[!ht]
\centering
\resizebox{0.5\textwidth}{!}{
\begin{tikzpicture}
 \draw[red] (2.5,-1)--(2.5,1) (3,-1)--(3,1) (3.5,-1)--(3.5,1) (5,-1)--(5,1) (6.5,-1)--(6.5,1) (7,-1)--(7,1) (7.5,-1)--(7.5,1);
   \draw[blue]  (3.8,-0.1)--(4.8,0.9) (5.2,-0.1)--(6.2,0.9);
   \draw (6.05,0.75)--(7,0.75) (5.9,0.6)--(6.5,0.6); 
   \draw  (3.5,0.15)--(4.05,0.15) (3,0)--(3.9,0); 
   \draw (2.5,-0.5)--(7.5,-0.5); 
\end{tikzpicture}
}
\caption{The brane setup after the HW moves and after we sent to infinity the D3-branes suspended between the two D5's.}
\label{tikzfig:step2}
\end{figure}
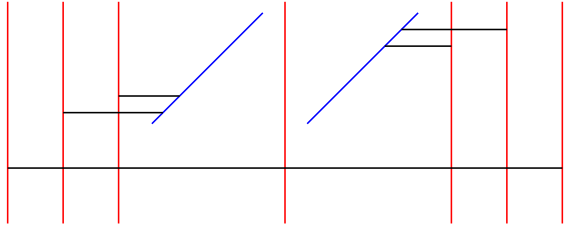

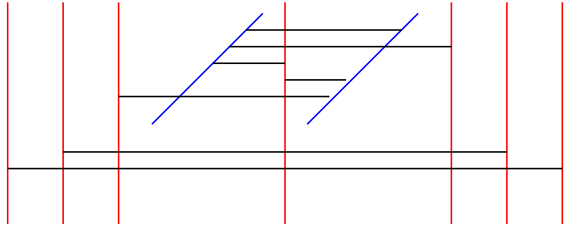
\begin{figure}[!ht]
\centering
\resizebox{0.5\textwidth}{!}{
\begin{tikzpicture}
 \draw[red] (2.5,-1)--(2.5,1) (3,-1)--(3,1) (3.5,-1)--(3.5,1) (5,-1)--(5,1) (6.5,-1)--(6.5,1) (7,-1)--(7,1) (7.5,-1)--(7.5,1);
   \draw[blue]  (3.8,-0.1)--(4.8,0.9) (5.2,-0.1)--(6.2,0.9);
   \draw (4.65,0.75)--(6.05,0.75) (4.5,0.6)--(6.5,0.6) (4.35,0.45)--(5,0.45); 
   \draw  (5,0.3)--(5.55,0.3) (3.5,0.15)--(5.4,0.15); 
   \draw (2.5,-0.5)--(7.5,-0.5) (3,-0.35)--(7,-0.35); 
\end{tikzpicture}
}
\caption{The brane setup after we suspended one more D3-brane between the two NS5's.}
\label{tikzfig:step3}
\end{figure}

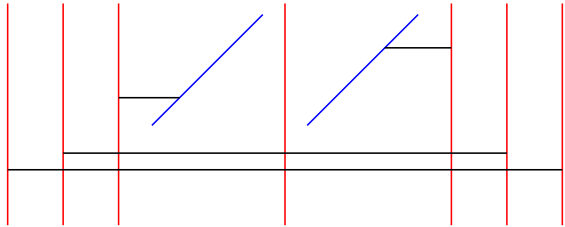
\begin{figure}[!ht]
\centering
\resizebox{0.5\textwidth}{!}{
\begin{tikzpicture}
 \draw[red] (2.5,-1)--(2.5,1) (3,-1)--(3,1) (3.5,-1)--(3.5,1) (5,-1)--(5,1) (6.5,-1)--(6.5,1) (7,-1)--(7,1) (7.5,-1)--(7.5,1);
   \draw[blue]  (3.8,-0.1)--(4.8,0.9) (5.2,-0.1)--(6.2,0.9);
   \draw (5.9,0.6)--(6.5,0.6); 
   \draw (3.5,0.15)--(4.05,0.15); 
   \draw (2.5,-0.5)--(7.5,-0.5) (3,-0.35)--(7,-0.35); 
\end{tikzpicture}
}
\caption{The brane setup describing the mirror dual of $U(2)$ with 7 flavors.}
\label{tikzfig:step4}
\end{figure}

In order to find the other frames we start over from the setup on Figure \ref{tikzfig:step1} and suspend another D3-brane between NS5’s. This leads to the configuration in Figure \ref{tikzfig:step3}.
\\
Close to the D5-branes the theory looks like the magnetic dual of $U(4)$ with 3 flavors and moving along the Higgs branch of this model we get the brane system in Figure \ref{tikzfig:step4}. 
This becomes the brane system describing the mirror dual of $U(N_f+1-N_c)$ SQCD with $N_f$ flavors, corresponding to the frame with $n=M+\epsilon$. 

\begin{figure}[!ht]
\centering
\resizebox{0.5\textwidth}{!}{
\begin{tikzpicture}
 \draw[red] (2.5,-1)--(2.5,1) (3,-1)--(3,1) (3.5,-1)--(3.5,1) (5,-1)--(5,1) (6.5,-1)--(6.5,1) (7,-1)--(7,1) (7.5,-1)--(7.5,1);
   \draw[blue]  (3.8,-0.1)--(4.8,0.9) (5.2,-0.1)--(6.2,0.9);
   \draw (4.65,0.75)--(6.05,0.75) (4.35,0.45)--(5,0.45); 
   \draw  (5,0.3)--(5.55,0.3) (4,0.15)--(5.4,0.15); 
   \draw (2.5,-0.5)--(7.5,-0.5) (3,-0.35)--(7,-0.35)  (3.5,-0.2)--(6.5,-0.2); 
\end{tikzpicture}
}
\caption{The brane setup describing the mirror dual of $U(3)$ with 7 flavors.}
\label{tikzfig:step5}
\end{figure}
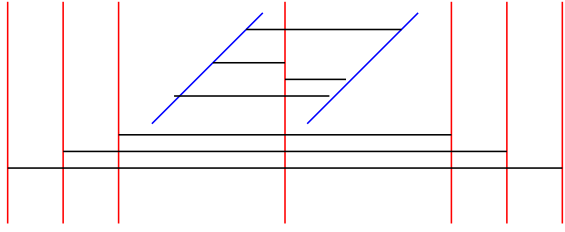

Now we can go back to Figure \ref{tikzfig:step3} and by suspending even more D3-branes between NS5’s we can find all the other frames up to $U(\lfloor N_f/2\rfloor)$ SQCD with $N_f$ flavors. In the case at hand we have just one further step to consider, starting from Figure \ref{tikzfig:step3}, which is shown in Figure \ref{tikzfig:step5}.   

Close to the D5-branes the theory locally looks like the magnetic dual of $U(3)$ SQCD with one flavor. By moving to a generic point of its Higgs branch we recover the brane system describing the mirror dual of $U(3)$ SQCD with 7 flavors, which is the $n=0$ frame. 

So far we have neglected the free hypermultiplets, which are present in all the frames we have discussed above. These arise upon going to a generic point of the Higgs branch of the magnetic theory. For example, in Figure \ref{tikzfig:step5} we have close to the D5-branes the magnetic dual of $U(3)$ SQCD with one flavor, whose moduli space is known to be smooth and the low-energy theory at each point is given by a collection of three hypermultiplets. 

Notice that in all the frames, except the last one with no delta distribution associated, we have at least one D3-suspended between the two D5’s which cannot end on any of the NS5-branes without violating the s-rule (see Figure \ref{tikzfig:step3} and \ref{tikzfig:step5}). This tells us that only in the last frame, with no delta distribution associated, we are allowed to turn on a FI deformation, which in the magnetic configuration we are considering involves moving vertically one of the D5-branes. We therefore recover the expected result that in all the frames, except for the last frame, the FI deformation is frozen. This is indeed consistent with the delta functions appearing in the partition function of the theory. Equivalently, we can go back to Figure \ref{tikzfig:step0} and notice that in presence of a non-zero FI deformation the two D5-branes are not at the same position in the vertical direction and therefore with a sequence of Hanany--Witten moves we can directly bring the brane system to the form Figure \ref{tikzfig:Itamar}. In this case we do not find any other frames and we get to the conclusion that the bad SQCD theory and its Seiberg-like dual are infrared equivalent once a FI deformation has been activated.

\acknowledgments
We would like to thank Chris Beem, Julius Grimminger and Zhenghao Zhong for useful discussions. CH is supported by the National Natural Science Foundation of China under Grant No.~12247103.
FM is partially supported by STFC under Grant No.~ST/V507118 - Maths 2020 DTP. 
SP is partially supported by the MUR-PRIN grant No. 2022NY2MXY.
MS is partially supported by the ERC Consolidator Grant \#864828 “Algebraic Foundations of Supersymmetric Quantum Field Theory (SCFTAlg)” and by the Simons Collaboration for the Nonperturbative Bootstrap under grant \#494786 from the Simons Foundation.

\appendix

\section{The good SQCD mirror pair}
\subsection{The 4d mirror pair}\label{The 4d mirror pair}
We report here the index of the  SQCD with $N_c$ colors and $N_f$ flavors, which is shown in Figure \ref{fig:good_SQCD_Sdualised_4d}: 
\begin{align}
	& \mathcal{I}_{\text{SQCD}(N_c,N_f)}(\vec x;y_1,y_2;t;c) = \nonumber\\
	& \quad =
	\oint d\vec{z}_{N_c}\Delta_{N_c}\left(\vec{z};t\right)
	\prod_{j=1}^{N_c}\prod_{a=1}^{N_f}\Gamma_e\left( (pq)^{\frac{1}{2}}t^{-\frac{1}{2}} z_j^{\pm} x_a^{\pm} \right) \nonumber\\
	& \quad\qquad\times
	\prod_{j=1}^{N_c} \Gamma_e\left( c z_j^{\pm} y_1^{\pm} \right)
	\prod_{j=1}^{N_c} \Gamma_e\left( t^{\frac{N_f}{2}-N_c+1} c^{-1} z_j^{\pm} y_2^{\pm} \right) \nonumber\\
	& \quad\qquad\times
	\prod_{a=1}^{N_f} \Gamma_e\left((pq)^{\frac{1}{2}} t^{N_c-\frac{N_f}{2}-\frac{1}{2}}c x_a^{\pm}y_2^{\pm} \right)
	\prod_{j=1}^{N_c} \Gamma_e\left( pqt^{1-j}c^{-2} \right) \,.
\end{align}
If the SQCD is good (namely $N_f \geq 2N_c$), its mirror dual is unique and its index is the following:
\begingroup\allowdisplaybreaks
\begin{align}
	& \mathcal{I}_{\widehat{\text{SQCD}}(N_c,N_f\geq 2N_c)}(\vec x;y_1,y_2;t;c) = \nonumber\\
	& \quad =
	\oint 
	\prod_{k=1}^{N_c} \left( d\vec{z}^{\,(k)}_{k}\Delta_{k}\left(\vec{z}^{\,(k)};pqt^{-1}\right) \right) 
	\prod_{k=N_c+1}^{N_f-N_c-1} \left( d\vec{z}^{\,(k)}_{N_c}\Delta_{N_c}\left(\vec{z}^{\,(k)};pqt^{-1}\right) \right) 
	\nonumber\\
	& \quad\qquad\times
	\prod_{k=N_f-N_c}^{N_f-1} \left( d\vec{z}^{\,(k)}_{N_f-k}\Delta_{N_f-k}\left(\vec{z}^{\,(k)};pqt^{-1}\right) \right)
	\nonumber\\
	& \quad\qquad\times
	\prod_{k=1}^{N_c-1}\prod_{j=1}^{k}\prod_{l=1}^{k+1} \Gamma_e\left( t^{\frac{1}{2}} z_j^{(k)\,\pm}z_l^{(k+1)\,\pm} \right)
	\prod_{k=N_c}^{N_f-N_c-1}\prod_{j=1}^{N_c}\prod_{l=1}^{N_c} \Gamma_e\left( t^{\frac{1}{2}} z_j^{(k)\,\pm}z_l^{(k+1)\,\pm} \right)
	\nonumber\\
	& \quad\qquad\times
	\prod_{k=N_f-N_c}^{N_f-2}\prod_{j=1}^{N_f-k}\prod_{l=1}^{N_f-k-1} \Gamma_e\left( t^{\frac{1}{2}} z_j^{(k)\,\pm}z_l^{(k+1)\,\pm} \right)
	\nonumber\\
	& \quad\qquad\times
	\prod_{j=1}^{N_c}\Gamma_e\left( t^{\frac{1}{2}} z_j^{(N_c)\,\pm} y_1^{\pm} \right)\Gamma_e\left( t^{\frac{1}{2}} z_j^{(N_f-N_c)\,\pm} y_2^{\pm} \right)
	\nonumber\\
	& \quad\qquad\times
	\prod_{k=1}^{N_c}\prod_{j=1}^{k} \Gamma_e\left( (pq)^{\frac{1}{2}}t^{\frac{k-N_c}{2}}c^{-1} z_j^{(k)\,\pm}x_k^{\pm} \right)
	\prod_{k=N_c+1}^{N_f-N_c-1}\prod_{j=1}^{N_c} \Gamma_e\left( (pq)^{\frac{1}{2}}t^{\frac{k-N_c}{2}}c^{-1} z_j^{(k)\,\pm}x_k^{\pm} \right)
	\nonumber\\
	& \quad\qquad\times
	\prod_{k=1}^{N_f-N_c}\prod_{j=1}^{N_f-k} \Gamma_e\left( (pq)^{\frac{1}{2}}t^{\frac{k-N_c}{2}}c^{-1} z_j^{(k)\,\pm}x_k^{\pm} \right)
	\nonumber\\
	& \quad\qquad\times
	\prod_{k=1}^{N_c}\prod_{j=1}^{k} \Gamma_e\left( (pq)^{\frac{1}{2}}t^{\frac{N_c-k-2}{2}}c z_j^{(k)\,\pm}x_{k+1}^{\pm} \right)
	\prod_{k=N_c+1}^{N_f-N_c-1}\prod_{j=1}^{N_c} \Gamma_e\left( (pq)^{\frac{1}{2}}t^{\frac{N_c-k-2}{2}}c z_j^{(k)\,\pm}x_{k+1}^{\pm} \right)
	\nonumber\\
	& \quad\qquad\times
	\prod_{k=N_f-N_c}^{N_f-1}\prod_{j=1}^{N_f-k} \Gamma_e\left( (pq)^{\frac{1}{2}}t^{\frac{N_c-k-2}{2}}c z_j^{(k)\,\pm}x_{k+1}^{\pm} \right)
	\nonumber\\
	& \quad\qquad\times
	\prod_{k=1}^{N_c} \Gamma_e\left( (pq)^{\frac{1}{2}} t^{-\frac{1}{2}}c x_k^{\pm}y_1^{\pm} \right)
	\prod_{k=1}^{N_f-N_c} \Gamma_e\left( (pq)^{\frac{1}{2}} t^{N_c-\frac{N_f}{2}-\frac{1}{2}}c x_k^{\pm}y_2^{\pm} \right) 
	\nonumber\\
	& \quad\qquad\times
	\prod_{j=1}^{N_c}\Gamma_e\left( t^{N_f-N_c+2-j}c^{-2} \right) \,.
\end{align}
\endgroup
Therefore, we have
\begin{equation}
	\mathcal{I}_{\text{SQCD}(N_c,N_f\geq 2N_c)}(\vec x;y_1,y_2;t;c) = \mathcal{I}_{\widehat{\text{SQCD}}(N_c,N_f\geq 2N_c)}(\vec x;y_1,y_2;t;c) \,.
\end{equation}

\subsection{The 3d mirror pair}\label{3dgoodres}

If we take the 3d limit of Figure \ref{fig:good_SQCD_Sdualised_4d}, we have what is shown in Figure \ref{fig:good_SQCD_Sdualised_3d}, namely
\begin{align}
	& \left(e^{2\pi i Y_2 \sum_{a=1}^{N_f}X_a}\right) 
	\mathcal{Z}^{3d}_{\text{SQCD}_U(N_c,N_f\geq 2N_c)}\left(\vec{X};Y_1-Y_2\right) 
	= \nonumber\\
	& \qquad =
	\left(
	e^{2\pi i Y_1 \sum_{a=1}^{N_c} X_a} \times
	e^{2\pi i Y_2 \sum_{a=1}^{N_f-N_c} X_a}
	\right)
	\mathcal{Z}^{3d}_{\widehat{\text{SQCD}}_U(N_c,N_f\geq 2N_c)}\left(Y_1,Y_2;X_{k+1}-X_{k}\right) \,,
\end{align}
where we defined
\begin{align}
	\mathcal{Z}^{3d}_{\text{SQCD}_U(N_c,N_f\geq 2N_c)}\left(\vec{X};Y_1-Y_2\right) & = 
	\int \udl{\vec{Z}_{N_c}}\Delta_{N_c}^{(3d)}\left(\vec{Z};m_A \right) 
	\times e^{2\pi i (Y_1-Y_2) \sum_{j=1}^{N_c}Z_j} 
	\nonumber\\
	& \qquad\quad\times
	\prod_{j=1}^{N_c}\prod_{a=1}^{N_f} s_b\left( \frac{iQ}{2}-m_A\pm(Z_j-X_a) \right)
	\label{eq:def_Z3d_SQCD_electric}
\end{align}
and
\begingroup\allowdisplaybreaks
\begin{align}
	&\mathcal{Z}^{3d}_{\widehat{\text{SQCD}}_U(N_c,N_f\geq 2N_c)}\bigg(Y_1,Y_2;X_{k+1}-X_{k}\,(k=1,\dots,N_f-1)\bigg) =\nonumber\\
	& \quad = 
	\int 
	\prod_{k=1}^{N_c}\left(\udl{\vec{Z}_{k}^{(k)}}\Delta_k^{(3d)}\left(\vec{Z}^{(k)};i\frac{Q}{2}-m_A \right) \right)
	\prod_{k=N_c+1}^{N_f-N_c-1}\left(\udl{\vec{Z}_{N_c}^{(k)}}\Delta_{N_c}^{(3d)}\left(\vec{Z}^{(k)};i\frac{Q}{2}-m_A \right) \right)
	\nonumber\\
	& \qquad\times
	\prod_{k=N_f-N_c}^{N_f-1}\left(\udl{\vec{Z}_{N_f-k}^{(k)}}\Delta_{N_f-k}^{(3d)}\left(\vec{Z}^{(k)};i\frac{Q}{2}-m_A \right) \right)
	\prod_{k=1}^{N_f-1}e^{-2\pi i(X_k-X_{k+1})\sum_{j=1}^{N_k}Z_j^{(k)}}
	\nonumber\\
	& \qquad\quad\times
	\prod_{k=1}^{N_c-1}\prod_{j=1}^{k}\prod_{l=1}^{k+1} s_b\left( m_A \pm\left(Z_j^{(k)}-Z_l^{(k+1)}\right) \right)
	\nonumber\\
	& \qquad\quad\times
	\prod_{k=N_c}^{N_f-N_c-1}\prod_{j=1}^{N_c}\prod_{l=1}^{N_c} s_b\left( m_A \pm\left(Z_j^{(k)}-Z_l^{(k+1)}\right) \right)
	\nonumber\\
	& \quad\qquad\times
	\prod_{k=N_f-N_c}^{N_f-2}\prod_{j=1}^{N_f-k}\prod_{l=1}^{N_f-k-1} s_b\left( m_A \pm\left(Z_j^{(k)}-Z_l^{(k+1)}\right) \right)
	\nonumber\\
	& \quad\qquad\times
	\prod_{j=1}^{N_c}s_b\left( m_A \pm\left(Z_j^{(N_c)}- Y_1\right)\right)s_b\left( m_A \pm\left(Z_j^{(N_f-N_c)}- Y_2 \right)\right) \,,
\end{align}
\endgroup
where we defined, as in  \cite{Comi:2022aqo},
\begin{align}
    \udl{\vec{Z}_n}\Delta_n^{(3d)}\big( \vec{Z};\alpha \big) 
    & = 
    \frac{\prod_{j=1}^n dZ_j}{n!} 
    \frac{\prod_{j,l=1}^n s_b\left( -i\frac{Q}{2}+2\alpha + (Z_j-Z_l) \right)}{\prod_{j<l}^n s_b\left( i\frac{Q}{2} \pm (Z_j-Z_l) \right)} \,.
\end{align}
Note that the rank of each gauge node, $U(N_k)_{Z^{(k)}}$, is given by
\begin{align}
	N_k = 
	\begin{cases}
		k \qquad\qquad\,\,\, \text{for } k=1,\dots,N_c-1 \text{ (increasing ramp)}\\
		N_c \qquad\qquad\! \text{for } k=N_c,\dots,N_f-N_c \text{ (plateau)} \\
		N_f-k \qquad \text{for } k=N_f-N_c+1,\dots,N_f-1 \text{ (decreasing ramp)}
	\end{cases}
	\,.
\end{align}
We also find it convenient to shift all of the magnetic gauge variables by $Y_1$, so that also on this mirror side the integral depends on the combination $Y_2-Y_1$. We get
\begin{align}
	& \left(e^{\pi i (Y_2 -Y_1)\sum_{a=1}^{N_f}X_a}\right) 
	\mathcal{Z}^{3d}_{\text{SQCD}_U(N_c,N_f\geq 2N_c)}\left(\vec{X};Y_1-Y_2\right) 
	= \nonumber\\
	& \qquad =
		e^{2\pi i (Y_2-Y_1) \left(\sum_{a=1}^{N_f-N_c} X_a-\frac{1}{2}\sum_{a=1}^{N_f}X_a\right)}
	\mathcal{Z}^{3d}_{\widehat{\text{SQCD}}_U(N_c,N_f\geq 2N_c)}\left(0,Y_2-Y_1;X_{k+1}-X_{k}\right) \,.
\end{align}

%
%
%
%
%
%
%

\afterpage{
\begin{landscape}
\begin{figure}[!ht]
\centering
    \includegraphics[scale=.4]{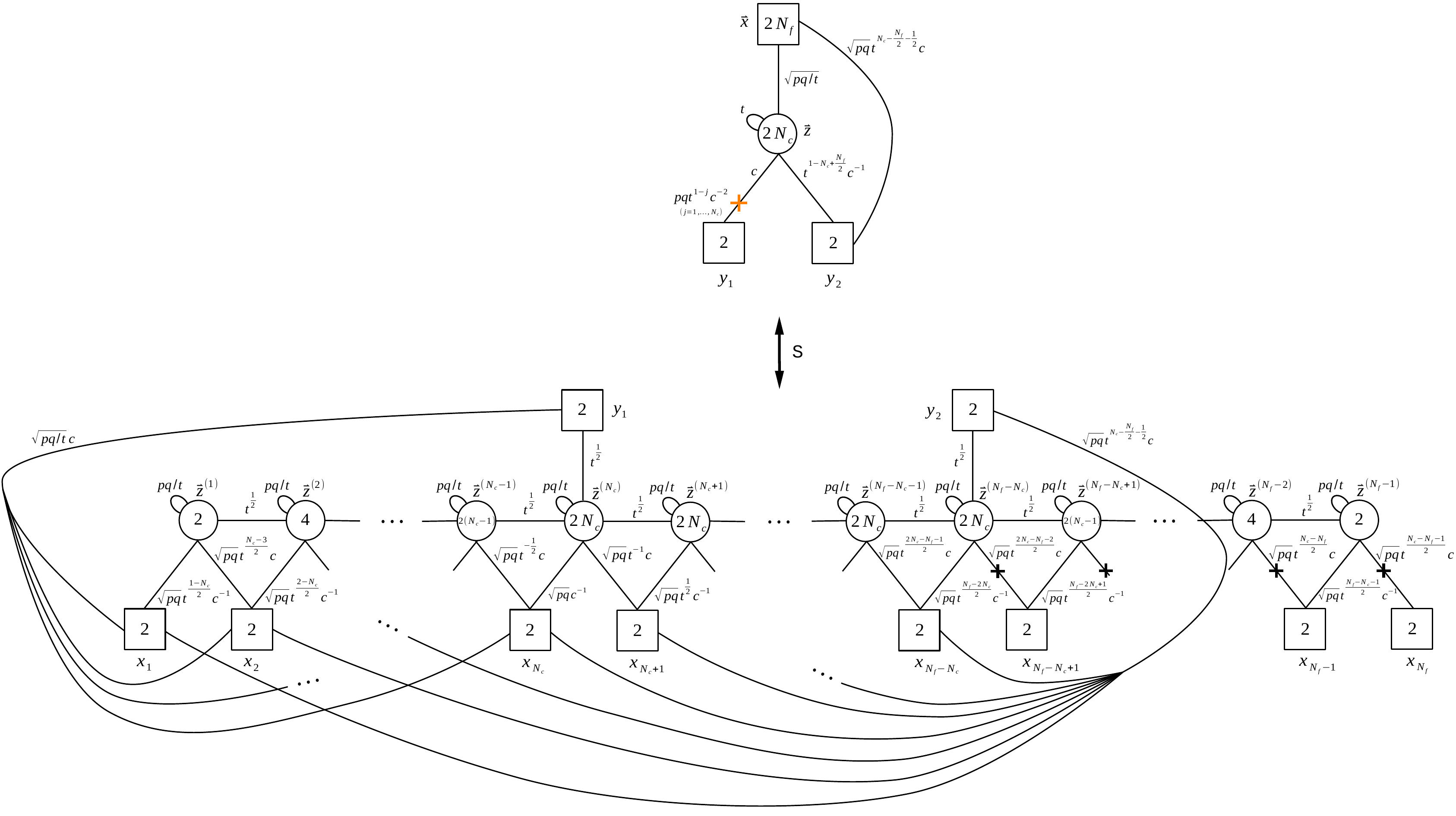}
    \caption{The 4d mirror pair for a generic good SQCD.}
    \label{fig:good_SQCD_Sdualised_4d}
\end{figure}
\end{landscape}
}

\afterpage{
\begin{landscape}
\begin{figure}[!ht]
\centering
    \includegraphics[scale=.4]{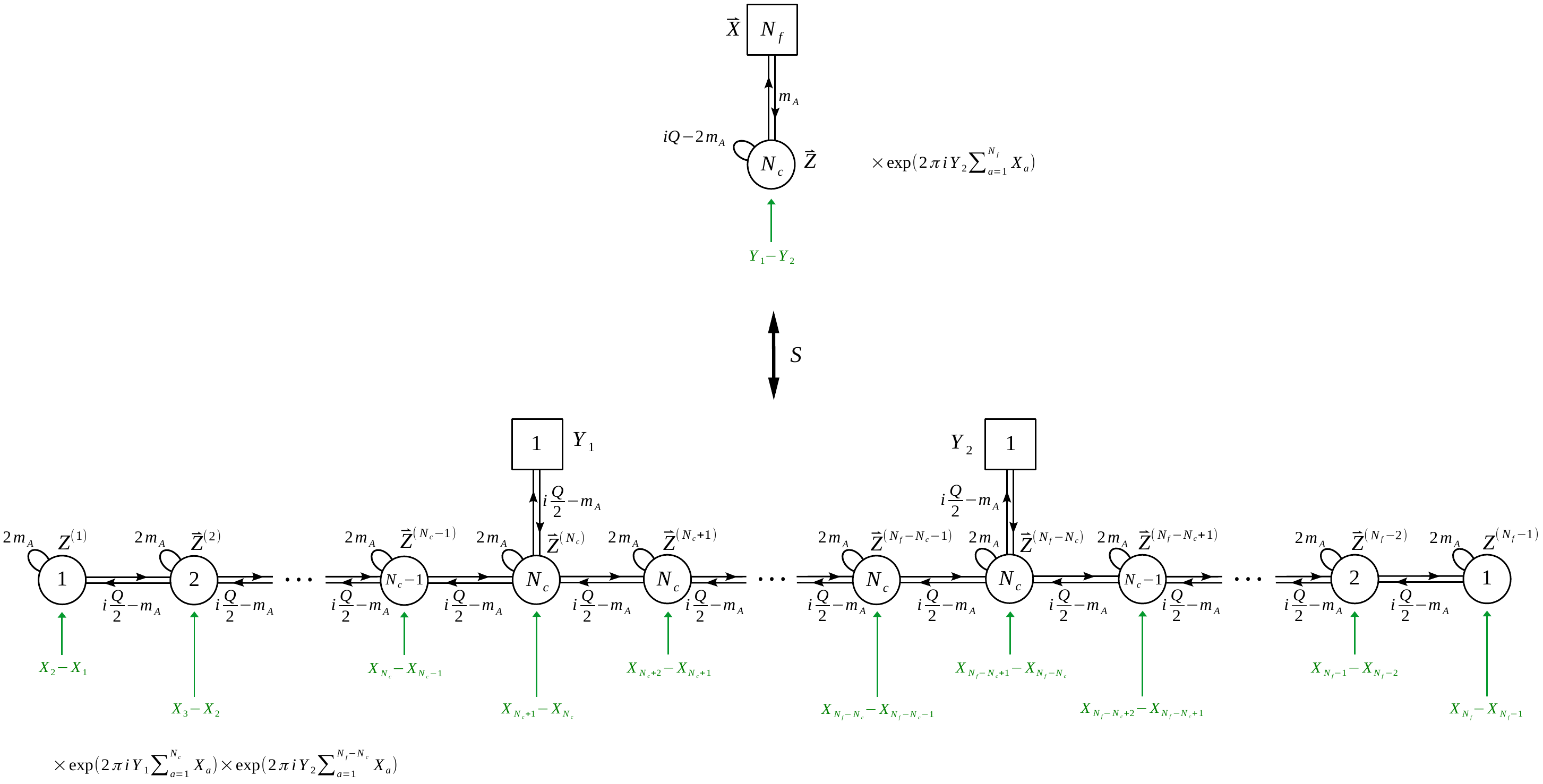}
    \caption{The 3d mirror pair for a generic good SQCD. The FI terms are denoted in green (with an arrow that shows which gauge node they are referred to): they have to be intended as exponentiated with also a $2\pi i$ normalization. In addition, we wrote aside the background BF couplings.}
    \label{fig:good_SQCD_Sdualised_3d}
\end{figure}
\end{landscape}
}

\section{Comments on the possible frames for the $N_f=N_c=4$ case}
\label{moreframe}

In this appendix, we comment more on the criterion used in Section \ref{sec:4dbadSQCD} to decide which identifications in the colliding Identity-walls lead to physical frames. 
We will focus on the example of $N_c=N_f=4$ discussed in Subsubsection \ref{44sec}, although what we are going to say can be applied in general. 

In this case, we recall that the colliding Identity-walls lead to the identifications of the two following sets of variables:
\be\label{eq:setfugNc4Nf4}
\vec{f}&=&\left\{u_1,u_2,y_1t^{\frac{1}{2}},y_1t^{-\frac{1}{2}}\right\}\,,\nn\\
\vec{h}&=&\left\{v_1,v_2,y_2t^{\frac{1}{2}},y_2t^{-\frac{1}{2}}\right\}\,.
\ee
The criterion we used in the main text, which we want to further justify in this subsection, is that we should consider all possible identifications of the sets of variables $\vec{f}$ and $\vec{h}$ in such a way that after the identifications the variables $f_i$ satisfy the following conditions:
\begin{enumerate}
\item $f_i\neq f_j^{\pm1}$ for any pair of $i$ and $j$;
\item $f_i\neq 1$ for any $i$.
\end{enumerate}
These conditions exclude in particular the fixed points of the action of the $USp(8)$ Weyl group on the variables $\vec{f}$. 

\begin{figure}[!ht] 
	\centering
	\includegraphics[width=.75\textwidth]{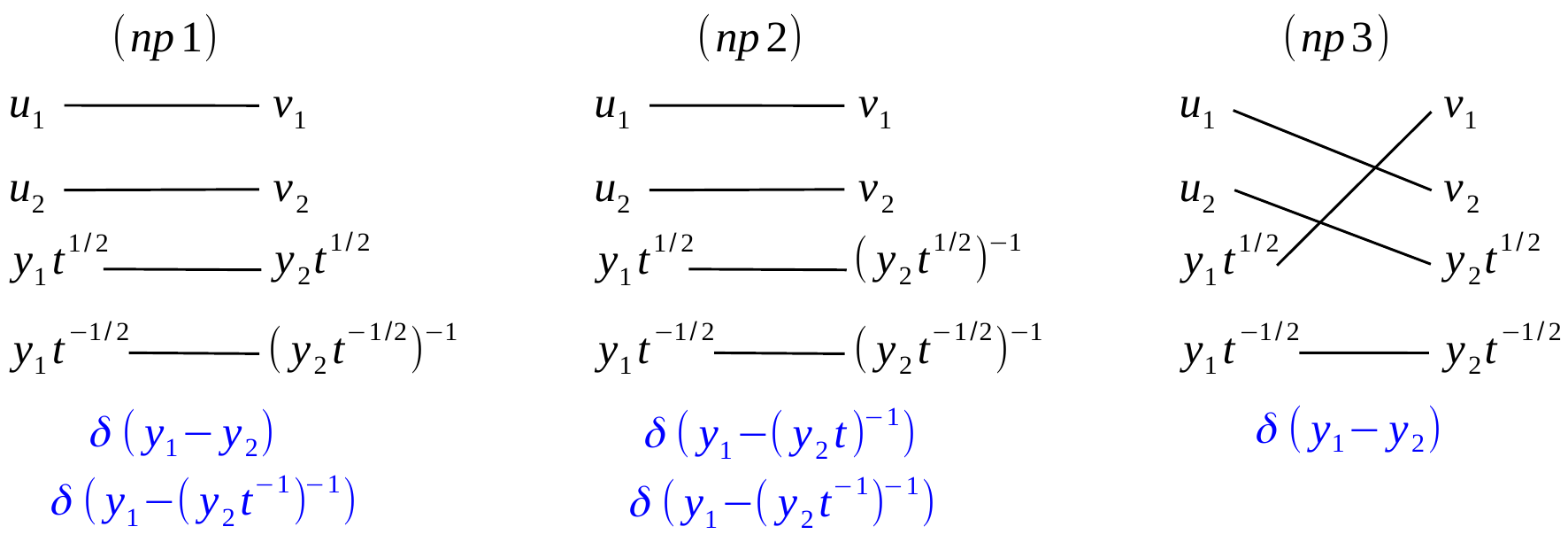}
	\caption{Non-physical identifications for the case $N_c=N_f=4$.}
	\label{n44id}
\end{figure}

Some examples of identifications that violate one of the above conditions are shown in Figure \ref{n44id}.

\noindent Let us discuss why these three possibilities do not satisfy our criterion in turn:
\begin{itemize}
\item $(np1)$: in this case the identifications lead to $y_1=y_2=t^{\frac{1}{2}}$, which after plugging back into \eqref{eq:setfugNc4Nf4} gives
\be
\vec{f}=\left\{u_1,u_2,y_1,1\right\}\,.
\ee
Since the last entry is equal to 1, this possibility should be discarded.
\item $(np2)$: in this case the identifications lead to $y_1=y_2^{-1}$ and $t=1$, which gives
\be
\vec{f}=\left\{u_1,u_2,y_1,y_1\right\}\,,
\ee
so again this possibility should be discarded since the last two entries are identical.
\item $(np3)$: in this case the identifications lead to $y_1=y_2$ and $u_2=y_2t^{\frac{1}{2}}=y_1t^{\frac{1}{2}}$, which gives
\be
\vec{f}=\left\{u_1,y_1t^{\frac{1}{2}},y_1t^{\frac{1}{2}},y_1t^{-\frac{1}{2}}\right\}\,,
\ee
so again this possibility should be discarded since the second and the third entries are identical.
\end{itemize}

We will provide some arguments that  indeed we need to drop these identifications in two ways: first we will study explicitly the Higgsing without using the HW move and then we will use the HW move to propagate the VEV but in a different way than what we did in Subsubsection \ref{44sec}. In both cases, we will see that there are no frames corresponding to the these identifications.

\subsection{Higgsing}\label{Higgsing}

Instead of using the HW moves, we can also study the index of a bad theory by examining the propagation of the VEV following the discussion of Section 3.2
in \cite{Comi:2022aqo}.

\begin{figure}[!ht] 
	\centering
	\includegraphics[width=\textwidth]{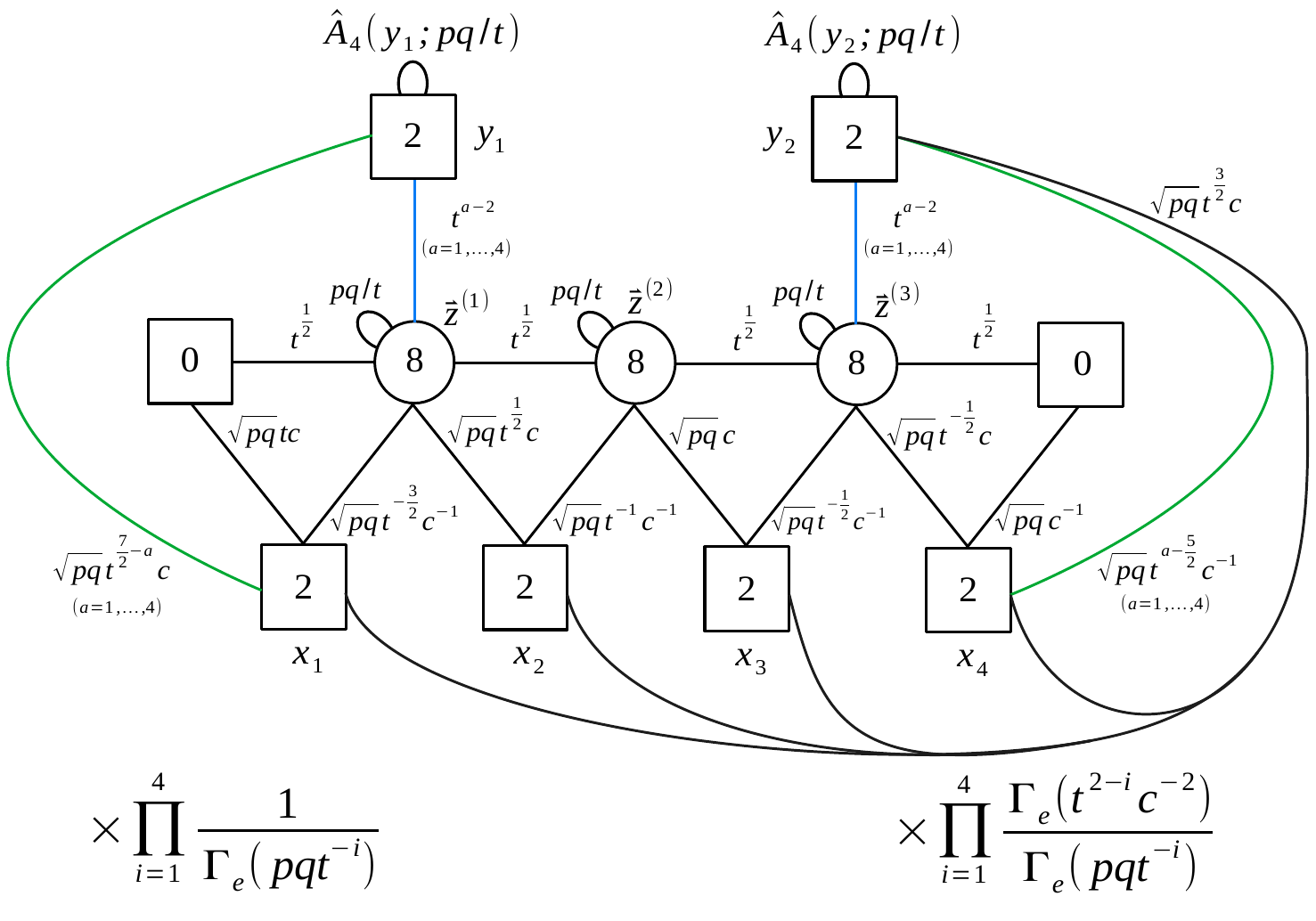}
	\caption{$\mathsf{S}$-dualized SQCD for $N_c=N_f=4$ after implementing the Identity-walls, which is the starting point of the VEV propagation. Each blue line denotes a set of $4$ chirals in the fundamental representation of $USp(8)_{\vec{z}_1}$ and $USp(8)_{\vec{z}_3}$ respectively with different $t$ charges, which indicate that some of them are taking VEVs and their index contributions lead to colliding poles.}
	\label{fig:SQCD_Nc=Nf=4_dualised_AsymmIdentitiesCollapsed}
\end{figure}

\begin{figure}[!ht] 
	\centering
	\includegraphics[width=\textwidth]{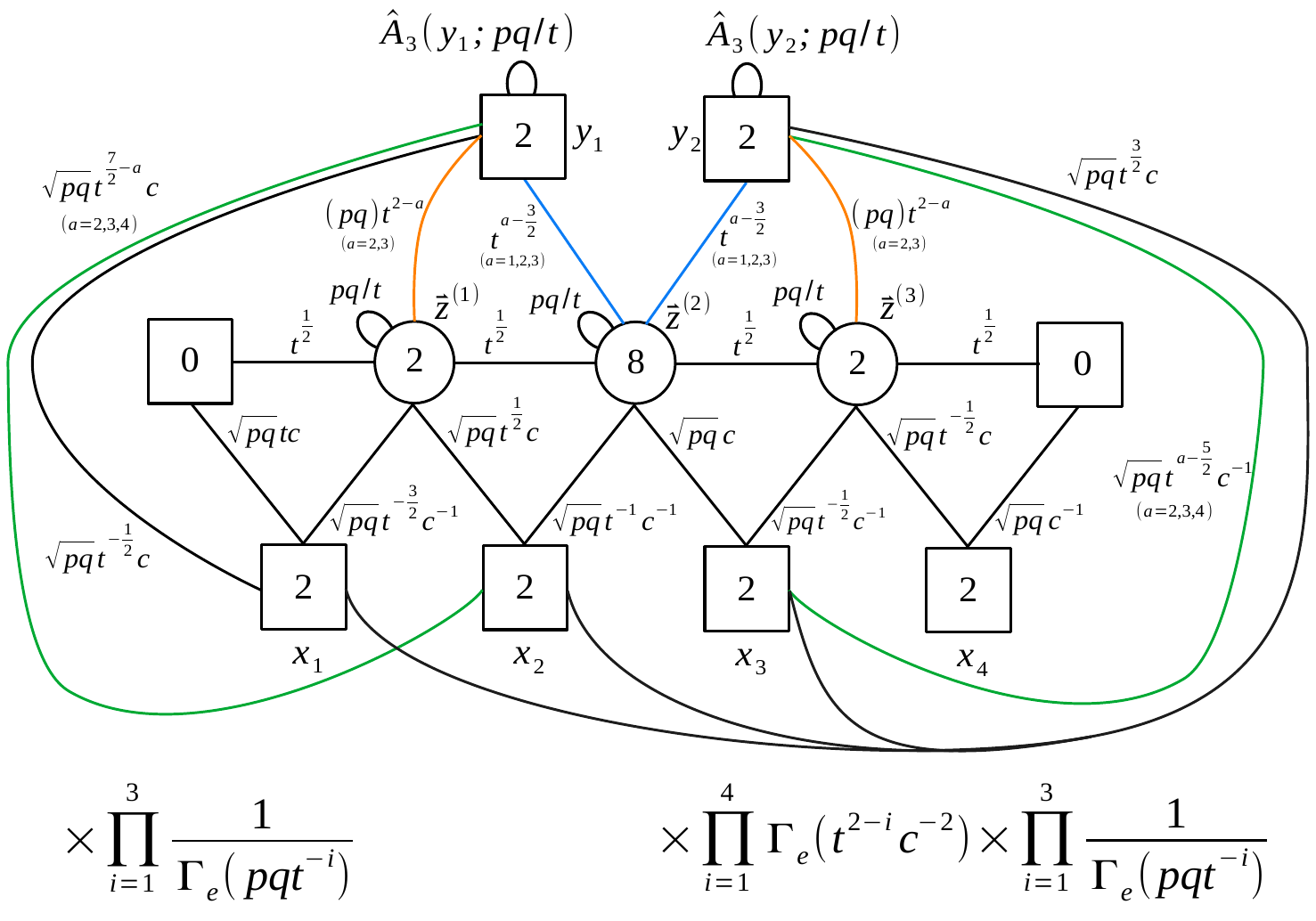}
	\caption{Result of the Higgsing of the leftmost and of the rightmost gauge nodes. Again, some of the blue fields lead to colliding poles in the index.}
	\label{fig:SQCD_Nc=Nf=4_dualised_Higgsing_setI}
\end{figure}

We start with the quiver in Figure  \ref{fig:SQCD_Nc=Nf=4_dualised_AsymmIdentitiesCollapsed}.
Here we observe that some of  the mesons constructed with the $USp(2)_{y_1}\times USp(8)_{z^{(1)}}$ and $USp(2)_{y_2}\times USp(8)_{z^{(3)}}$   bifundamentals denoted in blue acquire VEVs, which initiate a sequential Higgsing. In the first step, the
outmost $USp(8)$ nodes are Higgsed down to $USp(2)$ taking us to the quiver in Figure \ref{fig:SQCD_Nc=Nf=4_dualised_Higgsing_setI}.
Again some of the mesons built with blue bifundamentals acquire VEVs, and one can keep proceeding. However, as mentioned in Sections \ref{sec:4dbadSQCD} and \ref{sec:higgs}, this procedure yields a singular factor $\Gpq{1}$ and its inverse $\Gpq{pq}$, demanding a regularized version of the analysis; for this reason, we introduce refined fugacities for the blue lines as follows:
\begin{align}
\left\{y_1^\pm t^\frac32, y_1^\pm t^\frac12, y_1^\pm t^{-\frac12}\right\} = t^\frac12 \left\{\left(y_1 t\right)^\pm,y_1^\pm,\left(y_1 t^{-1}\right)^\pm\right\} \quad &\longrightarrow \quad t^\frac12 \left\{\left(a_1 t\right)^\pm,a_2^\pm,\left(a_3 t^{-1}\right)^\pm\right\} \,, \\
\left\{y_2^\pm t^\frac32, y_2^\pm t^\frac12, y_2^\pm t^{-\frac12}\right\} = t^\frac12 \left\{\left(y_2 t\right)^\pm,y_2^\pm,\left(y_2 t^{-1}\right)^\pm\right\} \quad &\longrightarrow \quad t^\frac12 \left\{\left(b_1 t\right)^\pm,b_2^\pm,\left(b_3 t^{-1}\right)^\pm\right\} \,,
\end{align}
where $a_i$ and $b_i$ will be sent to $y_1$ and $y_2$, respectively, at the end of the analysis. {For generic $a_i$ and $b_i$ this refinement restore the asymmetric $\mathsf S$-wall to the symmetric one.} Accordingly, $\hat A_3(y_{1,2};pq/t)$ should be replaced by
\begin{align}
\hat A_3(y_1;pq/t) = A_3(y_1 t, y_1, y_1 t^{-1};pq/t) \quad &\longrightarrow \quad A_3(a_1 t, a_2, a_3 t^{-1};pq/t) \,, \\
\hat A_3(y_2;pq/t) = A_3(y_2 t, y_2, y_2 t^{-1};pq/t) \quad &\longrightarrow \quad A_3(b_1 t, b_2, b_3 t^{-1};pq/t) \,.
\end{align}
These two $A_3(\dots;pq/t)$ include the following zero factors when $a_i \rightarrow y_1$ and $b_i \rightarrow y_2$:
\begin{gather}
\label{eq:vanishing factors}
\Gpq{pq a_1 a_2^{-1}} \,, \qquad \Gpq{pq a_2 a_3^{-1}} \,, \qquad \Gpq{pq b_1 b_2^{-1}} \,, \qquad \Gpq{pq b_2 b_3^{-1}} \,,
\end{gather}
which make the index vanish unless they are compensated by other diverging factors.

Some of such divergent factors are as follows:
\begin{gather}
\Gpq{t^{-\frac12} a_1^{-1} z^{(2)}_i{}^\pm} \,, \quad \Gpq{t^\frac12 a_2 z^{(2)}_i{}^\pm} \,, \quad \Gpq{t^\frac12 a_2^{-1} z^{(2)}_i{}^\pm} \,, \quad \Gpq{t^{-\frac12} a_3 z^{(2)}_i{}^\pm} \,, \\
\Gpq{t^{-\frac12} b_1^{-1} z^{(2)}_i{}^\pm} \,, \quad \Gpq{t^\frac12 b_2 z^{(2)}_i{}^\pm} \,, \quad \Gpq{t^\frac12 b_2^{-1} z^{(2)}_i{}^\pm} \,, \quad \Gpq{t^{-\frac12} b_3 z^{(2)}_i{}^\pm} \,,
\end{gather}
which can be rewritten as
\begin{gather}
\Gpq{a_1^{-\frac12} a_2^\frac12 \left(t^\frac12 a_1^\frac12 a_2^\frac12\right)^\pm z^{(2)}_i{}^\pm} \,, \qquad \Gpq{a_2^{-\frac12} a_3^\frac12 \left(t^{-\frac12} a_2^\frac12 a_3^\frac12\right)^\pm z^{(2)}_i{}^\pm} \,, \\
\Gpq{b_1^{-\frac12} b_2^\frac12 \left(t^\frac12 b_1^\frac12 b_2^\frac12\right)^\pm z^{(2)}_i{}^\pm} \,, \qquad \Gpq{b_2^{-\frac12} b_3^\frac12 \left(t^{-\frac12} b_2^\frac12 b_3^\frac12\right)^\pm z^{(2)}_i{}^\pm} \,.
\end{gather}
These are part of the contributions of the blue lines in Figure \ref{fig:SQCD_Nc=Nf=4_dualised_Higgsing_setI} and yield colliding poles in the limit $a_i \rightarrow y_1$ and $b_i \rightarrow y_2$, which can combine with the vanishing factors in \eqref{eq:vanishing factors} to form several delta distributions. As we explain shortly, we get a different frame depending on how many deltas in $z^{(2)}_i$ are formed in this way.

\paragraph{Frame zero.}
Let us first consider the case where the following factors combine into delta distributions:\footnote{One might wonder if we also need to consider other distributions such as $\tilde \delta\left(z^{(2)}_i,t^\frac12 b_1 \right)$ and $\tilde \delta\left(z^{(2)}_j,t^{-\frac12} b_2 \right)$ originating from similar factors written in terms of $b_i$ instead of $a_i$. However, as we will see shortly, the final result is proportional to $\delta(y_1-y_2) = \delta(a_i-b_j)$, so when evaluated on this delta, the $b_i$ factors yield \eqref{eq:colliding poles 1} and \eqref{eq:colliding poles 2}. Thus, both factors in $a_i$ and $b_i$ correspond to the same singularity, and we don't need to consider the delta in $b_i$ separately. 
}
\begin{align}
\Gpq{pq a_1 a_2^{-1}} \Gpq{a_1^{-\frac12} a_2^\frac12 \left(t^\frac12 a_1^\frac12 a_2^\frac12 z^{(2)}_i{}^{-1}\right)^\pm} \quad &\longrightarrow \quad \tilde \delta\left(z^{(2)}_i,t^\frac12 a_1 \right) \,, \label{eq:colliding poles 1} \\
\Gpq{pq a_2 a_3^{-1}} \Gpq{a_2^{-\frac12} a_3^\frac12 \left(t^{-\frac12} a_2^\frac12 a_3^\frac12 z^{(2)}_j{}^{-1}\right)^\pm} \quad &\longrightarrow \quad \tilde \delta\left(z^{(2)}_j,t^{-\frac12} a_2 \right) \label{eq:colliding poles 2}
\end{align}
in the limits $a_2 \rightarrow a_1$ and $a_3 \rightarrow a_2$, respectively. Here we use
 \begin{align}
    \lim_{s \rightarrow 1} \Gamma_e\left(pq s^2\right) \Gamma_e\left(s^{-1} x y^{-1} \right)\Gamma_e\left(s^{-1} x^{-1} y \right)  = 
    \tilde \delta(x,y) \,,
\end{align}
where
\begin{align}
\tilde \delta(x,y) = \frac{2 \pi i x}{(p;p)_\infty (q;q)_\infty} \delta(x-y) \,.
\end{align}
However, we should not take those limits right away because we still need the refined variables to avoid a strict zero in the remaining steps. Instead, we focus on the residue at one of the colliding poles on the left hand side and take the limit at the end of the computation to keep the refined variables until then. After taking the limit, this residue will give the value of the integral with the delta distributions evaluated.

The colliding poles in this case are given by
\begin{align}
&\left\{\begin{array}{l}
z^{(2)}_i = t^\frac12 a_1 \,, \\
z^{(2)}_i = t^\frac12 a_2\,,
\end{array}\right. \qquad \text{for \eqref{eq:colliding poles 1},} \\
&\left\{\begin{array}{l}
z^{(2)}_j = t^{-\frac12} a_2 \,, \\
z^{(2)}_j = t^{-\frac12} a_3\,,
\end{array}\right. \qquad \text{for \eqref{eq:colliding poles 2}.}
\end{align}
Therefore, choosing $i = 3$ and $j = 4$ without loss of generality, we evaluate the residue at the following pole:
\begin{align}
\begin{aligned}
\label{eq:pole}
z^{(2)}_3 &= t^\frac12 a_1 \,, \\
z^{(2)}_4 &= t^{-\frac12} a_2 \,,
\end{aligned}
\end{align}
which Higgses $USp(8)_{z^{(2)}}$ down to $USp(4)_{z^{(2)}}$. After taking the residue, we have two zero factors remaining in \eqref{eq:vanishing factors}:
\begin{align}
\Gpq{pq b_1 b_2^{-1}} \,, \quad \Gpq{pq b_2 b_3^{-1}}
\end{align}
and one new zero from $A_4(\vec z^{(2)};pq/t)$:
\begin{align}
\label{eq:extra zero}
\left.\Gpq{pq t^{-1} z^{(2)}_3 z^{(2)}_4{}^{-1}}\right|_{\substack{z^{(2)}_3 = t^\frac12 a_1 \,, \\
z^{(2)}_4 = t^{-\frac12} a_2 }} \quad = \quad \Gpq{pq a_1 a_2^{-1}} \,.
\end{align}
Those three zeros need to be canceled appropriately to have the non-vanishing index. It turns out the contribution of the blue bifundamentals in Figure \ref{fig:SQCD_Nc=Nf=4_dualised_Higgsing_setI} includes the following factors once evaluated at the pole \eqref{eq:pole}:
\begin{align}
&\Gpq{a_1 b_1^{-1}} \,, \quad \Gpq{a_1^{-1} b_2} \,, \quad \Gpq{a_2 b_2^{-1}} \,, \quad \Gpq{a_2^{-1} b_3} \,, \label{eq:diverging factors 01} \\
&\Gpq{a_2^{-1} b_1^{-1}} \,, \quad \Gpq{a_1^{-1} b_2^{-1}} \,, \quad \Gpq{a_2 b_2} \,, \quad \Gpq{a_1 b_3} \,. \label{eq:diverging factors 02}
\end{align}
First looking at \eqref{eq:diverging factors 01}, we find that
\begin{align}
\begin{aligned}
\label{eq:delta ratio}
&\lim_{a_2 \rightarrow a_1} \Big(\Gpq{pq a_1 a_2^{-1}} \Gpq{a_1^{-1} b_2} \Gpq{a_2 b_2^{-1}} \Big) \\
&\quad \times \Big(\Gpq{pq b_1 b_2^{-1}} \Gpq{a_1 b_1^{-1}}\Big) \times \Big(\Gpq{pq b_2 b_3^{-1}} \Gpq{a_2^{-1} b_3}\Big) \\
&=\tilde \delta\left(a_1,b_2\right) \times \Big(\Gpq{pq b_1 b_2^{-1}} \Gpq{a_1 b_1^{-1}}\Big) \times \Big(\Gpq{pq b_2 b_3^{-1}} \Gpq{a_1^{-1} b_3}\Big) \\
&=\tilde \delta\left(a_1,b_2\right) \,.
\end{aligned}
\end{align}
Also sending the remaining $a_i$ and $b_i$ to $y_1$ and $y_2$, respectively, we obtain $\tilde \delta(y_1,y_2)$, and the resulting quiver is exactly the one in Figure \ref{fig:SQCD_Nc=Nf=4_ACdual}. Similarly, we can also pair the zero factors with \eqref{eq:diverging factors 02}, leading to $\tilde \delta\left(y_1,y_2^{-1}\right)$ instead.

Interestingly, one can also express \eqref{eq:delta ratio} in terms of two delta distributions using
\begin{align}
\Gpq{pq b_1 b_2^{-1}} \Gpq{a_1 b_1^{-1}} \Gpq{a_1^{-1} b_2} \quad &\longrightarrow \quad \tilde \delta\left(y_1,y_2\right) \,, \\
\Gpq{pq b_2 b_3^{-1}}\Gpq{a_2 b_2^{-1}} \Gpq{a_2^{-1} b_3} \quad &\longrightarrow \quad \tilde \delta\left(y_1,y_2\right)\,,
\end{align}
with $a_i \rightarrow y_1$ and $b_i \rightarrow y_2$.
Combining with $\Gpq{pq a_1 a_2^{-1}} = \Gpq{pq}$ in \eqref{eq:extra zero}, we get
\begin{align}
\label{eq:alternative}
\left(\tilde \delta\left(y_1,y_2\right)\right)^2 \Gpq{pq} = \frac{2 \pi i y_1 \delta\left(y_1-y_2\right)}{(p;p)_\infty (q;q)_\infty} \times \frac{2 \pi i y_1 \delta\left(0\right)}{(p;p)_\infty (q;q)_\infty} \times \Gpq{pq} \,,
\end{align}
which is the expression we encounter in Subsubection \ref{44sec}. The analysis in this appendix shows that the last two factors are canceled out because \eqref{eq:alternative} must be the same as \eqref{eq:delta ratio}.

\paragraph{Frame one.}
Next, let us consider the case where we get three deltas in $z^{(2)}_i$ as follows:
\begin{align}
\Gpq{pq a_1 a_2^{-1}} \Gpq{a_1^{-\frac12} a_2^\frac12 \left(t^\frac12 a_1^\frac12 a_2^\frac12 z^{(2)}_i{}^{-1}\right)^\pm} \quad &\longrightarrow \quad \tilde \delta\left(z^{(2)}_i,t^\frac12 a_1\right) \,, \\
\Gpq{pq a_2 a_3^{-1}} \Gpq{a_2^{-\frac12} a_3^\frac12 \left(t^{-\frac12} a_2^\frac12 a_3^\frac12 z^{(2)}_j{}^{-1}\right)^\pm} \quad &\longrightarrow \quad \tilde \delta\left(z^{(2)}_j,t^{-\frac12} a_2\right) \,, \\
\Gpq{pq b_1 b_2^{-1}} \Gpq{b_1^{-\frac12} b_2^\frac12 \left(t^\frac12 b_1^\frac12 b_2^\frac12 z^{(2)}_k{}^{-1}\right)^\pm} \quad &\longrightarrow \quad \tilde \delta\left(z^{(2)}_k,t^\frac12 b_1\right) \label{eq:colliding poles 3}
\end{align}
in the limit $a_2 \rightarrow a_1$, $a_3 \rightarrow a_2$ and $b_2 \rightarrow b_1$, respectively. As before, to implement such deltas, we take the residue at the pole
\begin{align}
\begin{aligned}
z^{(2)}_2 &= t^\frac12 a_1 \,, \\
z^{(2)}_3 &= t^{-\frac12} a_2 \,, \\
z^{(2)}_4 &= t^{\frac12} b_1 \,,
\end{aligned}
\end{align}
which Higgses $USp(8)_{z^{(2)}}$ down to $USp(2)_{z^{(2)}}$. After taking the residue, we have one zero factor remaining in \eqref{eq:vanishing factors}
\begin{align}
\Gpq{pq b_2 b_3^{-1}}
\end{align}
and one new zero from $A_4(\vec z^{(2)};pq/t)$
\begin{align}
\left.\Gpq{pq t^{-1} z^{(2)}_2 z^{(2)}_3{}^{-1}}\right|_{\substack{z^{(2)}_2 = t^\frac12 a_1 \,, \\
z^{(2)}_3 = t^{-\frac12} a_2 }} \quad = \quad \Gpq{pq a_1 a_2^{-1}} \,.
\end{align}
The diverging factors cancelling these zeros are
\begin{align}
&\Gpq{t a_1 b_2^{-1}} \,, \quad \Gpq{t^{-1} a_1^{-1} b_3} \,, \quad \Gpq{t a_2 b_3^{-1}} \,, \label{eq:diverging factors 11}\\
&\Gpq{t a_2^{-1} b_2^{-1}} \,, \quad \Gpq{t a_1^{-1} b_3^{-1}} \,, \quad \Gpq{a_2 b_3} \label{eq:diverging factors 12}
\end{align}
in the contribution of the blue bifundamentals in Figure \ref{fig:SQCD_Nc=Nf=4_dualised_Higgsing_setI}. From \eqref{eq:diverging factors 11} and \eqref{eq:diverging factors 12}, respectively, we obtain
\begin{align}
&\lim_{a_2 \rightarrow a_1} \Big(\Gpq{pq a_1 a_2^{-1}} \Gpq{t^{-1} a_1^{-1} b_3} \Gpq{t a_2 b_3^{-1}} \Big) \times \Big(\Gpq{pq b_2 b_3^{-1}} \Gpq{t a_1 b_2^{-1}}\Big) = \tilde \delta\left(a_1,t^{-1} b_3\right) \,, \\
&\lim_{a_2 \rightarrow a_1} \Big(\Gpq{pq a_1 a_2^{-1}} \Gpq{t a_1^{-1} b_3^{-1}} \Gpq{t^{-1} a_2 b_3} \Big) \times \Big(\Gpq{pq b_2 b_3^{-1}} \Gpq{t a_2^{-1} b_2^{-1}}\Big) = \tilde \delta\left(a_1,t b_3^{-1}\right) \,.
\end{align}
Sending the remaining $a_i$ and $b_i$ to $y_1$ and $y_2$, respectively, we obtain $\tilde \delta\left(y_1,\left(t^{-1} y_2\right)^\pm\right)$, and the resulting quiver is exactly the one in Figure \ref{fig:SQCD_Nc=Nf=4_INTdual}.

Similarly, one can also consider another case with
\begin{align}
\Gpq{pq b_2 b_3^{-1}} \Gpq{b_2^{-\frac12} b_3^\frac12 \left(t^{-\frac12} b_2^\frac12 b_3^\frac12 z^{(2)}_k{}^{-1}\right)^\pm} \quad &\longrightarrow \quad \tilde \delta\left(z^{(2)}_k,t^{-\frac12} b_2\right) 
\end{align}
instead of \eqref{eq:colliding poles 3}, leading to $\tilde \delta\left(y_1,\left(t y_2\right)^\pm\right)$.

\paragraph{Frame two.}
The last case is the one where we get four deltas in $z^{(2)}_i$
\begin{align}
\Gpq{pq a_1 a_2^{-1}} \Gpq{a_1^{-\frac12} a_2^\frac12 \left(t^\frac12 a_1^\frac12 a_2^\frac12 z^{(2)}_i{}^{-1}\right)^\pm} \quad &\longrightarrow \quad \frac{2 \pi i z^{(2)}_i \delta\left(z^{(2)}_i-t^\frac12 a_1\right)}{(p;p)_\infty (q;q)_\infty} \,, \\
\Gpq{pq a_2 a_3^{-1}} \Gpq{a_2^{-\frac12} a_3^\frac12 \left(t^{-\frac12} a_2^\frac12 a_3^\frac12 z^{(2)}_j{}^{-1}\right)^\pm} \quad &\longrightarrow \quad \frac{2 \pi i z^{(2)}_j \delta\left(z^{(2)}_j-t^{-\frac12} a_2\right)}{(p;p)_\infty (q;q)_\infty} \,, \\
\Gpq{pq b_1 b_2^{-1}} \Gpq{b_1^{-\frac12} b_2^\frac12 \left(t^\frac12 b_1^\frac12 b_2^\frac12 z^{(2)}_k{}^{-1}\right)^\pm} \quad &\longrightarrow \quad \frac{2 \pi i z^{(2)}_k \delta\left(z^{(2)}_k-t^\frac12 b_1\right)}{(p;p)_\infty (q;q)_\infty} \,, \\
\Gpq{pq b_2 b_3^{-1}} \Gpq{b_2^{-\frac12} b_3^\frac12 \left(t^{-\frac12} b_2^\frac12 b_3^\frac12 z^{(2)}_l{}^{-1}\right)^\pm} \quad &\longrightarrow \quad \frac{2 \pi i z^{(2)}_l \delta\left(z^{(2)}_l-t^{-\frac12} b_2\right)}{(p;p)_\infty (q;q)_\infty}
\end{align}
in the limit $a_2 \rightarrow a_1$, $a_3 \rightarrow a_2$, $b_2 \rightarrow b_1$, and $b_3 \rightarrow b_2$, respectively. We take the residue at the following pole to implement these deltas:
\begin{align}
\begin{aligned}
\label{eq:pole 2}
z^{(2)}_1 &= t^\frac12 a_1 \,, \\
z^{(2)}_2 &= t^{-\frac12} a_2 \,, \\
z^{(2)}_3 &= t^{\frac12} b_1 \,, \\
z^{(2)}_4 &= t^{-\frac12} b_2 \,,
\end{aligned}
\end{align}
which completely Higgses $USp(8)_{z^{(2)}}$.

While we have no remaining zero factor in \eqref{eq:vanishing factors} after taking this residue,
we get new ones from $A_4(\vec z^{(2)};pq/t)$
\begin{align}
\begin{aligned}
\label{eq:extra zero 2}
\left.\Gpq{pq t^{-1} z^{(2)}_1 z^{(2)}_2{}^{-1}}\right|_{\substack{z^{(2)}_1 = t^\frac12 a_1 \,, \\
z^{(2)}_2 = t^{-\frac12} a_2 }} \quad = \quad \Gpq{pq a_1 a_2^{-1}} \,, \\
\left.\Gpq{pq t^{-1} z^{(2)}_3 z^{(2)}_4{}^{-1}}\right|_{\substack{z^{(2)}_3 = t^\frac12 b_1 \,, \\
z^{(2)}_4 = t^{-\frac12} b_2 }} \quad = \quad \Gpq{pq b_1 b_2^{-1}} \,.
\end{aligned}
\end{align}
Unlike the previous cases, these zeros cannot be canceled by the contribution of the blue bifundamentals. Instead, the nodes $USp(2)_{z^{(1)}}$ and $USp(2)_{z^{(3)}}$ have the following fundamental chirals after taking the residue at \eqref{eq:pole 2}:
\begin{align}
&\Gpq{b_1^{-1} {z^{(1)}}^\pm} \,, \quad \Gpq{b_2 {z^{(1)}}^\pm} \,, \quad \Gpq{t b_1 {z^{(1)}}^\pm} \,, \quad \Gpq{t b_2^{-1} {z^{(1)}}^\pm} \,, \\
&\Gpq{a_1^{-1} {z^{(3)}}^\pm} \,, \quad \Gpq{a_2 {z^{(3)}}^\pm} \,, \quad \Gpq{t a_1 {z^{(3)}}^\pm} \,, \quad \Gpq{t a_2^{-1} {z^{(3)}}^\pm} \,,
\end{align}
where the first two factors in each line provide colliding poles. Thus, the extra zeros in \eqref{eq:extra zero 2} combine with these colliding poles to form delta distributions that completely Higgs $USp(2)_{z^{(1)}}$ and $USp(2)_{z^{(2)}}$, leading to the quiver theory in Figure \ref{fig:SQCD_Nc=Nf=4_ITAdual}.

Those are the all possible ways to compensate the vanishing factors in \eqref{eq:vanishing factors} up to equivalent choices, which exactly lead to the frames obtained by the HW moves following the rules as explained in Subsubection \ref{44sec}. Especially, there is no way to obtain the non-physical frames corresponding to the identifications shown in Figure \ref{n44id}, justifying the rules we gave at the beginning of Appendix \ref{moreframe}. 

\subsubsection{Flips and a single frame duality}
\label{appsingleframe}
 We argued in Section \ref{sec:single frame duality} that an appropriate choice of singlets flipping some of the mesons kills the multiple dual frames but the last one so that we find a more conventional single frame duality for the flipped bad theories. Let us examine the $N_c = N_f = 4$ case in detail.

On the original SQCD side, one can construct the following mesons out of the fields in the saw dressed $j$ times with the antisymmetric field:
\begin{align}
\label{eq:singlets}
\prod_{j = 0}^3 \Gpq{t^{j-1} y_1^\pm y_2^\pm} .
\end{align}
In particular, we have seen in Section \ref{sec:index higgsing} that the VEVs of the mesons dressed with the antisymmetric field once and twice lead to the dual frame zero and frame one.
We want to couple those mesons with extra singlets
\begin{align}
\label{eq:extra singlets}
\prod_{k = 0}^3 \Gpq{p q t^{j-2} y_1^\pm y_2^\pm} ,
\end{align}
which forbid the VEVs of the mesons so that the aforementioned dual frames are all lifted, leaving the last dual frame only.

As explained in this appendix, from the index perspective, the multiple dual frames arise from separate singularities giving rise to different delta distributions, either in gauge fugacities or flavor ones. It turns out the extra singlets we introduce provide extra zeros canceling such singularities, leaving the one leading to the last dual frame. To see this, let us refine the index contribution \eqref{eq:extra singlets} of the extra singlets as follows:
\begin{align}
\begin{aligned}
\label{eq:refined extra singlets}
\Gpq{p q t^{-2} y_1^\pm y_2^\pm} \quad &\longrightarrow \quad \Gpq{p q t^{-2} a_3 b_2^\pm} \Gpq{p q t^{-2} a_1^{-1} b_2^\pm} , \\
\Gpq{p q t^{-1} y_1^\pm y_2^\pm} \quad &\longrightarrow \quad \Gpq{p q t^{-1} a_1 b_3} \Gpq{p q t^{-1} a_1 b_1^{-1}} \Gpq{p q t^{-1} a_3^{-1} b_3} \Gpq{p q t^{-1} a_3^{-1} b_1^{-1}} , \\
\Gpq{p q y_1^\pm y_2^\pm} \quad &\longrightarrow \quad \Gpq{p q a_1 b_2^\pm} \Gpq{p q a_3^{-1} b_2^\pm} , \\
\Gpq{p q t y_1^\pm y_2^\pm} \quad &\longrightarrow \quad \Gpq{p q t a_1 b_1} \Gpq{p q t a_1 b_3^{-1}} \Gpq{p q t a_3^{-1} b_1} \Gpq{p q t a_3^{-1} b_3^{-1}} ,
\end{aligned}
\end{align}
which are tuned in a way to couple them to the refined mesons.

Now let us first examine frame zero, accompanied by $\tilde \delta(y_1,y_2^\pm)$. In \eqref{eq:delta ratio}, the delta distribution $\tilde \delta(y_1,y_2)$ originates from the colliding poles of
\begin{align}
\label{eq:delta poles}
\Gpq{a_1^{-1} b_2} , \qquad \Gpq{a_2 b_2^{-1}}
\end{align}
in the limit $a_2 \rightarrow a_1$, which compensate the zero of $\Gpq{pq a_1 a_2^{-1}}$. Once we introduce the extra singlets in \eqref{eq:refined extra singlets}, those colliding poles are canceled by
\begin{align}
\Gpq{pq a_1 b_2^{-1}} , \qquad \Gpq{pq a_3^{-1} b_2}
\end{align}
in the limit $a_i \rightarrow y_1$, leaving the vanishing factor $\Gpq{pq a_1 a_2^{-1}}$. Similarly, $\tilde \delta(y_1,y_2^{-1})$ originates from the colliding poles of
\begin{align}
\Gpq{a_1^{-1} b_2^{-1}} , \qquad \Gpq{a_2 b_2} ,
\end{align}
which are canceled by the zeros of
\begin{align}
\Gpq{pq a_1 b_2} , \qquad \Gpq{pq a_3^{-1} b_2^{-1}}
\end{align}
in \eqref{eq:refined extra singlets} in the limit $a_i \rightarrow y_1$, again leaving the vanishing factor $\Gpq{pq a_1 a_2^{-1}}$. Thus, there is no singularity leading to frame zero anymore in the presence of the extra singlets.

One can check that the other singularities leading to frame one are also cancelled by the extra singlet contributions in \eqref{eq:refined extra singlets}. On the other hand, frame two arises from the pole \eqref{eq:pole 2} in $z_i^{(2)}$, which never cancels out with the singlet contributions in \eqref{eq:refined extra singlets}. Instead, the extra singlets couple to some of the original singlets in frame two shown in \eqref{eq:singlets}, and their contributions are completely cancelled out. Thus, we end up with the unique dual frame, given by frame two where the singlets in \eqref{eq:singlets} are flipped by the extra singlets \eqref{eq:extra singlets}.

\subsection{Alternative HW collisions}
\label{app:alternativeHWcollisions}

\begin{figure}[!ht]
    \includegraphics[width=1\textwidth]{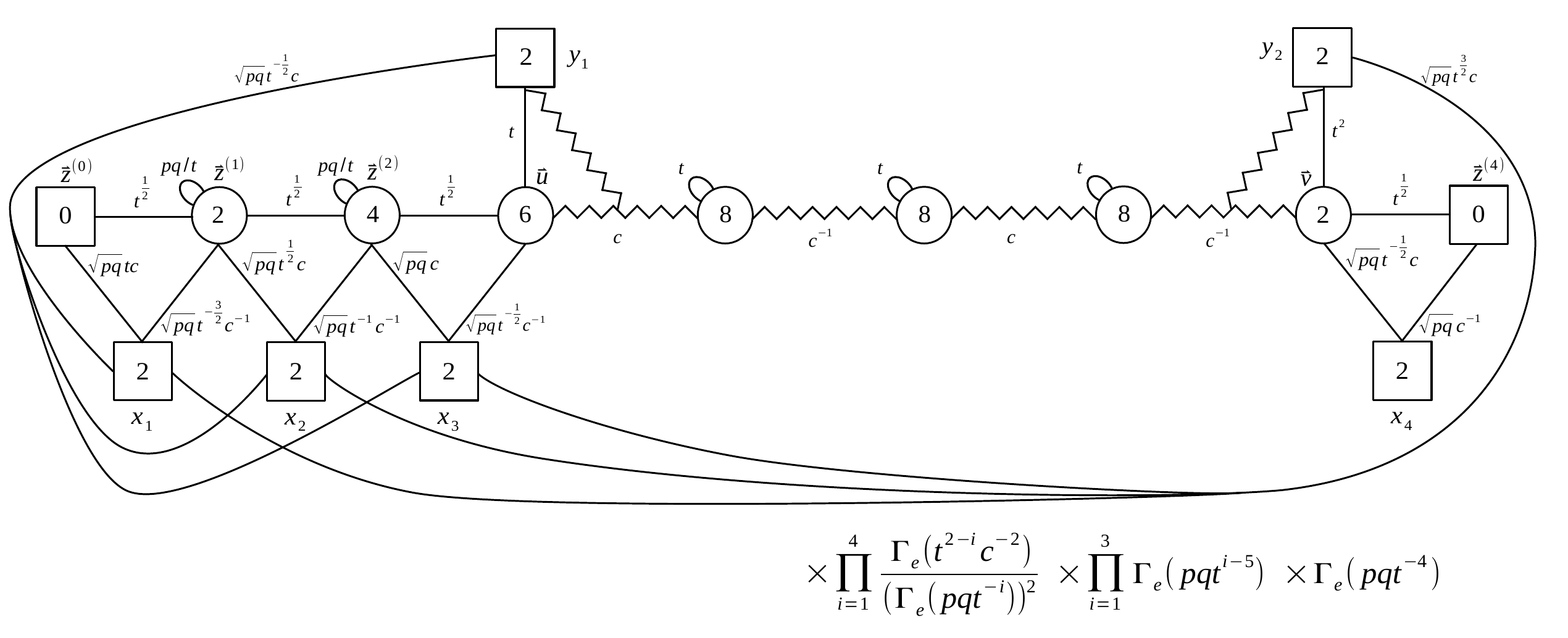}
    \caption{Different application of the HW moves in the $\mathsf{S}$-dualized SQCD for $N_c=N_f=4$.}
    \label{fig:SQCD_Nc=Nf=4_dual_HW_setII}
\end{figure}

We can also consider other ways of implementing the HW moves, all of which must give the same result. For example, we can perform 3 HW move on the l.h.s.~and one HW move on the r.h.s.~to obtain the theory in Figure \ref{fig:SQCD_Nc=Nf=4_dual_HW_setII}.

\begin{figure}[!ht] 
	\centering
	\includegraphics[width=.9\textwidth]{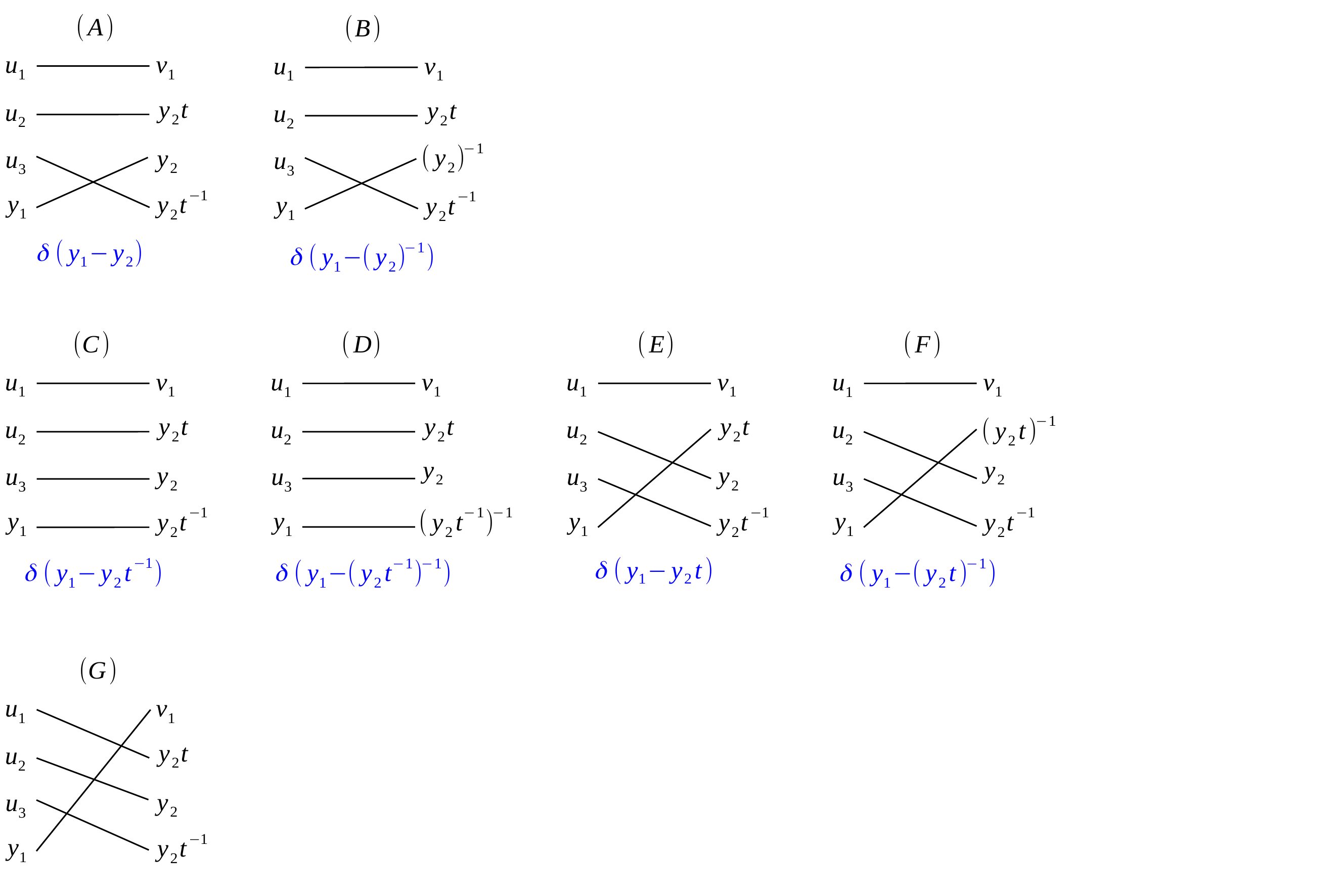}
	\caption{Possible inequivalent identifications of fugacities enforced by the colliding Identity-walls for $N_c=N_f=4$ after using the different application of the HW moves as in Figure \ref{fig:SQCD_Nc=Nf=4_dual_HW_setII}.}
	\label{44bisid}
\end{figure}

Now the two colliding  asymmetric Identity-walls enforce the identification of the two fugacity vectors:
\begin{align}
\vec f &= \left\{ u_1, u_2, u_3, y_1 \right\} \,, \\
\vec h &= \left\{ v_1, y_2t, y_2, y_2t^{-1} \right\}
\end{align}
up to the Weyl group actions of $USp(4)$, consisting of the permutations and the inversion. With these HW moves, it is much simpler to classify all possible identifications. Since the permutations and the inversion of the dynamical variables do not affect the result, we only need to focus on the possible ways to pair  $y_1$ from the first vector $\vec f$ to an element of the second vector $\vec h$ as shown in Figure \ref{44bisid}. Now the identifications  $(A), (B)$ lead to frame zero, $(C),(D),(E),(F)$ lead to frame one and $(G)$ to frame two. Thus, in this case we directly see the three expected  frames.

We could consider the case where we perform all the four HW moves from left to right (or right to left), whose analysis is slightly more subtle though. In this case, naively one would find a single dual frame because one fugacity vector includes all dynamical fugacities while the other includes global fugacities only. However, one can show that the index of this theory still has some singularities leading to a sequential Higgsing, which need to be taken into account by a careful inspection of the poles structure. One can carry out the analysis in the same way as done in the previous subsection and recover the expected frames.

\section{Comments on the derivation for generic $N_c$ and $N_f$}
\label{app:general}

In this appendix, we sketch the derivation of the general formulas for the index of the 4d bad SQCD that we presented in Subsections \ref{subsubsec:genNfeven} and \ref{subsubsec:genNfodd}. Again we distinguish our analysis between the cases of $N_f$ even and odd.

\subsection{Even $N_f$}
\label{app:generalNfeven}

We first consider the general case of arbitrary $N_c$ and arbitrary even $N_f=2k$ with $k< N_c$. We will only focus on the gauge groups of the quiver and on the matter fields that transform under them to demonstrate that we indeed get the structure of the frames given by the good $USp(2r)$ SQCD for $r=N_f-N_c,\cdots,\frac{N_f}{2}$, while we will ignore the gauge singlet fields.

\begin{figure}[!ht]
\includegraphics[width=\textwidth]{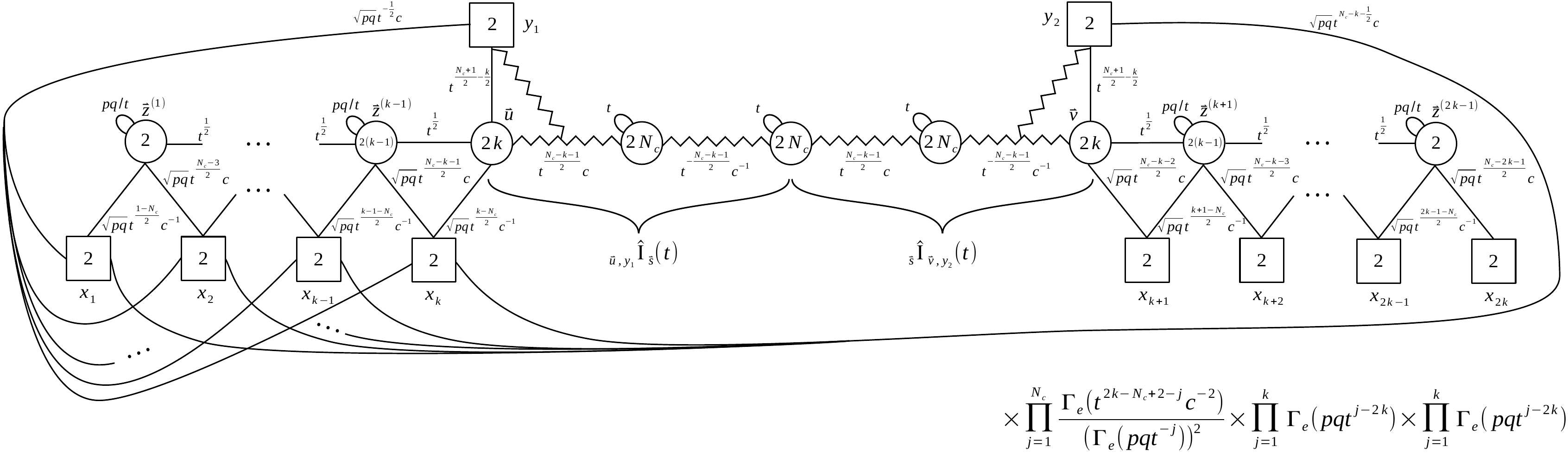}
\caption{$\mathsf{S}$-dualized SQCD for $N_f=2k$ even after having applied the HW moves.}
\label{eclash}
\end{figure}

We start from the $\mathsf{S}$-dualized theory in Figure \ref{SQCDone} and we perform $k$ HW moves on the left side of the quiver and $k$ HW moves on the right side. We then arrive at the situation depicted in Figure \ref{eclash}, where two asymmetric Identity-walls collide, yielding a product of delta functions that identify the fugacity vectors on the two sides

\begin{eqnarray}
	\vec{f}&=& \Big\{ u_1,\cdots u_k,y_1t^{\frac{N-k-1}{2}},y_1t^{\frac{N-k-3}{2}}, \cdots, y_1t^{-\frac{N-k-1}{2}} \Big\} \,,\\ \nonumber
	\vec{h}&=& \Big\{ v_1,\cdots v_k,y_2t^{\frac{N-k-1}{2}},y_2t^{\frac{N-k-3}{2}}, \cdots ,y_2t^{-\frac{N-k-1}{2}} \Big\} \,.
\end{eqnarray}
The physical identifications we consider are those that do not correspond to fixed points of the Weyl action of $USp(2N_c)$, that is
\begin{enumerate}
\item $f_i\neq f_j^{\pm1}$ for any pair of $i$ and $j$;
\item $f_i\neq 1$ for any $i$.
\end{enumerate}
Such physical identifications yield $N-k+1=N-\frac{N_f}{2}+1$ independent frames and their twins obtained by taking the inverse of each element of the second set $\vec{h}$.

As we have seen in explicit examples in Section \ref{sec:4dbadSQCD}, the delta conditions once implemented might determine some VEVs. By studying the Higgs mechanism triggered by each of these VEVs we find an interacting part consisting of the good SQCD and some singlets, which as we said we will ignore in this general discussion. Let us explain the procedure after implementing the delta conditions for each frame. 

\begin{figure}[!ht]
\centering
\includegraphics[width=.9\textwidth]{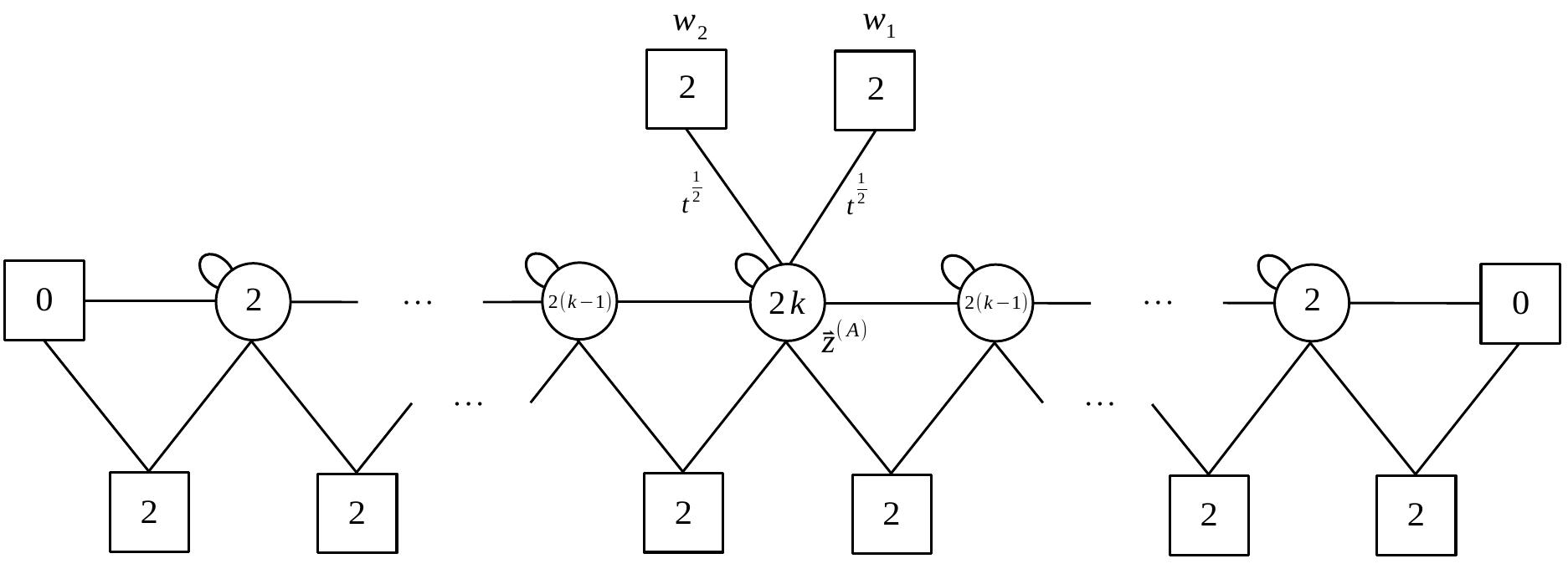}
\caption{The mirror dual of frame zero of the bad SQCD for $N_f=2k$ even.}
\label{fig:bad_SQCD_even_Nf_frame0}
\end{figure}

\paragraph{Frame zero.} The first set of conditions is
\begin{align}
\label{eq:condition 1}
\left\{\begin{array}{rll}
u_i &= v_i \,, &\qquad i = 1\,,\, \dots \,,\, k \,,\\
y_1 &= y_2 \,.
\end{array}\right.
\end{align}
The resulting theory is shown in Figure \ref{fig:bad_SQCD_even_Nf_frame0}, which is the mirror of the good $USp(2k)=USp(N_f)$ SQCD.\footnote{In the drawing we are representing the flavor symmetries under which the fundamental chirals transform as $SU(2)_{w_1}\times SU(2)_{w_2}$ or $USp(4)$ with $SU(2)^2$ subgroup, although generically there are extra singlet fields which we are ignoring here that do not form nice $SU(2)_{w_1}\times SU(2)_{w_2}$ representations. Nevertheless, after all the manipulations that we will perform the $SU(2)_{w_1}\times SU(2)_{w_2}$ symmetry will become manifest in the final result also for the singlet fields. For the rest of this subsection we will ignore this fact since we will not take into account the singlets and always draw $SU(2)$ flavor boxes in the quivers to simplify the drawings.}
After implementing the delta conditions, the gauge node labelled by A in Figure \ref{fig:bad_SQCD_even_Nf_frame0}, which comes from the $k$-th gauge node from the left in Figure \ref{eclash} that has been identified with the $(k+4)$-th node, has several chirals in its fundamental representation, which come from the original flavors attached to the $k$-th node and the $(k+4)$-th node and the remaining contributions of the collapsed Identity-walls. However, most of them couple to each other and become massive. As a result, one can verify that only two doublets remain massless, whose index contribution is
\begin{align}
\Gpq{t^\frac12 u_i^\pm \left(y_2 t^\frac{N-k}{2}\right)^\pm} \quad &\equiv \quad \Gpq{t^\frac12 u_i^\pm w_2{}^\pm} \,, \\
\Gpq{t^\frac12 u_i^\pm \left(y_1 t^{-\frac{N-k}{2}}\right)^\pm} \quad &\equiv \quad \Gpq{t^\frac12 u_i^\pm w_1{}^\pm} \,,
\end{align}
where we have defined
\begin{align}
\left\{\begin{array}{rl}
w_1 &= y_1 t^{-\frac{N-k}{2}} \,, \\
w_2 &= y_2 t^\frac{N-k}{2} \,.
\end{array}\right.
\end{align}

\begin{figure}[!ht]
\centering
\includegraphics[width=.9\textwidth]{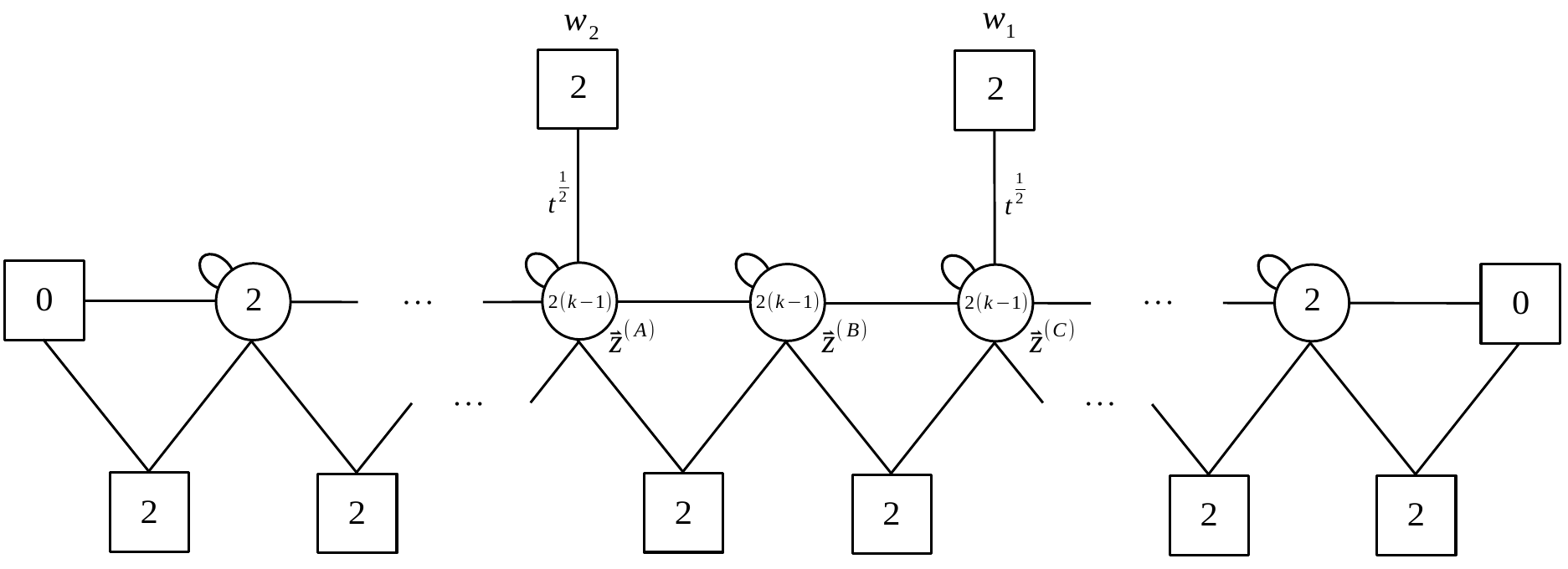}
\caption{The mirror dual of frame one of the bad SQCD for $N_f=2k$ even.}
\label{fig:bad_SQCD_even_Nf_frame1}
\end{figure}

\paragraph{Frame one.} The second set of conditions is
\begin{align}
\label{eq:condition 2}
\left\{\begin{array}{rll}
u_i &= v_{i+1} \,, &\qquad i = 1\,,\, \dots \,,\, k-1 \,, \\
u_k &= y_2 t^\frac{N-k-1}{2} \,, \\
v_1 &= y_1 t^{-\frac{N-k-1}{2}} \,, \\
y_1 &= y_2 t^{-1} \,.
\end{array}\right.
\end{align}
The resulting theory is shown in Figure \ref{fig:bad_SQCD_even_Nf_frame1}, which is the mirror of the good $USp(2k-2)=USp(N_f-2)$ SQCD.
The gauge node A, which comes from the $(k-1)$-th node from the left in Figure \ref{eclash}, has one fundamental doublet, originating from the broken part of the bifundamental field between the $(k-1)$-th node and the $k$-th node, which is now called gauge node B. This is because the $k$-th node is broken from $USp(2 k)$ due to the constraint \eqref{eq:condition 2} on $u_k$
\begin{align}
\Gpq{t^\frac12 z^{(A)}_i{}^\pm u_k^\pm}=\Gpq{t^\frac12 z^{(A)}_i{}^\pm \left(y_2 t^\frac{N-k-1}{2}\right)^\pm} \quad &\equiv \quad \Gpq{t^\frac12 z^{(A)}_i{}^\pm w_2{}^\pm}\,.
\end{align}
Similarly, the gauge node C, which comes from the $(k+5)$-th node from the left in Figure \ref{eclash}, has a fundamental doublet, originating from the broken part of the bifundamental field between the $(k+4)$-th gauge node and the $(k+5)$-th gauge node
\begin{align}
\Gpq{t^\frac12 z^{(C)}_i{}^\pm v_1^\pm}=\Gpq{t^\frac12 z^{(C)}_i{}^\pm \left(y_1 t^{-\frac{N-k-1}{2}}\right)^\pm} \quad &\equiv \quad \Gpq{t^\frac12 z^{(C)}_i{}^\pm w_1{}^\pm}\,.
\end{align}
In both of the previous expressions we defined
\begin{align}
\left\{\begin{array}{rl}
w_1 &= y_1 t^{-\frac{N-k-1}{2}} \,, \\
w_2 &= y_2 t^\frac{N-k-1}{2} \,.
\end{array}\right.
\end{align}
On the other hand, the gauge node B, coming from the $USp(2k-2)$ part of the $USp(2k)$ of the $k$-th node from the left in Figure \ref{eclash} that has been identified by \eqref{eq:condition 2} with the $USp(2k-2)$ part of the $USp(2k)$ of the $(k+4)$-th node, doesn't have any residual fundamental chiral since they all become massive. Therefore, the final configuration is as shown in Figure \ref{fig:bad_SQCD_even_Nf_frame1}.

\begin{figure}[tbp]
\centering
\includegraphics[width=\textwidth]{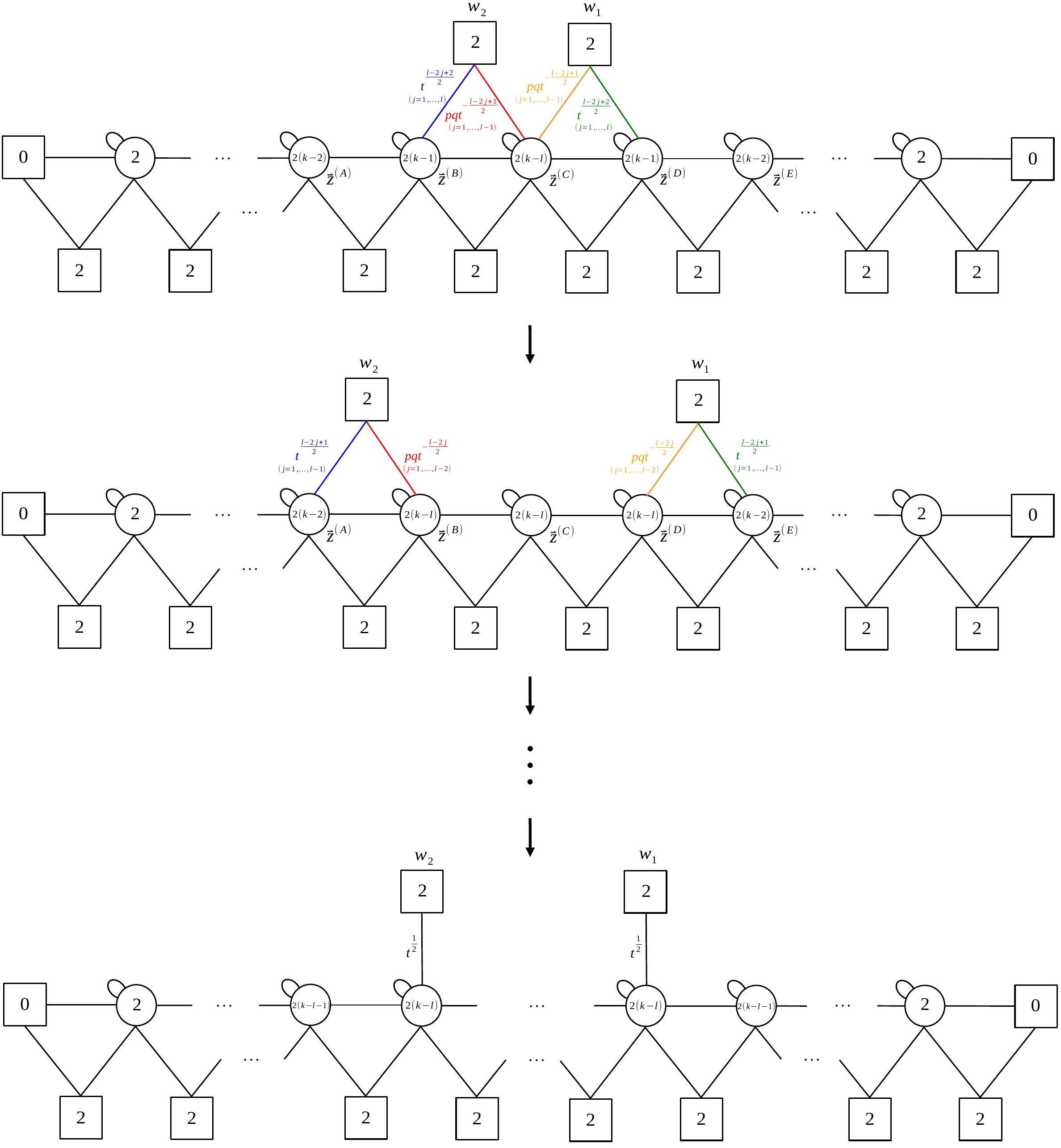}
\caption{Sequential Higgsing that results in the mirror dual of frame $l$ of the bad SQCD with $N_f=2k$ even.}
\label{fig:bad_SQCD_even_Nf_frameL}
\end{figure}

\paragraph{Frame \boldmath$l$.} We now consider the general frame $l$ with $1 < l < N-k$. The condition we impose in this case is
\begin{align}
\label{eq:condition 3}
\left\{\begin{array}{rll}
u_i &= v_{i+l} \,, &\qquad i = 1 \,,\, \dots \,,\, k-l \,, \\
u_{k-j+1} &= y_2 t^\frac{N-k-2 l+2 j-1}{2} \,, &\qquad j = 1\,,\, \dots \,,\, l \,, \\
v_j &= y_1 t^{-\frac{N-k-2 j+1}{2}} \,, &\qquad j = 1 \,,\, \dots \,,\, l \,, \\
y_1 &= y_2 t^{-l} \,.
\end{array}\right.
\end{align}
The first quiver shown in Figure \ref{fig:bad_SQCD_even_Nf_frameL} is the theory after collapsing the Identity-walls with the condition \eqref{eq:condition 3}.
As frames zero and one above, one can work out the fundamental flavors attached to the gauge nodes B, C, D, where the node C comes from the $USp(2k-2l)$ subgroup of the $USp(2k)$ of the $k$-th node from the left in Figure \ref{eclash} that has been identified with the $USp(2k-2l)$ subgroup of the $USp(2k)$ of the $(k+4)$-th node in Figure \ref{eclash} by the condition \eqref{eq:condition 3}. Such fundamental flavors are as follows:
\begin{itemize}
\item for node B, they come from the bifundamentals connecting the $(k-1)$-th node in Figure \ref{eclash} and the $USp(2l)$ subgroup of the $k$-th node in Figure \ref{eclash} whose fugacities has been identified with global fugacities by \eqref{eq:condition 3}, which we can repackage as
\begin{align}
\label{eq:B}
\prod_{j=1}^l\Gpq{t^{\frac{1}{2}}z_i^{(B)}{}^\pm u_{k-j+1}^\pm}=\prod_{j=1}^l\Gpq{t^{\frac{l-2 j+2}{2}} z_i^{(B)}{}^\pm \left(y_2 t^\frac{N-k-l}{2}\right)^\pm} \equiv \prod_{j=1}^l\Gpq{t^{\frac{l-2 j+2}{2}} z_i^{(B)}{}^\pm w_2^\pm} \,;
\end{align}
\item for node D, they come from the bifundamentals connecting the $(k+5)$-th node in Figure \ref{eclash} and the $USp(2l)$ subgroup of the $(k+4)$-th node in Figure \ref{eclash} whose fugacities has been identified with global fugacities by \eqref{eq:condition 3}, which we can repackage as
\begin{align}
\label{eq:D}
\prod_{j=1}^l\Gpq{t^{\frac{1}{2}}z_i^{(D)}{}^\pm v_{j}^\pm}=\prod_{j=1}^l\Gpq{t^{\frac{l-2 j+2}{2}} z_i^{(D)}{}^\pm \left(y_1 t^{-\frac{N-k-l}{2}}\right)^\pm} \equiv \prod_{j=1}^l\Gpq{t^{\frac{l-2 j+2}{2}} z_i^{(D)}{}^\pm w_1^\pm} \,;
\end{align}
\item finally, for node C one has to collect various contributions coming from the original fundamentals attached to the $k$-th and the $(k+4)$-th node in Figure \ref{eclash}, the components of their antisymmetrics and vectors for the Higgsed part of the gauge group and some of residual contribution from the collapsed Identity-walls, which one can verify that overall give
\begin{align}
\label{eq:C}
\prod_{j=1}^{l-1}\Gpq{t^{-\frac{l-2 j+1}{2}} z_i^{(C)}{}^\pm \left(y_2 t^\frac{N-k-l}{2}\right)^\pm} \equiv \prod_{j=1}^{l-1}\Gpq{t^{-\frac{l-2 j+1}{2}} z_i^{(C)}{}^\pm w_2^\pm} \,, \\
\prod_{j=1}^{l-1}\Gpq{t^{-\frac{l-2 j+1}{2}} z_i^{(C)}{}^\pm \left(y_1 t^{-\frac{N-k-l}{2}}\right)^\pm} \equiv \prod_{j=1}^{l-1}\Gpq{t^{-\frac{l-2 j+1}{2}} z_i^{(C)}{}^\pm w_1^\pm} \,.
\end{align}
\end{itemize}
Again, in the previous expressions we have defined
\begin{align}
\left\{\begin{array}{rl}
w_1 &= y_1 t^{-\frac{N-k-l}{2}} \,, \\
w_2 &= y_2 t^\frac{N-k-l}{2} \,.
\end{array}\right.
\end{align}

The crucial fact at this point is that the chirals attached to the gauge nodes B and D have non-zero VEVs, which we can see in the index since the contributions in \eqref{eq:B} and \eqref{eq:D} yield poles that pinch the integration contour at, say,
\begin{align}
z_{k-i}^{(B)} &= y_2 t^\frac{N-k-2 i}{2} \,, \qquad i = 1, \dots, l-1 \,, \\
z_{k-i}^{(D)} &= y_1 t^{-\frac{N-k-2 i}{2}} \,, \qquad i = 1, \dots, l-1 \,.
\end{align}
These VEVs then Higgs the B and D gauge nodes from $USp(2k-2)$ to $USp(2k-2l)$. As usual, we can study the result of the Higgsing at the level of the index by taking the residues at the poles \cite{Gaiotto:2012xa}. This allows us to keep track of all the remaining massless fields, which we summarize in the second quiver of Figure \ref{fig:bad_SQCD_even_Nf_frameL}.

One can notice that the VEV has not been extinguished completely yet. In particular, some of the chirals attached now to the nodes A and E still have VEVs which trigger another Higgsing. By iteration we observe a sequential Higgsing of the quiver, which eventually leads to the last quiver in Figure \ref{fig:bad_SQCD_even_Nf_frameL}. This is the mirror of the good $USp(2k-2l)=USp(N_f-2l)$ SQCD, as expected.

\subsection{Odd $N_f$}
\label{app:generalNfodd}

We now consider the general case of arbitrary $N_c$ and arbitrary odd $N_f=2k+1$ with $k< N_c-1$. As we did in our previous discussion of the even $N_f$ case, we will only focus on the gauge groups of the quiver and on the matter fields that transform under them to demonstrate that we indeed get the structure of the frames given by the good $USp(2r)$ SQCD for $r=N_f-N_c,\cdots,\frac{N_f}{2}$, while we will ignore the gauge singlet fields.

\begin{figure}[!ht]
\includegraphics[width=\textwidth]{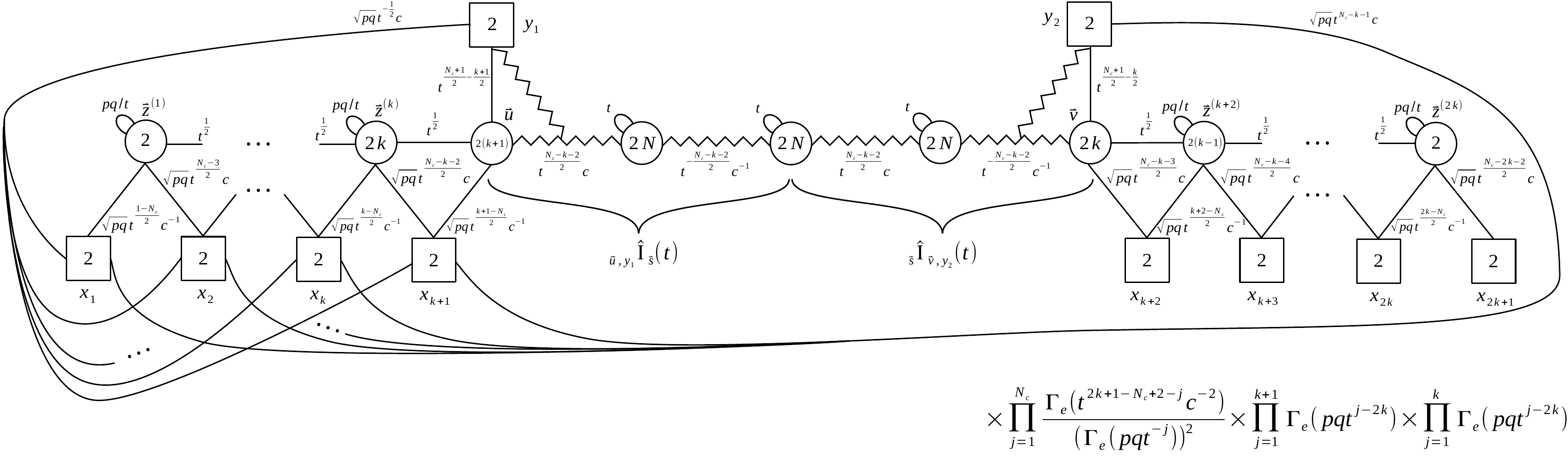}
\caption{$\mathsf{S}$-dualized SQCD for $N_f=2k+1$ odd after having applied the HW moves.}
\label{odd}
\end{figure}

We start from the $\mathsf{S}$-dualized theory in Figure \ref{SQCDone} and we perform $k+1$ HW moves on the left side of the quiver and $k$ HW moves on the right side. We then arrive at the situation depicted in Figure \ref{odd}, where two asymmetric Identity-walls collide, yielding a product of delta functions that identify the fugacity vectors on the two sides

As we have seen explicitly in the example $N_c=6$ and $N_f=7$ in Subsubsection \ref{67sec}, the delta conditions once implemented might determine some VEVs. By studying the Higgs mechanism triggered by each of these VEVs we find an interacting part consisting of the good SQCD and some singlets, which as we said we will ignore in this general discussion but they can be traced exactly in explicit examples. Let us describe the procedure after implementing the delta conditions for each frame. 

\begin{figure}[!ht]
\centering
\includegraphics[width=.9\textwidth]{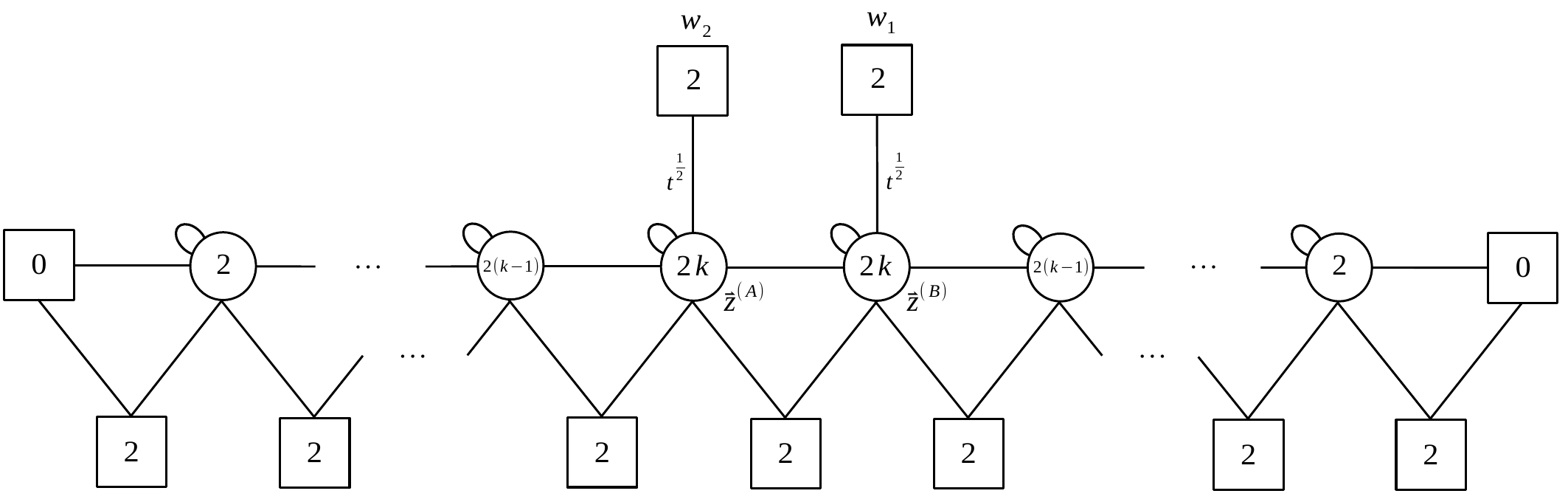}
\caption{The mirror dual of frame zero of the bad SQCD for $N_f=2k+1$ odd.}\label{fig:bad_SQCD_odd_Nf_frame0}
\end{figure}

\paragraph{Frame zero.} The first set of conditions is
\begin{align}
\label{eq:condition 1_odd}
\left\{\begin{array}{rll}
u_i &= v_i \,, & \qquad i = 1\,,\, \dots \,,\, k \,,\\
u_{k+1} &= y_2 t^\frac{N-k-1}{2} \,, \\
y_1 &= y_2 t^{-\frac12} \,.
\end{array}\right.
\end{align}
The resulting theory is shown in Figure \ref{fig:bad_SQCD_odd_Nf_frame0}, which is the mirror of the good $USp(2k)=USp\left(\left\lfloor\frac{N_f}{2}\right\rfloor\right)$ SQCD.
The gauge node A, which comes from the $k$-th gauge node from the left in Figure \ref{odd}, now has one fundamental doublet:
\begin{align}
\Gpq{t^\frac12 z^{(A)}_i{}^\pm (y_2 t^\frac{N-k-1}{2})^\pm} \equiv \Gpq{t^\frac12 z^{(k)}_i{}^\pm w_2{}^\pm} \,, \qquad 1 \leq i \leq k\,.
\end{align}
This doublet originates from the broken part\footnote{Note that the $(k+1)$-th gauge node is broken from $USp(2 k+2)$ to $USp(2 k)$ after implementing delta.} of the bifundamental field between the $k$-th gauge node and the $(k+1)$-th gauge node, which is now identified as the gauge node B.

On the other hand, the gauge node B has several fundamental contributions after implementing delta functions, which originate either from the original fundamental flavors attached to the $(k+1)$-th node and the $(k+5)$-th node in Figure \ref{odd}, now identified by the Identity-walls, or from the remaining contributions of the collapsed Identity-walls. However, most of them are massive, and only one massless doublet remains
\begin{align}
\Gpq{t^\frac12 u_i^\pm (y_2 t^{-\frac{N-k}{2}})^\pm} = \Gpq{t^\frac12 u_i^\pm (y_1 t^{-\frac{N-k-1}{2}})^\pm} \equiv \Gpq{t^\frac12 u_i^\pm w_1^\pm}\,.
\end{align}
In both of the previous expressions we defined
\begin{equation}
\begin{cases}
w_1 = y_1 t^{-\frac{N-k-1}{2}}\,,\\
w_2 = y_2 t^\frac{N-k-1}{2}\,.
\end{cases}
\end{equation}
Overall, we find the quiver in Figure \ref{fig:bad_SQCD_odd_Nf_frame0}.

\begin{figure}[!ht]
\centering
\includegraphics[width=.9\textwidth]{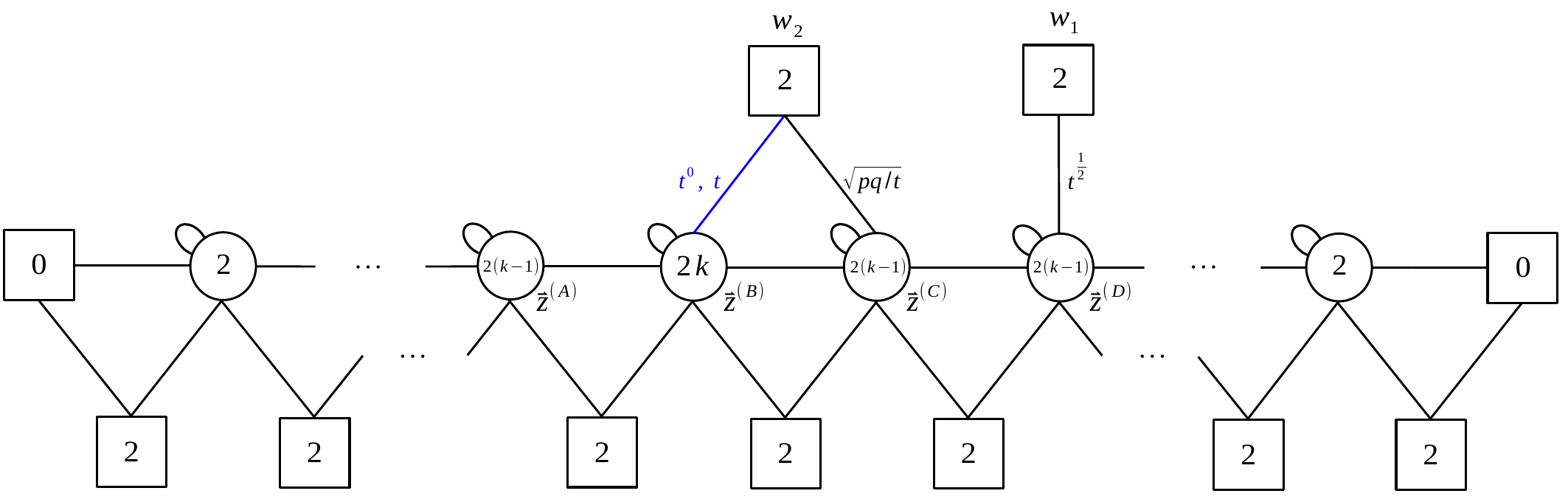}
\caption{The quiver for frame one of the bad SQCD with $N_f=2k+1$ odd. The blue line indicates a pair of fields, one of which is taking a VEV.}\label{fig:bad_SQCD_odd_Nf_frame1_beginning}
\end{figure}

\paragraph{Frame one.} The second set of conditions is
\begin{align}
\left\{\begin{array}{rll}
u_i &= v_{i+1} \,, & \qquad i = 1\,,\, \dots \,,\, k-1 \,, \\
u_k &= y_2 t^\frac{N-k-1}{2} \,, \\
u_{k+1} &= y_2 t^\frac{N-k-3}{2} \,, \\
v_1 &= y_1 t^{-\frac{N-k-2}{2}} \,, \\
y_1 &= y_2 t^{-\frac32} \,.
\end{array}\right.
\end{align}
The resulting theory is shown in Figure \ref{fig:bad_SQCD_odd_Nf_frame1_beginning}, which is the mirror of the good $USp(2k-2)$ SQCD.
The gauge node B, which comes from the $k$-th gauge node from the left in Figure \ref{odd}, has two fundamental doublets originating from the broken part of the bifundamental between the $k$-th node and the $(k+1)$-th node because the $(k+1)$-th node is broken from $USp(2 k+2)$ to $USp(2 k-2)$
\begin{align}
\label{eq:doublet B}
\Gpq{z^{(B)}_i{}^\pm \left(t^\frac{N-k-2}{2} y_2\right)^\pm} \,, \qquad \Gpq{t z^{(B)}_i{}^\pm \left(t^\frac{N-k-2}{2} y_2\right)^\pm} \,, \qquad 1 \leq i \leq k \,.
\end{align}
The gauge node C, which comes from the $(k+1)$-th node in Figure \ref{odd} that has been identified with the $(k+5)$-th node, has one fundamental doublet originating from the remaining contributions of the collapsed Identity-walls
\begin{align}
\Gpq{pq t^{-\frac12} z^{(C)}_i{}^\pm \left(t^\frac{N-k-2}{2} y_2\right)^\pm} \equiv \Gpq{pq t^{-\frac12} z^{(C)}_i{}^\pm w_2^\pm} \,, \qquad 1 \leq i \leq k-1 \,.
\end{align}
Lastly, the gauge node D, which comes from the $(k+6)$-th node in Figure \ref{odd}, has one fundamental doublet originating from the broken part of the bifundamental between the $(k+5)$-th node and the $(k+6)$-th node because the $(k+5)$-th node is broken from $USp(2 k)$ to $USp(2 k-2)$
\begin{align}
\Gpq{t^{\frac12} z^{(D)}_i{}^\pm \left(t^{-\frac{N-k-2}{2}} y_1\right)^\pm} \equiv \Gpq{t^{\frac12} z^{(D)}_i{}^\pm w_1^\pm} \,, \qquad 1 \leq i \leq k-1\,.
\end{align}
We have defined
\begin{equation}
\begin{cases}
w_1 = t^{-\frac{N-k-2}{2}} y_1\,,\\
w_2 = t^\frac{N-k-2}{2} y_2\,.
\end{cases}
\end{equation}

\begin{figure}[!ht]
\centering
\includegraphics[width=.9\textwidth]{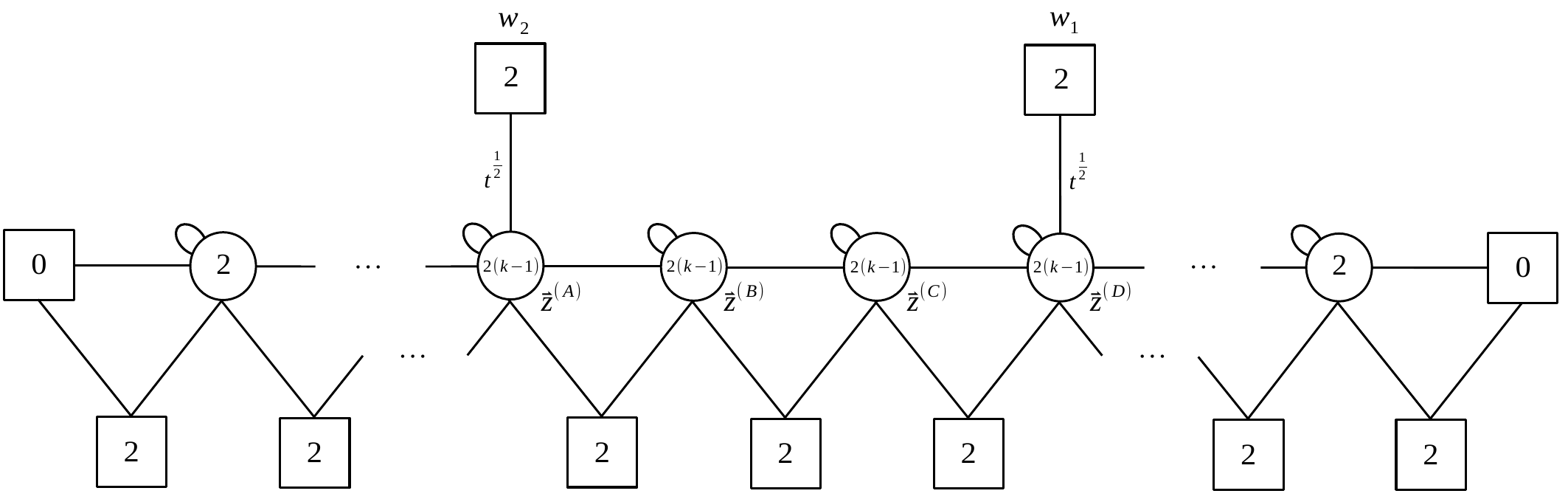}
\caption{The result of the Higgsing that gives the mirror dual of frame one of the bad SQCD for $N_f=2k+1$ odd.}\label{fig:bad_SQCD_odd_Nf_frame1_end}
\end{figure}

At this point we observe that the chirals attached to the node B are getting a non-trivial VEV. Indeed, from their index contribution in \eqref{eq:doublet B}, we can see that the poles pinch the integration contour at, say,
\begin{align}
z^{(B)}_k = t^\frac {N-k-2}{2} y_2 \,.
\end{align}
As usual, we should evaluate the residue at this pole, which is equivalent to Higgsing the gauge node B down to $USp(2 k-2)$. 
As a result, the final theory is as shown in Figure \ref{fig:bad_SQCD_odd_Nf_frame1_end}.

\begin{figure}[tbp]
\centering
\includegraphics[width=.8\textwidth]{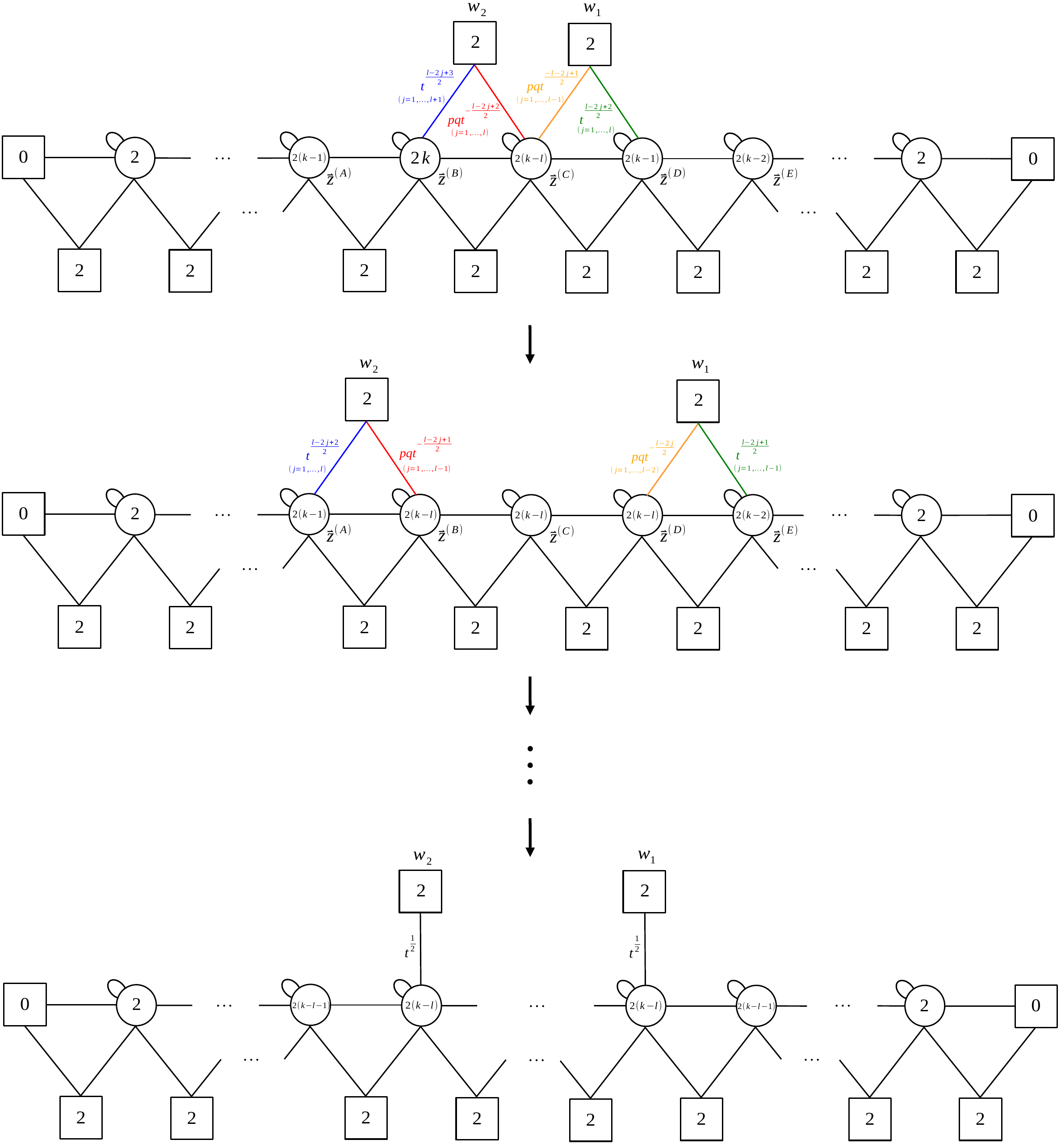}
\caption{Sequential Higgsing that results in the mirror dual of frame $l$ of the bad SQCD for $N_f=2k+1$ odd.}
\label{fig:bad_SQCD_odd_Nf_frameL} 
\end{figure}

\paragraph{Frame $l$.} We consider now the general frame $l$ with $1 < l < N-k$. The associated conditions are
\begin{align}
\left\{\begin{array}{rll}
u_i &= v_{i+l} \,, & \qquad i = 1 \,,\, \dots \,,\, k-l \,, \\
u_{k-l+j} &= y_2 t^\frac{N-k-2 j+1}{2} \,, & \qquad j = 1\,,\, \dots \,,\, l+1 \,, \\
v_j &= y_1 t^{-\frac{N-k-2 j}{2}} \,, & \qquad j = 1 \,,\, \dots \,,\, l \,, \\
y_1 &= y_2 t^{-\frac{2 l+1}{2}} \,.
\end{array}\right.
\end{align}
Implementing this condition, we obtain the first quiver in Figure \ref{fig:bad_SQCD_odd_Nf_frameL}.
The gauge nodes B and D have $l+1$ and $l$ fundamental doublets, respectively, which come from the broken part of bifundamental fields. The gauge node C has $2l-1$ fundamental doublets originating from the remaining contributions of the Identity-walls.

Some of the doublets of the nodes B and D acquire VEVs which Higgs these nodes from $USp(2k)$ down to $USp(2k-2l)$. Again this can be seen from their index contribution, which give sets of poles that pinch the integration contour at, say,
\begin{align}
z^{(B)}_{k-l+j} &= t^\frac {N-k-2 j}{2} y_2 \,, \qquad 1 \leq j \leq l \,, \\
z^{(D)}_{k-l+j} &= t^{-\frac {N-k-2 l+2 j-1}{2}} y_1 \,, \qquad 1 \leq j \leq l-1 \,.
\end{align}
Evaluating the residue at this set of poles implements the Higgsing of the nodes B and D at the level of the index. The result is shown in the second quiver in Figure \ref{fig:bad_SQCD_odd_Nf_frameL}, where $w_1$ and $w_2$ are defined by
\begin{equation}
\begin{cases}
w_1 = t^{-\frac{N-k-l-1}{2}} y_1\,,\\
w_2 = t^\frac{N-k-l-1}{2} y_2\,.
\end{cases}
\end{equation}

One can notice that the VEV has not been extinguished completely yet. In particular, some of the chirals attached now to the nodes A and E still have VEVs which trigger another Higgsing. By iteration we observe a sequential Higgsing of the quiver, which eventually leads to the last quiver in Figure \ref{fig:bad_SQCD_odd_Nf_frameL}. This is the mirror of the good $USp(2k-2l)$ SQCD, as expected.

\section{Connecting UV fugacities to IR fugacities}
\label{app:UVIRfug} 

In this appendix, we derive the generic mapping between the UV and IR fugacities in the bad SQCD that we found in Section \ref{sec:4dbadSQCD}, but from the perspective of the Higgsing implemented by the delta constraints which we discussed in Section \ref{sec:higgs}.

We first remind that for each frame, which we label by an integer $n$, the constraints coming from the delta functions are\footnote{For $\epsilon=n=0$ we only have two constraints $y_1=y_2^{\pm1}$}
\be
y_1=\left(y_2^{\pm1} t^{\epsilon-n}\right)^{\pm1},\qquad n=0,\cdots,\Big\lfloor\frac{N_f}{2}\Big\rfloor-(N_f-N_c)-1\,,
\ee
where
\be
\epsilon=\begin{cases}0&N_f\text{ even}\,,\\-\frac{1}{2}&N_f\text{ odd}\,.\end{cases}
\ee
Focusing on the constraint $y_1=y_2t^{\epsilon-n}$ for definiteness, in Section \ref{sec:higgs} we interpreted this as a VEV for an operator, which is a meson built from the fields $P_{L,1}$ and $P_{R,2}$ dressed with the antisymmetric $A$ to the power $n-\epsilon+M$ (or equivalently $k-1=n-\epsilon+M$). This can be implemented by giving the VEV \eqref{vevsol} to $A$, $P_{L,1}$ and $P_{R,2}$, which Higgses the $USp(2N_c)$ gauge group to $USp(N_f-2n+2\epsilon)$. Denoting again the locked  fugacities associated with the broken $USp$ Cartan generators as $z_i$ with $i=1,\dots, n-\epsilon+M+1$, let us recall \eqref{Aconstr2} and \eqref{Pconstr2}: The VEV for the antisymmetric $A$ in \eqref{vevsol} imposes the constraints 
\be\label{Aconstr} z_{i+1}=tz_i\,,\quad i=1,\dots, n-\epsilon+M\,,\ee 
and the VEV for $P_{L,1}$ and $P_{R,2}$ lead instead to 
\be\label{Pconstr} y_1cz_{n-\epsilon+M+1}=1\,,\quad y_2cz_1=t^{-M}\,.\ee 
These together imply the delta constraint $y_1=y_2t^{\epsilon-n}$. These are the points \eqref{eq:pinchpoints} where the integration contours are pinched that we found in the index analysis, up to a trivial permutation of the $USp(2N_c)$ fugacities.

Let us now proceed with the mapping between UV and IR fugacities. It is convenient to start with the singlets $\beta_i$. As we have explained, those with $i\leq n-\epsilon+M$ end up becoming massive while those with $i>n-\epsilon+M$ become the $\beta_i$ singlets of the infrared $USp(N_f-2n+2\epsilon)$ gauge theory. We therefore find at the level of fugacities the equation 
\be\label{fug1} \frac{pq}{c^2t^{n-\epsilon+M+1+j}}=\frac{pq}{(c’)^2(t’)^j},\quad 0\leq j\leq N-n+\epsilon-M-1\,,\ee
where the primed fugacities denote those of the infrared theory. Provided that $n-\epsilon+M<N-1$ (so that $j$ in \eqref{fug1} can take more than one value), we immediately conclude that \eqref{fug1} can hold for every $j$ only if $t’=t$ and from now on we will assume this holds true. Notice that this is also consistent with the requirement that the unlocked components of the antisymmetric multiplet $A$ map to the corresponding components of the antisymmetric field in the IR. We also find from \eqref{fug1} the relation 
\be\label{fug2} c’=ct^{\frac{n-\epsilon+M+1}{2}}\,.\ee 
This tells us that the index of our frame appears in the index of the bad theory as a function of $c t^{\frac{n-\epsilon+M+1}{2}}$ only and this is precisely what was found in Section \ref{sec:4dbadSQCD}. 

Let us now analyze the fields $P_{L,1}$ and $P_{R,2}$ of the infrared theory. As we have explained in Section \ref{sec:higgs}, these come from the locked components of the antisymmetric field $A$ with fugacities $1/z_1$ and $z_{n-\epsilon+M+1}$ respectively. We therefore find the relations 
\be\label{constr1} \frac{t}{z_1}=c’w_2\ee and 
\be\label{constr2} tz_{n-\epsilon+M+1}=\frac{t^{n-\epsilon+1}}{c’w_1}\,,\ee 
where we have set $w_2=(y_1)’$ and $w_1=(y_2)’$, to conform with the conventions of the previous sections. Combining these with \eqref{Pconstr} and \eqref{fug2} we find  
\be\label{fug3} w_1=y_1t^{\frac{n-\epsilon-M-1}{2}}\,,\quad w_2=y_2t^{\frac{M+1-n+\epsilon}{2}}\,,\ee 
which is precisely the relation derived in Section \ref{sec:4dbadSQCD}. Let us see explicitly how it works for $w_2$. From \eqref{constr1}  we find 
\be
\frac{t}{z_1}=\frac{ty_2c}{y_2cz_1}=t^{M+1}y_2c=t^{\frac{M+1-n+\epsilon}{2}}y_2c’=w_2c'\,,
\ee
where in the second equality we have used \eqref{Pconstr} and in the third we have exploited \eqref{fug2}. A similar manipulation of \eqref{constr2} leads to the equation for $w_1$ in \eqref{fug3}. This completes our derivation of the connection between UV and IR fugacities from the equations of motion. Notice that in the derivation we have made use of \eqref{Pconstr} only and not \eqref{Aconstr}, therefore we did not need to exploit the delta constraint.  Notice also that in terms of the fugacities $w_{1,2}$ the constraint $y_1=y_2t^{\epsilon-n}$ reads $w_1=w_2t^{-M-1}$ and is frame independent.

Let us now discuss some consistency checks of our analysis. As we have seen before, the $2N_c$-th component of $Q$ plays the role of the multiplet $S_1$ in the infrared. This leads to the equation  
\be\label{fug4} \frac{\sqrt{pq}}{\sqrt{t}z_{n-\epsilon+M+1}}=\sqrt{pq}t^{-n+\epsilon-\frac{1}{2}}w_1c’\,.\ee 
We can now check that, using \eqref{fug2} and \eqref{fug3}, \eqref{fug4} becomes equivalent to the first equation in \eqref{Pconstr} and is therefore automatically satisfied. Similarly, we have argued before that the unlocked components of $P_{L,2}$ and $P_{R,1}$ are mapped to the same fields in the infrared. This leads to the relations  
\be \frac{c}{y_1}=\frac{c’}{w_2}\,,\qquad \frac{t^{-M}y_2}{c}=\frac{t^{n-\epsilon+1}w_1}{c’}\,,\ee 
which become both equivalent to the delta constraint $y_1=y_2t^{\epsilon-n}$ once \eqref{fug2} and \eqref{fug3} are used. Regarding the locked components of $P_{R,1}$, using \eqref{Aconstr}, \eqref{Pconstr} and \eqref{fug3} we can rewrite them as follows: 
\be \frac{t^{-M}y_2}{cz_j}=t^{n-\epsilon+1-j}w_1w_2\,,\qquad  \frac{t^{-M}y_2z_j}{c}=\frac{t^{j-2M-1}}{c^2}\,, \quad j=1,\dots , n-\epsilon+M+1\ee
and we recognize above the singlets contributing to the index of the bad theory. They appear in the prefactor multiplying the index of the infrared frame. Notice that we have used both \eqref{Aconstr} and \eqref{Pconstr}, and therefore the delta constraint, in the derivation.  As for the locked components of $P_{L,2}$, we can reexpress them as follows: 
\be \frac{cz_j}{y_1}=\frac{t^{j-1-M}}{w_1w_2}\,,\qquad \frac{c}{y_1z_j}=c^2t^{n-\epsilon+M+1-j}\,,\quad j=1,\dots , n-\epsilon+M+1\,.\ee 
On the left we recognize another group of singlets contributing to the prefactor in the index and on the right we have singlets with the correct quantum numbers to combine with the $\beta_{i\leq n-\epsilon+M}$ fields into massive multiplets. This is precisely consistent with our analysis of the equations of motion.

\section{The $\mathcal N=2^*$ moduli space with non-trivial flavor masses}
\label{app:N=2}

In this appendix, we extend the discussion in Section \ref{sec:N=2}, where we examined the $\mathcal N=2^*$ moduli space with  vanishing flavor masses, to the following choice of flavor masses:
\begin{align}
\vec m = \left(m_1,\dots,m_{N_f}\right) = \left(\undermat{\frac{N_f}{2}}{\mathsf m_A,\dots,\mathsf m_A},\undermat{\frac{N_f}{2}}{-\mathsf m_A,\dots,-\mathsf m_A}\right) , \\
\nonumber
\end{align}
where we have assumed $N_f$ is even, for simplicity. If $N_f$ is odd, one can add $m_{N_f} = 0$ to ensure the condition $\sum_{a = 1}^{N_f} m_a = 0$. This set of flavor masses admits a moduli space more similar to that of the $\mathcal N = 4$ theory, making the comparison with the $\mathcal N=4$ case more straightforward.

Let us consider the Higgs branch solution with a generic value of the FI parameter first. Recall the vacuum equations
\begin{align}
\label{eq:F-term*}
Q \widetilde Q = 0 \,, \qquad \Phi Q = 0 \,, \qquad \widetilde Q \Phi = 0 \,,
\end{align}
\begin{align}
\label{eq:D-term*}
[\Phi^\dagger,\Phi]+Q Q^\dagger{}-\widetilde Q^\dagger \widetilde Q-\frac{\zeta_\text{eff}}{2 \pi} \, \mathbb I_{N_c}-\frac{k_\text{eff}}{2\pi} \Sigma = 0 \,,
\end{align}
\begin{align}
\begin{aligned}
 \label{eq:mass***}
M_{\text{eff}}(Q^i{}_a)&\equiv\left(\mathsf m_A-m_a+\sigma_i\right) Q^i{}_a = 0 \,, \\
M_{\text{eff}}(\widetilde{Q}^a{}_i)&\equiv\left(\mathsf m_A+m_a-\sigma_i\right) \widetilde Q^a{}_i = 0 \,, \\
M_{\text{eff}}(\Phi^i{}_j)&\equiv\left(-2 \mathsf m_A+\sigma_i-\sigma_j\right) \Phi^i{}_j = 0 \,.
\end{aligned}
\end{align}
From \eqref{eq:mass***}, the effective real masses of fundamental and anti-fundamental fields are given by
\begin{align}
M_{\text{eff}}(Q^i{}_a) &= \begin{cases}
\sigma_i \,, \qquad &1 \leq a \leq \frac{N_f}{2} \,, \\
2 \mathsf m_A+\sigma_i \,, \qquad &\frac{N_f}{2}+1 \leq a \leq N_f \,, \\
\end{cases} \\
M_{\text{eff}}(\widetilde Q^a{}_i) &= \begin{cases}
2 \mathsf m_A-\sigma_i \,, \qquad &1 \leq a \leq \frac{N_f}{2} \,, \\
-\sigma_i \,, \qquad &\frac{N_f}{2}+1 \leq a \leq N_f \,. \\
\end{cases}
\end{align}
One can see that $\sigma_i=0$ for some $i$ allows both $Q$ and $\widetilde Q$ to have non-zero VEVs simultaneously, yielding a non-compact Higgs branch as follows:
\begin{align}
\begin{aligned}
\label{eq:vev1}
Q  &=
\left (
\begin{array}{cccccc | ccc | ccc}
q_1 & & & & & & & & & & & \\
& \ddots & & & & & & & & & & \\
& & q_{N_f-N_c} & & & & & & & & & \\
& & & \sqrt{\zeta_\text{eff}} & & & & & & & & \\
& & & & \ddots & & & 0 & & & & \\
& & & & & \sqrt{\zeta_\text{eff}} & & & & & & \\
& & & & & & & & & \sqrt{\zeta'_\text{eff}} & & \\
& & & & & & & & & & \ddots & \\
\undermat{N_f-N_c}{\,\,\,\,\,\,&\,\,\,\,\,\,\,\,&\,\,\,\,\,\,\,\,\,\,\,\,\,\,\,\,\,\,\,} & \undermat{N_c-\frac{N_f}{2}}{\,\,\,\,\,\,\,\,\,\,\,\,\,\,& \,\,\,\,\,\,\,\,&\,\,\,\,\,\,\,\,\,\,\,\,\,\,} & \undermat{N_f-N_c}{\,\,\,\,\,\,\,\,\,\,\,\,\,&\,\,\,\,\,\,\,\,&\,\,\,\,\,\,\,\,\,\,\,\,\,} & \undermat{N_c-\frac{N_f}{2}}{\,\,\,\,\,\,\,\,\,\,\,\,\,\,& \,\,\,\,\,\,\,\,&\sqrt{\zeta'_\text{eff}}}
\end{array}
\right ) , \\
\\
\\
\widetilde{Q}^T &=
\left (
\begin{array}{ccclcc | ccc | ccc}
& & & & & & \tilde q_1 & & & & & \\
& & & & & & & \ddots & & & & \\
& & & & & & & & \tilde q_{N_f-N_c} & & & \\
& & & & & & & & & & & \\
& & & 0 & & & & & & & 0 & \\
& & & & & & & & & & & \\
& & & & & & & & & & & \\
& & & & & & & & & & & \\
\undermat{\frac{N_f}{2}}{\,\,\,\,\,\,&\,\,\,\,\,\,\,\,&\,\,\,\,\,\,\,\,\,\,\,\,\,\,\,\,\,\,\,&\,\,\,\,\,\,\,\,\,\,\,\,\,\,& \,\,\,\,\,\,\,\,&\,\,\,\,\,\,\,\,\,\,\,\,\,\,} & \undermat{N_f-N_c}{\,\,\,\,\,\,&\,\,\,\,\,\,\,\,&\,\,\,\,\,\,\,\,\,\,\,\,\,\,\,\,\,\,\,\,} & \undermat{N_c-\frac{N_f}{2}}{\,\,\,\,\,\,\,\,\,\,\,\,\,\,& \,\,\,\,\,\,\,\,&\,\,\,\,\,\,\,\,\,\,\,\,\,\,}
\end{array}
\right ) . \\
\\
\end{aligned}
\end{align}
\begin{align}
\label{eq:Sigma 2}
\Sigma = \mathrm{Diag} \left(\undermat{\frac{N_f}{2}}{0, \dots, 0},\undermat{N_c-\frac{N_f}{2}}{-2 \mathsf m_A,\dots,-2 \mathsf m_A}\right) , \\
\nonumber
\end{align}
where
\begin{align}
\label{eq:higgs}
q_i^2-\tilde q_i^2 = \zeta_\text{eff}
\end{align}
for $i = 1,\dots, N_f-N_c$, and $\Phi$ vanishes due to \eqref{eq:F-term*}. The value of $\Sigma$ in \eqref{eq:Sigma 2} is fixed to let $Q$ have the vanishing effective real mass and breaks the gauge group $U(N_c)$ into $U(N_f/2) \times U(N_c-N_f/2)$, allowing two separate effective FI parameters $\xi_\text{eff}$ and $\xi'_\text{eff}$. $Q$ and $\widetilde Q$ in \eqref{eq:vev1} satisfying \eqref{eq:higgs} are then fixed by the D-term equation \eqref{eq:D-term*} and eventually break the gauge group completely. Notice that, except the quantum correction to the FI parameter, \eqref{eq:vev1} is the same as the $\mathcal N = 4$ Higgs branch solution \eqref{eq:vev with real FI} up to a flavor permutation and renaming $\alpha$ to $\tilde q$.

If the FI parameter satisfies certain conditions such that its quantum corrected value vanishes, one can also find solutions with a flat direction of $\Sigma$, i.e., the mixed branch solutions. Such solutions are of the form
\begin{align}
\begin{aligned}
\label{eq:vev2}
Q  &=
\left (
\begin{array}{ccc | c | ccc | c}
q_1 & & & & & & & \\
& \ddots & & & & & & \\
& & q_r & 0 & & 0 & & 0 \\
& & & & & & & \\
& & & & & & & \\
\undermat{r}{\,\,\,\,\,\,&\,\,\,\,\,\,\,\,&\,\,\,\,\,\,} & \undermat{\frac{N_f}{2}-r}{\,\,\,\,\,\,\,\,\,\,\,\,\,\,\,\,\,\,\,\,\,} & \undermat{r}{\,\,\,\,\,\,&\,\,\,\,\,\,\,\,&\,\,\,\,\,\,} & \undermat{\frac{N_f}{2}-r}{\,\,\,\,\,\,\,\,\,\,\,\,\,\,\,\,\,\,\,\,\,}
\end{array}
\right ) , \,\,
\widetilde{Q}^T &=
\left (
\begin{array}{ccc | c | ccc | c}
& & & & \tilde q_1 & & & \\
& & & & & \ddots & & \\
& 0 & & 0 & & & \tilde q_r & 0 \\
& & & & & & & \\
& & & & & & &  \\
\undermat{r}{\,\,\,\,\,\,&\,\,\,\,\,\,\,\,&\,\,\,\,\,\,} & \undermat{\frac{N_f}{2}-r}{\,\,\,\,\,\,\,\,\,\,\,\,\,\,\,\,\,\,\,\,\,} & \undermat{r}{\,\,\,\,\,\,&\,\,\,\,\,\,\,\,&\,\,\,\,\,\,} & \undermat{\frac{N_f}{2}-r}{\,\,\,\,\,\,\,\,\,\,\,\,\,\,\,\,\,\,\,\,\,}
\end{array}
\right ) . \\
\\
\\
\end{aligned}
\end{align}
\begin{align}
\label{eq:Sigma 3}
\Sigma = \mathrm{Diag} \left(\undermat{r}{0, \dots, 0},\sigma, \dots, \sigma\right) , \\
\nonumber
\end{align}
for $i = 1,\dots, r \leq \frac{N_f}{2}$. Again, the value of $\Sigma$ is fixed to let $Q$ have vanishing effective real mass and breaks the gauge group $U(N_c)$ into $U(r) \times U(N_c-r)$ in this case. The solution \eqref{eq:vev2} is the same as the $\mathcal N=4$ mixed branch solution \eqref{quarkvev5} up to a flavor permutation.

A crucial difference from the $\mathcal N=4$ case is that the FI parameter now receives quantum corrections. While the mixed branch solution of the $\mathcal N=4$ case requires the bare FI parameter to vanish regardless of $r$, the $\mathcal N=2^*$ case has a different effective FI parameter for the $U(N_c-r)$ part for each value of $r$, which can be determined as follows. As in Section \ref{sec:N=2}, first we note that the $SU(N_c-r)$ part and the diagonal $U(1)$ part obtain different quantum corrections to $\xi_\text{eff}+k_\text{eff} \, \Sigma$ because the adjoint field is only charged under $SU(N_c-r)$ and neutral under $U(1)$. Since the effective real mass of the adjoint field $\Phi$ restricted to the $U(N_c-r)$ part is
\begin{align}
M_{\text{eff}}(\Phi) &= -2 \mathsf m_A \,,
\end{align}
it generates a CS term of level
\begin{align}
\label{eq:CS adjoint}
-\mathrm{sgn}(\mathsf m_A) (N_c-r)
\end{align}
only for $SU(N_c-r)$ after integrated out, as shown in \eqref{eq:CS}. On the other hand, it doesn't yield any shift of the FI parameter since it is neutral under the diagonal $U(1)$. Then the corrections to $\zeta_\text{eff}+k_\text{eff} \, \Sigma$ for the $U(N_c-r)$ part from the other fields read as follows:
\begin{align}
\label{eq:F*}
&\quad \frac12 \frac{N_f}{2} |\sigma|+\frac12 \frac{N_f}{2} |2 \mathsf m_A+\sigma| \nonumber \\
&-\frac12 \frac{N_f}{2} |2 \mathsf m_A-\sigma|-\frac12 \frac{N_f}{2} |-\sigma| \nonumber \\
&+ \frac12 \sum_{j = 1}^r \Big(|-2 \mathsf m_A+\sigma-\sigma_j|-|-2 \mathsf m_A+\sigma_j-\sigma|\Big)
\end{align}
with $\sigma_j = 0$ for $j = 1, \dots, r$, as shown in \eqref{eq:Sigma 3}. Each line is the contribution of respectively the fundamental, the anti-fundamental and the components of the original $U(N_c)$ adjoint field that are now in the bifundamental representation of $U(r) \times U(N_c-r)$. If $r = \frac{N_f}{2}$, \eqref{eq:F*} vanishes regardless of $\sigma$. Therefore, the effective theory in the IR consists of the $\mathcal N = 2$ $U(1)$ Maxwell theory with FI parameter $\zeta$ and the $\mathcal N=2$ $SU(N_c-r)$ CS theory with level $-(N_c-r) \, \mathrm{sgn}(\mathsf m_A)$. The latter CS theory has a unique supersymmetric vacuum \cite{Intriligator:2013lca,Witten:1999ds}, whereas the $U(1)$ theory has supersymmetric vacua only when the bare FI parameter $\zeta$ vanishes.

If $r < \frac{N_f}{2}$, \eqref{eq:F*} varies depending on the range of $\sigma$. For $|\sigma| > 2 |\mathsf m_A|$, it is evaluated as $\mathrm{sgn}(\sigma) (N_f-2 r) \mathsf m_A$, indicating the one-loop correction to the FI parameter but the absence of the contribution to the CS level. Hence, the effective theory consists of the $\mathcal N = 2$ $U(1)$ Maxwell theory now with the one-loop corrected FI parameter
\begin{align}
\zeta+\mathrm{sgn}(\sigma) (N_f-2 r) \mathsf m_A
\end{align}
and the $\mathcal N=2$ $SU(N_c-r)$ CS theory with level $-(N_c-r) \, \mathrm{sgn}(\mathsf m_A)$. As mentioned, the latter CS theory has a unique vacuum \cite{Intriligator:2013lca,Witten:1999ds}, whereas the $U(1)$ theory has supersymmetric vacua only when $\zeta_\text{eff}$ vanishes. This condition is exactly the delta constraint associated with the dual $U(r)$ theory with $\beta = -\mathrm{sgn}(\sigma)$ found in the partition function analysis (see \eqref{eq:3d pf}) once we identify\footnote{The shift by $-\frac{iQ}{2}$ appears due to our choice of the UV R-symmetry in our partition function computation.}
\begin{align}
\zeta &= Y_1-Y_2 \,, \\
\mathsf m_A &= m_A-\frac{iQ}{2} \,.
\end{align}
This free $U(1)$ vector multiplet can be dualized into a chiral multiplet, which corresponds to the chiral with vanishing charges represented by a delta distribution in the partition function.

Next let us check what happens when $|\sigma| < 2 |\mathsf m_A|$. In this range, \eqref{eq:F*} is evaluated as
\begin{align}
\mathrm{sgn}(\mathsf m_A) \left(\frac{N_f}{2}-r\right) \sigma \,,
\end{align}
indicating the shift of the $U(N_c-r)$ CS level by $\mathrm{sgn}(\mathsf m_A) \left(\frac{N_f}{2}-r\right)$. Together with the shift due to the adjoint field \eqref{eq:CS adjoint}, the $SU(N-r)$ part now has the CS level
\begin{align}
k = -\mathrm{sgn}(\mathsf m_A) \left(N_c-\frac{N_f}{2}\right)\,.
\end{align}
Since $|k| < N_c-r$, the theory has the vanishing Witten index and is supposed to have no supersymmetric vacuum \cite{Intriligator:2013lca,Witten:1999ds}.\footnote{For the $\mathcal N=2$ $SU(n)_k$ pure CS theory, the Witten index is given by
\begin{align}
I(k) = \frac{1}{(n-1)!} \prod_{j=1}^{n-1} (k-j) \,,
\end{align}
which vanishes when $|k| < n$ \cite{Intriligator:2013lca,Witten:1999ds}.} Therefore, we don't find an extra solution in the range $|\sigma| < 2 |\mathsf m_A|$.

One may also consider the case where $\sigma_i$ are not all equal. For example, one can slightly perturb the above solution such that all $\sigma_i$ are distinct but not very far apart; i.e., $|\sigma_i-\sigma_j| < 2 |\mathsf m_A|$. The gauge group $U(N_c-r)$ is now broken to $U(1)^{N_c-r}$, and $\Phi$ is decomposed into bifundamentals between these $U(1)$'s. A vacuum solution should satisfy\footnote{In addition to the semi-classical vacuum equation, the monopole instanton superpotential should also be taken into account to find the Coulomb branch solution breaking the non-abelian gauge group into its abelian subgroup \cite{Affleck:1982as}. We will show no such solution exists in our case, so we don't discuss the monopole instanton effect.}
\begin{align}
\label{eq:Coulomb}
0 = \zeta_i+\sum_j k_{ij} \sigma_j &= \zeta+\frac12 \left(\frac{N_f}{2}-r\right) \Big(|2 \mathsf m_A+\sigma_i|-|2 \mathsf m_A-\sigma_i|\Big) \nonumber \\
&\quad +\frac12 \sum_{j \neq i} \Big(|-2 \mathsf m_A+\sigma_i-\sigma_j|-|-2 \mathsf m_A+\sigma_j-\sigma_i|\Big)
\end{align}
for $i = r+1, \, \dots, \, N_c$, including the mixed CS term contributions from the bifundamentals between the $U(1)$'s, which can be computed by \eqref{eq:CS} with replacing $n_\alpha^2$ to $n^{(i)}_\alpha n^{(j)}_\alpha$, where $n^{(i)}_\alpha$ is the $U(1)_i$ charge of the $\alpha$-th field. Since we are looking at the region where $|\sigma_i| > 2 |\mathsf m_A|$ and $|\sigma_i-\sigma_j| < 2 |\mathsf m_A|$, \eqref{eq:Coulomb} is given by
\begin{align}
\zeta+\mathrm{sgn}(\sigma) \left(N_f-r\right) \mathsf m_A+(N_c-r-1) \sigma_i-\sum_{j \neq i} \sigma_j \,,
\end{align}
whose solution forces $\sigma_i$ equal to each other. Thus, we should go back to the previous analysis with equal $\sigma_i$.

Instead, let us consider the case where $\sigma_i$ are far apart from each other; i.e., $|\sigma_i-\sigma_j| > |2 \mathsf m_A|$. In this case, \eqref{eq:Coulomb} is given by
\begin{align}
\zeta+\mathrm{sgn}(\sigma) \left(N_f-r\right) \mathsf m_A-2 (N_c-2 i+r+1) \mathsf m_A \,,
\end{align}
which, given that $\mathsf m_A \neq 0$, cannot vanish simultaneously for all $i$. One can also imagine an intermediate type of a solution where $\sigma_i$ gather into several groups such that $U(N_c-r)$ is broken to $\prod_k U(n_k)$. Combining the previous analyses, the diagonal $U(1)$ of each $U(n_k)$ has a different value of the FI parameter and suffers from the same issue.

Hence, there is no solution other than the one with equal $\sigma_i$, which is the $\mathcal N=2^*$ counterpart of the mixed branch solution of the $\mathcal N=4$ case. Together with the Higgs branch solution \eqref{eq:vev1}, they exactly demonstrate the structure of the multiple dual frames associated with different values of the FI parameter, obtained from the partition function identity.

\bibliographystyle{JHEP}
\bibliography{ref}

\end{document}